%% file: Thesis.tex
\documentclass[g5paper,10pt,final,openright,coverpage,swedish]{thesis}
\pdfoutput=1
\usepackage[dvips]{graphicx}
\usepackage[section] {placeins}
\usepackage{rotating}
\usepackage{epsfig,amsmath,amssymb}
\usepackage{varioref}
\usepackage{graphicx}
\usepackage{verbatim,amsfonts}
\usepackage{array}
\usepackage{pstricks}
\usepackage{multirow}
\usepackage{multicol}
\usepackage{hhline}
\usepackage{epsfig}
\usepackage{rotate}
\usepackage{longtable}
\usepackage{rotating}
\usepackage{mathtools}

\newgray{dg}{0.3}
\newgray{mg}{0.8}
\newgray{lg}{0.8}

\title{Measurements of cosmic ray antiprotons with PAMELA and studies of propagation models}
\author{Juan Wu}
\date{April 2012}
\shortdate{2012}
\type{Doctoral Thesis in Physics}
\division{Particle and Astroparticle Physics}
\department{Department of Physics,}
\address{SE-106 91 Stockholm, Sweden}
\city{Stockholm}
\country{Sweden}
\publisher{Printed by Universitetsservice US-AB 2012}
\copyrightline{\copyright\ Juan Wu, April 2012}
\trita{TRITA-FYS 2012:22}
\issn{0280-316X}
\isrn{KTH/FYS/\mbox{-}\mbox{-12:22}\mbox{-}\mbox{-}SE}
\isbn{978-91-7501-330-5}

\cplogo{kth_svv_enginee_sciences}
\cplogonblines{1} 
\innerlogo{\includegraphics[width=32mm]{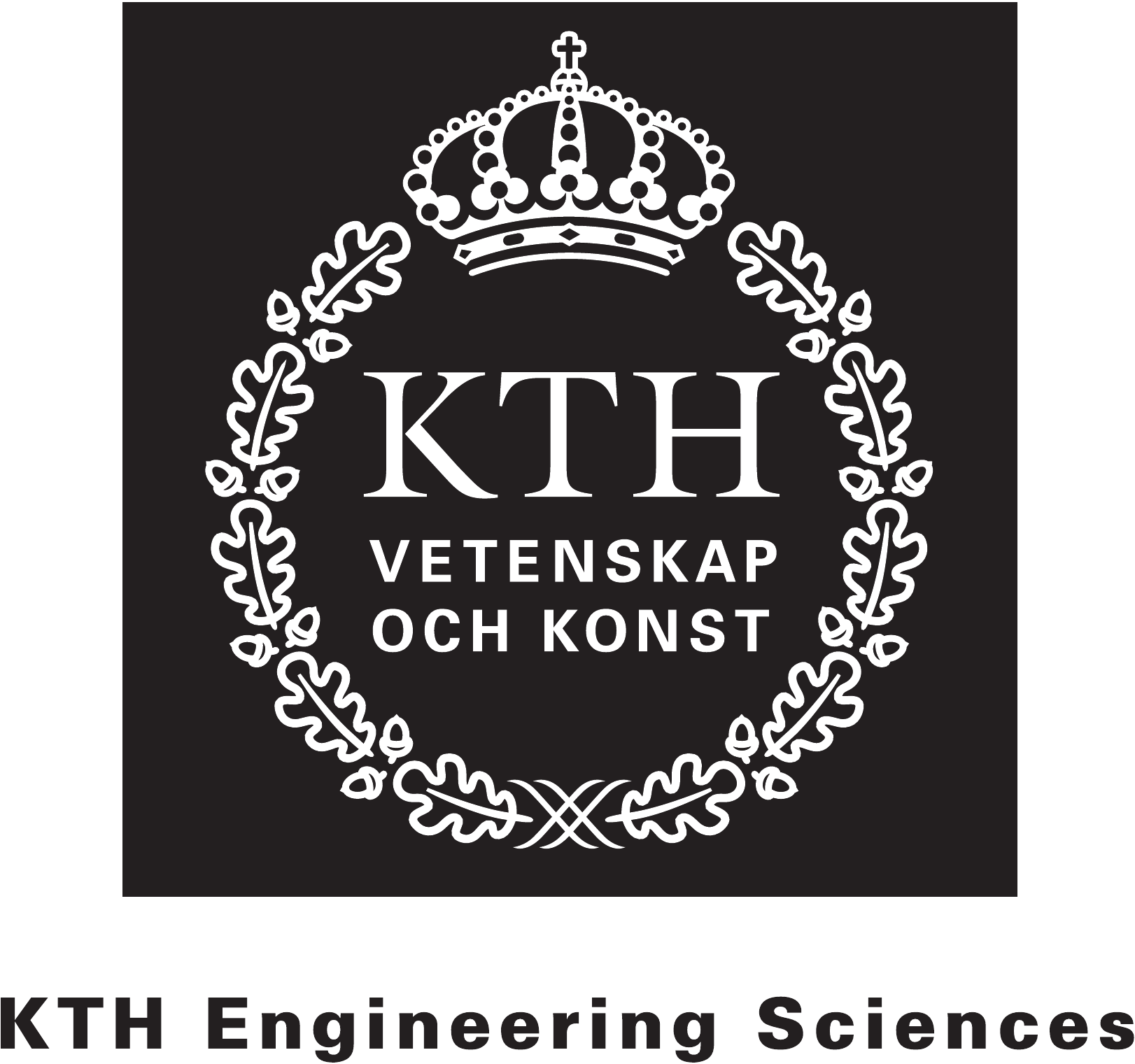}}
\centercomment{Cover illustration: The Milky Way galaxy. \\Taken from \texttt{http://www.universetoday.com/21563/milky-way/}.\\}
\foregincomment{Akademisk avhandling som med tillst{\aa}nd av Kungliga Tekniska H{\"o}gskolan i Stockholm framl{\"a}gges till offentlig granskning f{\"o}r avl{\"a}ggande av teknologie doktorsexamen fredagen den 1 juni 2012 kl 13:00 i sal FA31, AlbaNova Universitetscentrum, Roslagstullsbacken 21, Stockholm. \\ \\ Avhandlingen f{\"o}rsvaras p{\aa} engelska.}

\newcommand{\beq}{\begin{equation}}
\newcommand{\eeq}{\end{equation}}

\begin{document}

\maketitle

\cleardoublepage
\addcontentsline{toc}{chapter}{Abstract}
\include{Phd-Abstract}

\cleardoublepage
\tableofcontents

\mainmatter

\addcontentsline{toc}{chapter}{Introduction}
\include{Phd-Introduction}
\cleardoublepage

\include{Phd-Ch-Cosmic_Rays}

\cleardoublepage

\include{Phd-Ch-Propagation}

\cleardoublepage

\include{Phd-Ch-PAMELA_Experiment}
\cleardoublepage

\include{Phd-Ch-Antiproton}

\cleardoublepage

\include{Phd-Ch-Model_Constraints}
\cleardoublepage

\include{Phd-Ch-Outlook}

\cleardoublepage

\addcontentsline{toc}{chapter}{Acknowledgements}
\include{Phd-Acknowledgements}
\cleardoublepage


\addcontentsline{toc}{chapter}{List of figures}
\listoffigures
\cleardoublepage

\addcontentsline{toc}{chapter}{List of tables}
\listoftables
\cleardoublepage

\include{Phd-References}

\bibliographystyle{unsrt}


\end{document}

%% file: Phd-Abstract.tex
\chapter*{Abstract}
\label{chapt:abstract}

Studying the acceleration and propagation mechanisms of Galactic cosmic rays can provide information regarding astrophysical sources, the properties of our Galaxy, and possible exotic sources such as dark matter. To understand cosmic ray acceleration and propagation mechanisms, accurate measurements of different cosmic ray elements over a wide energy range are needed. The PAMELA experiment is a satellite-borne apparatus which allows different cosmic ray species to be identified over background. 

Measurements of the cosmic ray antiproton flux and the antiproton-to-proton flux ratio from 1.5 GeV to 180 GeV are presented in this thesis, employing the data collected between June 2006 and December 2008. Compared to previous experiments, PAMELA extends the energy range of antiproton measurements and provides significantly higher statistics. During about 800 days of data collection, PAMELA identified approximately 1300 antiprotons including 61 above 31.7 GeV. A dramatic improvement of statistics is evident since only 2 events above 30 GeV are reported by previous experiments. The derived antiproton flux and antiproton-to-proton flux ratio are consistent with previous measurements and generally considered to be produced as secondary products when cosmic ray protons and helium nuclei interact with the interstellar medium.

To constrain cosmic ray acceleration and propagation models, the antiproton data measured by PAMELA were further used together with the proton spectrum reported by PAMELA, as well as the B/C data provided by other experiments. Statistical tools were interfaced with the cosmic ray propagation package GALPROP to perform the constraining analyses.

Different diffusion models were studied. It was shown in this work that only current PAMELA data, i.e. the antiproton-to-proton ratio and the proton flux, are not able to place strong constraints on propagation parameters. Diffusion models with a linear diffusion coefficient and modified diffusion models with a low energy dependence of the diffusion coefficient were studied in the $\chi^{2}$ study. Uncertainties on the parameters and the goodness of fit of each model were given. Some models are further studied using the Bayesian inference. Posterior means and errors of the parameters base on our prior knowledge on them were obtained in the Bayesian framework. This method also allowed us to understand the correlation between parameters and compare models. 

Since the B/C ratio used in this analysis is from experiments other than PAMELA, future PAMELA secondary-to-primary ratios (B/C, $^{2}$H/$^{4}$He and $^{3}$He/$^{4}$He) can be used to avoid the data sets inconsistencies between different experiments and to minimize uncertainties on the solar modulation parameters. More robust and tighter constraints are expected. The statistical techniques have been demonstrated useful to constrain models and can be extended to other observations, e.g. electrons, positrons, gamma rays etc. Using these channels, exotic contributions from, for example, dark matter will be further investigated in future.

%% file: Phd-Introduction.tex
\chapter*{Introduction}
\label{chapt:introduction}

\section*{Outline of the thesis}
This thesis presents measurements of cosmic ray antiprotons performed with the PAMELA\footnote[1]{a Payload for Antimatter Matter Exploration and Light-nuclei Astrophysics.} satellite experiment. Cosmic ray propagation models are studied by using the antiproton and proton data measured by PAMELA and measurements of the B/C ratio from other experiments. An overview of cosmic rays is given in chapter 1, including the acceleration and transport mechanisms of cosmic rays, detection techniques for cosmic rays, and knowledge we obtain from cosmic ray studies. Chapter 2 further details the possible processes during cosmic ray propagation in our Galaxy and summarizes the current status of previous studies on cosmic ray propagation. Chapter 3 describes the PAMELA experiment. The scientific objectives of the experiment are illustrated. The design and identification capabilities of all the sub-detectors are detailed. Chapter 4 identifies antiprotons from a large background of various cosmic ray species and reconstructs the antiproton flux and the antiproton-to-proton ratio by estimating the selection efficiencies as well as other correction factors. In chapter 5, by employing the antiproton and proton data from PAEMLA and the B/C ratio data from other experiments, the $\chi^{2}$ minimization method and the Bayesian inference are used to constrain cosmic ray propagation models. Finally, some discussion and outlook are given in chapter~6.

\section*{The author's contribution}

My work on PAMELA started in September 2007 when I started my PhD position in the group of Particle and Astroparticle Physics at KTH. The first year as a PhD student was mainly focused on familiarization of the PAMELA experiment and the data analysis framework. A few months were spent on an analysis to study the separation capability of particles with equal charge but different mass in the calorimeter by using a $p\beta$ (momentum-velocity) method based on the multiple scattering effect. The analysis was documented as a Collaboration note, but not described further in this thesis. 

From the second year, I took part in the analysis of antiproton measurements. Building on the antiproton selection criteria developed by the PAMELA Collaboration, I started working on estimating the antiproton selection efficiencies and reconstructing antiproton flux. The efficiencies were derived by using different methods to fully understand the detector performance and possible systematic effects. This analysis also provided some fundamental information for the analysis of proton measurements, which is performed elsewhere in the PAMELA Collaboration. This work was presented in my licentiate thesis in April 2010, entitled ``Measurements of cosmic ray antiprotons with PAMELA''. A part of this doctoral thesis concerning about the antiproton measurements, i.e. chapter 4, is selected from the licentiate thesis. 

After doing the data analysis on antiproton measurements, I focused on studying cosmic ray propagation models by using statistical methods. The source and propagation parameters charactering the injection primary cosmic ray spectrum and different propagation processes were constrained under the framework of different propagation models. The GALPROP package which solves the transport equation numerically was used in my work to simulate cosmic ray propagation. I interfaced GALPROP with the statistical tools MINUIT and MULTINEST to perform a $\chi^{2}$ minimization analysis and a Bayesian analysis, respectively. The constraining capability of current antiproton and proton data from PAMELA data were demonstrated, as well as the upcoming PAMELA B/C data. Furthermore, in the Bayesian analysis I also produced the credible intervals on the parameters as well as on the predicted cosmic ray fluxes and flux ratios to understand the statistical uncertainties on parameters and predicted fluxes as well as the correlation between parameters. 

My work has been presented at several PAMELA Collaboration meetings and international conferences and has been discussed in several publications. 

\section*{Publications}
\begin{itemize}
\item J.~Wu, ``Measurements of cosmic-ray antiprotons with PAMELA'', Royal Institute of Techology Licentiate Thesis (2010), ISBN: 978-91-7415-585-3.
\item J.~Wu on behalf of the PAMELA collaboration, ``Measurements of cosmic-ray antiprotons with PAMELA'', Astrophys. Space Sci. Trans. 7 (2011) 225-228.
\item J.~Wu et al., ``Constraints on cosmic-ray propagation and acceleration models from recent data'', Proceedings of 32nd International Cosmic Ray Conference (2011). 
\item O.~Adriani et al., ``PAMELA Results on the Cosmic-Ray Antiproton Flux from 60 MeV to 180 GeV in Kinetic Energy'', Physical Review Letters 105 (2010) 121101.
\item O.~Adriani et al., ``PAMELA Measurements of Cosmic-Ray Proton and Helium Spectra'', Science 332 (2011) 69-72.
\item O.~Adriani et al., ``Cosmic-Ray Electron Flux Measured by the PAMELA Experiment between 1 and 625~GeV'', Physical Review Letters 106 (2011) 201101.
\item O.~Adriani et al., ``The Discovery of Geomagnetically Trapped Cosmic-ray Antiprotons'', The Astrophysical Journal Letters, 737 (2011) L29.
\item O.~Adriani et al., ``Observations of the 2006 December 13 and 14 Solar Particle Events in the 80~MeV\,n$^{-1}$ - 3~GeV\,n$^{-1}$ Range from Space with the PAMELA Detector'', The Astrophysical Journal 742 (2011) 102.
\item O.~Adriani et al., ``A statistical procedure for the identification of positrons in the PAMELA experiment'', Astroparticle Physics 34 (2010) 1-11. 
\item O.~Adriani et al., ``Measurements of quasi-trapped electrons and positron fluxes with PAMELA'', Journal of Geophysical Research 114 (2009) A12218.
\end{itemize}

\section*{Presentations} 
\begin{itemize}
\item 21st Nordic Conference in Particle Physics, Sp\aa tind, Norway. January 3-7, 2010. Contributed talk, ``The cosmic ray antiproton flux between 1 GeV and 180 GeV measured by PAMELA''.
\item A workshop on cosmic ray backgrounds for dark matter searches, Oscar Klein Center, Stockholm, Sweden. January 25-27, 2010. Contributed talk, ``The cosmic-ray antiproton flux measured by PAMELA''.
\item 22nd European Cosmic Ray Symposium in Turku, Finland. August 3-6, 2010. Contributed talk, ``Measurements of cosmic-ray antiprotons with PAMELA''.
\item 7th TeV Particle Astrophysics Conference, Stockholm, Sweden. August 1-5, 2011. Contributed talk, ``Constraints on cosmic-ray propagation and acceleration models from recent data''.
\item 32nd International Cosmic Ray Conference, Beijing, China. August 11-18, 2011. Contributed talk, ``Constraints on cosmic-ray propagation and acceleration models from recent data''.
\end{itemize}

%% file: Phd-Ch-Cosmic_Rays.tex
\chapter{Cosmic rays} \label{chapt:cosmic_rays}

A general picture about the basics of cosmic rays is given in this chapter. In section \ref{sec:CRintroduction} we review the characteristics  of cosmic rays and discuss the questions concerning the hypotheses on cosmic ray production and propagation which remain unclear. Section \ref{sec:CRdetection} focuses on  detection techniques for cosmic rays, from high altitude balloon-borne experiments or space missions to the ground based detection of ultra high energy particles. Finally, section \ref{sec:CRtools} presents the use of cosmic rays as a tool to study aspects of astrophysics, dark matter, and the matter-antimatter asymmetry in the Universe.

\section{Introduction to cosmic rays} \label{sec:CRintroduction}

Cosmic rays are energetic charged particles from outer space that travel at nearly the speed of light and impinge on Earth from all directions. They are composed mainly of ionized nuclei, roughly 90$\%$ protons, 10$\%$ helium nuclei, and slightly under 1$\%$ heavier elements as well as electrons. In 1912 Victor Hess found that an electroscope discharged faster as he ascended in a balloon to altitudes up to 5~km~\cite{Berezinskii1990}. He therefore concluded that cosmic rays arrived from outside our atmosphere and did not originate from decaying radioactive isotopes in the ground. Since their discovery, cosmic ray nuclei and electrons have been studied extensively \cite{Gaisser1990} with a special emphasis on their main characteristics: the energy spectrum as well as their elementary composition and abundances. 

%

\begin{figure}[tbhp]
\begin{center}
\includegraphics[width=0.85\textwidth]{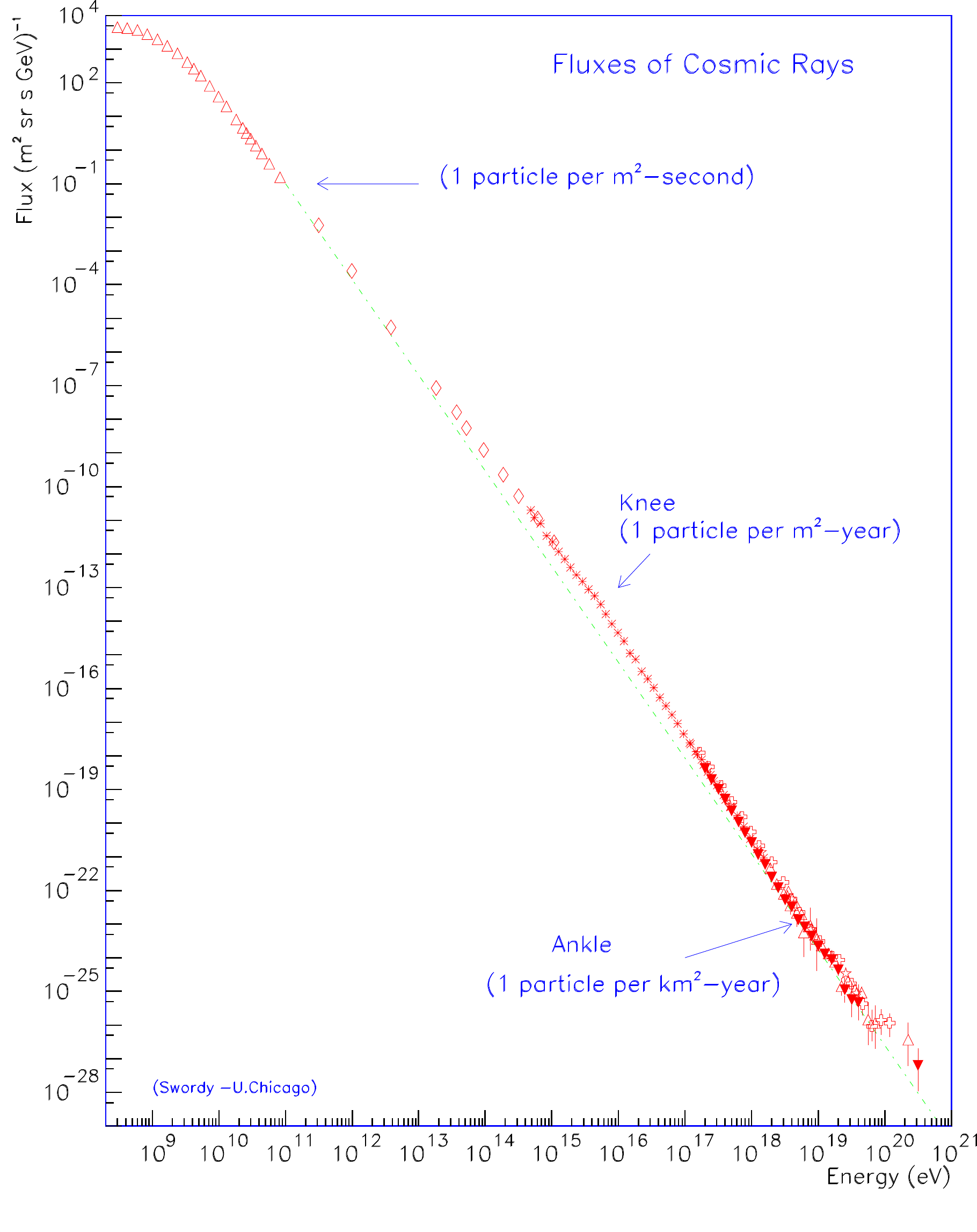}
\end{center}
\caption[The energy spectra of cosmic rays]{{\footnotesize The energy spectra of cosmic rays (taken from \cite{SwordyChicago}). Above $10^{10}$ eV the spectrum shows a power-law behaviour. An obvious change in the slope is observed at the knee ($4 \times 10^{15}$ eV) and at the ankle ($5 \times 10^{18}$ eV).}}
\label{fig:powerlawspectra}
\end{figure}

The overall energy spectrum of cosmic ray for energies $> 10^{10}$~eV, where solar effects are negligible, is well described by an inverse power law with evident features: the ``knee'', at $4\times10^{15}$ eV, a not so evident second knee at $\sim 10^{17}$ eV, and a flatter supposedly extragalactic component at energies larger than about $5 \times 10^{18}$ eV (figure \ref{fig:powerlawspectra}). Beyond this energy, where data become sparse, another steepening appears above $5\times 10^{19}$~eV possibly due to the Greisen-Zatsepin-Kuzmin limit (GZK limit), which is expected as cosmic ray protons with energies above $5\times 10^{19}$~eV interact with cosmic microwave background photons to produce pions:
\begin{equation*}
p+\gamma_{CMB} \longrightarrow \Delta^{+} \rightarrow 
\begin{array}{rl}
p+\pi^{0}&\\
n+\pi^{+}
\end{array}. 
\end{equation*}


\begin{figure}[tbh]
\begin{center}
\includegraphics[width=0.85\textwidth]{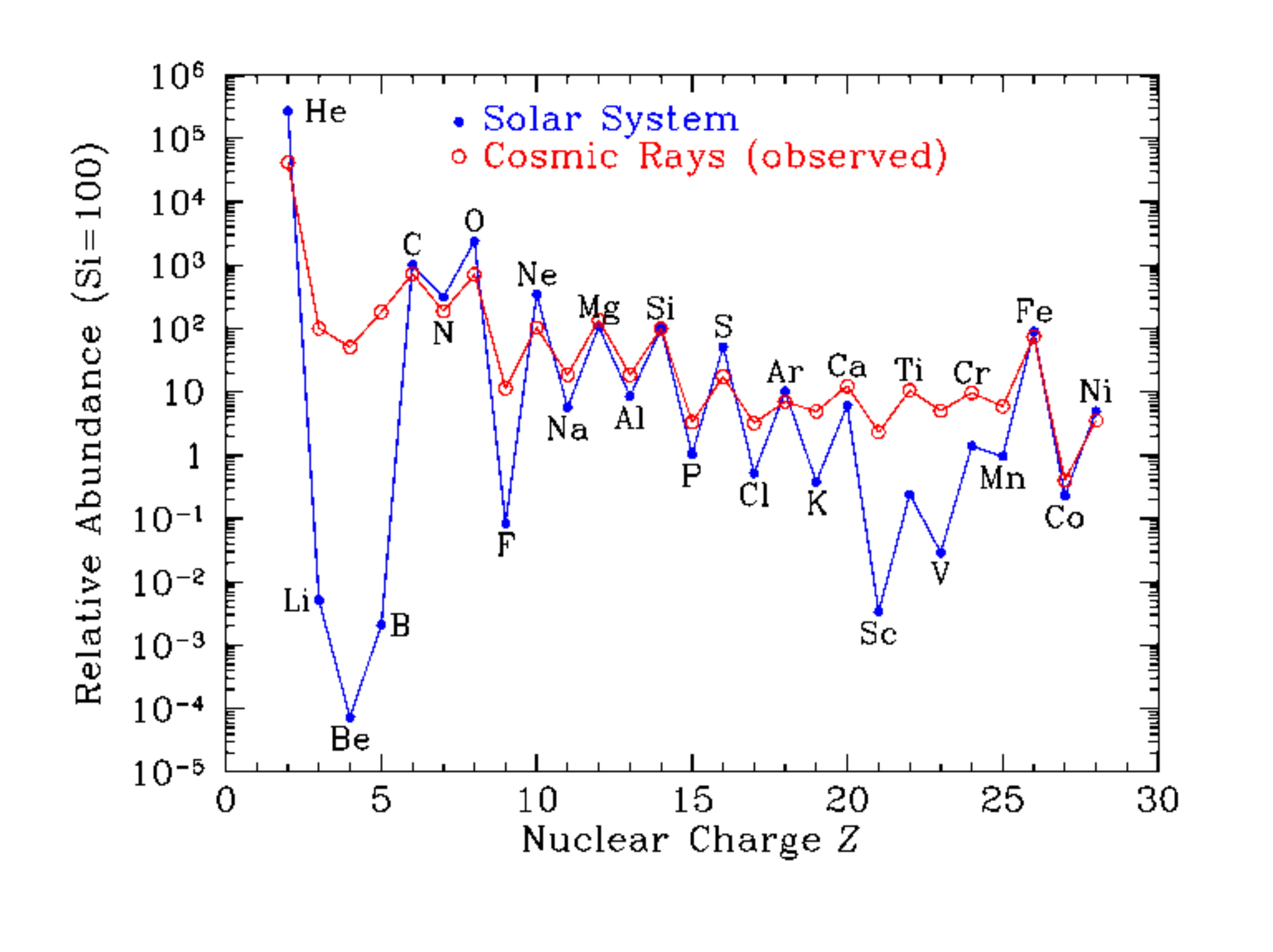}
\end{center}
\caption[The cosmic ray abundances] {\footnotesize The relative abundances of cosmic rays measured at Earth compared to the Solar system abundances (taken from \cite{AltasAstroLec}, normalized to Si=100).}
\label{fig:CRabundances}
\end{figure}

The main features of the cosmic ray composition at low energies ($<10^{14}$~eV) were known by 1950 and still remain unclear at higher energies. The relative abundances of cosmic rays are similar to the abundances of common elements in the Solar system, as shown in figure \ref{fig:CRabundances}. This consistency indicates that the composition of cosmic ray material injected into the interstellar medium (ISM) is very similar to that of the nebula that formed the Solar system. However, a striking difference can be seen between these two compositions. Chemical elements including Li, Be, B, F, Sc, Ti, V, Cr, Mn which are rare in the Solar system are many orders of magnitude more abundant in the cosmic rays. Since these elements are essentially absent as end products of stellar nucleosynthesis, they are generated as spallation products of abundant cosmic rays interacting with hydrogen or helium nuclei in the interstellar gas. For instance, Li, Be, B isotopes are mainly created from fragmented progenitors C, N and O nuclei. An example of the reaction contributing to boron production is $^{12}\text{C}\,+\,\text{p}\, \longrightarrow \, ^{10}\text{B}\,+\,^{3}\text{He}$.


Although a great amount of information on the composition and the energy spectrum of cosmic rays on Earth has been gathered, fundamental questions concerning the origin of these particles, the mechanism through which they are accelerated to high energies, and the processes they undergoing before they arrive at the Earth remain unanswered. 

\subsection{Cosmic ray sources and acceleration}

Since the 1960s, supernova remnants (SNRs) -- the tattered, gaseous remains of supernovae -- have been discussed as the breeding ground of Galactic cosmic rays for energies up to $10^{15}$~eV. On average, about one supernova occurs in our Galaxy every 30 years, releasing 10$^{44}$ J in the form of kinetic energy in the ejecta. Therefore supernovae have enough power to energize the Galactic cosmic ray population at the observed level if there exists a mechanism for converting about 10$\%$ of the mechanical energy into relativistic particles. The knee in the cosmic ray energy spectrum presumably indicates a limit for the acceleration of cosmic ray protons by SNRs. It is argued that type II supernova surrounded by a dense stellar wind may be responsible for accelerating cosmic ray heavy nuclei up to energies about 10$^{18}$~eV \cite{Hillas2005}. Above $\sim10^{19}$~eV, the Galactic magnetic field would not be able to trap effectively even the heaviest elements of cosmic rays and an extra-galactic origin is required. Candidates such as external shocks in jets of active galactic nuclei and long gamma ray bursts have been proposed \cite{Torres2004, Hillas2006, Berezinsky2008} as sources of ultra high energy cosmic rays.


Once created in the sources, cosmic rays need to be accelerated and injected into the ISM. Diffusive shock acceleration (DSA) operating at expanding supernova shells is the most-favored mechanism for the production and acceleration of Galactic cosmic rays. The supernova remnants expand into the surrounding interstellar gas, compressing both the interstellar gas and magnetic field, producing a shock front. As the fast-moving charged particles move through the shocked gas, they diffuse by scattering on the contorted magnetic fields. Particles gain energy by bouncing between converging upstream and downstream regions around the shock front. This process naturally generates power-law energy spectra $N\left(E\right) \propto E^{-\alpha}$ which is the striking characteristic of cosmic rays, with $\alpha = 2$ for large sonic Mach numbers \cite{Drury1983, Blanford1987, Jones1991, Malkov2001, Hillas2005}. In a more realistic picture, cosmic rays being accelerated can cause streaming instabilities and generate hydromagnetic waves which make the acceleration a non-linear process. A deviation from $\alpha = 2$ can then occur. However, the injection spectrum remains poorly known since it does not only depend on the instantaneous spectrum of particles being accelerated at a shock, but also relates to how and when accelerated particles are released into the Galaxy, as well as the details of the interplay between accelerated particles, magnetic field amplification and shock dynamics. Calculations of non-linear DSA (NLDSA) models can either predict a hard injection spectrum with a spectral index less than $\simeq 2.1-2.15$ \cite{Malkov2001, Berezhko2007, Caprioli2011_Aph}, or produce steeper spectrum up to $\alpha \sim 2.5$ \cite{Caprioli2011_JCAP}. 

Despite the appeal of the SNR conjecture, verification from observational evidence is needed. The problem is that cosmic rays are deflected and isotropized by the Galactic magnetic field and as a result the actual position of acceleration sites can not be extrapolated from their arrival direction. Thus some other tools are required to test the supernova paradigm. Significant progress has been achieved in recent years by keV X-ray and GeV to TeV gamma ray observations of young SNRs, providing very useful information about cosmic ray acceleration by supernova shocks. Since the acceleration of cosmic rays in SNRs must be accompanied by copious gamma ray emission due to the decay of neutral pions produced in nuclear collisions between relativistic nuclei and the background gas atoms, gamma ray detection is a good tracer of cosmic ray accelerators. Young SNRs which have strong shocks and can actively accelerate particles to the highest energies are usually chosen to be targets to investigate acceleration processes. 

\begin{figure}[tbhp]
\begin{center}$
\begin{array}{cc}
\includegraphics[width=5cm]{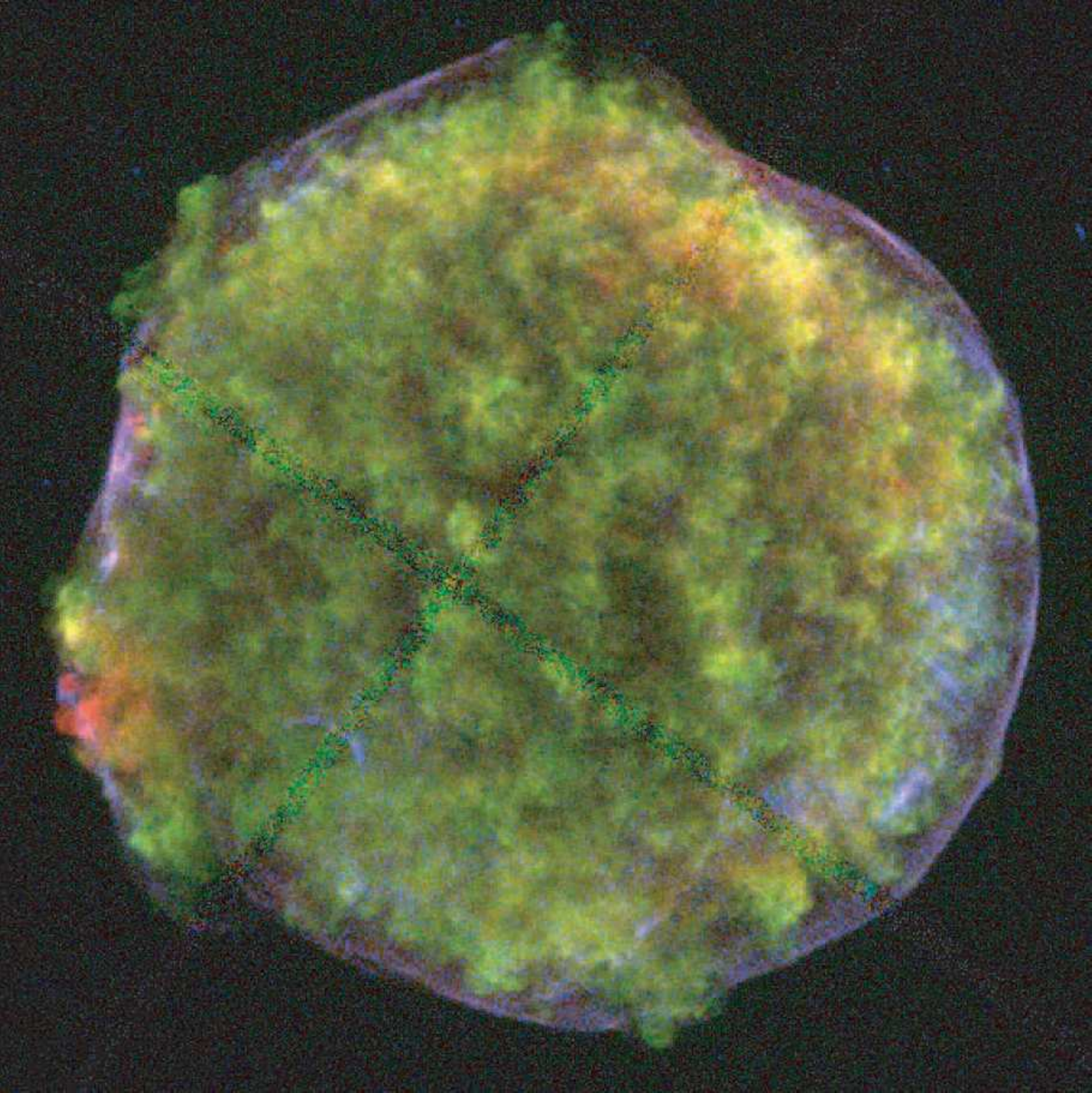}&
\includegraphics[width=5cm]{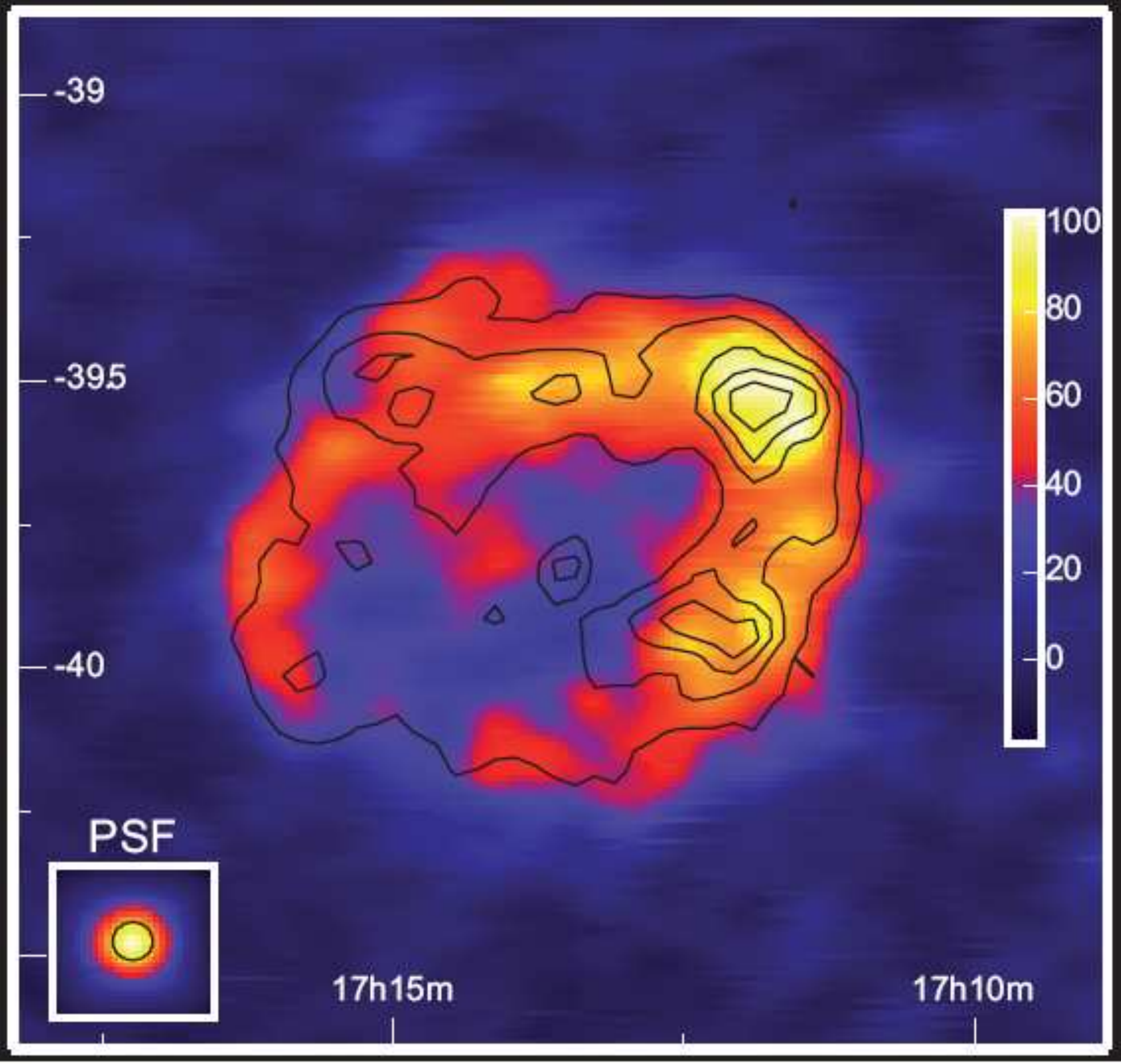}
\end{array}$
\end{center}
\caption[The observed shell-like SNRs Tycho and RX J1713.7-3946]{\footnotesize Left: Three-color composite image of Tycho's SNR observed by Chandra (taken from \cite{Warren2005}: 0.95-1.26 keV emitted from Fe L-shell (red), 1.63-2.26 keV emitted from Si K-shell (green) and 4.1-6.1 keV continuum (blue). Right: RX J1713.7-3946 as seen by HESS (colors) and by ASCA in the 1-3keV energy band (contours) (taken from \cite{Aharonian2007_AA}). The image is smoothed with a Gaussian of 2Ô and the linear color scale is in units of excess counts per smoothing radius.}
\label{fig:SNRcontours}
\end{figure}

\begin{figure}[tbhp]
\begin{center}
\includegraphics[width=0.85\textwidth]{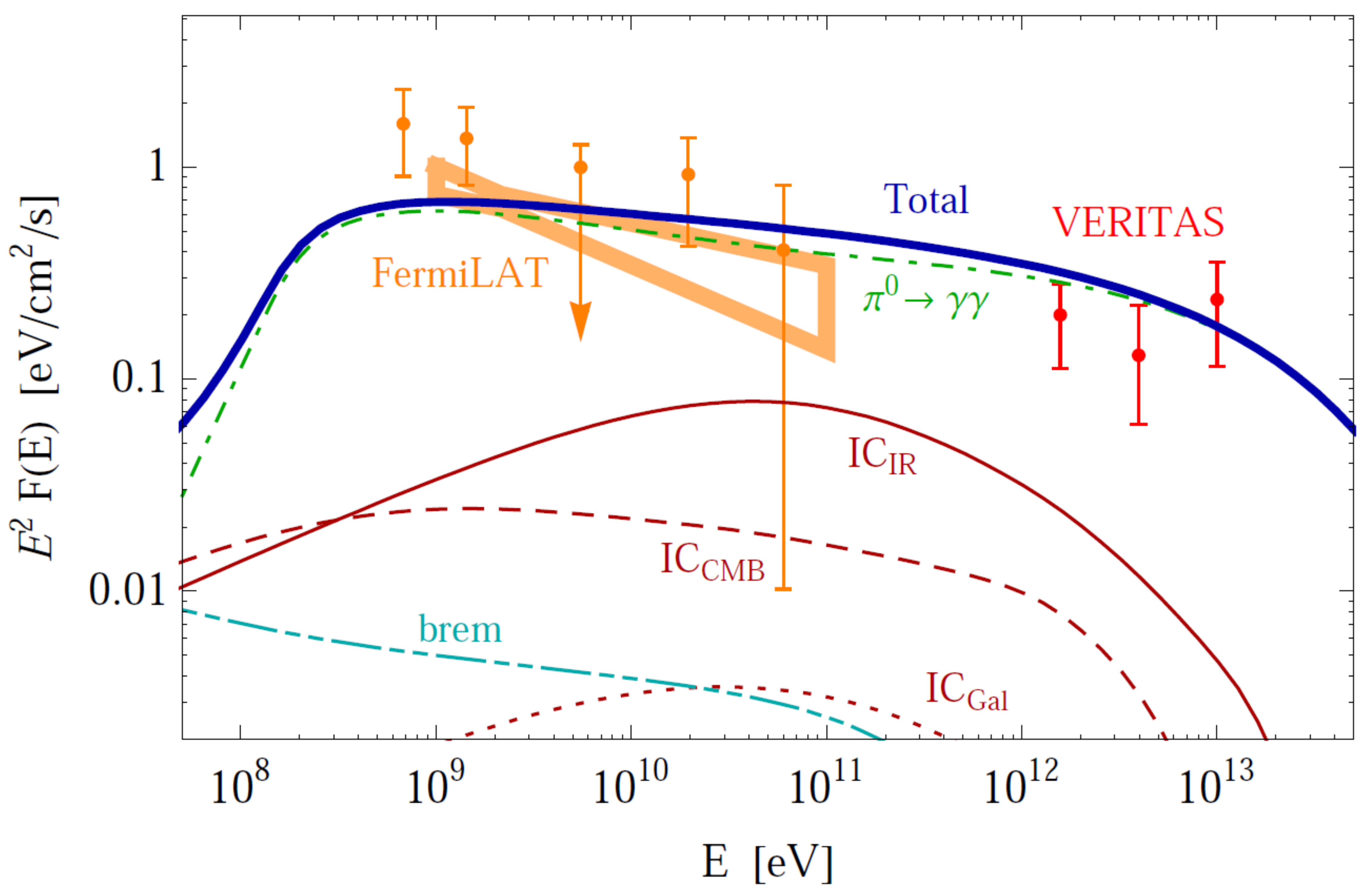}
\end{center}
\caption[Tycho spectrum]{\footnotesize The spectrum of gamma ray emission from Tycho's SNR measured by Fermi-LAT and VERITAS, compared with different theoretical contributions. Taken from \cite{Morlino2011}.}
\label{fig:SNR_TychoFlux}
\end{figure}

Many young SNRs exhibit shell-like morphologies at different wavelength bands. Examples of Tycho \cite{Warren2005} and RX J1713.7-3946 \cite{Aharonian2004, Aharonian2007_AA} are shown in figure \ref{fig:SNRcontours}. While the non-thermal X-rays detected in the shells of SNRs are generated by electrons via synchrontron processes, the mechanism responsible for the gamma ray emission is still under debate. Two scenarios have been proposed. Hadronic models connect gamma rays with neutral pion decay following proton-proton interactions while the leptonic models suggest gamma rays are generated through inverse Compton scattering by the same populations of electrons interpreting the X-ray emission \cite{Gabici2008, Berezhko2008, Morlino2009}. Very recently, GeV gamma ray emission from RX J1713.7-3946 was measured by Fermi-LAT \cite{Abdo2011} and disfavors the $\pi^{0}$-decay mechanism. A very hard photon spectrum was observed which agrees well with the prediction of leptonic origin. However, the observed photon flux could still be reasonable in hadronic models considering low-density hot bubbles around the SNR shocks \cite{Fang2011} or interaction between shocks and interstellar clouds \cite{Inoue2011, Fukui2011}. Another young SNR, Tycho, newly detected in GeV energies by Fermi-LAT \cite{Giordano2011} and in TeV energies by VERITAS \cite{Acciari2011}, strongly supports the hadronic scenario, as shown in figure \ref{fig:SNR_TychoFlux}.  However, both scenarios could be adapted to the experimental data under the assumption of a SNR environment with non-uniform magnetic fields \cite{Atoyan2011}. Neutrino observations with km$^{3}$-class detectors such as IceCube \cite{Achterberg2007, Halzen2008} or KM3NeT \cite{Sapienza2011} may improve our confidence in the hadronic mechanism, since high energy neutrinos are created mainly in the decay of charged pion mesons produced in collisions of cosmic ray protons with nuclei in the ambient gas.

\subsection{The journey of cosmic rays from the source to the Earth} \label{sec:CRjourney}

No matter where and how the cosmic rays were produced and accelerated, they subsequently propagate through the Galaxy before reaching Earth. A common view derived from observations hints that cosmic rays travel in a confinement volume with an average residence time $\sim10^{7}$ years. The amount of matter traversed by cosmic rays is estimated to be less than the density of the disk, indicating that cosmic rays are trapped mostly in low-density regions. Several complex phenomena might occur during propagation. It is believed that during their residence time in the Galaxy, cosmic rays diffuse randomly due to the irregularities in the Galactic magnetic field. The diffusion process was proposed to explain cosmic ray confinement in the Galaxy and the observed isotropy. The details of the Galactic magnetic field structure is not well understood. Assuming the average magnetic field strength is $B$, a particle scattering on magnetic field irregularities with weak random fluctuations $\delta B << B$ can be treated in the quasi-linear theory of plasma turbulence. Theoretically, different types of spectral energy density of interstellar turbulence have been proposed. The favored ones are Kolmogorov-type \cite{Goldreich1995} and Kraichnan-type \cite{Yan2004} spectra. However, current data do not allow us to distinguish between these different turbulence types. 

In addition to diffusion, other processes could also play a role in cosmic ray transport. Cosmic rays could possibly be convected if the medium responsible for diffusion is moving away from the disc, i.e. a galactic wind is present. The scattering of cosmic ray particles on magnetized plasmas in the ISM causing stochastic acceleration could also happen, but cannot serve as the main mechanism of cosmic ray acceleration. Moreover, when charged cosmic ray nuclei travel in the ISM, they undergo nuclear destruction due to fragmentation and unstable nuclei decay to stable nuclei. Through these processes, secondary cosmic rays are created as spallation products of primary progenitors. Additionally, energy losses arise from interactions such as ionization and Compton scattering, which dominate for cosmic ray nuclei. Cosmic ray electrons, however, do not only lose energy by virtue of interactions with the ISM but also with the Galactic magnetic field or the interstellar radiation field. A more detailed description of the propagation processes in the Galaxy will be a focus of Chapter \ref{chapt:propagation}.

\begin{figure}[tbh]
\begin{center}
\includegraphics[width=0.85\textwidth]{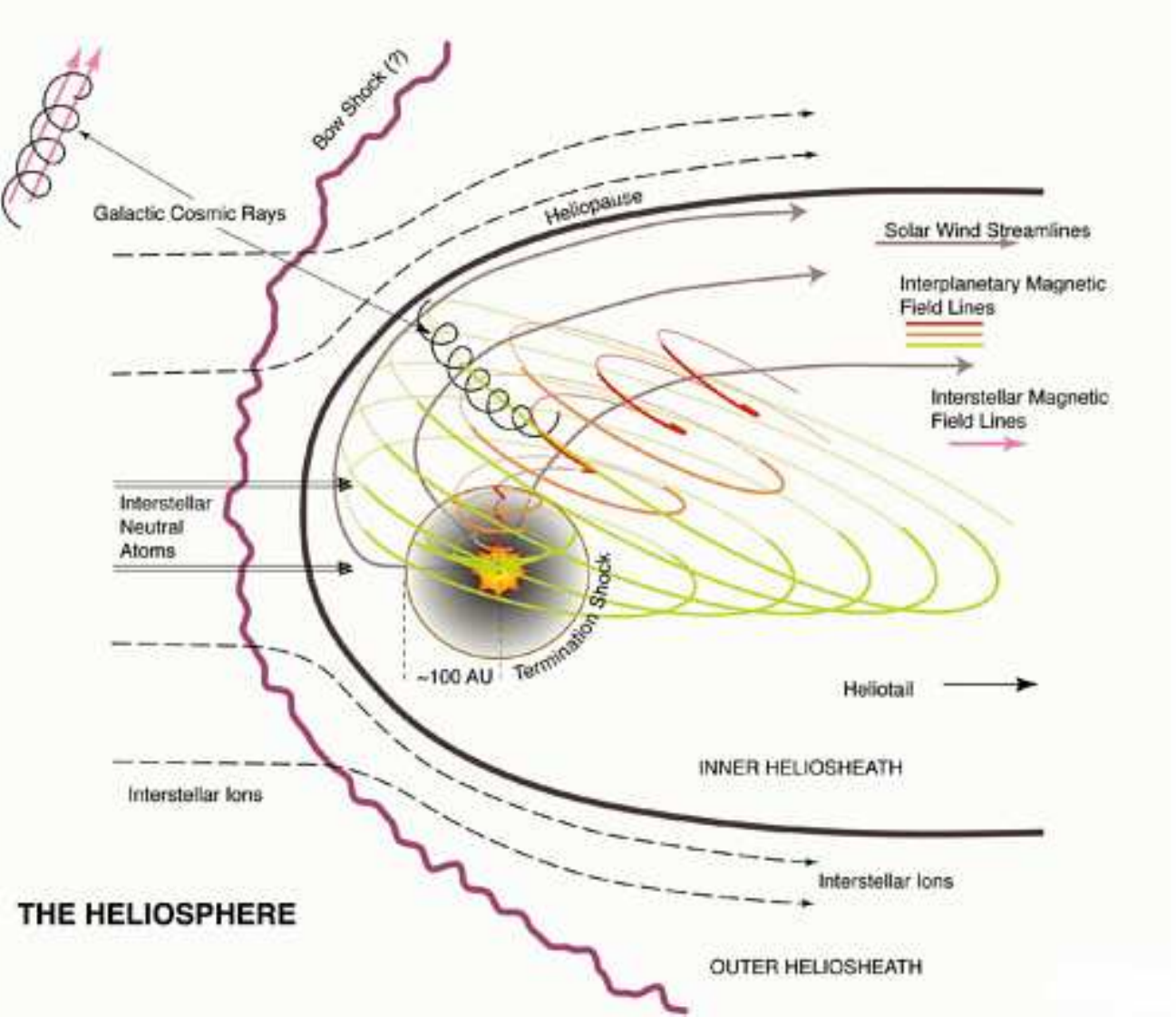}
\end{center}
\caption[The heliosphere] {\footnotesize A schematic diagram of the heliosphere. Taken from \cite{HelioNap2004}.}
\label{fig:heliosphere}
\end{figure}

Finally, before arriving at Earth, cosmic rays are affected by the outstreaming particles ejected from Sun and the geomagnetic field. The Sun emits low energy particles in the form of a fully ionized plasma called the solar wind, dominating in a cavity known as heliosphere as shown in figure \ref{fig:heliosphere}. The solar wind has a supersonic speed of about 400-800 km/s, flows outward and decreases to subsonic flow at the termination shock. Beyond this, the solar wind which carries the spiraling interplanetary magnetic field is turned toward the heliotail. At larger radial distances, a surface called the heliopause is reached, separating the solar material and the solar magnetic fields from the interstellar material and the interstellar magnetic fields. Interstellar ions are diverted around the heliosphere. An outward pointing bow shock may also be formed beyond the heliosphere.

\begin{figure}[tbh]
\begin{center}
\includegraphics[width=0.85\textwidth]{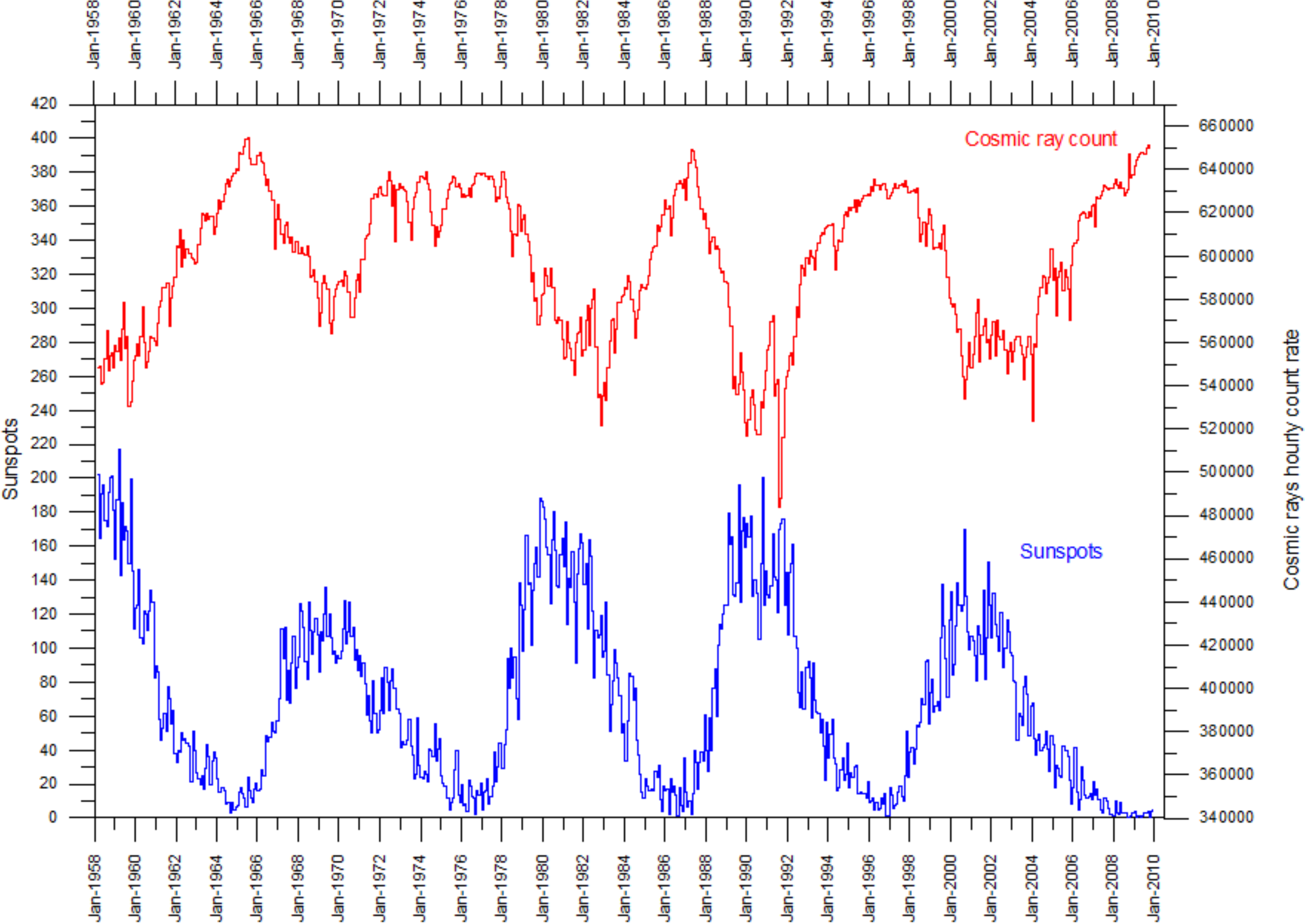}
\end{center}
\caption[The anticorrelation between cosmic ray intensity and the solar activity] {\footnotesize Variation of cosmic ray neutron intensity and the solar activity represented by the sunspot numbers. High cosmic ray intensity corresponds to low sunspot activity, and vice versa. Taken from \cite{climate4you}.}
\label{fig:CRintensity_sunspot}
\end{figure}

The solar wind prevents low energy cosmic rays from penetrating the heliosphere and modulates the cosmic ray energy spectra. This phenomenon is called solar modulation and was developed originally by Parker \cite{Parker1965}, varying according to the 11 year solar cycle. The greater the solar activity, fewer cosmic ray particles can get into the heliosphere, as shown in figure \ref{fig:CRintensity_sunspot}. The solar modulation is determined by four mechanisms, including convection by the outward solar wind flow, diffusion in a turbulent heliospheric magnetic field (HMF) carried by the wind, drift due to the gradients, curvature and current sheet of the HMF and adiabatic energy changes. A simple but frequently used model is the force-field approximation~\cite{Gleeson1968} which depends on a single parameter, the modulation potential $\Phi$. For a nucleus with charge $Z$, mass $m$ and atomic number $A$, its interstellar flux $J_{IS}$ is modulated to the top-of-atmosphere flux $J_{TOA}$ by the relation
\begin{equation} \label{eq:forcefield}
J_{TOA}(E)= \frac{E^{2}-m^{2}}{(E+|Z|\Phi)^{2}-m^{2}} J_{IS}(E+|z|\Phi),
\end{equation}
where $E$ is the total energy of the nucleus. The modulation potential $\Phi$ is determined by fitting the observed spectrum above the atmosphere with the assumed interstellar spectrum. Although the force-field approximation is useful in most cases, it is worth to note that this approximation cannot consider any charge-sign dependence of solar modulation indicated in experimental data \cite{Mitchell2007, Gast2009}. Drift models are suggested in literature \cite{LeRoux1995, Langner2004, ValdesGalicia2005, Potgieter2008} which can produce a clear charge-sign-dependent modulation. For example, during the $A<0$ polarity cycles, i.e. when the HMF is directed toward the Sun in the northern hemisphere, the negatively charged particles will drift inward primarily through the polar regions of the heliosphere and the positively charged particles will drift primarily through the equatorial regions of the heliosphere.
Nevertheless, the realistic time-dependent modulation could be very complex and needs further investigation \cite{Potgieter2008}.  Cosmic ray nuclei with energies larger than about 10~GeV/n are not sensitive to the solar activity~\cite{Moskalenko2002}. 

After penetrating the heliosphere, low energy charged particles are deflected by the geomagnetic field, which is the last obstacle for cosmic rays on their way to the top of Earth's atmosphere. A charged particle traverses this magnetic field in a curved path and a minimum rigidity (momentum per unit charge) referred to as the \textit{cutoff} is required to penetrate the geomagnetic field. The cutoff rigidity, varying with the geomagnetic position and the approaching direction of cosmic ray particles, was first treated by Stoermer who approximated the Earth's magnetic field as a dipolar field \cite{Stoermer1911}. For particles incident vertically towards the center of the magnetic dipole, the Stoermer vertical cutoff (SVC) can be written as \cite{Longair1992}
\begin{equation}
p \geq 14.9 Z \cos^{4}{\lambda} \; \mathrm{GeV/c},
\end{equation}
where $\lambda$ is the geomagnetic latitude. Consequently, the detected cosmic ray intensity will be lower at the magnetic equator and higher at the magnetic pole as the geomagnetic cutoff value is largest at the equator and diminishes closer to the poles. The SVC is often the reference quantity calculated and is used as an effective average over all arrival directions. However, the SVC has limited accuracy because the realistic geomagnetic field does not obey an ideal dipole geometry but is offset by some 400~km from Earth's center and has higher order components. Furthermore, Stoermer's theory allows the penetration of charged particles with trajectories that would go through Earth and generally underestimates the cutoffs. This problem is called the Earth's shadow and was first addressed by Vallarta \cite{Vallarta1948}, who showed that a range of magnetic rigidities exist above the Stoermer cutoff where the penumbral shadow of Earth casts a broken pattern of allowed or forbidden bands of magnetic rigidity. More reliable and precise determination of the geomagnetic field can be done by tracing trajectories of cosmic rays in higher order geomagnetic field models \cite{Smart2005, Smart2009}.

\section{Detection techniques}\label{sec:CRdetection}

In order to understand the nature of cosmic rays, many experiments have been performed during the last century, producing a large amount of observational data. Different kinds of detectors are used to detect cosmic rays depending on the energy of interest. Direct detection experiments record cosmic rays directly, while indirect ones measure the secondary showers initiated from the incident cosmic rays interacting with atmosphere.

Direct detection is used to study particles below 10$^{15}$~eV for which the flux of particles is sufficiently large that individual primary nuclei can be studied by instruments carried in high-altitude balloons or in space. The purpose of direct detection is to discriminate the incoming cosmic ray particles and to measure their abundances and energies. Various types of detector are utilized, such as magnetic spectrometers, calorimeters, transition radiation detectors, scintillators or solid state detectors, Cherenkov counters and time-of-flight systems. A number of these detectors are appropriately assembled as a package either in high-attitude balloon experiments such as MASS91 \cite{Hof1996}, CAPRICE \cite{Barbiellini1996, Bergstrom2001}, TRACER \cite{Ave2008}, ATIC \cite{Guzik2004}, BESS \cite{Mitchell2004} and CREAM \cite{Ahn2007}, or in space-based experiments, for example Spacelab 2 \cite{Swordy1990}, HEAO3-C2 \cite{Engelmann1990}, ACE-CRIS \cite{Stone1998}, AMS \cite{Ahlen1994, Lubelsmeyer2011} and PAMELA \cite{Picozza2007}. In general, balloon-borne experiments allow multiple flights with a moderate budget and can provide a prototype test which can be further employed in space. However, the exposure time they can provide is up to 42 days (CREAM-1 performed in 2004 \cite{Ahn2007}), which is restricted mostly by the wind and limited resources on-board. Space missions are more expensive and risky, but highly increase the statistics benefiting from much longer exposure time and reduce the systematic uncertainties caused by the interference of cosmic rays with the residual atmosphere above balloons. A part of this thesis focuses on the data analysis of the satellite-borne experiment PAMELA, which will be described in more detail in chapter \ref{chapt:PAMELA}. 

Very high energy (above $\sim$10$^{14}$~eV) cosmic rays are extremely rare, for example the flux is only 1~km$^{-2}$sr$^{-1}$year$^{-1}$ above 10$^{19}$~eV (see figure \ref{fig:powerlawspectra}), and only ground-based experiments with huge effective areas and long exposure times can hope to acquire a significant statistical sample. The ground-based experiments exploit the atmosphere as a giant calorimeter. An incident cosmic ray particle interacts with air molecules, mainly oxygen and nitrogen, and produces a cascade of lighter particles, spreading out over large areas, called an extensive air shower. Rather than detecting the primary cosmic rays directly, ground-based detectors detect the remnants of the atmospheric cascades of particles initiated by the primary particle. Composition and energy information of incident particle species can be derived from the EAS properties based on hadronic models. Several techniques are used in current instruments, ranging from direct sampling of secondary particles in the shower to measurements of fluorescence from atmospheric nitrogen excited by the charged particles, and radio emission emanating from the air shower. Some experiments employing one or more of these techniques, are AGASA \cite{Chiba1992}, HiRes \cite{AbuZayyad1999}, Auger \cite{Abraham2004}, KASCADE \cite{Antoni2003} and TA \cite{Kawai2008}. 

\section{Cosmic rays as observational tools} \label{sec:CRtools}

From the various elemental cosmic ray data, measured by different experiments, the principle astrophysical issues concerning cosmic ray acceleration and propagation mechanisms can be investigated. Moreover, cosmic ray observations provide potential to help us understand topics such as the nature of dark matter and the apparent matter-antimatter asymmetry in the Universe. 

\subsection{Astrophysics}

The energy spectra of cosmic rays, extending over a wide energy range from tens of MeV/n to EeV/n, provide a useful means to probe the properties of cosmic ray sources, acceleration mechanisms, propagation processes in the Galactic halo and the interstellar environment itself. At energies higher than tens of GeV, the observed abundances are affected by the injection spectrum from the sources, the diffusion in the Galactic magnetic field and the nuclear interactions in the Galaxy. The low energy tail, however, also has a contribution from other phenomena, such as convection, reacceleration and heliospheric physics. Therefore, tracing back from cosmic rays observed at Earth, we can effectively investigate the processes happening before cosmic rays reach Earth. In addition, the properties of the ISM and the structure of the Galactic magnetic field can then be better understood.

\begin{figure}[tbh]
\begin{center}
\includegraphics[width=0.85\textwidth]{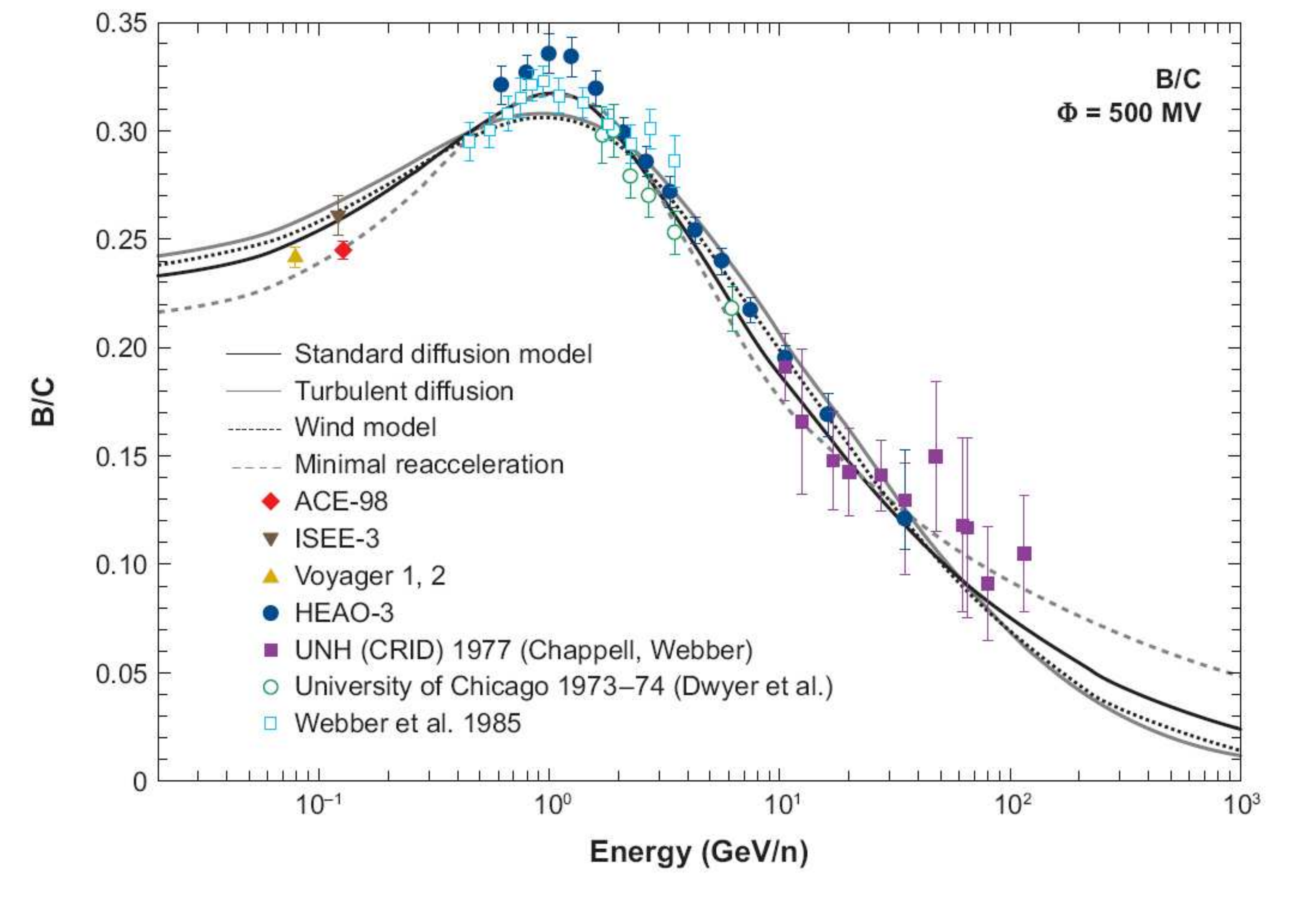}
\end{center}
\caption[Measured B/C Ratio compared with models]{{\footnotesize Measured boron-to-carbon ratio (B/C) compared with four models, which are modulated with the solar modulation potential $\Phi=500$~MV. Taken from {\cite{Strong2007}}.}}
\label{fig:BCRatio}
\end{figure}

As indicated in figure \ref{fig:CRabundances}, elements like Li, Be, B are secondary nuclei produced by primary cosmic rays interacting with the interstellar gas. Therefore, the relative abundances of secondary nuclei shed light on the properties of matter in the Galaxy. Measurements of secondary-to-primary ratios are useful probes of cosmic ray transport since they mainly depend on the mean amount of interstellar matter that primaries have encountered before reaching Earth rather than the source spectrum of the progenitors. The B/C ratio has been considered as one of the most important quantities for decades, as B is entirely secondary and its main progenitors C and O are primaries directly produced in the SNR. The B/C ratio is also the best measured secondary-to-primary ratio since it depends on the elemental separation capability of the detector but not on the isotopic separation capability which are important for the ratios like $^{2}$H/$^{4}$He. Figure \ref{fig:BCRatio} shows the measured B/C ratio compared to some models, which cannot be distinguished using the data listed here alone. Apart from B/C data, other quantities such as $^{2}$H/$^{4}$He, $^{3}$He/$^{4}$He and $\bar{\text{p}}$/p are also secondary-to-primary ratios which are useful to probe cosmic ray transport processes \cite{Coste2011}. 

The other major constraint on propagation models comes from radioactive species, which are unstable nuclei undergoing radioactive decays such as $\beta$ decay and electron capture. Especially useful species are secondary radioactive isotopes for which no extra contribution from sources need to be accounted for. The ratios of unstable to stable isotopes of secondary nuclei tell us the global properties of the Galaxy through the surviving fraction of unstable isotopes in the Galaxy. A combination of secondary-to-primary ratios and radioactive isotope ratios allows one to derive the size of Galaxy halo. The most notable unstable nucleus is $^{10}$Be, which is best measured and has a lifetime of $\sim 3.9\times10^{6}$~years for $\beta$ decay, comparable with the escape time of cosmic rays in the Galaxy. Other long-lived radioactive nuclei such as $^{14}$C, $^{26}$Al, $^{36}$Cl and $^{54}$Mn also provide constraints on cosmic ray propagation \cite{Webber1998, Strong2001}.

The spectra observed on Earth are affected by a combination of acceleration and propagation. Generally, while the propagation processes can be understood using secondary cosmic rays, the information on the acceleration can be derived from the primary cosmic ray spectra \cite{Putze2011}. Propagation and source parameters are degenerate. Simultaneously fitting secondary-to-primary ratios as well as primary fluxes allows us to explore both source and propagation mechanisms \cite{Trotta2011}. 

\subsection{Dark Matter} \label{sec:darkmatter}

The existence of dark matter (DM) is motivated by a wealth of observational evidence, including galactic rotation curves, gravitational lensing, the anisotropies of the cosmic microwave background, and primordial light element abundances. The approximate distribution of DM, which constitutes about a quarter of the mass of the Universe, can be deduced from its gravitational effects, but its nature and microphysical properties remain one of the great unsolved problems of physics \cite{Trimble1987, Ashman1992}. The lack of observation of DM particles indicate that DM particles are primarily non-baryonic which only interact through the weak force and gravity. In addition, cold DM is necessary to explain structure formation in Universe, since relativistic DM moves too quickly to clump together on small scale of galaxies. One of the most common proposed candidates is referred to as a Weakly Interacting Massive Particle (WIMP) \cite{Jungman1996, Feng2010}. DM signals are possible to be detected directly by creating DM particles in accelerators or searching for the scattering of DM particles off atomic nuclei within a detector, and indirectly by gathering information from WIMP annihilation products.

Indirect detection of DM is based on the search for anomalous features in energy spectrum in cosmic rays due to WIMP annihilation in the Galactic halo, on the top of the expectation from standard astrophysics background. Since matter dominates cosmic rays, antimatter ($\bar{\text{p}}$, $\bar{\text{D}}$, e$^{+}$), gamma ray and neutrino channels have better potential to probe dark matter. The spectral distortion of these components over the astrophysical background may give evidence for dark matter.

The DM signal prediction depends on the models in which the properties of DM particles and their interaction strength with Standard Model (SM) states are assumed. There are numerous models with a WIMP DM candidate. One popular framework is the Minimal Supersymmetric extension to the Standard Model (MSSM), in which the lightest supersymmetry particle, known as ``neutralino'', is stable and therefore provides a good candidate. Another widely discussed scenario is universal extra dimensions introducing a tower of Kaluza-Klein partners for every SM particle in which the lightest Kaluza-Klein particle is stable and considered as a good DM candidate. For any particular model, the annihilation products of DM candidates can be predicted. A robust and accurate estimation of the cosmic ray background contribution as well as the propagation of DM annihilation products is required to clarify DM models and consequently help us to understand the properties of DM particles. 

\begin{figure}[tbh]
\begin{center}
\includegraphics[width=0.80\textwidth]{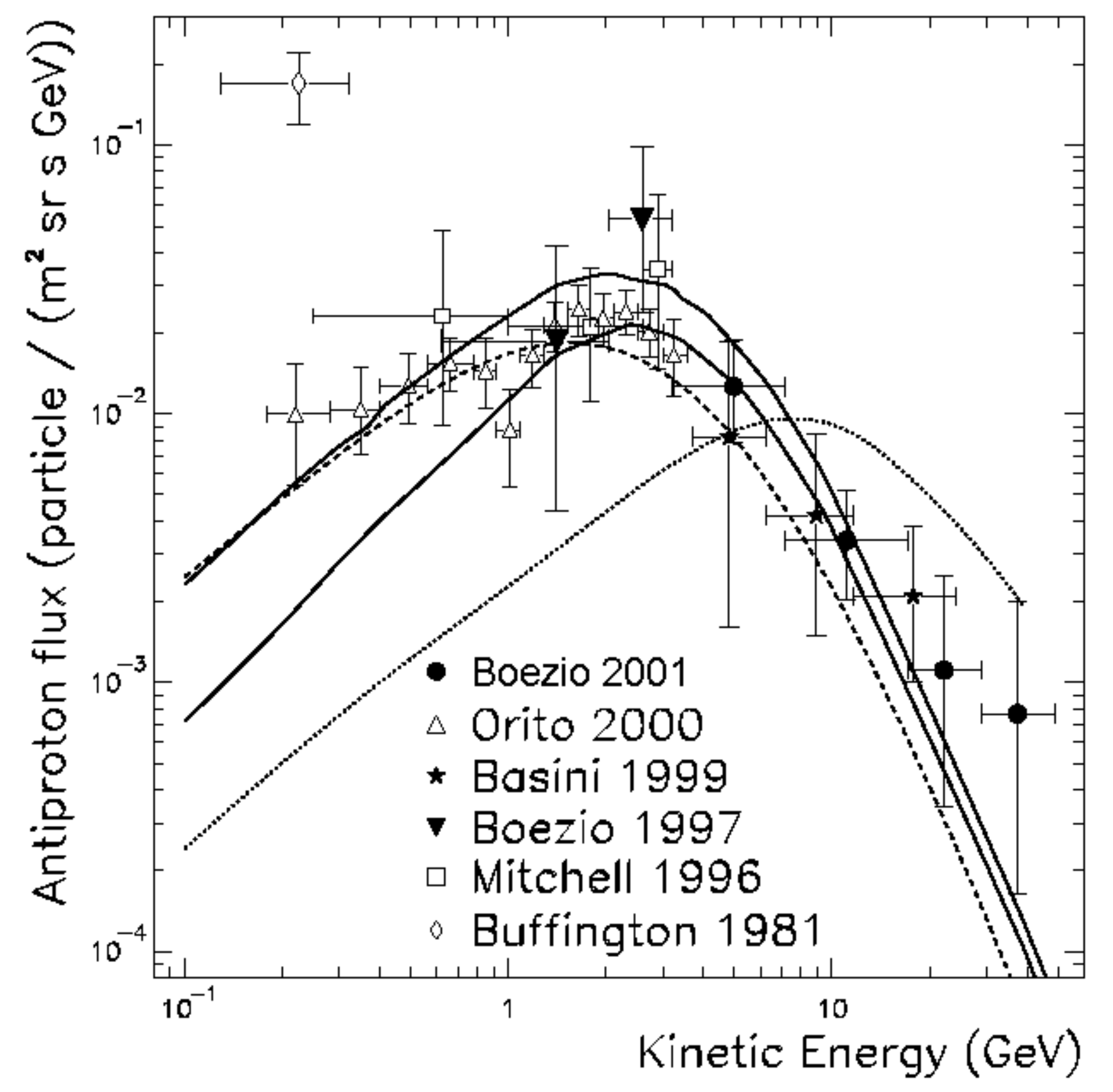}
\end{center}
\caption[The antiproton flux predicted by a neutralino model]{{\footnotesize A primary antiproton flux assuming the annihilation from neutralino from MSSM with a mass of 964 GeV (dotted line) compared with experiment results. The solid lines show the upper and lower limit of calculated flux of interstellar secondary antiprotons by Simon el al. The dashed line shows the theoretical calculation of interstellar secondary antiprotons by Bergstr\"om \& Ullio. All the references for the experiment results and theoretical models can be found in {\cite{Boezio2001}}.}}
\label{fig:Ulliomodel}
\end{figure}

Recently, the positron fraction first reported by PAMELA \cite{Adriani2009_positron} and then confirmed by Fermi-LAT \cite{Ackermann2012} show an unexpected excess above 10~GeV over the prediction of propagation models, triggering many theoretical interpretations including a contribution from DM (see, e.g. \cite{Bergstrom2008, Hooper2009a, Hooper2009b, Hooper2009c, Okada2009, Yin2009}). The antiproton flux is also an interesting means for indirect DM detection as they are inevitably produced whenever it is kinematically possible and the final states of DM annihilation contain quarks or gauge bosons. One example is shown in figure \ref{fig:Ulliomodel}, which shows that the antiproton flux at energies of tens of GeV resulting from annihilation of high-mass neutralinos could be more than an order of magnitude above the flux of secondary antiprotons and thus could be observed. The PAMELA measurements of antiproton flux presented in this thesis, which extends to an energy of about 180~GeV provides an important test for these models. 

\subsection{Matter-antimatter asymmetry}

The matter-antimatter asymmetry in the Universe is one of the most important and puzzling questions indicated from cosmic rays observation. So far, no sizable amounts of antimatter has been observed. The asymmetry is also inferred from the baryon-to-photon ratio in the cosmic microwave background radiation, i.e. $ \eta_{B} \approx 10^{-9}$~\cite{Bergstrom2004}. While fundamental theories for elementary particles predict the same laws for matter and antimatter, CP-violation and baryon number non-conservation have been proposed to explain the depletion of antimatter \cite{Sakharov1967}. However, particle experiments do not support large levels of violations \cite{Nakamura2010}. Domains of antimatter in our Universe are suggested in \cite{Fargion2003, Bambi2007, Dolgov2009}. 

The detection of antinuclei with charge $|Z|>=2$ would constitute a smoking gun if they can be found in future experiments since the secondary production of antinuclei is negligible due to the extremely small production probability in the ISM through spallation (e.g. 10$^{-13}$ for $^{3}\overline{\text{He}}$). Inferred from cosmic ray matter composition, cosmic ray antihelium are the most possible detectable antinuclei compared to other species. So far, no $\overline{\text{He}}$ has been detected and only an upper limit on $\overline{\text{He}}$/He has been reported by experiments. Cosmic ray antiprotons and positrons, which are measured with much higher statistics than antihelium nuclei, could also provide signals on primordial antimatter sources.  However, studies could become complicated since contributions from such as non-standard astrophysical sources and dark matter may also give an excess on the cosmic ray antiproton spectrum or positron spectrum. 

\vspace{10mm}
In summary, this chapter reviews the fundamental issues of cosmic rays. A lot of questions about their nature, origin and propagation are still unanswered. Accurate measurements of individual cosmic ray elements are necessary to study the cosmic ray acceleration and propagation phenomena. Additionally, dark matter and matter-antimatter asymmetry can be inferred from cosmic antimatter measurements. As outlined in this chapter, the study of cosmic ray propagation plays a key role in understanding the processes occurring in our Galaxy. The propagation processes briefly described here will be elaborated in chapter \ref{chapt:propagation}.


%% file: Phd-Ch-Propagation.tex
\chapter{Cosmic ray propagation}
\label{chapt:propagation}

This chapter discusses general questions related to the propagation of cosmic rays with energies up to $10^{15}$~eV. It starts with an overview of the basics of cosmic ray propagation, such as energy losses/gains and nuclear interactions of cosmic rays with the interstellar medium. In section \ref{sec:prop_equation} the transport equation is constructed in the form of the continuity equation and Fick's law. All relevant parameters used to characterize the cosmic ray propagation processes are described in section \ref{sec:paradescription}. Section \ref{sec:prop_approach} discusses different approaches to solve the transport equation. The final section summarizes the current status of studies on cosmic ray propagation.

\section{Basics of cosmic ray propagation}
\label{sec:prop_basics}

As mentioned in chapter \ref{chapt:cosmic_rays}, due to insufficient information on the properties of the ISM and on the structure of the Galactic magnetic field, the specific mechanisms of cosmic ray propagation are not known yet. All our knowledge is developed and constructed semi-empirically based on cosmic ray observational results. So far it is generally supposed that the diffusion process with possible reacceleration and convection can be crucial during cosmic ray propagation.

\subsection{Diffusion}
From cosmic ray observations, especially secondary-to-primary ratios (e.g., B/C, $\bar{\text{p}}$/p) and the unstable-to-stable isotope ratios of secondary nuclei (e.g., $^{10}$Be/$^{9}$Be, $^{26}$Al/$^{27}$Al), the mean amount of matter (referred to as grammage $X$) traversed by cosmic rays and their escape time $\tau_{esc}$ from our Galaxy can be established. The grammage X is found to be about 5~g/cm$^{2}$ and the escape time $\tau_{esc}$ is estimated to be tens of million years. This suggests that cosmic rays travel in a confinement volume with average gas density $\rho_{gas}$ about 0.3~protons/cm$^{3}$, deduced from the relation $X =\int \upsilon \rho_{gas} \tau_{esc}$ where $\upsilon$ is the particle velocity. Since the Galactic plane has an average gas density about 1~proton/cm$^{3}$, it appears that cosmic rays must spend most of their time in low-density regions of the ISM, which could be either a hot coronal phase of the ISM and/or refer to a Galactic halo surrounding the disk with low gas density. Radio measurements support the halo hypothesis since significant amounts of synchrotron radiation are detected far away from the galactic disk and may be emitted from cosmic ray electrons (see \cite{Berezinskii1990} and references therein). 

A process which can confine cosmic rays inside the Galaxy and can send them back to the disk from the halo is needed. A natural hypothesis is that cosmic rays scatter on turbulence in the magnetic field. Cosmic rays form a plasma of ionized particles. On the microscopic level, the cosmic ray particles interact with the magnetohydrodynamic (MHD) waves arising in magnetized plasmas. If the interaction is resonant, particles are scattered by the waves leading to diffusion. The diffusion process can explain not only the cosmic rays' long travel time but also their highly isotropic distribution in the Galaxy. If there was no scattering, due to the particular location of the Solar system there should be more cosmic rays from sources towards the Galactic center and we would expect a strong anisotropy towards this direction. Such an anisotropy is not seen for cosmic rays with energies less than $10^{15}$~eV which apparently is destroyed by multiple scattering of the cosmic rays on their path from the sources to us. 

Locally, cosmic ray diffusion occurs along the magnetic field lines and thus can be quite anisotropic. However, on scales larger than 100~pc,  the wandering of cosmic rays on irregular magnetic field can make the diffusion isotropic and randomize the trajectories of particles. 

\subsection{Energy losses and gamma ray production} \label{sec:Eloss}
During propagation, cosmic ray nuclei and electrons interact with other constituents of the Galaxy and continuously lose energy. Meanwhile, electromagnetic radiation is produced in some interactions. By observing the emissions from radio to gamma ray frequencies, it is possible to probe cosmic ray propagation.

For relativistic cosmic ray electrons, considering the content and structure of the Galaxy, the following interactions may occur:
\begin{itemize}
\item ionization of neutral interstellar matter.
\item Coulomb scattering of individual plasma electrons in the fully ionized plasma.
\item Bremsstrahlung in the neutral and ionized medium. An electron is deflected in the electrostatic potential of an atom, ion or molecule, losing energy by emitting a $\gamma$-ray photon.
\item Inverse Compton scattering of the interstellar radiation field. In this interaction the target photon is scattered to higher frequencies by receiving part of the kinetic energy transferred from the relativistic cosmic ray electron.
\item Synchrotron radiation in magnetic fields. Electromagnetic radiation is emitted when a charged relativistic particle travels in a magnetic field that is uniform on scales much larger than the gyroradius of the particle.
\end{itemize}

For cosmic ray nucleons, the cross sections for electromagnetic interactions of cosmic ray nuclei are much smaller than those of electrons. Therefore all the electromagnetic processes responsible for electron energy losses can be neglected. Interactions of cosmic ray nuclei with cosmic photons only becomes important at energies $>10^{17}$~eV and will not be considered here. The remaining contributions are interactions with ISM:
\begin{itemize}
\item Ionization of atoms and molecules in the ISM.
\item Coulomb interactions with the ionized plasma.
\end{itemize}

\begin{figure}[tb!h]
\begin{center}$
\begin{array}{cc}
\includegraphics[width=6cm]{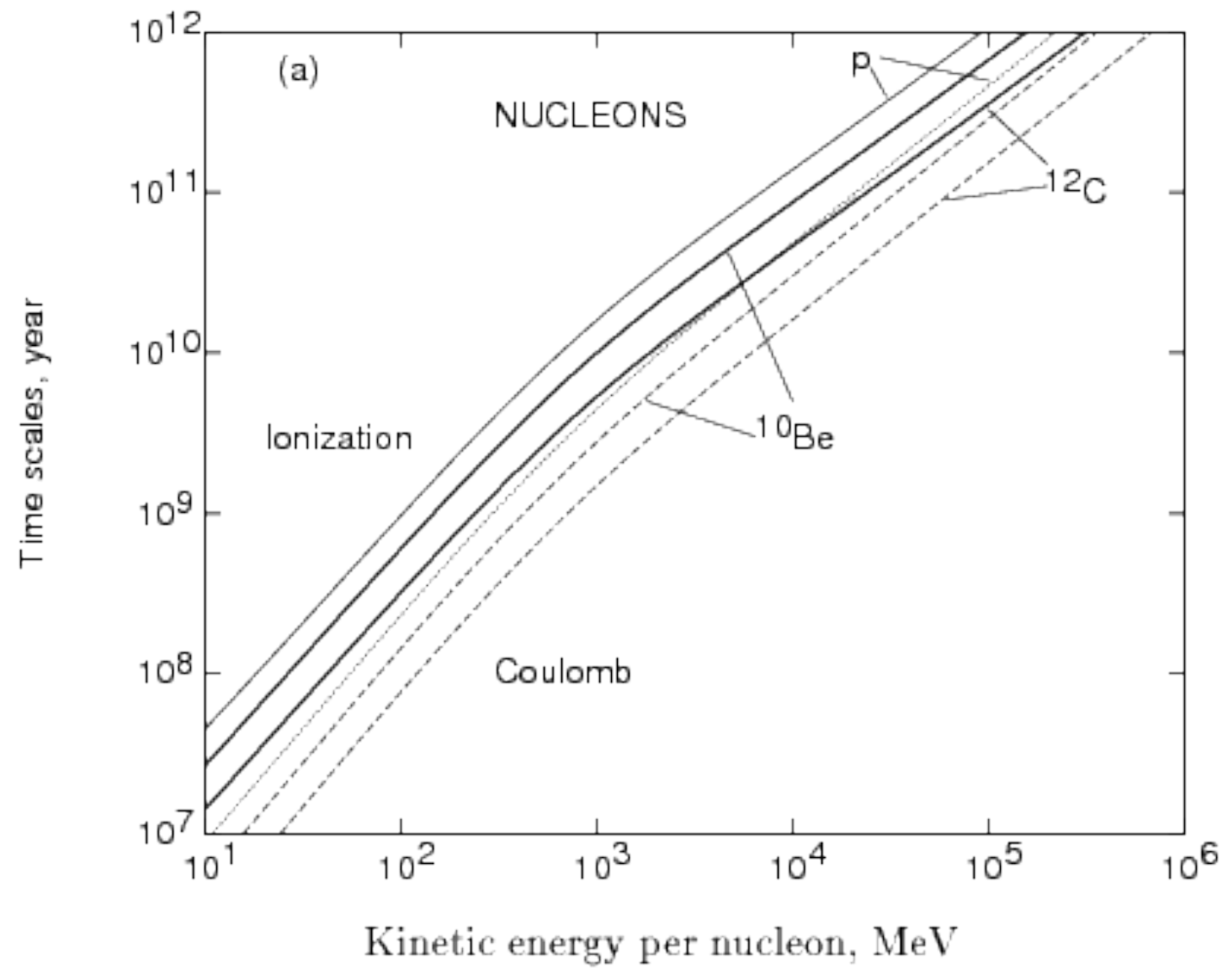}&
\includegraphics[width=6cm]{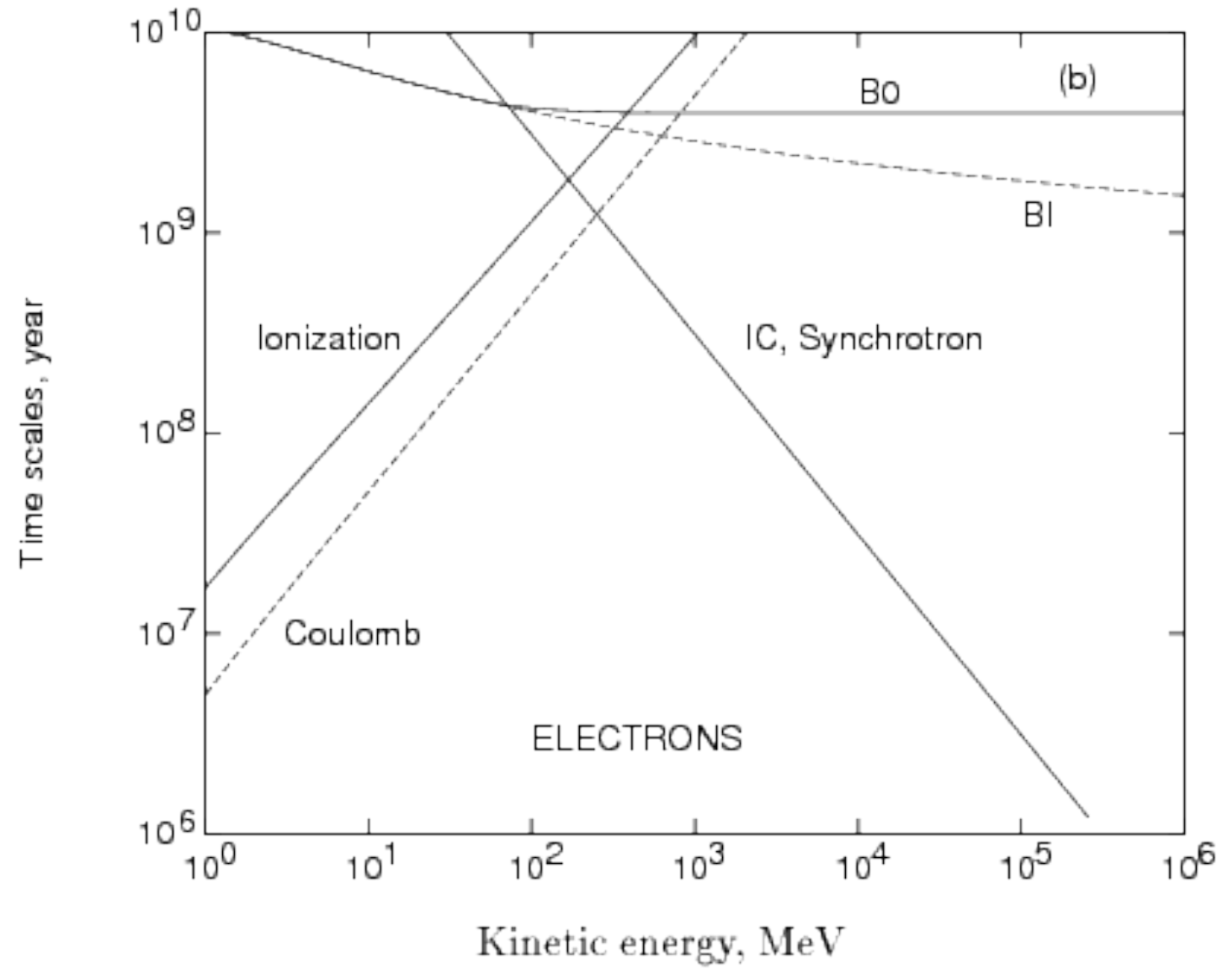}
\end{array}$
\end{center}
\caption[Energy-loss timescales of nucleons and electron in ISM] {\footnotesize Energy-loss timescales of nucleons (left) and electrons (right) in neutral and ionized hydrogen. In the left figure, solid lines show ionization losses and dashed lines show Coulomb losses. In the right figure, B0 (BI) means the Bremsstrahlung losses in the neutral gas (ionized gas). The curves are calculated based on the assumption of equal neutral and ionized gas number densities (0.01 cm$^{-3}$), and equal energy densities of photons and magnetic field (1 eV cm$^{-3}$). Taken from \cite{Strong1998}.}
\label{fig:elosstimescale}
\end{figure}

When these interactions occur, the original cosmic ray energy spectrum and propagation processes are affected. The contributions of different interactions to the energy loss are energy dependent. The energy loss timescales, which are associated to the inverse of energy loss rates, are shown in figure  \ref{fig:elosstimescale}. For nucleons and low energy electrons, the most important processes responsible for the energy loss are Coulomb scattering and ionization. But for electrons with energies higher than around 1~GeV, synchrotron losses become dominant. A complete summary of energy losses can be found in \cite{Strong1998}. 

\subsection{Nuclear interactions} \label{sec:Nint}

When cosmic rays traverse the ISM, they may interact with an interstellar hydrogen or helium nucleus and initiate nuclear reactions. For a specific type of cosmic ray nucleus, several kinds of nuclear interactions can be discussed: 

\begin{itemize}
\item Inelastic scattering of cosmic ray nuclei with ISM atoms and molecules which results in destruction of the given species governed by the total reaction cross section of that species.
\item Spallation determined by the formation rate from each parent element.
\item Radioactive decay of unstable cosmic ray nucleons.
\item Radioactive spallation generated via the decay of parent unstable isotopes. 
\end{itemize}
 

\section{The transport equation} \label{sec:prop_equation}

Taking into account the decisive role played by diffusion as well as other possible interactions, the cosmic ray transport equation can be built by incorporating the continuity equation with Fick's law \cite{Fick1855}. 

The fundamental continuity equation can be written as:
\begin{equation}
\frac{\partial N}{\partial t} = - \nabla \cdot \vec{J}+q,
\end{equation}
where $N$ is the number density, $\vec{J}$ is its current generated due to a spatial gradient in the density $N$ and $q$ is the source term. Assuming that the diffusing particles obey Fick's law, $\vec{J}=-\hat{D} \nabla N$, where $\hat{D}$ is the diffusion tensor, the continuity equation leads to the diffusion approximation:
\begin{equation}
\frac{\partial N}{\partial t} = \nabla \cdot \left( \hat{D}\nabla N \right) +q.
\end{equation}


A rather general equation for cosmic ray species $i$ can be constructed by taking into account all the relevant processes in addition to diffusion:
\begin{itemize}
\item Continuous energy losses
\begin{equation}
\frac{\partial N_{i}}{\partial t} = - \frac{\partial }{\partial E} \left( b_{i} N_{i}\right),
\end{equation}
where $b_{i}=\mathrm{d}E/\mathrm{d}t$ is the first order of energy loss;

\item Nuclear destruction
\begin{equation}
\frac{\partial N_{i}}{\partial t} = - nv\sigma_{i}N_{i},
\end{equation}
where $n$ is the density of the interstellar gas, $v$ is the particle velocity and $\sigma_{i}$ is the inelastic scattering cross section of a nucleus of type $i$ with nuclei of the interstellar gas;

\item Spallation from heavier nuclei
\begin{equation}
\frac{\partial N_{i}}{\partial t}=\sum_{m_{j}>m_{i}} nv\sigma_{ij} N_{j},
\end{equation}
where $\delta_{ij}$ is the production cross section of nuclei  of type $i$ from heavier nuclei of type $j$;

\item Radioactive decay
\begin{equation}
\frac{\partial N_{i}}{\partial t}=-\frac{1}{\tau_{i}} N_{i}, 
\end{equation}
where $\tau_{i}$ is the lifetime of a nucleus of type $i$; 

\item Radioactive spallation
\begin{equation}
\frac{\partial N_{i}}{\partial t}= \sum_{m_{j}>m_{i}} \frac{1}{\tau_{ij}}N_{j},
\end{equation}
where $\tau_{ij}$ is the lifetime of a nucleus of type j decaying radioactively to a nucleus of type i;
\end{itemize}

Adding up all these terms, the diffusion equation can be written as:
\begin{equation} \label{eq:simple_PropagateEquation}
\begin{split}
\frac{\partial N_{i}}{\partial t} & = q_{i} \\
&+ \sum_{m_{j}>m_{i}} \left( nv\sigma_{ij} + \frac{1}{\tau_{ij}}\right)N_{j} + \nabla \cdot \left( \hat{D} \nabla N_{i} \right) - \left(nv\sigma_{i}+\frac{1}{\tau_{i}}\right)N_{i}  - \frac{\partial }{\partial E} \left( b_{i} N_{i}\right).
\end{split}
\end{equation}


\subsection{Transport of cosmic rays by convection}
While diffusion is necessary to explain the high degree of isotropy and confinement in the Galaxy, other processes may also be of importance. Particularly, it is very likely that in our Galaxy there is large-scale motion of the interstellar gas with a ``frozen'' magnetic field, caused by the stellar activity and the energetic phenomena associated with the late stage of stellar evolution. This is referred to as \textit{convection} or \textit{galactic wind}. Cosmic rays are carried by the wind ``as a whole'' with some velocity $\vec{V}_{c}$  outwards from the Galactic plane. Galactic winds are found in many galaxies {\cite{Breitschwerdt2000}}. It is natural to propose that supernovae also power a similar wind in the Milky Way \cite{Jokipii1976, Owens1977, Jones1978, Jones1979, Bloemen1993}. 

Observational support comes from Galactic diffuse soft X-ray emission measured by ROSAT, which can be interpreted by assuming the presence of a strong galactic wind in our Galaxy \cite{Everett2008}. Convection adds a term  -$\left( \nabla \cdot  \vec{V}_{c} \right)N_{i}$ to the diffusion equation (equation \ref{eq:simple_PropagateEquation}) and causes adiabatic energy losses,  of the form $\frac{\partial}{\partial E} \left(\frac{\nabla \cdot \vec{V}_{c}}{3} \frac{p^{2}}{E} N_{i} \right)$.

\subsection{Transport of cosmic rays by reacceleration}
Though cosmic rays lose energy during propagation in the Galactic environment, they may also gain energy via stochastic acceleration. Above a few GeV/n, the fraction of secondary nuclei decrease as energy increases, which indicates that higher energy cosmic rays traverse less amount of matter (i.e. spend shorter time) in the Galaxy than lower energy ones. If acceleration only occurred together with fragmentation, the expected time spent to accelerate cosmic rays to higher energies would be longer. Hence, cosmic rays are mainly accelerated before their propagation, as discussed in chapter \ref{chapt:cosmic_rays} where diffusive shock acceleration at the shock-wave fronts in SNRs are proposed as the main mechanism of cosmic ray acceleration in the Galaxy. However, this does not exclude the possibility that cosmic rays might experience some additional acceleration after being injected from sources. Due to relativistic cosmic rays scattering on magnetic turbulence in the interstellar hydrodynamical plasma, some weak stochastic acceleration is almost unavoidable. This can be called reacceleration. 

Reacceleration may be significant at low energies to explain the peaks of B/C ratio around 1~GeV/n as shown in figure \ref{fig:BCRatio}, but should only slightly distort the ratios above few GeV. Reacceleration leads to a second order energy gain, which adds a term $\frac{\partial}{\partial E} \beta^{2} D_{pp} \frac{\partial N_{i}}{\partial E}$ to the transport equation, where $D_{pp}$ is the diffusion coefficient in momentum space.

\subsection{A full transport equation}
\begin{figure}[tb!h]
\begin{center}
\includegraphics[width=0.85\textwidth]{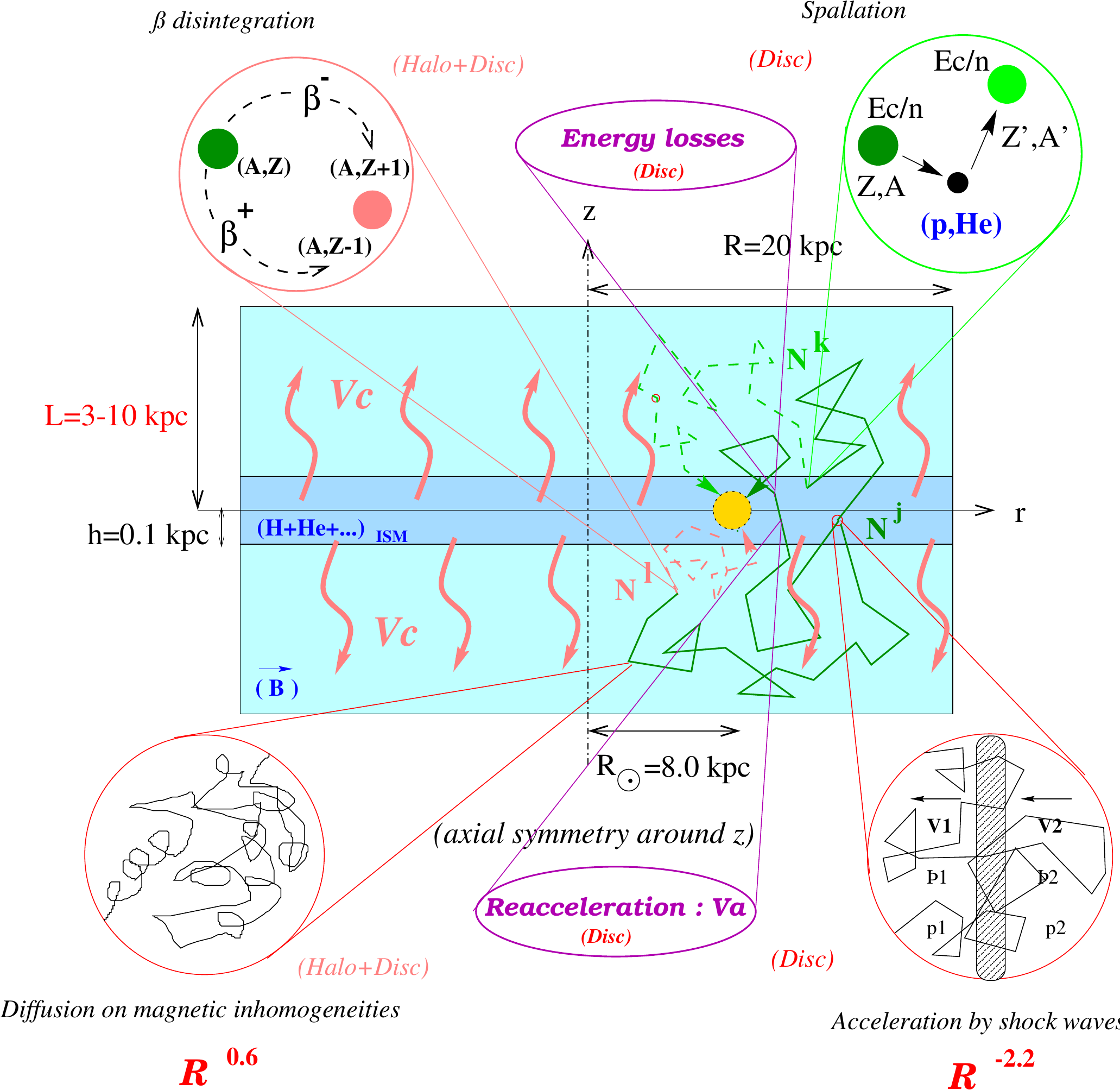}
\end{center}
\caption[Schematic view of cosmic ray propagation]{\footnotesize Schematic view of the propagation of cosmic rays in our Galaxy. After accelerated by SNR shock waves, cosmic rays suffer a combination of propagation processes in our Galaxy, including diffusion, reacceleration, convection, nuclear interaction, radioactive decay and energy losses. Taken from \cite{Maurin2002}.}
\label{fig:propagation}
\end{figure}

A schematic view of cosmic ray transport including all the most important propagation steps is illustrated in figure \ref{fig:propagation}. By adding the convection and reacceleration terms, the full transport equation can be written as:

\begin{equation} \label{eq:prop_full_equation}\begin{split}
\frac{\partial N_{i}}{\partial t} &= q_{i} + \sum_{m_{j}>m_{i}} \left( nv\sigma_{ij} + \frac{1}{\tau_{ij}}\right)N_{j} + \nabla \cdot \left[ \hat{D} \nabla N_{i} - \vec{V}_{c} N_{i}\right] \\
& - \left(nv\sigma_{i}+\frac{1}{\tau_{i}}\right)N_{i}+ \frac{\partial}{\partial E} \beta^{2} D_{pp} \frac{\partial N_{i}}{\partial E}- \frac{\partial }{\partial E} \left( b_{i} N_{i}-\frac{\nabla \cdot \vec{V}_{c}}{3}\frac{p^{2}}{E} N_{i}\right).
\end{split}\end{equation} 

\section{Parameter description} \label{sec:paradescription}

In the framework of the diffusion equation, it is possible to use observational data to study the properties of cosmic ray composition, abundances, anisotropy and to determine the composition at the sources. By combining numerous experimental facts within certain models, the propagation parameters which will be further described in this section can be investigated.  This allow us to better understand the related transport processes.

\subsubsection{Spatial diffusion}

Diffusion is a result of cosmic ray particles interacting with MHD waves. While our knowledge on the structure of the Galactic magnetic field is limited, we generally assume that diffusion mainly takes place along the mean magnetic field direction. Charged cosmic rays scatter mainly on resonant magnetic field fluctuations and the diffusion coefficient is estimated to follow a rigidity power law according to the quasi-linear theory \cite{Ptuskin2006, Strong2007}. The diffusion coefficient is usually represented by $D_{xx}$ and assumed to have the form: 

\begin{equation} \label{eq:diff_coeffieient}
D_{xx}=D_{0}\beta \left( \frac{\rho}{\rho_{0}}\right)^{\delta},
\end{equation}
where $D_{0}$ is the normalization at reference rigidity $\rho_{0}$, linked to the fluctuation level of the hydromagnetic turbulence; the factor $\beta=\upsilon/c$ is the particle velocity and $\delta$ the spectral index of diffusion coefficient related to the spectral index of turbulence spectrum. The rigidity $\rho$ is usually used as the kinematic variable instead of momentum $p$. The free parameters concerning diffusion are $D_{0}$ and $\delta$. 


\subsubsection{Reacceleration}

The energy gain through reacceleration is a result of diffusion in momentum space. The associated diffusion coefficient in momentum space $D_{pp}$ is taken from the model of minimal reacceleration by interstellar turbulence and is correlated to the velocity of disturbances in the hydrodynamical plasma, called the \textit{Alfv\'{e}n velocity}.  $D_{pp}$ is related to the spatial diffusion coefficient $D_{xx}$ with the expression \cite{Seo1994}:
\begin{equation}
D_{pp}=\frac{4v_{A}^{2}p^{2}}{3\delta\left(4-\delta^{2}\right)\left(4-\delta\right)D_{xx}},
\end{equation}
where $v_{A}$ is the Alfv\'en velocity - the main free parameter related to reacceleration. 

\subsubsection{Convection}

Considering the convection mechanism, through which cosmic rays can be transported in bulk away from the Galactic plane, the convection velocity $V_{c}(z)$ is the quantity used to describe the convective wind. It is usually assumed that the velocity varies linearly with the distance from the Galactic plane $z = 0$ as 
\begin{equation}
V_{c}\left(z\right)=V\left(0\right)+\frac{\mathrm{d}V}{\mathrm{d}z}z,
\end{equation}
in which $V(0)$ is usually taken to be zero for a simplicity and $\mathrm{d}V/\mathrm{d}z$ is the main free parameter. Besides, some studies \cite{Webber1992, Maurin2002} assume a constant velocity in order to make the transport equation analytically solvable. Nevertheless, detailed information on the convection velocity is still unknown. 

\subsubsection{Source term}

As well as transport processes, the source term is indispensable in order to describe cosmic ray data. In section \ref{sec:CRintroduction} it was stated that SNRs are believed to be the main sources of primary nuclei. For a cosmic ray species the injected density is assumed to be a power law in momentum $p$ (or rigidity $\rho$) as expected from diffusive shock acceleration theory:
\begin{equation}
q_{i}\left( p \right) \propto  p^{-\nu} \propto \rho^{-\nu}.
\end{equation}

The general form of the average source term depends not only on the point-source injection spectrum but also on the spatial distribution of sources $f (R,z)$ in the Galaxy:
\begin{equation} \label{eq:source}
q_{i}\left( p, \vec{r} \right) = N_{ i} f (R,z) \rho^{-\nu}, 
\end{equation}
where $N_{i}$ is the normalisation abundance for the cosmic ray species $i$. The free parameters related to the source terms are the normalisation abundance $N_{i}$ and 
the injection index $\nu$.

The SNR distribution in the Galaxy is very poorly determined by radio surveys due to the small sample available and selection effects \cite{Case1996, Case1998}. Pulsars could be a useful tracer of the SNR distribution since they are born in the core collapse of supernovae. Large samples of pulsars can be obtained, but could be biased by distance and interstellar dispersion uncertainties \cite{Lorimer2004}. Moreover, the distributions of SNRs and pulsars as a function of galactocentric radius are both steeper than the distribution of cosmic ray sources chosen to reproduce the EGRET $\gamma$-ray data \cite{Strong1998}. An enhancement of molecular gas in the outer Galaxy was proposed in \cite{Strong2004} to moderate this $\gamma$-ray gradient problem but is disfavored by Fermi-LAT data \cite{Abdo2010}.

\begin{figure}[tb!h]
\begin{center}
\includegraphics[width=0.85\textwidth]{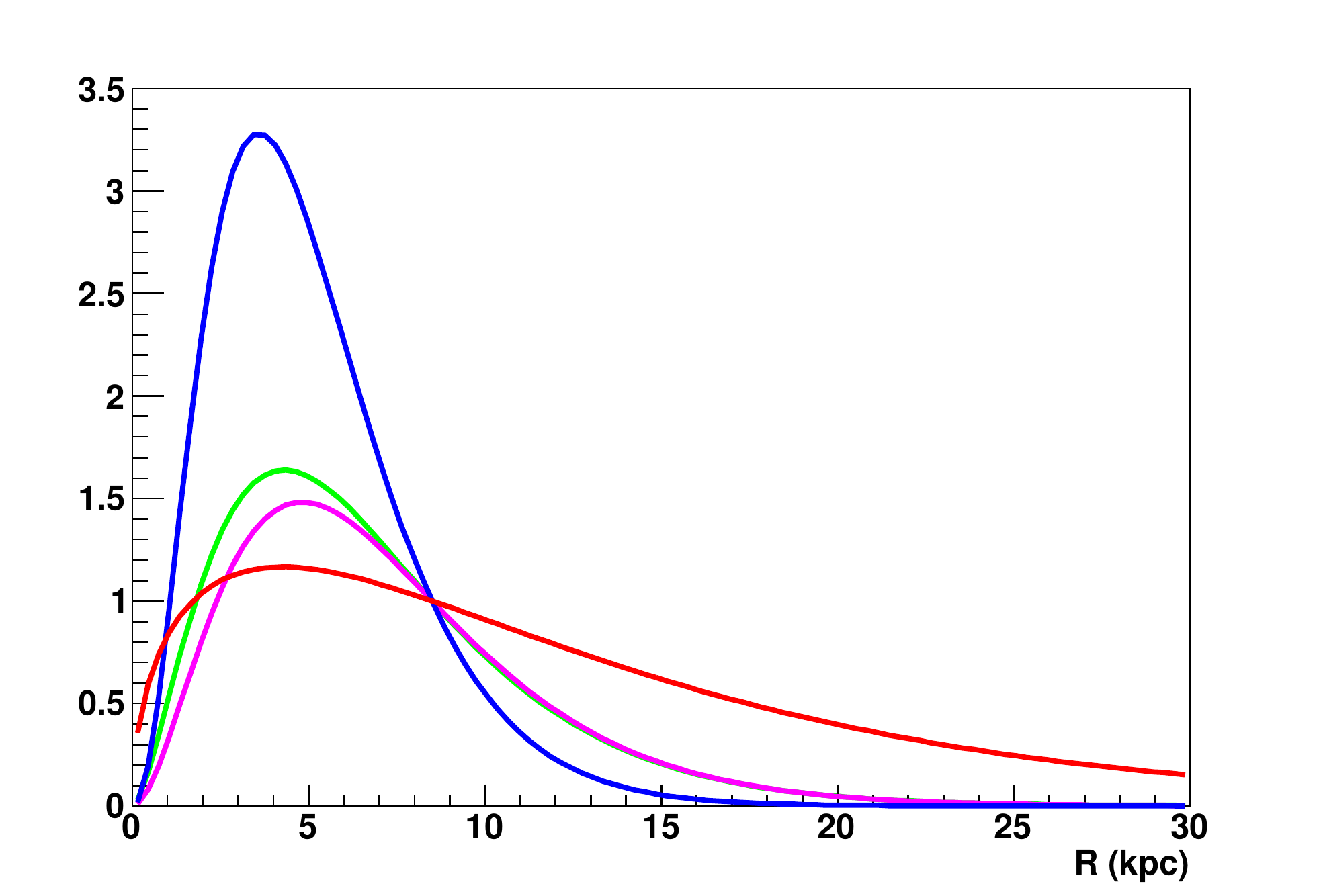}
\end{center}
\caption[Cosmic ray source density as function of galactocentric radius $R$.] {\footnotesize Cosmic ray source density as function of galactocentric radius $R$, normalized at the position of the Sun with $R = 8.5$~kpc. The references of these curves are \cite{Case1996} in green, \cite{Case1998} in purple, \cite{Lorimer2004} in blue and \cite{Strong1996} in red.}
\label{fig:sourcedist}
\end{figure}

As suggested in \cite{Stecker1977}, the radial dependence of the SNR distribution can have the form:
\begin{equation}
f(R)=\left(\frac{R}{R_{\odot}}\right) ^{\alpha} \exp{\left(-\beta \frac{R-R_{\odot}} {R_{\odot}}\right)},
\end{equation}
where $R_{\odot}$ is the radial distance of the Sun from the Galactic center. Different values of the parameters $\left(\alpha, \beta \right)$ have been adopted in the literature. For instance, (1.69, 3.33) was found in \cite{Case1996} and (2.00, 3.53) in \cite{Case1998} to model the SNR distribution. The combination (2.35, 0.654) was obtained in \cite{Lorimer2004} to fit the pulsar distribution whereas (0.5, 1.0) was determined to reproduce observed EGRET $\gamma$-ray based gradient \cite{Strong1996}. These distributions are shown in figure \ref{fig:sourcedist}. More sets of favored values can be found in \cite{Delahaye2011} and references therein.


\section{Propagation models}\label{sec:prop_approach}

Various approaches are useful to solve the transport equation. The simplest approximation is the so called ``leaky-box'' model (LBM) which was employed in pioneering studies. In the leaky-box approximation, the Galaxy is described as a finite and homogeneous volume with uniform gas density. Each nucleus escapes from this volume with a probability $1/\tau_{esc}$. The diffusion term $\nabla \cdot \left( \hat{D} \nabla N_{i} \right) $ can be expressed as $-N_{i}/\tau_{esc}$. The LBM can be considered as a diffusion model in two limiting cases:
\begin{itemize}
\item cosmic rays diffuse rapidly in the Galaxy and reflect at the Galaxy halo boundary with little leakage from the system;
\item the Galaxy halo is much flatter than the radius of the Galaxy with thin source and gas disks \cite{Ginzbury1976}. 
\end{itemize}
The leaky-box model has been used successful to explain most observed cosmic ray fluxes of stable nuclei, however, it cannot deal with the complexities such as spatially dependent source distributions, etc. More complete treatment of all the relevant processes and more realistic description of the Galactic ingredients use other techniques to solve equation \ref{eq:prop_full_equation} explicitly, in which two main ones have been employed to date: analytical (or semi-analytical) models and purely numerical models. Several software packages have been developed for this purpose. For example, USINE is a commonly used code employing the analytical approach but no public version has been released yet \cite{Maurin2001, Maurin2002}, GALPROP \cite{GALPROPWebsite} is a publicly available numerical code used widely not only for cosmic ray nuclei, electrons but also for photons \cite{Strong1998, Moskalenko2002}. An overview of these models will be presented in this section.



\subsection{Description of the Galaxy}
Any solution of the propagation equation is based on some fundamental assumptions regarding the Galaxy, including its geometry, its matter content as well as the magnetic fields. Cosmic rays are thought to diffuse in some containment volume beyond which they freely stream out. The density outside the boundary drops to zero. Radio observations of galactic halos indicate that the shape of the confinement volume might radially follow the galactic disc, but with a greater thickness. Commonly the Galaxy halo is modeled with cylindrical symmetrically with radius $R=20$~kpc and half-height $z_{h}$ whose value is still unknown but is reasonably believed to be greater than a few kpc. The density of cosmic rays satisfies the boundary condition $N(r=R, z)=N(r, z=\pm z_{h})=0$. The Galactic disk is embedded in the halo with half-height $h \sim 100$~pc. 

\subsubsection{The gas density distribution}

As discussed in section \ref{sec:Eloss} and section \ref{sec:Nint}, cosmic rays may interact with interstellar gas causing secondary production of particles and energy losses. Therefore, the gas density is a basic ingredient to affect these processes. 

\begin{figure}[tb!h]
\begin{center}
\includegraphics[width=0.85\textwidth]{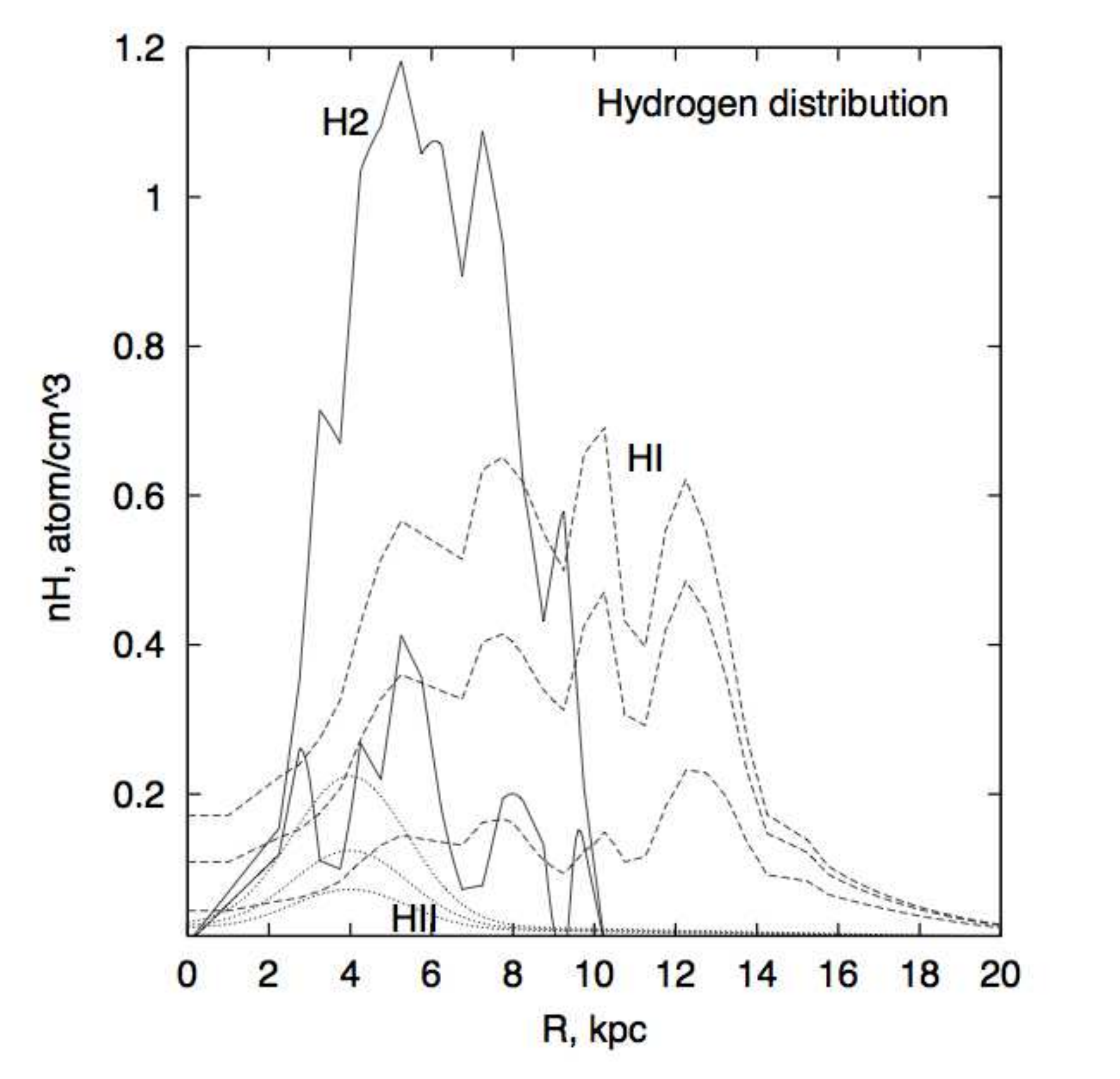}
\end{center}
\caption[The hydrogen number density distribution.] {\footnotesize The number density distribution for HI (dashed lines), HII (dotted lines) and H$_{2}$ (solid lines) in the Galaxy, taken from \cite{Moskalenko2002}. Curves for a specific gas type are arranged with decreasing density for $z=$0, 0.1 and 0.2~kpc ($n_{H_{2}}$ at $z=0.2$~kpc is too low to be shown in this figure).} 
\label{fig:gasdistribution}
\end{figure}

The ISM is a mixture of neutral atomic hydrogen HI, ionized hydrogen HII, molecular hydrogen H$_{2}$ and helium components. The densities of these components vary with the radial distance $r$. In some cases, for example in the analytical code USINE, it is taken as a simplified average gas density in the disk of $\sim 1$~proton/cm$^{3}$. In numerical code GALPROP, a more realistic gas distribution is used instead of a constant gas density. The H$_{2}$ density is calculated from the CO volume emissivity \cite{Bronfman1988} and the conversion factor from CO emissivity to $n_{\text{H}_{2}}$ is taken as $1.9\times 10^{20}$~molecules\,cm$^{-2}$\,(K\,km\,s$^{-1}$)$^{-1}$ \cite{Strong1996}. 
The HI distribution is taken from the model in \cite{Gordon1976} but renormalized to agree with the density distribution perpendicular to the Galactic plane \cite{Dickey1990} and \cite{Cox1986}. The HII distribution is taken from a cylindrically symmetric model \cite{Cordes1991}. The hydrogen number density distributions are plotted in figure \ref{fig:gasdistribution} for height $z=$0, 0.1 and 0.2~kpc.  The helium number density fraction in the gas is taken as 0.11 \cite{Moskalenko2002}.

\subsubsection{The interstellar radiation field and the Galactic magnetic field}

The interstellar radiation field (ISRF) and the magnetic field have a strong influence on electron energy losses, $\gamma$-ray production from inverse Compton scattering and synchrotron radiation. These aspects are taken into account in numerical codes but can only be treated by assuming mean values in analytical codes.

The Galactic interstellar radiation field (ISRF) results from emission from stars,  and the scattering, absorption, re-emission of absorbed star light by dust in the ISM. Therefore, the estimation of the ISRF distribution is difficult and relies on the luminosity distribution from the stellar populations of the Galaxy,  the dust distribution and the description of the absorption,  scattering of star light and re-radiation processes. A recent calculation of ISRF can be found in \cite{Moskalenko2006, Porter2006, Porter2008}. 

The fine structure of the Galactic magnetic field is far from being fully understood. An assumption regarding the magnetic field is only used to calculate the electron synchrotron losses. A spatially dependent model adjusted to match the 408~MHz synchrotron longitude and latitude distributions \cite{Strong2000}, is used in GALPROP.

\subsection{Analytical approach}

Based on all the assumptions made on the Galaxy geometry, since the disk is much thinner than the halo size, the disk is considered as infinitely thin for practical purpose in the analytical approach. Cosmic ray sources and their interactions with the ISM are confined to this thin disk, and reacceleration is also assumed to take place in the disk. As a consequence, a factor $2h\delta \left( z \right)$ is added to the terms relating to the source and fragmentation processes, as well as the terms relating to energy losses and gains. The diffusion which occurs throughout the disk and the halo is assumed to have the same strength but does not have any spatial dependence. 

Assuming steady-state, the transport equation can be rewritten as a Laplace equation in a cylindrical geometry, by replacing $\nabla \cdot \left[ \hat{D} \nabla N_{i} -\left( \nabla \cdot  \vec{V}_{c} \right)N_{i}\right]$  in equation \ref{eq:prop_full_equation} by
\begin{equation}
D\left[  \frac{\partial^{2}}{\partial z^{2}} + \frac{1}{r} \frac{\partial}{\partial r \left( r \frac{\partial}{\partial r}  \right)} \right] - V_{c} \frac{\partial}{\partial z}.
\end{equation}

The density can be obtained by solving the equation using a Bessel expansion method. One can expand all the quantities over the orthogonal set of Bessel functions $\left[J_{0}\left( \zeta_{k} \frac{r}{R}\right)\right]^{k=1,...,\infty}$ in which k is the order of the Bessel decomposition and $\zeta_{k}$  are the successive zeros of function $J_{0}$:
\begin{equation}
N\left( r,z \right) = \sum_{k=1}^{\infty} N_{k}(z) J_{0}\left(  \zeta_{k} \frac{r}{R}  \right), 
\end{equation}

\begin{equation}
q\left( r \right)= \sum_{k=1}^{\infty} \hat{q}_{k} J_{0}\left(  \zeta_{k} \frac{r}{R}  \right).
\end{equation}

The solution of the cosmic ray density $N\left( R,z \right)$ comprises contributions from the disk and from the halo. The contribution from the disk involves the primary sources and the spallation production concentrated in the disk 
as well as the energy losses and the diffusive reacceleration. 
The contribution from the halo involves the products from radioactive decay in the whole halo. For each contribution, the detailed expression of  the solution can be found in \cite{Maurin2001, Maurin2002}. 

The analytical approach has the advantage of showing a direct relationship between propagation parameters. Another benefit is that the computation is fast. However, the analytical solution is only based on some simplified assumptions and thus cannot extend to more complicated cases. For example, anisotropic diffusion proposed in some literatures \cite{Gebauer2009} is not able to obtain analytical solutions. It is a challenge to use analytical methods to treat electron energy losses and photon production since information on ISRF and magnetic fields are difficult to be imported in analytical codes.

\subsection{Numerical approach}

The numerical solution of the transport equation is based on a implicit scheme in the GALPROP code. Terms such as diffusion, reacceleration,  convection and energy loss in equation \ref{eq:prop_full_equation} can all be finite-differenced for each coordinate ($R$, $z$, $p$) or ($x$, $y$, $z$, $p$) in the form
\begin{equation}
\frac{\partial N_{i}}{\partial t}= \frac{N_{i}^{t+\Delta t}- N_{i}^{t}}{\Delta t}= \frac{ \alpha_{1} N_{i-1}^{t+\Delta t} -\alpha_{2} N_{i}^{t+\Delta t} +\alpha_{3} N_{i+1}^{t+\Delta t}  }{\Delta t} + q_{i},
\end{equation}
 where all terms are functions of ($R$, $z$, $p$) or ($x$, $y$, $z$, $p$). 
 
To ensure stability for large time step $\Delta t$, the Crank-Nicolson method \cite{Press1992} is used in GALPROP, which is second-order accurate in time since in this method all the terms are alternatively finite-differenced in the form
 \begin{equation}\begin{split}
\frac{\partial N_{i}}{\partial t} &= \frac{N_{i}^{t+\Delta t}- N_{i}^{t}}{\Delta t}\\
& = \frac{ \alpha_{1} N_{i-1}^{t+\Delta t} -\alpha_{2} N_{i}^{t+\Delta t} +\alpha_{3} N_{i+1}^{t+\Delta t}  }{2\Delta t} + \frac{ \alpha_{1} N_{i-1}^{t} -\alpha_{2} N_{i}^{t} +\alpha_{3} N_{i+1}^{t}  }{2\Delta t} +q_{i}.
\end{split}\end{equation} 
 The detailed derivation and expression of coefficients $\alpha_{1}$, $\alpha_{2}$ and $\alpha_{3}$ for each term can be found in \cite{Strong1998}. 

Compared to analytical programs, numerical approaches need a heavier computation effort. Numerical methods have been developed to deal with both two and three dimensional spatial models, allowing us to handle more complicated and more realistic models involving spatially varying quantities, for example, anisotropic diffusion coefficients and electron energy losses. The production of photons can be treated more completely in numerical codes while only the hadronic production of photons can be dealt with in analytical programs until recently. 

\section{Current status} \label{sec:CRstatus}

Investigations of cosmic ray propagation have been addressed using both analytical and numerical methods (e.g., most recently \cite{Putze2010, Maurin2010, Putze2011, Bernardo2010, Trotta2011}), applied to experimental data including stable and unstable nuclei, electrons and gamma rays. In a given propagation model, stable secondary-to-primary ratios can be used to determine the ratio of the halo size to the diffusion coefficient while the radioactive isotopes allow us to break the degeneracy between these two parameters. Moreover, the source spectrum can be accessed from the propagated fluxes of primary nuclei (mainly protons) and electrons.

\begin{figure}[tb!h]
\begin{center}
\includegraphics[width=0.85\textwidth]{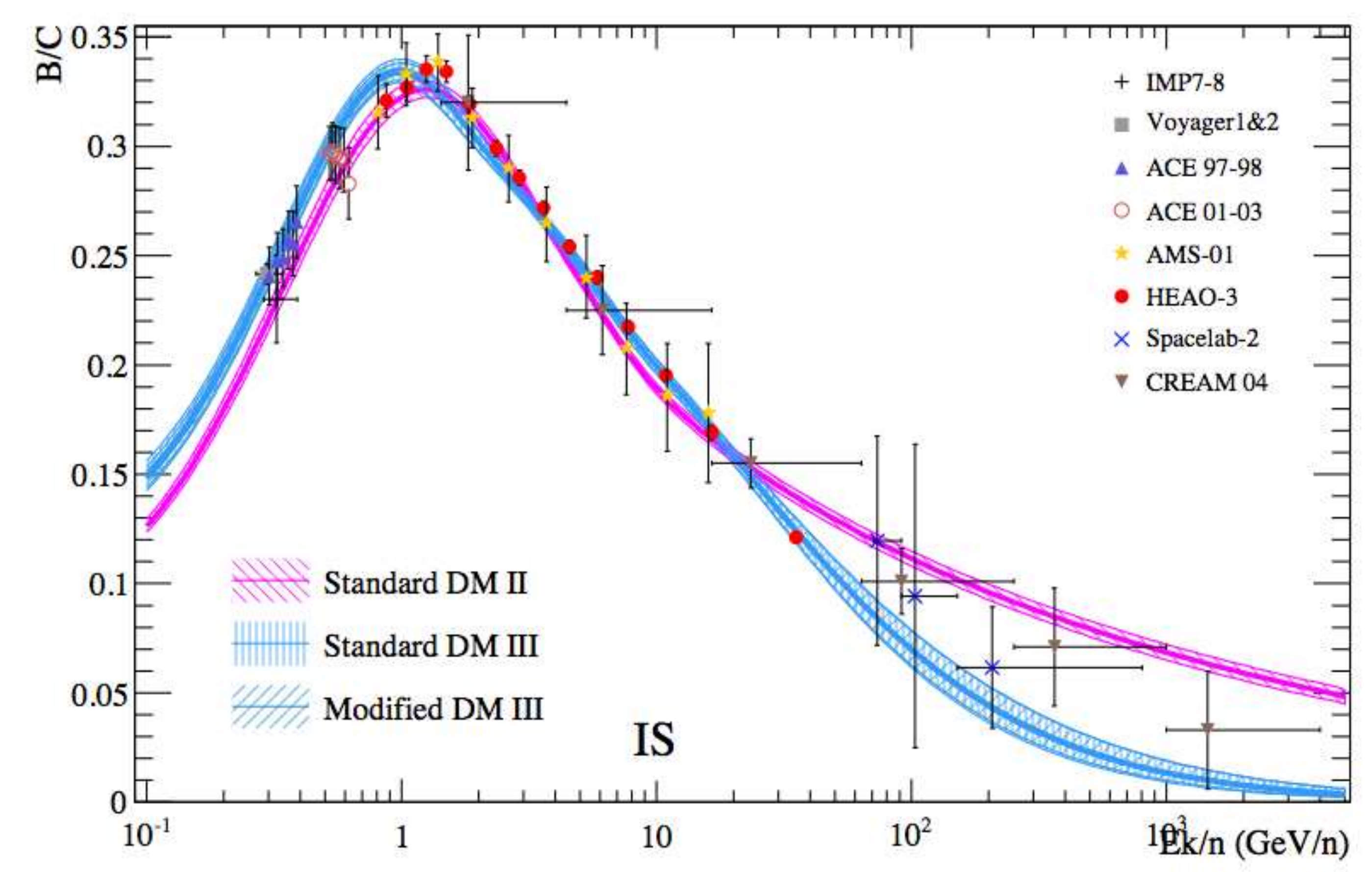}
\end{center}
\caption[BC compared with best-fit models by using USINE.] {\footnotesize B/C ratio compared with the ratio from best-fit DR (red) and DRC (blue) models. The shade areas are the 68\% confidence level. Taken from \cite{Putze2011}.} 
\label{fig:PutzeModels}
\end{figure}

\begin{figure}[tbhp]
\begin{center}$
\begin{array}{cc}
\includegraphics[width=6cm]{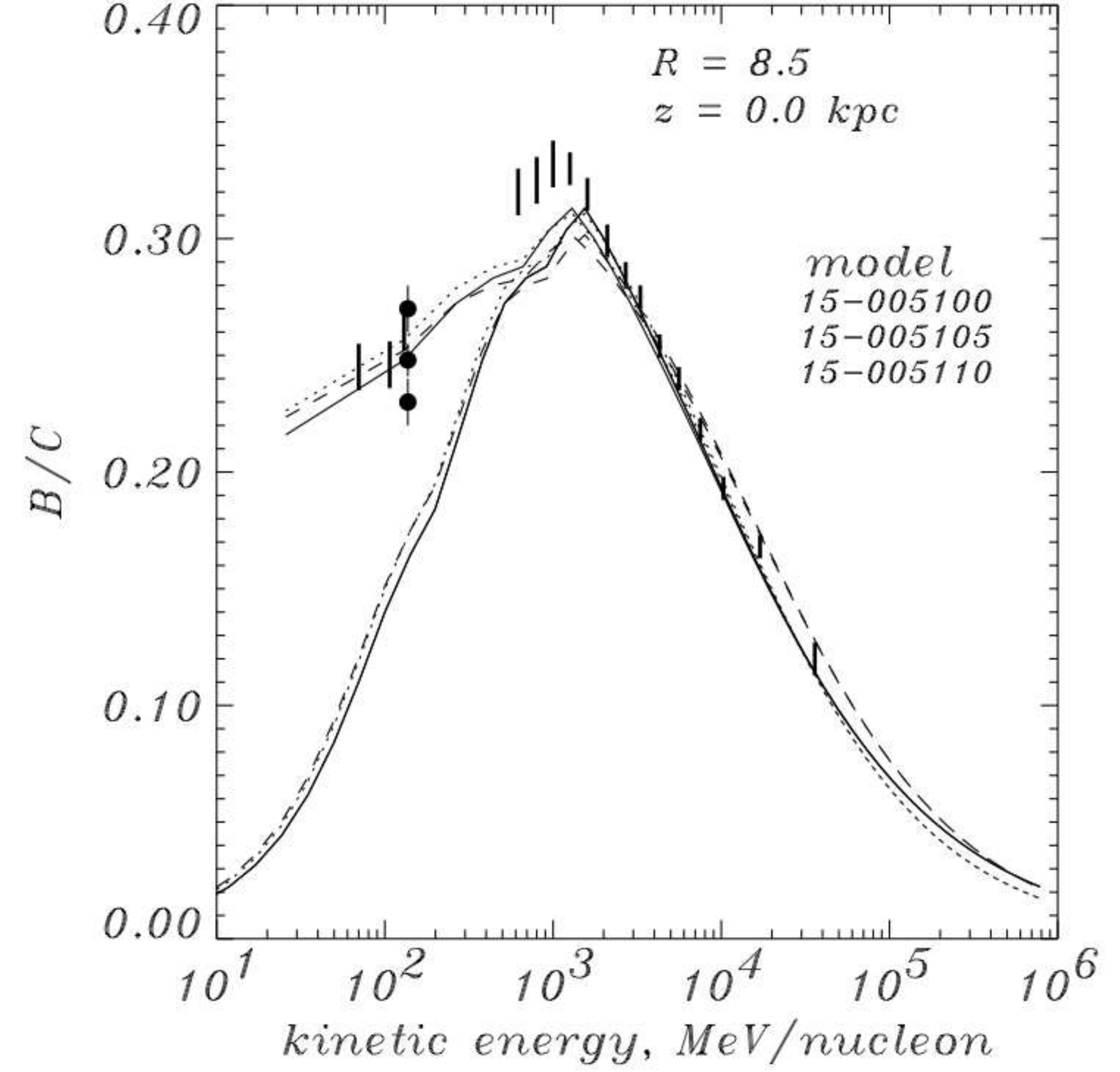}&
\includegraphics[width=6cm]{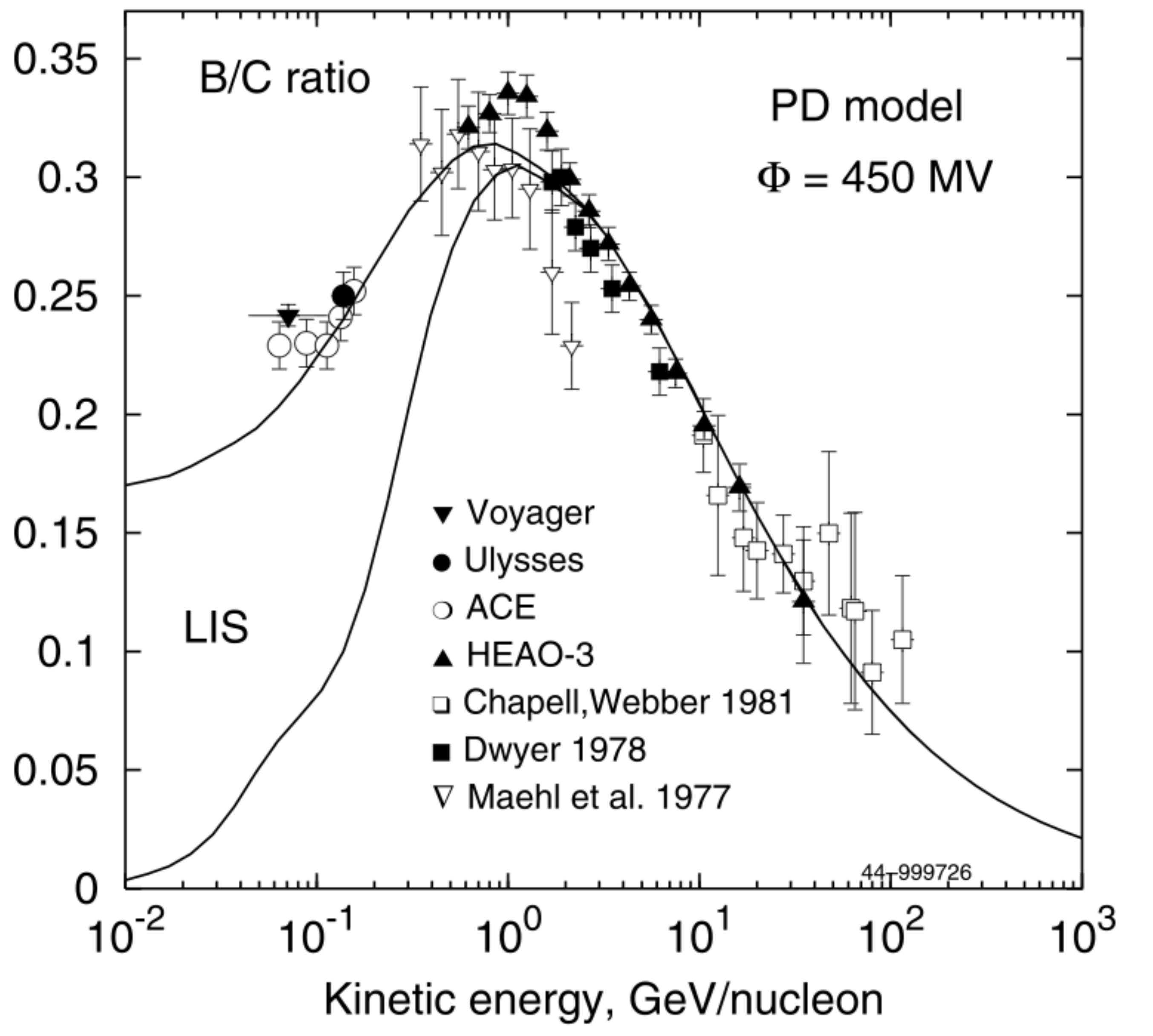}
\end{array}$
\end{center}
\caption[B/C ratio compared with models]{\footnotesize Left: B/C ratio for DC models with a break in the diffusion coefficient for convection velocity $\mathrm{d}V/\mathrm{d}z = 0$ (solid lines), 5 (dotted lines) and 10 (dashed lines)~km s$^{-1}$ kpc$^{-1}$ and $z_{h}$=5~kpc. From \cite{Strong1998}. Right:  B/C ratio for a PD model with both a break in the diffusion coefficient and an additional $\beta^{3}$ factor. Taken from \cite{Ptuskin2006}.}
\label{fig:BC_PD_DC}
\end{figure}

From the inference of cosmic ray isotropy and confinement, diffusion should inevitably be involved in the propagation processes. The existence of reacceleration and convection is not proved definitively. Therefore different models were studied in literature, such as plain diffusion (PD) models, diffusion reacceleration (DR) models, diffusion convection (DC) models and diffusion reacceleration convection (DRC) models. No specific model can be considered as the best one so far to explain all the observations as well as being physically reasonable. The most favored model claimed in \cite{Putze2010, Maurin2010, Putze2011} is the DRC model as shown in figure \ref{fig:PutzeModels}. It points to $\delta$ higher than 0.8 which is highly disfavored by the cosmic ray anisotropy problem, i.e. too large anisotropy is predicted compared to the observed one at highest energies $>10^{14}$~eV. DR models can explain quite well the sharp peaks observed in the secondary-to-primary ratios (e.g., B/C, [Sc+Ti+V]/Fe) at energies around 1~GeV/n but they cannot reproduce the proton and helium fluxes unless a break is introduced around 10~GeV in the injection spectra \cite{Moskalenko2002}. PD and DC models require a break in the rigidity dependency of the diffusion coefficient $D$, i.e. defining $D$ as $\beta D_{0}(\rho/\rho_{0})^{\delta_{1}}$ and as $\beta D_{0}(\rho / \rho_{0})^{\delta_{2}}$ below and above the reference rigidity $\rho_0$ respectively, or/and an additional factor $\beta^{\eta}$, as shown in figure \ref{fig:BC_PD_DC}. Both the break in the diffusion coefficient and the break in the injection spectra are arbitrary and not physically motivated. The factor $\beta^{\eta}$, that only has an effect on non-relativistic particles,  may be related to nonlinear MHD waves \cite{Ptuskin2006}.

Even in frameworks involving the same processes, the derived values of parameters vary in different studies. The published results do not always present consistent answers on the best-fit values of propagation parameters, especially the most relevant one, $\delta$. For example, for PD models, the typical diffusion coefficient slope $\delta$ (means $\delta_{2}$ if there is a break in $\delta$) obtained varies from 0.4 to 0.6 (e.g. \cite{Strong1998, Ptuskin2006,  Bernardo2010}. For DR models, the best-fit $\delta$ was determined to be about 0.3 in \cite{Strong1998, Moskalenko2002, Trotta2011} but  about 0.5 in \cite{Bernardo2010} and about 0.2 in \cite{Putze2010, Maurin2010}. The errors on $\delta$ are usually not larger than $\pm0.05$.

Systematic uncertainties on quantities such as gas density, cross sections and data bias,  are discussed in \cite{Maurin2010}. The gas density has a small impact on the diffusion slope but has very strong effects on other parameters like the normalization of the diffusion coefficient $D_{0}$, the Alfv\'en speed $v_{A}$ and the convection velocity $V_{c}$. Nuclear cross sections are crucially related to the destruction of primary cosmic rays and production of secondary cosmic rays. Their influences are model dependent, however. The typical variance in the best-fit values is a factor of 2 for $D_{0}$, $\sim$10\% for $\delta$, $\sim$50\% for $v_{A}$ and $\sim$5\% for $V_{c}$. Another important uncertainty arises from data bias since errors might be underestimated by a given experiment. The value of $\delta$ determined by using data sets from different experiments can vary by more than 0.3 \cite{Putze2010, Maurin2010}. Therefore, to improve the reliability of parameter determination, more realistic gas density distributions and cross sections should be employed. 
Furthermore, the systematic uncertainties could also be reduced by using spectra or spectrum ratios for all the necessary species provided by a single experiment to avoid data sets inconsistency arising from using data from different experiments. Therefore using PAMELA data exclusively to constrain propagation models is motived in my work and will be further discussed in chapter \ref{chapt:model_constraints}.

%% file: Phd-Ch-PAMELA_Experiment.tex
\chapter{The PAMELA experiment} \label{chapt:PAMELA}

The PAMELA experiment is a satellite-borne apparatus designed to identify and measure charged particles and especially antiparticles in the cosmic radiation {\cite{Picozza2007}}. PAMELA is installed inside a pressurized container attached to a Russian Resurs-DK1 Earth-observation satellite which was launched into Earth orbit by a Soyuz-U rocket on June 15th 2006 from the Baikonur cosmodrome in Kazakhstan. The container of PAMELA is connected to the satellite body with a mechanical arm which can move the container from the parked position with downward orientation in which it is kept during launch to the position with upward orientation kept during data acquisition mode. 

\begin{figure}[!htb]
\begin{center}
\includegraphics[width=0.8\textwidth]{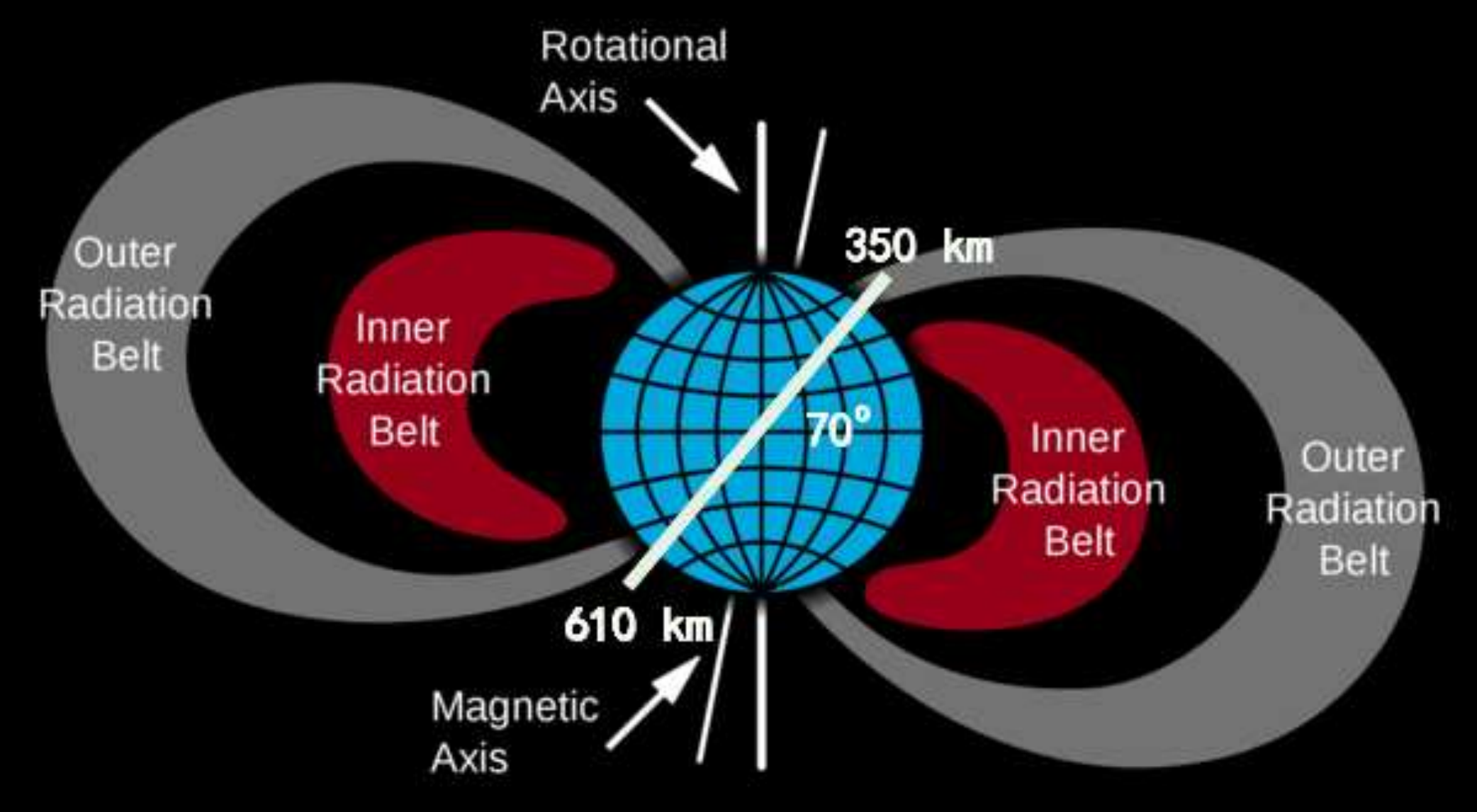}
\end{center}
\caption[Orbit of the Resurs-DK1 satellite]{{\footnotesize Orbit of the Resurs-DK1 satellite. The satellite is travelling around Earth along an elliptical orbit, at an altitude ranging between 350 km and 610 km with an inclination of $70^{\circ}$.}}
\label{fig:satelliteorbit}
\end{figure}

Until now the instrument has been traveling around Earth along an elliptical and semi-polar orbit for almost six years, with an altitude varying between 350 km and 600 km, at an inclination of 70 degrees. The trajectory thus goes through regions with varying geomagnetic cutoff, which effects the incident cosmic ray flux, and also passes the outer electron belt and the South Atlantic Anomaly~(SAA) (figure~\ref{fig:satelliteorbit}).

In this chapter, the scientific objectives of PAMELA will be presented in section \ref{sec:sci_obj} and each sub-detectors of the instrument will be described in section \ref{sec:sub_detectors}.

\section{Scientific objectives} \label{sec:sci_obj}

The design goal for PAMELA performance is to measure particle and nuclei fluxes over a wide energy range, and with unprecedented precision as a long exposure is achieved and no residual overburden of atmosphere needs to be compensated. Particularly, compared to previous experiments, PAMELA extends the energy range of antiprotons and positrons to both higher and lower energies. The statistics exceeds previous experiments by more than one order of magnitude after three years data taking.

Table \ref{tab:performance} shows the nominal design goals for PAMELA performance.

\begin{table}[h]
\scalebox{0.85}{
\centering
\begin{tabular}{c c c}
\hline
{\bf } & {\bf Energy range} & {\bf Statistics (3 years)} \\
\hline
{ Antiprotons } & { 80 MeV - 190 GeV} & { $10^{4}$} \\
{ Positrons } & { 50 MeV - 270 GeV} & { $10^{5}$} \\
{ Positrons+Electrons } & { Up to 2 TeV (from calorimeter only)} & { } \\
{  } & { } & { } \\
{ Electrons } & { 50 MeV - 400 GeV} & { $10^{6}$} \\
{ Protons } & { 80 MeV - 700 GeV} & { $10^{8}$} \\
{  } & { } & { } \\
{ Light nuclei (up to Z=6) } & { 100 MeV/n - 250 GeV/n He/Be/C } & { $10^{7/4/5}$} \\
{  } & { } & { } \\
{ Antinuclei search } & \multicolumn {2}{c} { Sensitivity of O ($10^{-7}$) for Antihelium/helium}  \\
\hline
\end{tabular}
}
\caption[PAMELA design goals]{PAMELA design goals.}
\label{tab:performance}
\end{table}

The PAMELA mission mainly focuses on the precise measurement of antiprotons and positrons. By studying the antimatter component of the cosmic radiation, the following themes will be addressed:

\begin{itemize}
\item To search for evidence of dark matter particle annihilations by precisely measuring the antiparticle (antiproton and positron) energy spectrum. 
\item To search for primordial antinuclei (e.g. antihelium) and to study low energy particles (e.g. trapped particles in the Earth's magnetic field and solar flare particles).
\item To test cosmic ray propagation models through precise measurements of the antiparticle energy spectrum and precision studies of light nuclei.
\end{itemize}

Besides, a reconstruction of the cosmic ray electron energy spectrum up to 2~TeV may give a hint for possible contribution from local sources.

\section{The detectors} \label{sec:sub_detectors}
\begin{figure}[tb!h]
\begin{center}
\includegraphics[width=0.9\textwidth]{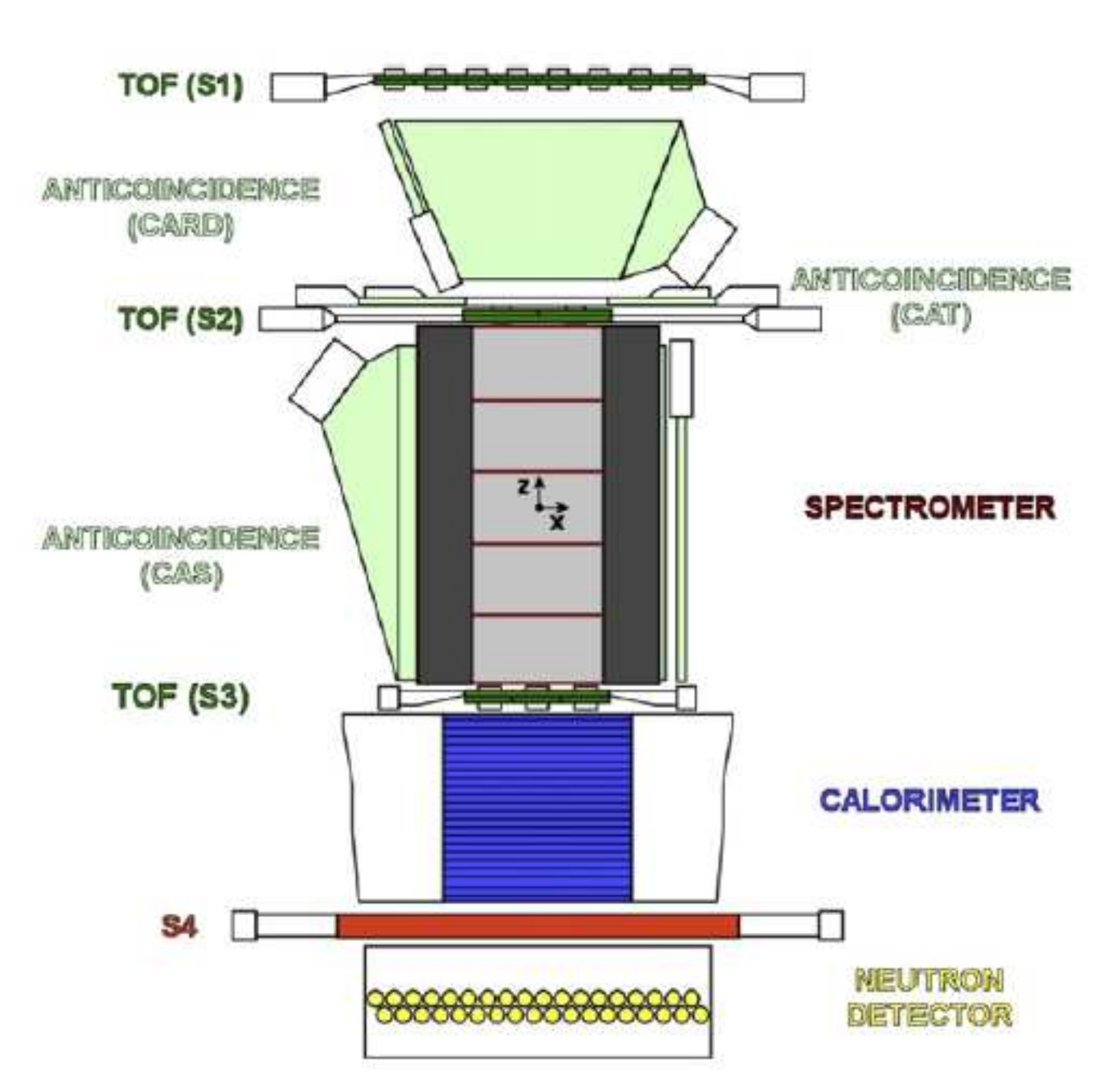}
\end{center}
\caption[A schematic overview of PAMELA instrument]{{\footnotesize A schematic overview of PAMELA instrument. The apparatus is $\sim 1.3$~m tall, with a mass of 470~kg. Taken from {\cite{Picozza2007}}.}}
\label{fig:detectors}
\end{figure}

In order to reach the design goals of performance, PAMELA comprises several subdetectors, each providing an independent measurement of the incident particles. Figure \ref{fig:detectors} presents a schematic overview of the PAMELA instrument and shows the location of each subdetector.

The core detector of PAMELA is a 0.43 Tesla permanent magnet spectrometer (tracker) equipped with 6 planes of double-sided silicon detectors, allowing the sign, absolute value of charge and momentum of traversing charged particles to be determined. The spectrometer geometry and dimensions define the overall acceptance of the experiment which is 21.5~cm$^{2}$sr. The maximum detectable rigidity is found to be $\sim 1$ TV from test beams. Spillover effects limit the upper detectable antiparticle momentum to $\sim 190$ GeV/c ($\sim 270$ GeV/c) for antiprotons (positrons). The spectrometer is surrounded by a plastic scintillator veto shield which can be used to reject particles not cleanly entering the acceptance. An electromagnetic calorimeter mounted below the spectrometer measures the energy of incident electrons and allows topological discrimination between electromagnetic and hadronic showers, or non-interacting particles. Planes of plastic scintillator mounted above and below the spectrometer form a time-of-flight system. It provides the primary experimental trigger, identifies albedo particles, measures the absolute charge of traversing particles and also allows proton-electron separation below $\sim 1$ GeV/c. The volume between the upper two time-of-flight planes is bounded by an additional plastic scintillator anticoincidence system. A plastic scintillator system mounted beneath the calorimeter aids in the identification of high energy electrons and is followed by a neutron detection system for the discrimination of high energy electrons and hadrons which shower in the calorimeter but do not necessarily pass through the spectrometer.


The PAMELA instrument is 1.3 m tall, with a mass of 470 kg. The average power consumption of PAMELA is 355 W, which is provided by the solar panels or batteries of the host satellite. Data are down-linked a few times per day to the mass memory of the satellite during acquisition, and radio-linked down to Earth when passing the ground center in Moscow, NTsOMZ. The average volume of data transmitted per day is about 15 GBytes, corresponding to $\sim 2$ million collected events.

\subsection{The Magnetic Spectrometer}\label{sec:spec} 
The magnetic spectrometer {\cite{Adriani2003}} is designed to give a precise measurement of momentum and charge (with sign) of the incident particle, as well as satisfying the requirements of the mission imposed by the satellite specifications. A compact mechanical assembly has been chosen and tested to withstand the stresses during the launch phase. The spectrometer is composed of a permanent magnet with an internal rectangular cavity, and a tracking system with six planes of double-sided silicon microstrip detectors, uniformly positioned along the cavity. Each plane independently measures both the X and Y coordinates of the crossing point of an incoming ionizing particle. 

The reconstruction of the trajectory is based on the impact points and the resulting determination of the curvature due to the Lorentz force. The equation of motion describing a charged particle (with mass $m$ and charge $q$) moving in a magnetic field $\vec{B}$ is:

\begin{equation}\label{motion}
m \gamma \frac{d^{2} \vec{r}}{dt^{2}}=q \left( \frac{d\vec{r}}{dt} \times \vec{B}\right),
\end{equation}
where $ \vec{r}$ is the position of the charge particle and $\gamma = 1 / \sqrt{1-v^{2}/c^{2}}$.
 
Introducing the path length $l=\beta ct$ and using $p=m\gamma \beta c$, equation \eqref{motion} can be rewritten as:

\begin{equation}\label{motion2}
\frac{d^{2}\vec{r}}{dl^{2}}=\frac{q}{m\gamma \beta c} \left( \frac{d\vec{r}}{dl} \times \vec{B} \right)=\eta \left( \frac{d\vec{r}}{dl}\times \vec{B} \right),
\end{equation}

where $\beta=v/c$ and the magnetic deflection $\eta$ is defined as the inverse of the rigidity $R$ ($R = p/q$):

\begin{equation}\label{eq:deflection}
\eta=\frac{1}{R}=\frac{q}{p}.
\end{equation}

Equation \eqref{motion2} can be solved by numerical methods, thus the deflection $\eta$ of the particle is derived by looking for the set of initial conditions which best fit the measured coordinates of the particle trajectory. The spectrometer can also measure the absolute value of the charge since the ionization energy loss deposited in the sensitive areas of one plane is proportional to the square of the charge of the particle.

\begin{figure}[!h]
\begin{center}
\includegraphics[width=0.7\textwidth]{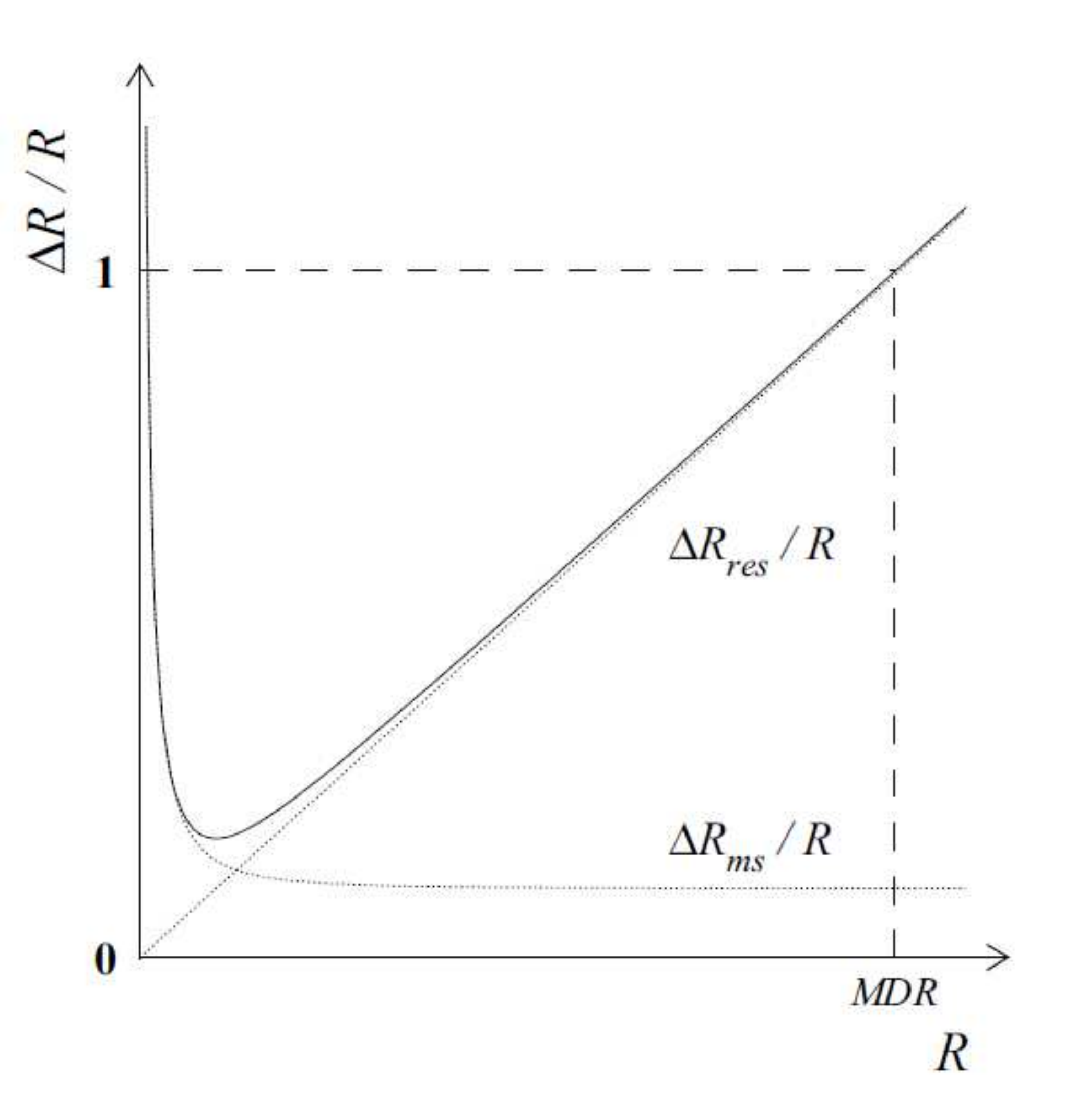}
\end{center}
\caption[Spectrometer resolution as a function of rigidity]{{\footnotesize Spectrometer resolution as a function of rigidity. The dotted lines present the rigidity relative error due to the finite spatial resolution $\Delta R_{res}/R$ and the error due to the multiple scattering $\Delta R_{ms}/R$ . The solid line shows the quadratic sum of the two. Taken from \cite{BongiPhd}.}}
\label{fig:rigidityerr}
\end{figure}

The upper limit of the detectable energy range PAMELA can achieve is constrained by the spectrometer bending power, expressed by the so-called \textit{Maximum Detectable Rigidity} (MDR). MDR is defined as the measured rigidity which corresponds to 100\% uncertainty. This feature conflicts with the detector acceptance, expressed as a \textit{Geometrical Factor} (GF), which is defined as the factor of proportionality between the detector counting rate and the intensity to isotropic radiation. While the acceptance grows with the cross section of the cavity, the bending power improves for a longer cavity and a larger magnitude of $\vec{B}$. Given a constraint on $\vec{B}$, a longer magnetic cavity enhances the MDR while lowering GF, and conversely, a wider acceptance increases GF but worsens the MDR since it is more difficult to maintain a high field over a larger area. The geometric design should give the best compromise between these two features. Since extending the measurement of antiprotons and positrons to higher energy is a main objective, the MDR has been preferred in designing the spectrometer. The measurement error for momentum $p$ depends on two contributions, the finite spatial resolution of the tracking system $ \sigma$ and multiple Coulomb scattering of the particles crossing the spectrometer, whose relative weight varies with momentum. The errors from these two contributions, expressed as $\Delta p_{res}$ and $\Delta p_{ms}$ can be derived as:
\begin{equation}
\frac{ \Delta p_{res}}{p} \propto \frac{\sigma}{BL^{2}}p,
\end{equation}
and
\begin{equation}
\frac{ \Delta p_{ms}}{p} \propto \frac{1}{\beta}= \sqrt{ 1+\left(\frac{mc^{2}}{pc} \right)^{2}}, 
\end{equation}
where $p$ is the momentum of the particle, $L$ is the length of the track when projected on the bending plane, and $\sigma$ is the spatial resolution. The measured spatial resolution with test beams is $\left(3.0 \pm 0.1 \right)$~$\mu$m and $\left(11.5 \pm 0.6 \right)$~$\mu$m in the bending and non-bending views, respectively {\cite{Picozza2007}}. Since the rigidity $R = p/q$, for a particle with a certain charge $q$, the relative error on rigidity is $\Delta R / R = \Delta p / p$ (see figure \ref{fig:rigidityerr}). While spatial resolution plays a crucial role at high energy, multiple Coulomb scattering cause the main uncertainty of measured rigidity at low energy. Consequently, a long magnetic cavity with a strong magnetic field provides a good spatial resolution at high energy, while a minimal amount of material along the path of the particles reduce the scattering effect at low energy. The expected MDR is about 1~TV/c and the computed GF for high-energy particles with straight tracks is about 21.5~cm$^{2}$sr \cite{SergioPhd}. 

\begin{figure}[!h]
\begin{center}
\includegraphics[width=0.7\textwidth]{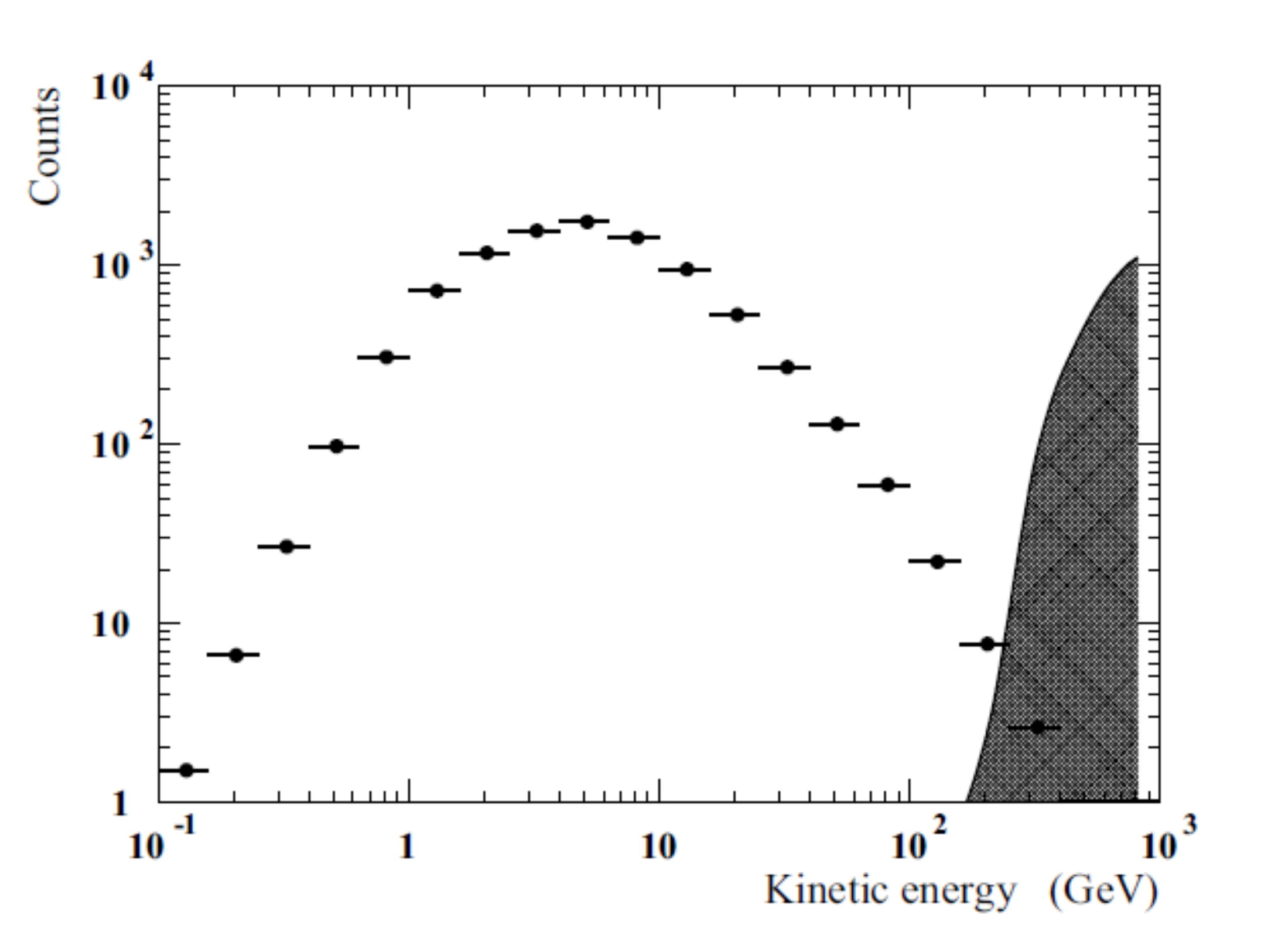}
\end{center}
\caption[Simulated spillover effect in the antiproton flux measurement]{{\footnotesize Simulated spillover effect in the antiproton flux measurement. The points show the expected antiproton flux in three years measurements with PAMELA according to pure secondary production. The shaded area shows the simulated proton spillover in the antiproton sample. Taken from \cite{BongiPhd}.}}
\label{fig:spillover}
\end{figure}

The upper rigidity limit for particles like protons, nuclei and electrons is directly connected to the MDR, but this is not the case for the antiparticles due to their rarity in cosmic rays. When the energy of particles increase, the tracks get closer and closer to a straight line. The finite spatial resolution makes it difficult to properly determine the charge sign, which is used to distinguish antiparticles from particles. This effect, called \textit{spillover}, causes a non-negligible background when measuring antiparticles at high energy, especially for antiprotons since the number of protons is much larger than the number of antiprotons (about a factor of $10^4$ at 10 GeV). A simulation of proton spillover into a sample of antiprotons implies the detection of antiprotons is limited to about 190 GeV (see figure \ref{fig:spillover}). 

\subsubsection{The magnet}
\begin{figure}[!h]
\begin{center}$
\begin{array}{cc}
\includegraphics[width=5cm]{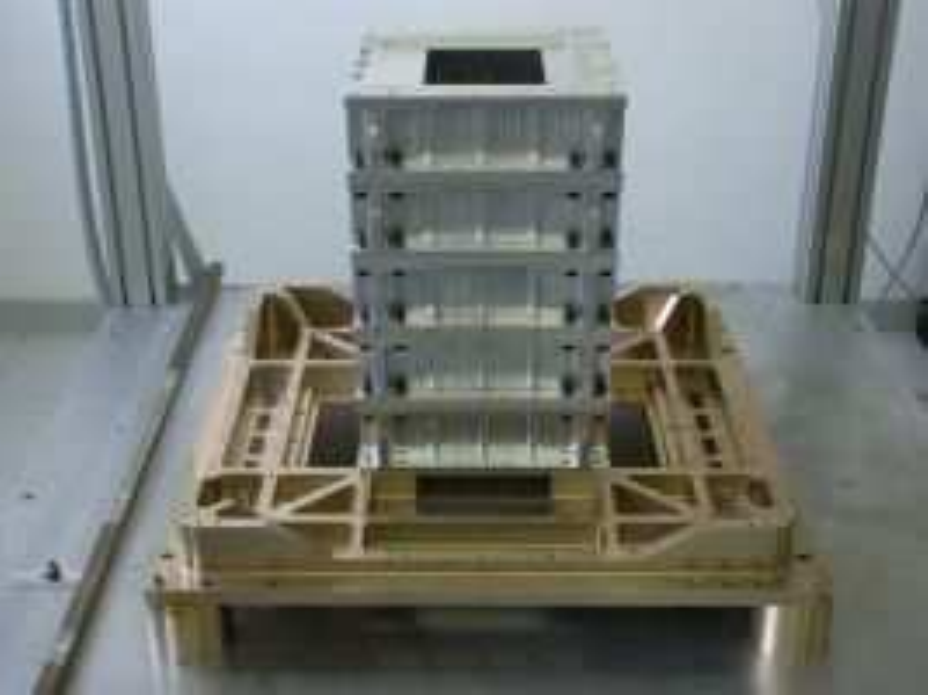}&
\includegraphics[width=5cm]{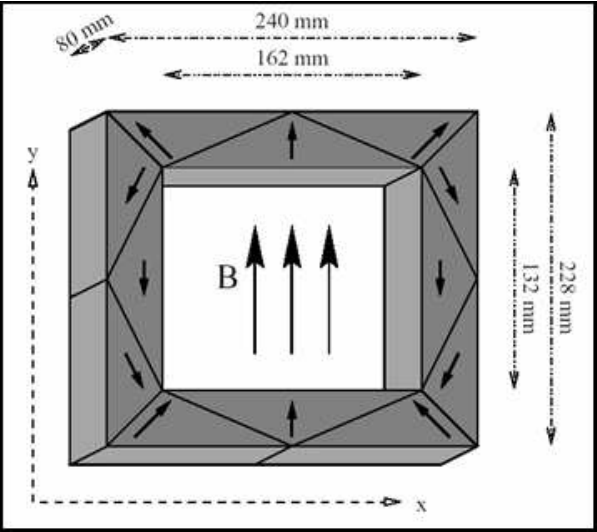}
\end{array}$
\end{center}
\caption[The magnet tower and a prototype of one magnet module]{{\footnotesize The magnet tower (left). Sketch of a prototype of one magnet module (right). Taken from \cite{PamelaHomepage}.}}
\label{fig:magnet}
\end{figure}

The magnet is a 43.66 cm high tower, formed by five identical modules with a central rectangular cavity (16.14 cm $\times$ 13.14 cm). Each module is composed of 12 Nd-Fe-B alloy elements with high residual magnetization ($\approx$ 1.32 T). A picture of the entire configuration is shown in figure \ref{fig:magnet}. 

\begin{figure}[!h]
\begin{center}
\includegraphics[width=0.95\textwidth]{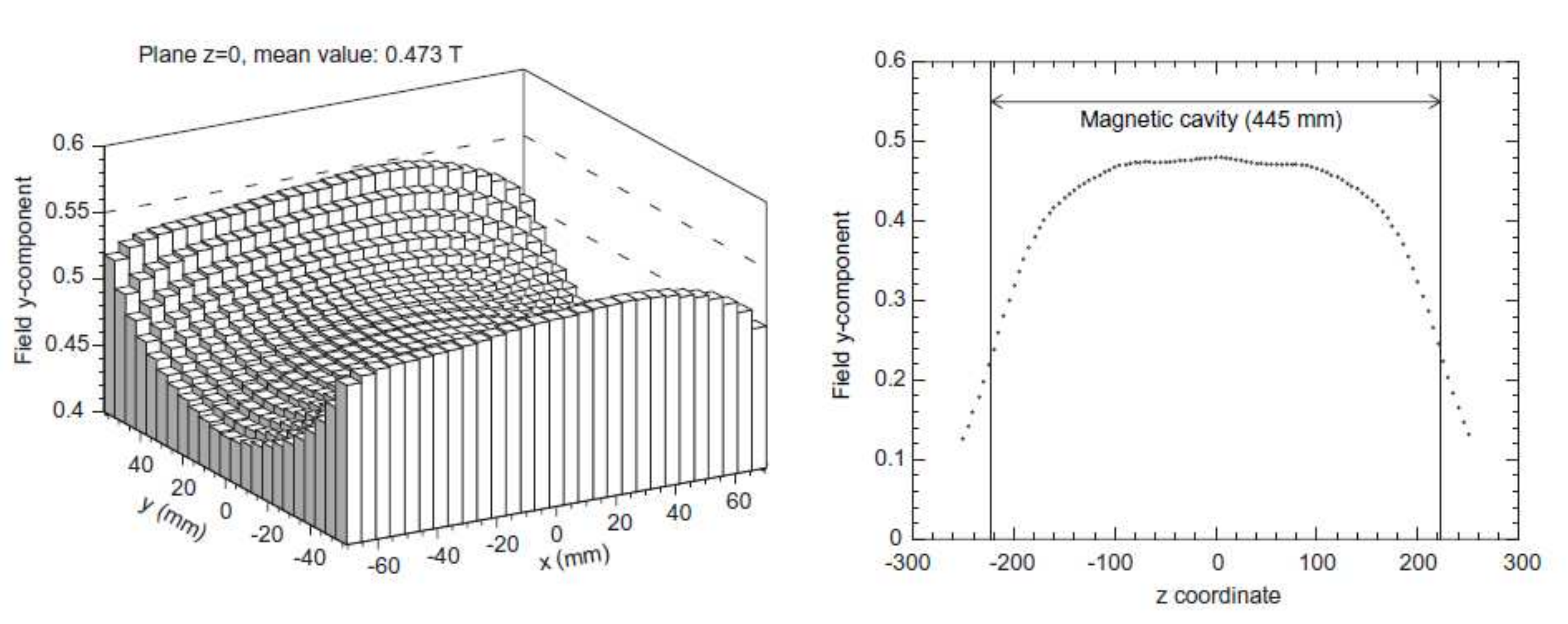}
\end{center}
\caption[The main magnetic field component]{{\footnotesize The main magnetic field component ($B_y$) plotted for the central plane ($z=0$) of the cavity (left) and the ($B_y$) plotted as a function of z coordinate along the central axis ($x=0$, $y=0$) of the cavity (right). Taken from \cite{PamelaHomepage}.}}
\label{fig:magneticfield}
\end{figure}

This configuration has been chosen to have very high and uniform field strength inside the cavity and the lowest possible field intensity outside. To protect the magnetic material from chemical attacks, a 500 $\mu$m aluminum layer covers all the free surfaces of the magnet. The field inside the cavity is almost uniform, with practically all the strength along the negative Y-direction {\cite{Adriani2003b}}. As a consequence, particles are bent in the XZ plane within the cavity, due to the Lorentz force $ \vec{F} = q \vec{v} \times \vec{B} $. The magnetic field has been mapped by means of an FW-Bell Gaussmeter equipped with a three-axis Hall probe mounted on an automatic positioning device. The measured main field component ($B_y$) is plotted for the central plane of the cavity and plotted for the central axis of the cavity in figure \ref{fig:magneticfield}. In the center of the cavity (z=0) the value reaches 0.48 T and remains nearly constant across a wide region. The measured average magnetic field is about 0.43 T. 

\subsubsection{Silicon tracking system}
\begin{figure}[!h]
\begin{center}
\includegraphics[width=0.7\textwidth]{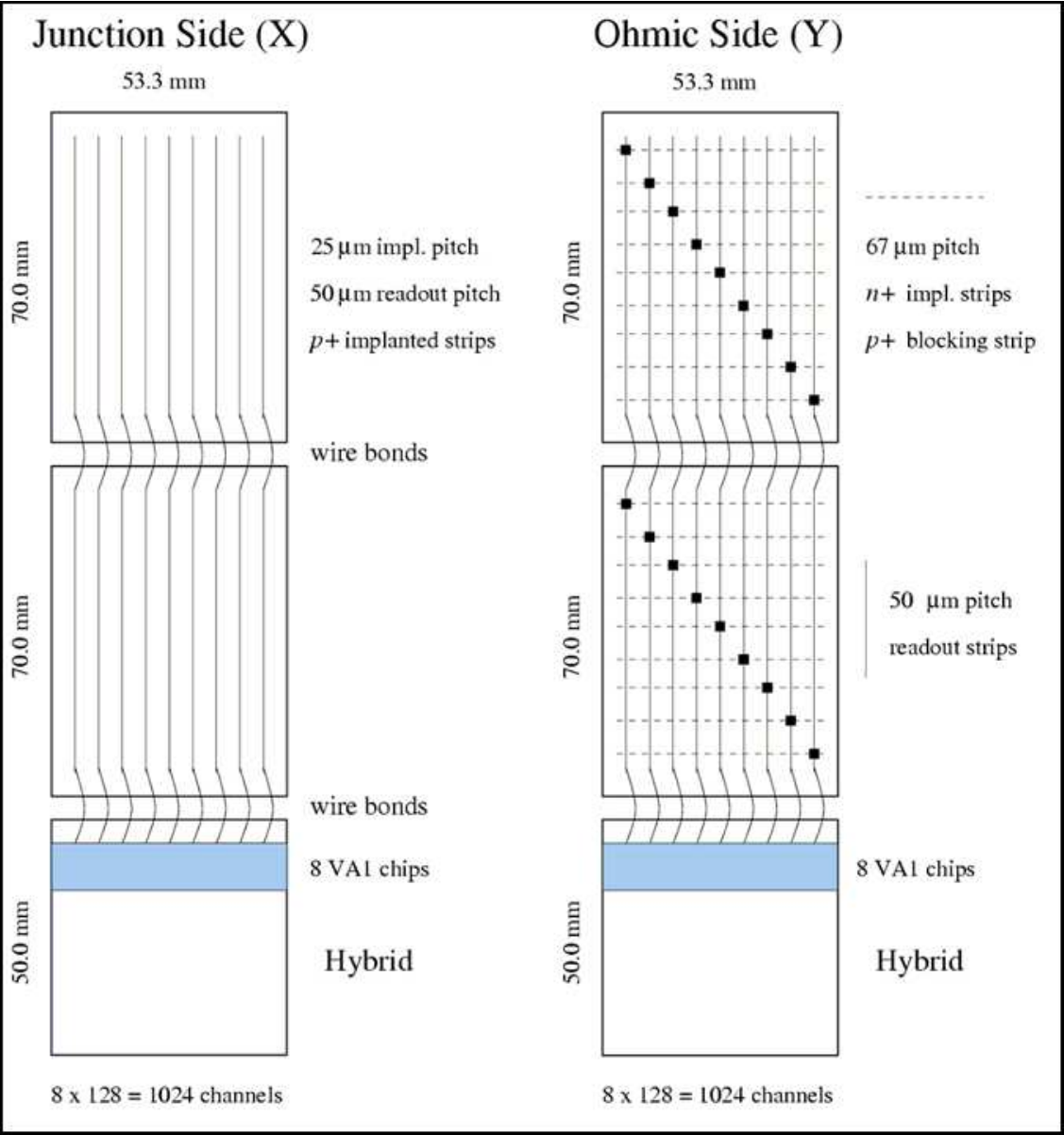}
\end{center}
\caption[The ladder structure]{{\footnotesize The sketch of the strips layout on the junction side and ohmic side of the ladder. Both the read-out electrodes perpendicular to the $n^{+}$ strips and their connections on the diagonal of the sensor are shown on the ohmic side. Taken from \cite{PamelaHomepage}.}}
\label{fig:ladder}
\end{figure}

The tracking system is composed of 6 planes of high-precision silicon microstrip detectors, placed between the five magnetic modules and above and below the openings of the magnetic tower, with an uniform vertical spacing of 8.9 cm. Each plane, housed in an aluminum frame, consists of 3 independent sections (ladders) along the X axis. Each ladder is formed by 2 rectangular double-sided n-type silicon sensors with dimensions 53.33 mm $\times$ 70.00 mm $\times$ 30 $\mu$m and a hybrid circuit which houses the front-end electronics (figure \ref{fig:ladder}). On the junction side 2035 $p^+$ type microstrips are implanted with a pitch of 25.5 $\mu$m and on the ohmic side 1024 $n^+$ type microstrips are implanted with a pitch of 66.5 $\mu$m. Strips on opposite sides are orthogonal and the spatial information of the impact point of the incident particle can be measured by looking at which strip collected the ionization charge on junction (X) view and ohmic (Y) view.

\subsection{The time of flight system}\label{sec:tof}
The time of flight system (ToF) is designed to fulfill several goals {\cite{Barbarino2008}}:
\begin{itemize}
\item provide a fast signal for triggering data acquisition of the whole instrument;
\item measure the flight time of particles crossing its planes; once this information is integrated with the measurement of the trajectory length through the instrument, the particle velocity $\beta$ can be derived. This feature
enable also the rejection of particles entering the apparatus from below, called $albedo$ particles;
\item determine the absolute value of the charge $Z$ of incident particles through the multiple measurement of the
ionization energy loss $dE/dx$ in the scintillator counters. 
\end{itemize}

Additionally, segmentation of each detector layer in strips can provide a rough tracking of particles, thus helping to reconstruct their trajectory outside the magnet volume.

\begin{figure}[tb!h]
\begin{center}
\includegraphics[width=0.7\textwidth]{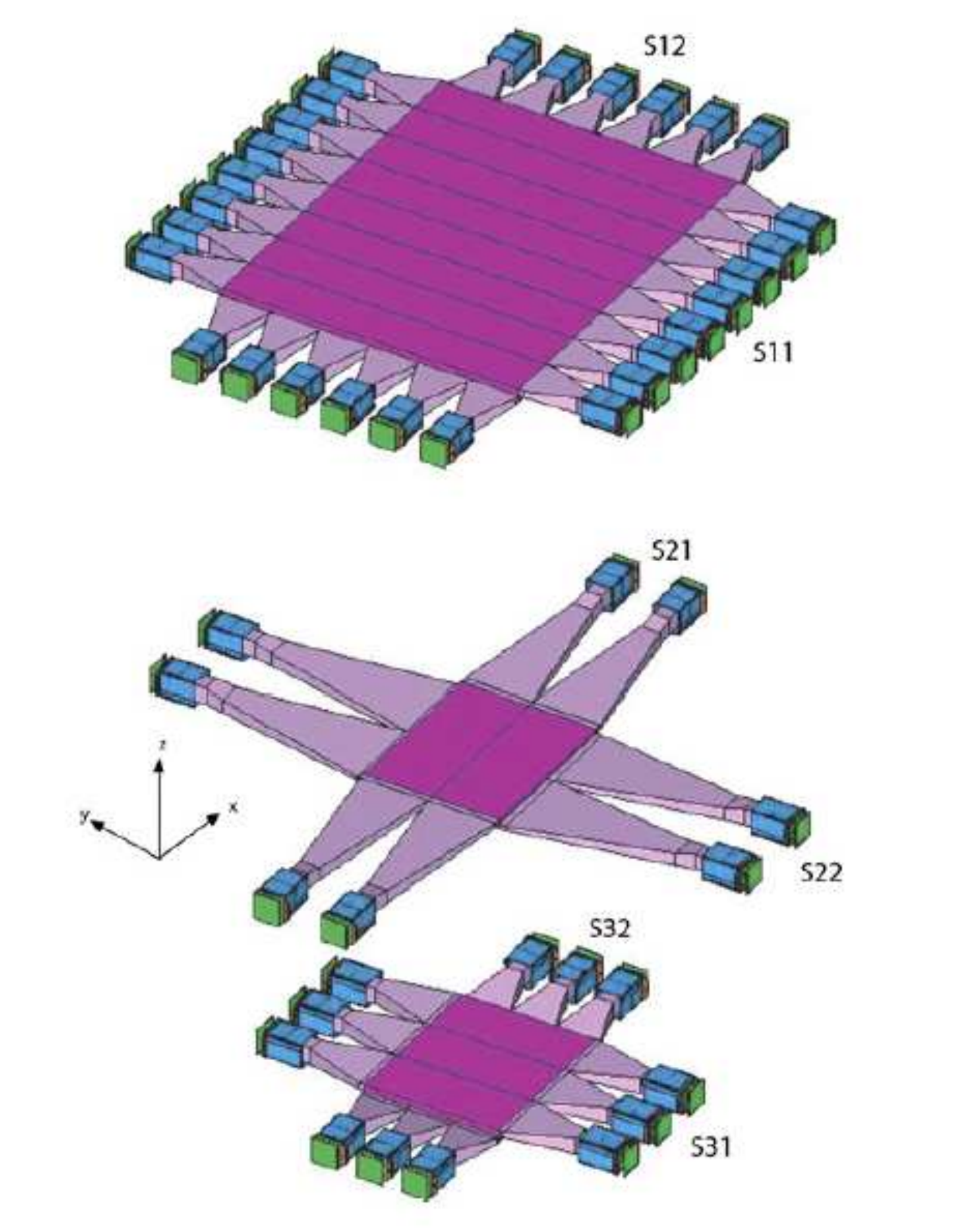}
\end{center}
\caption[The ToF system]{{\footnotesize The ToF system. The sensitive areas are  $330\times408$~mm$^{2}$ for S1 and $150\times180$~mm$^{2}$ for S2 and S3. The distance of S1 and S3 planes is 77.3~cm. Taken from \cite{PamelaHomepage}.}}
\label{fig:tof}
\end{figure}

The ToF (see figure \ref{fig:tof}) is arranged in three planes, referred to as S1, S2 and S3, and each composed of double layers of segmented plastic scintillators to improve the reconstruction efficiency for crossing particles. S1 is placed on top of the experiment, with eight 330 mm $\times$ 51 mm paddles forming its first layer S11 and six 408 mm $\times$ 55 mm paddles form its second layer S12. The overall sensitive area of each layer of S1 is 330 $\times$ 408 mm$^{2}$. S2 and S3 are placed above and below the spectrometer respectively. The two layers of S2, S21 and S22, are both divided into two paddles, with dimension 180 mm $\times$ 75 mm for S21 and 150 mm $\times$ 90 mm for S22, resulting in a sensitive area 150 $\times$ 180 mm$^{2}$ for each layer. For S3, the first layers S31 is divided into three 150 mm $\times$ 60 mm paddles, while the second layer S32 is divided into three 180 mm $\times$ 50 mm paddles. The overall sensitive area of each layer of S3 is 150 $\times$ 180 mm$^{2}$. For each plane, the paddles of the upper layer are orthogonal to those of the lower layer, therefore allowing a two dimensional coordinate measurement of the impact points of charged particles. BC-404 manufactured by Bicron company was chosen for the scintillator material, characterized by a rise time 0.7 ns and a decay time 1.8 ns. The two ends of each paddle are read-out by a Hamamatsu R5900 PMT, which can achieve an amplification of about $4 \times 10^{6}$ at 900 V. Since the core of the PAMELA apparatus is a permanent magnet, all the PMTs have been shielded with a 1 mm thick $\mu$-metal screen to avoid the influence of any residual magnetic field.

The anode pulse of each PMT is converted both in charge and time, by connecting to an analog-to-digital converter (ADC) and a time-to-digital converter (TDC) respectively. When a charged particle crosses a layer, the ADC measures the ionization energy loss, and the TDC provides the relative time. The charge identification capabilities of ToF were evaluated during a test with particle beams performed at the GSI laboratory in Germany, which indicates the measured charge uncertainty is less than 0.1 for protons and 0.16 for carbon {\cite{Campana2009}}. The combined TDC information of all the ToF planes is used to generate the main PAMELA trigger and determine the flight time of the incoming particle. The standard trigger configuration requires the coincidence of at least one TDC signal from each of the three planes. In the radiation belts and inside the SAA the requirement on S1 is removed as S1 is saturated by low energy particles. Moreover, TDC information can establish the incident particle direction by checking the order in which the layers have been hit. This is important since the sign of charge can be determined by the deflection only if the direction of motion is known.

The velocity $\beta$ ($v/c$) of an incident particle can be calculated by measuring the time needed for the particle traveling from S1 to S3. For a particle with velocity $\beta$, momentum $p$ and mass $m$, it follows that:

\begin{equation}
\beta = \frac{1} {\sqrt{ 1 + \left( mc^{2}/ pc \right)^{2}}},
\end{equation}

which give the possibility to discriminate types of particles with different masses at low energy. The measured ToF time resolution about 250 ps allows electrons (positrons) to be separated from antiprotons (protons) up to 1 GeV/c \cite{Picozza2007}. 

\subsection{Anticoincidence system}\label{sec:acinstrument}
The primary aim of the Anticoincidence (AC) system designed and built at KTH is to identify events yielding "false" triggers, which might be generated by secondary particles produced in the mechanical structure of the experiment. It also can help to reject out of acceptance events.

\begin{figure}[!h]
\begin{center}
\includegraphics[width=5cm]{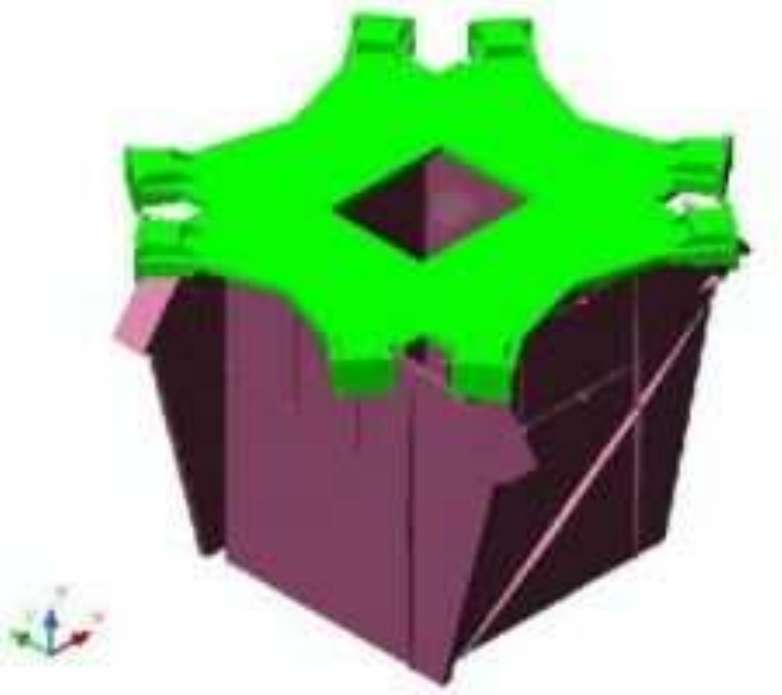}
\end{center}
\caption[A schematic view of CAS and CAT]{{\footnotesize A schematic view of CAS (purple) and CAT (green). The CAS scintillator is $\sim 40$~cm tall and 33~cm wide. The hole in the CAT scintillator is $\sim 22\times 18$~cm$^2$. Taken from \cite{PamelaHomepage}.}}
\label{fig:AC}
\end{figure}

The PAMELA experiment contains two AC systems. One of them consists of 4 plastic scintillators (CAS) covering the sides of the magnet each of which has an approximate rectangular shape, and 1 scintillator covering the top (CAT) which has a star shape with a rectangular hole in the center corresponding to the acceptance of the spectrometer (see figure \ref{fig:AC}). The other one consists of 4 plastic scintillators (CARD) surrounding the empty volume between S1 and S2. All the scintillators are Bicron BC-448M and each CAS and CARD detector is read out by two identical Hamamatsu R5900U PMTs in order to decrease the possibility of single point failure, while the CAT detector is read out by 8 PMTs for the same reason and also to cover the irregular shaped area. The scintillators and PMTs are housed in aluminum containers which provide light-tightness, allow fixation to the PAMELA superstructure and ensure that a reliable scintillator-PMT coupling is maintained. No additional magnetic shielding is required due to the small fringe field from the magnetic spectrometer at the position of the PMTs. A particle traversing an AC detector is registered as a hit if it deposits at least $\sim 0.8$ MeV energy in the scintillator. The detection efficiency for charged particles is measured to be 99.9\%. 

\subsection{Electromagnetic Calorimeter}\label{sec:calo}

The calorimeter is the key detector used to select antiprotons and positrons from like-charged backgrounds which are significant more abundant in cosmic rays. Antiprotons must be separated from a background of electrons that decreases from  $\sim 10^{3}$ times the antiproton component at 1 GeV/c to less than $10^{2}$ above 10 GeV/c, and positrons from a background of protons that increases from $\sim 10^{3}$ times the positron component at 1 GeV/c to $\sim 5\times 10^{3}$ at 10 GeV/c. This means that the PAMELA detectors have to separate electromagnetic from hadronic particles at a level of $10^{5} - 10^{6}$. Most of this separation is provided by the calorimeter.

\begin{figure}[!h]
\begin{center}
\includegraphics[width=8cm]{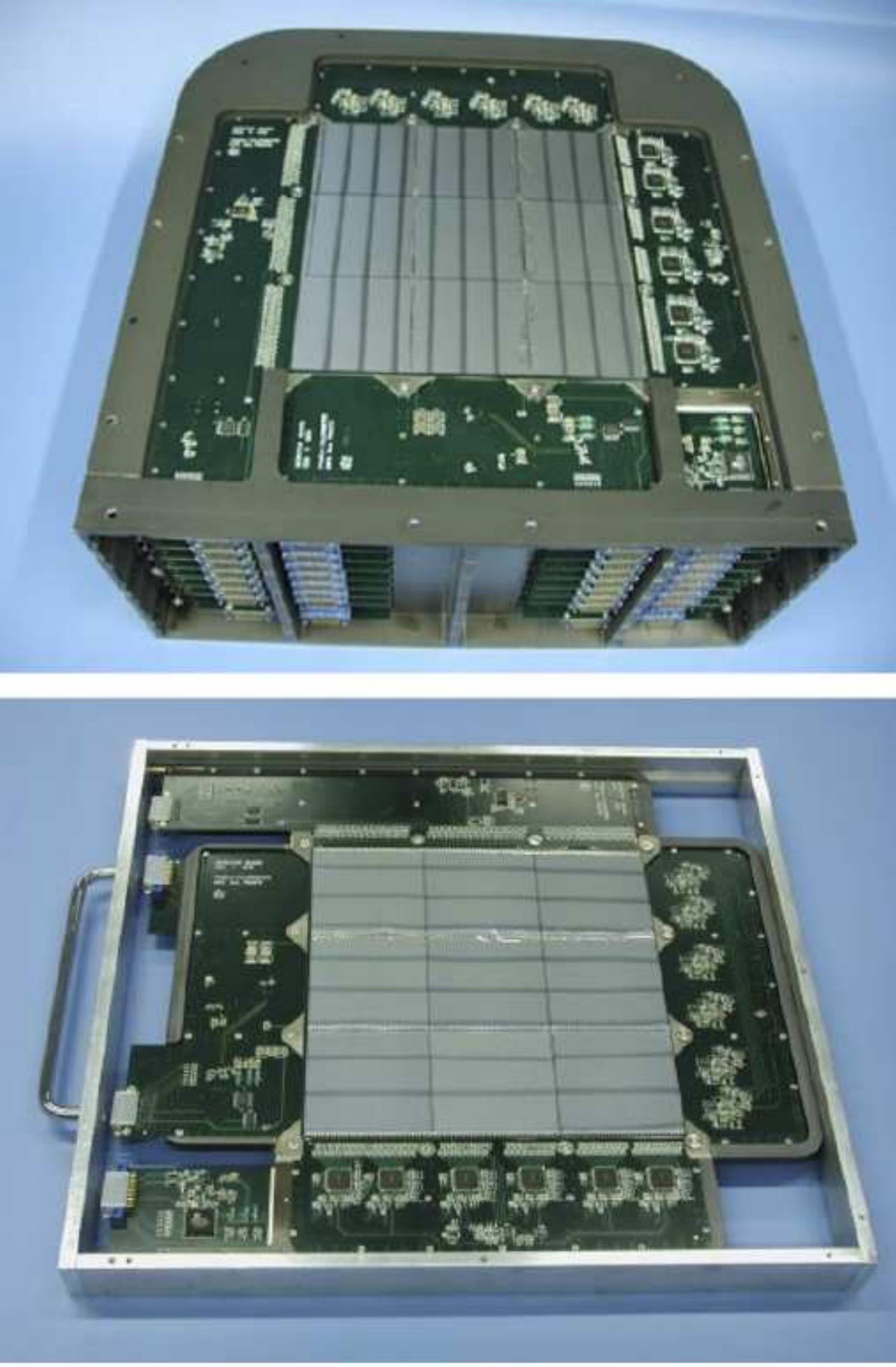}
\end{center}
\caption[The PAMELA electromagnetic calorimeter]{{\footnotesize The PAMELA electromagnetic calorimeter comprising 22 calorimeter modules. The device is $\sim 20$~cm tall and the active silicon layer is $\sim 24 \times 24$~cm$^2$ in cross-section. The bottom plot shows the detail of a single module consisting of a tungsten layer sandwiched between two silicon detector planes. Taken from \cite{Picozza2007}.}}
\label{fig:Calorimeter}
\end{figure}

The calorimeter is composed of 44 silicon sensor layers interleaved with 22 planes of tungsten absorbers (see figure \ref{fig:Calorimeter}). Each tungsten layer has a thickness of 26 mm which corresponds to 0.74 $X_{0}$ (radiation lengths), giving a total depth 16.3 $X_{0}$ or 0.6 $\lambda$ (interaction lengths). Each silicon plane consists of a $3\times 3$ matrix of $8\times 8$ cm$^{2}$ silicon detectors. The detectors are separated from each other by a $\sim 35$ mm wide non-sensitive area, called a \textit{dead area}. About 5\% area of one plane is covered by such dead areas. Each silicon detector is 380 $\mu$m thick and segmented into 32 strips with a pitch of 2.4 mm. Two consecutive sensor layers and one sandwiched tungsten absorber form a detector plane. The orientation of the strips of the two layers in a detector plane is orthogonal and therefore provides two-dimensional spatial information of a particle shower. 

\begin{figure}[!h]
\begin{center}
\includegraphics[width=10cm]{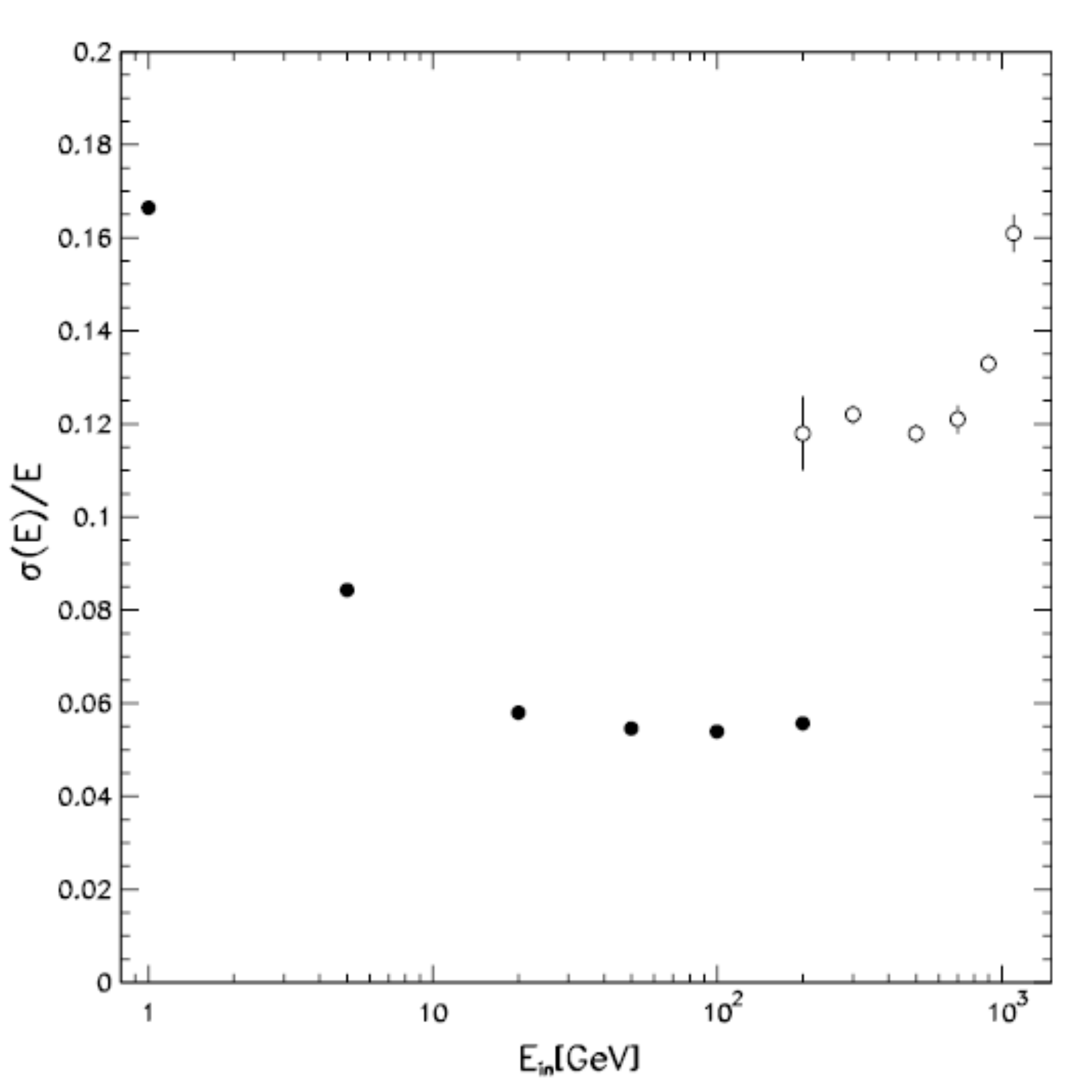}
\end{center}
\caption[The calorimeter energy resolution]{{\footnotesize The calorimeter energy resolution as a function of the incident electron energy $E_{in}$. The filled circles are for normal operation (experimental data) and the open circles are for the self-trigger mode (simulations). Taken from \cite{Picozza2007}. }}
\label{fig:caloenergyres}
\end{figure}

The longitudinal and transverse segmentation of the calorimeter, combined with the measurement of the particle energy loss in each silicon strip, allows a high rejection power of electrons in the antiproton sample and protons in the positron sample. A good agreement is found between simulated and experimental calorimeter data. Simulations demonstrate a rejection factor of about $10^{5}$ for electrons in antiproton measurements with 90\% antiproton identification efficiency \cite{Boezio2006}. The calorimeter is also used to reconstruct the energy of the electromagnetic showers, providing a measurement of the energy of the incident electrons independent from spectrometer, thus allowing a cross-calibration of two energy measurements. The calorimeter energy resolution has been measured as $\sim 5.5 \%$ up to several hundred GeV (shown in figure \ref{fig:caloenergyres}). In order to measure very high energy electrons ($\sim 300$ GeV to $>1$ TeV), calorimeter is equipped with a self-trigger capability. A self-trigger signal is generated when a specific energy distribution is detected predetermined planes within the lower half of the calorimeter. By requiring that self-triggering particles enter through one of the first four planes and cross at least 10 radiation lengths, the geometrical factor can achieve 600 cm$^2$sr, which is about 30 times larger than the default PAMELA geometrical factor defined by the magnetic spectrometer. Since the geometrical factor is highly increased in self-trigger mode, PAMELA has the capbility to measure the very-high energy electrons which are rare in the cosmic radiation. The calorimeter energy resolution in self-trigger mode is estimated through simulation to be $\sim 12 \%$ up to about 800 GeV, as shown in figure \ref{fig:caloenergyres}.

\subsection{Bottom Scintillator S4}
The bottom scintillator S4, referred to as the \textit{shower tail catcher}, is used to improve the PAMELA electron-hadron separation by measuring shower leakage from the calorimeter. S4, with a sensitive area of $482 \times 482$ mm$^{2}$ and a thickness of 10 mm, is located directly beneath the calorimeter and read out by six PMTs placed along the two opposite sides. The S4 detector detects showers not contained in the calorimeter. When the signal in S4 exceeds 10 MIPs (where 1 MIP is the most probable energy deposited by a normally incident minimum ionizing particle) and coincides with the main trigger signal, an on-board neutron detector is read out. 

\subsection{Neutron Detector}
The primary purpose of the neutron detector (ND) is to complement the electron-proton discrimination capabilities of the calorimeter. Also, combined analysis of calorimeter and ND information will expand the energy range for detected primary electrons up to 10 TeV.

\begin{figure}[!h]
\begin{center}
\includegraphics[width=10cm]{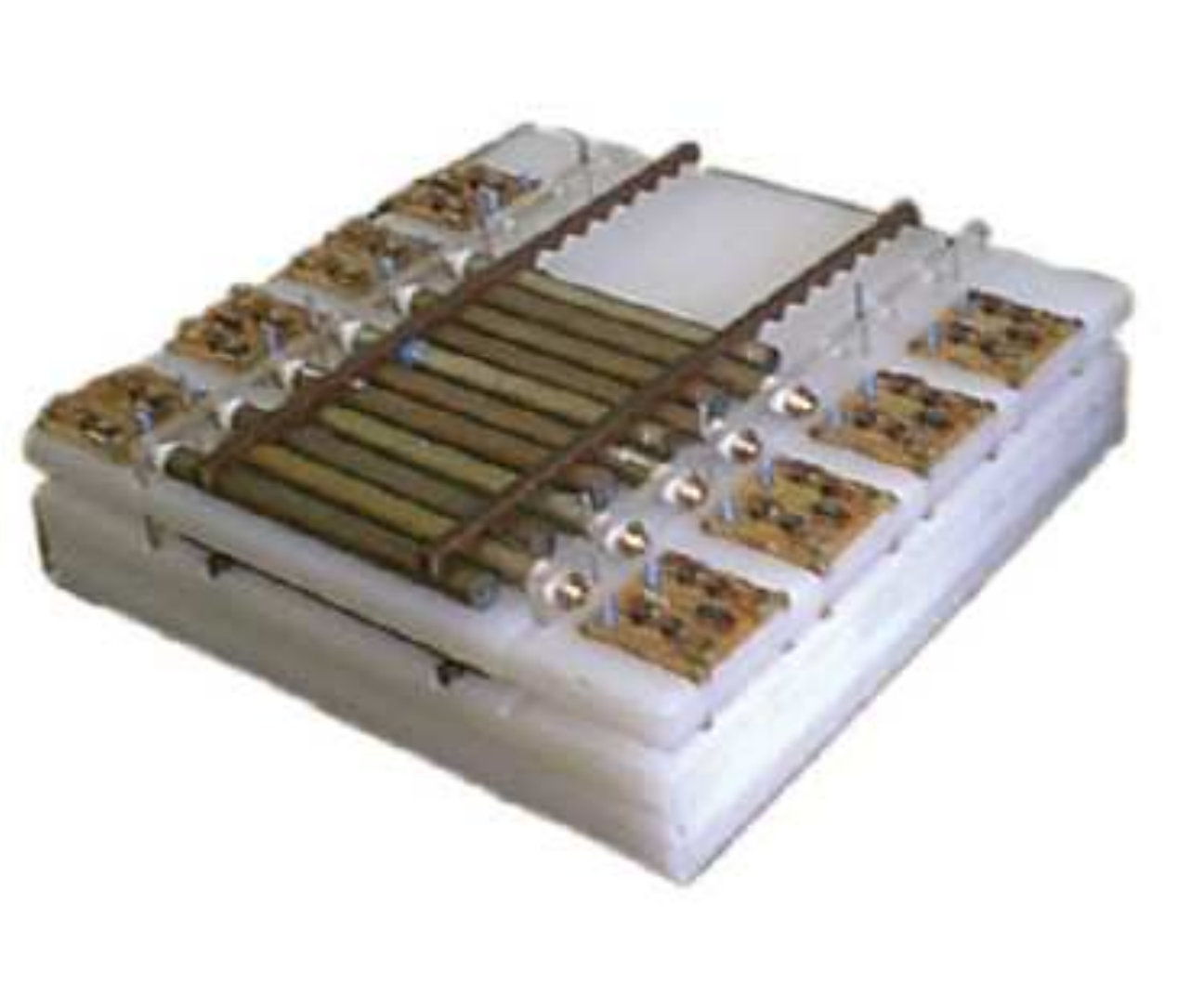}
\end{center}
\caption[The neutron detector partially equipped with $^{3}$He proportional counters]{{\footnotesize The neutron detector partially equipped with $^{3}$He proportional counters. The neutron detector covers an area of $60 \times 55$~cm$^{2}$. Taken from \cite{Picozza2007}.}}
\label{fig:ND}
\end{figure}

The detector is located below the S4 scintillator and consists of 36 gas proportional counters stacked in two planes of 18 counters each, oriented along the y-axis of the instrument. The size of the neutron detector is $600\times 550 \times 150$ ~mm$^{3}$. The counters are filled with $^{3}$He and surrounded by a polyethylene moderator enveloped in a thin cadmium foil to prevent thermal neutrons entering the detector from the sides and from below (see figure \ref{fig:ND}). 

When a high-energy hadron interacts inside the calorimeter, a large number of neutrons are produced by the decay of excited nuclei, while the number is 10-20 times lower if the primary particle is an electron and the neutrons are generated via photo-nuclear interactions. A part of these neutrons is thermalized by the polyethylene moderator and detected by the $^{3}$He counters.

\subsection*{}
To summarize, this chapter details the scientific objectives of the PAMELA experiment and describes all the sub-detectors including the spectrometer, the time-of-flight system, the anticounters, the calorimeter, the bottom scintillator and the neutron detector. Using a combination of these detectors, different cosmic ray species can be identified and their fluxes measured. In chapter \ref{chap:antiproton}, cosmic ray antiprotons will be selected with the help of PAEMLA instrument and the flux of antiprotons in the cosmic radiation as well as the antiproton-to-proton flux ratio will be reconstructed .

%% file: Phd-Ch-Antiproton.tex
\chapter{The antiproton flux and antiproton-to-proton flux ratio} \label{chap:antiproton}

Cosmic rays are dominated by protons and helium nuclei, with a small fraction of electrons, helium-3, deuterium and heavier nuclei, as well as rare antiparticles. In the negatively charged part, antiprotons constitute only a small fraction (about $10^{-3}$) compared to the main component, electrons. Besides primary cosmic rays, PAMELA also records particles created in the interactions of primary cosmic rays with the experiment materials, for example positive and negative pions, which complicates the identification of antiprotons. 

The procedure used to determine the antiproton flux and antiproton-to-proton flux ratio ($\bar{\text{p}}$/p) contains three steps:
\begin{enumerate}
\item Select a reliable antiproton sample. Section \ref{sec:selection} lists the selection criteria.
\item Calculate the efficiencies of the selection cuts, i.e. the probability that an antiproton will pass the selection criteria. This is presented in section \ref{sec:efficiencies}.
\item Correct for other factors such as geometrical factor, hadronic interaction losses, the live time of measurements and transmission through the geomagnetic field. All these corrections are discussed in section \ref{sec:corrections}.
\end{enumerate}
The final results of antiproton flux and $\bar{\text{p}}$/p ratio measured by PAMELA are shown in section \ref{sec:antiproton_results}.

\section{Antiproton selection} \label{sec:selection}

Before antiproton identification, non-corrupted data (i.e. data contain proper information from each sub-detector) are pre-selected. Events in the high radiation region such as the South Atlantic Anomaly (SAA), yielding high counting rate which may cause unstable performance of detectors, are rejected. 

To obtain a clean antiproton sample, selection cuts have been applied to several variables. Charge one hadrons entering the instrument from above with a well reconstructed track in the tracking system and a trajectory contained in the fiducial geometric acceptance have been selected. Among the surviving particles, negatively charged particles are selected as antiprotons, while a MDR cut is applied to remove the oppositely charged contamination due to the spillover effect. Since the number of antiprotons is about  $10^{-4}$ of the number of protons, the MDR cut is crucial for the selection of rare antiprotons in order to reduce significant proton contamination in the antiproton sample. In following sections all the selection criteria are classified by sub-detector and are described in more detail.

\subsection{Tracker criteria} \label{sec:tracker_sel}

The tracker cuts can be divided into four categories: (i) the basic tracker selection to reconstruct the track of the incident particle and to provide a reliable rigidity measurement; (ii) the geometrical selection to define the acceptance of the experiment; (iii) additional tracker cuts imposed to further clean the (anti)proton sample; (iv) a MDR cut to reduce the background of spillover protons in the antiproton sample at high energies due to the wrong assignment of charge sign.

\subsubsection{The basic tracker cuts}

The basic tracker selection cuts are:

\begin{itemize}
\item A single physical track reconstructed by the track fitting algorithm. Unreasonable events for which the fit routine does not converge or $\chi^{2}$ (the goodness of the fit) is less than zero are excluded. Most multiparticle events will be rejected by this cut.

\item Number of hits on planes in the x-view and y-view: $N_{x} \geq 4$, $N_{y} \geq 3$, and the lever-arm (defined as the distance (in planes) between the upper and lower impact points) in the x-view $ \geq 4$. This selection ensures a good quality of the track. The number of fit points is larger in the x-view than in the y-view since the rigidity construction is performed from the bending in the x-view. A larger number of fit points and a longer lever arm will give a better track reconstruction.

\item An upper limit on $\chi^{2}$, which is a comparison between measured and reconstructed impact points in the tracker planes 
\begin{equation}
\chi^{2} \leq 12.42+199.5 \times \eta^{2} + 153.1 \times  \eta^{4},
\end{equation}
where $\eta$ is the deflection defined in equation \ref{eq:deflection}. Multiple tracks or particles suffering multiple scattering when they cross the tracker planes usually yield rather high $\chi^{2}$. The upper limit on the $\chi^{2}$ is chosen so that the efficiency of this cut is about 95\% and constant with rigidity.
\end{itemize}

\subsubsection{The geometrical cut}

The geometrical selection requires that the particle track is contained inside the fiducial acceptance during its entire passage from S1 to S3. A 1.5 mm margin is subtracted from each side of the 8 tracker planes and 6 ToF planes which defines the geometrical acceptance.  Therefore, this fiducial acceptance is 92\% of the nominal acceptance mentioned in section \ref{sec:spec}. This cut is necessary to exclude particles crossing the magnet walls which might give incorrectly reconstructed tracks, and to avoid efficiency underestimation caused by bad tracking. For high energies, the fiducial acceptance is about 19.90 cm$^{2}$sr \cite{SergioPrivate}.

\subsubsection{Additional tracker cuts} \label{sec:addtrkcuts}

Additional tracker cuts contain:
\begin{itemize}
\item  A cut on the ionization energy loss in the tracker to select singly-charged particles. The mean values of  the $dE/dx$ measurements on the 12 tracker planes are contained inside the region bounded by the red lines shown in figure \ref{fig:dedxtrk}. This cut helps to exclude multiply charged particles, multi-particle events and particles interacting in the tracker material. 

\begin{figure}[!htbp]
\begin{center}
\includegraphics[width=0.85\textwidth]{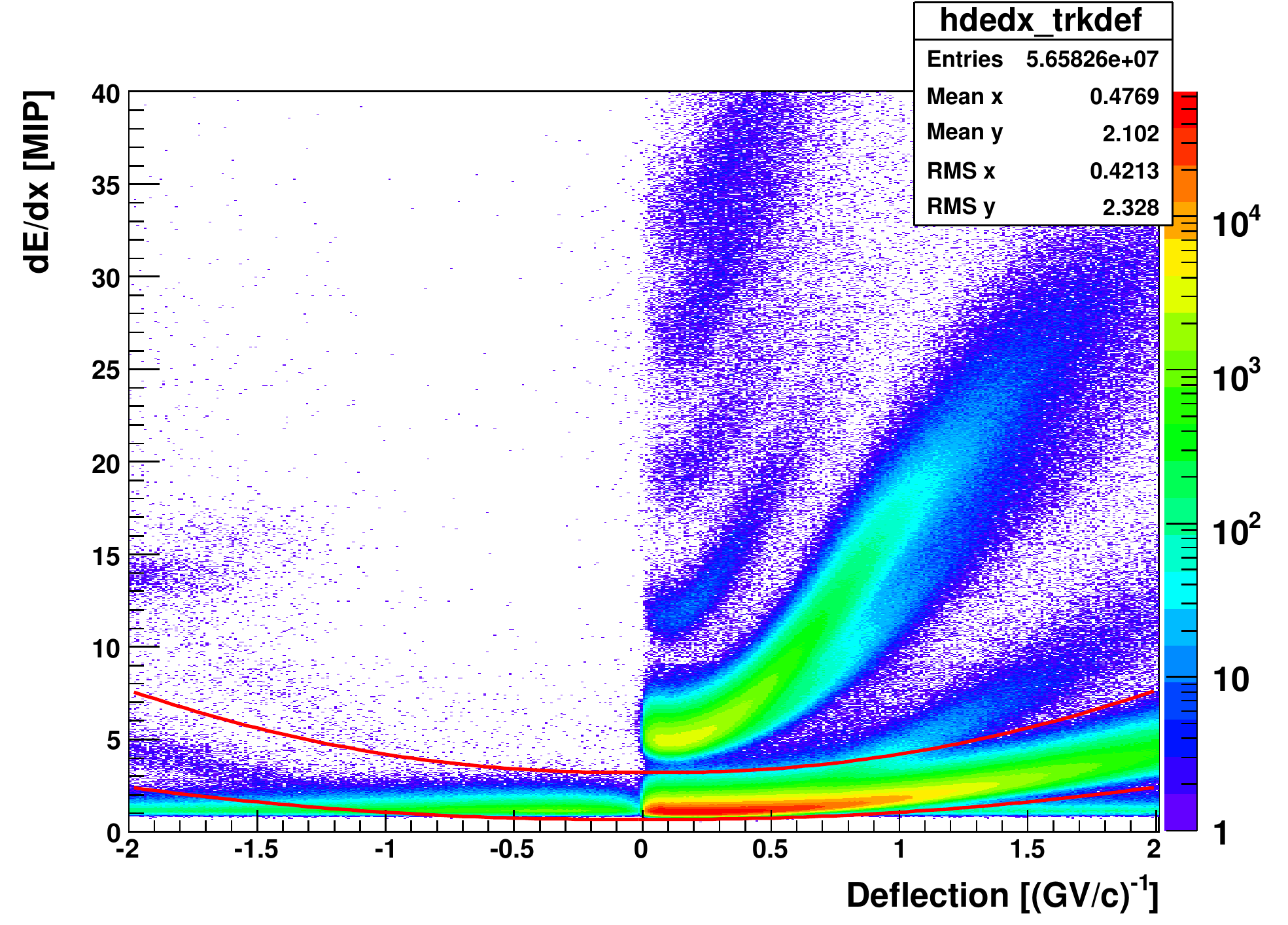}
\end{center}
\caption[Tracker $dE/dx$ selection]{{\footnotesize Tracker $dE/dx$ selection. Particles within the red lines are selected as proton (antiproton) candidates.}}
\label{fig:dedxtrk}
\end{figure}

\item A cut to further remove multiple tracks in the tracker (hereafter referred as TrkMultipleTracksCut). Events which fulfill one of the following conditions are excluded:
\begin{itemize}
\item[(a)] apart from the track, at least 3 hits located on the same side of the track in the x-view and 3 hits totally in the y-view; 
\item[(b)] apart from the track, at least 2 hits located on the same side of the track in the x-view and 2 hits totally in the y-view when hit PMTs are not associated to the track passing through S1 and S2. Information from the ToF is used since if there is at least one hit PMT outside the track on S1 and S2, this indicates a higher possibility of inelastic reaction occurred above the tracker and a lower limit should be put on the number of hits apart from the track.  
\end{itemize}

\item A $\chi^{2}$ cut imposed only for low energy particles: 
\begin{equation}
\chi^{2} \leq 5.99+131 \times \eta^{2} + 99.09 \times {\eta}^{4} 
\end{equation}

This selection, which is about 90\% efficient, places a stronger limit on the $\chi^{2}$ than the one used in the basic tracker cuts. As multiple scattering effects are significant at low energy, this cut is conservatively applied below 14.6~GV to reject particles scattering in the tracker system which might cause an unreliable track reconstruction. 

\item Requirements on high energy particles to clean the tracking position measurements, including a $\chi^{2}$ limitation, a cut placed on the maximum energy release on the first tracker plane and the maximum multiplicity to remove events with accompanying hits due to delta ray emissions, no bad strips and a spacial resolution less than 0.01~mm.

\end{itemize}

\subsubsection{The MDR cut}

\begin{figure}[!htb]
\begin{center}
\includegraphics[width=0.80\textwidth]{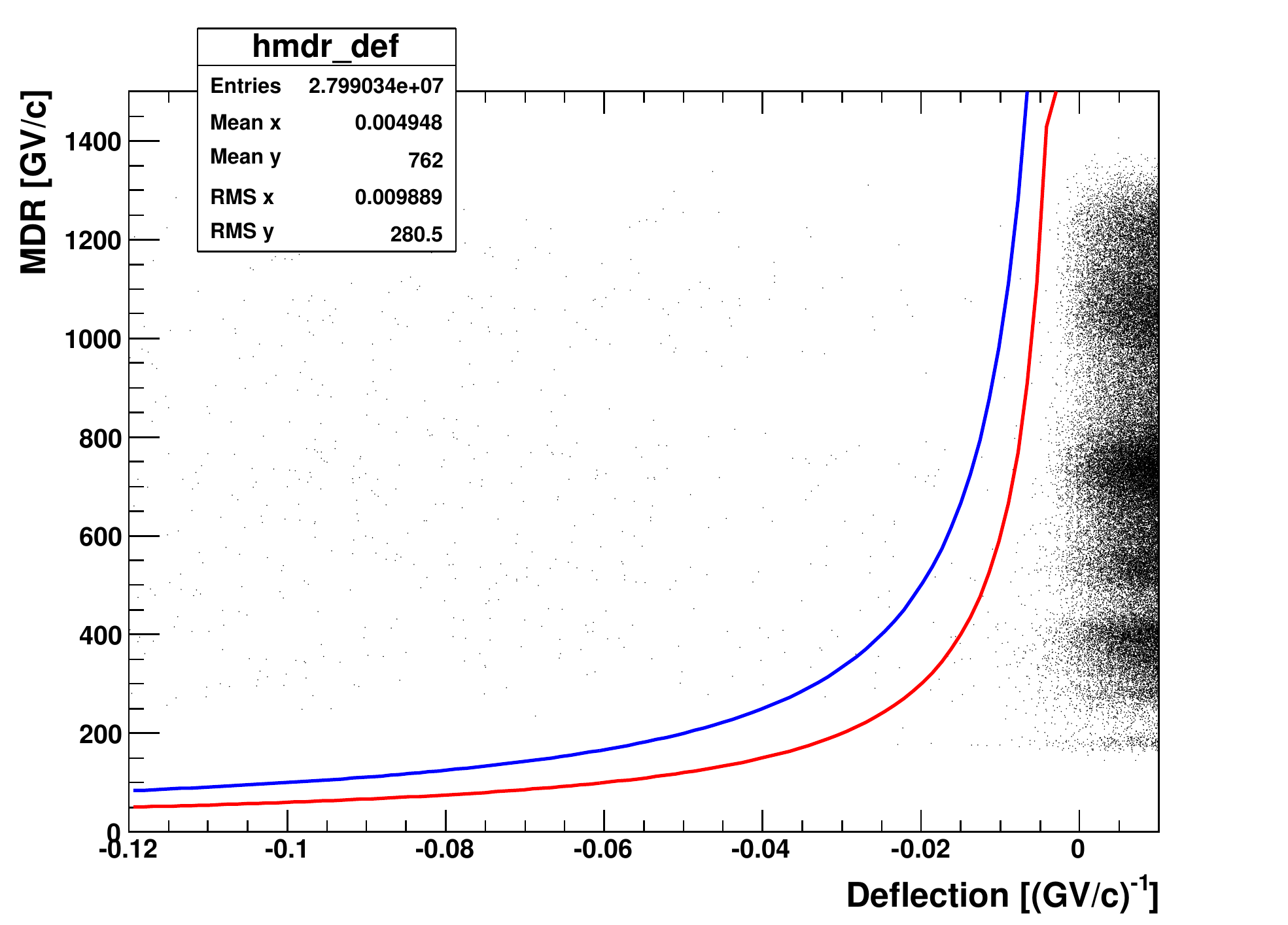}
\end{center}
\caption[The MDR distribution as a function of deflection]{{\footnotesize The MDR distribution as a function of deflection. The red line is $\text{MDR} =6  \times \text{rigidity}$ and the blue line is $\text{MDR} =10 \times \text{rigidity}$. The spillover protons can be clearly observed as the dense area which spills into the negatively-charged side. Events below the curve are rejected.}}
\label{fig:mdrdef}
\end{figure}

A further MDR cut is imposed due to the finite spectrometer resolution. The MDR, which is evaluated for each event during the fitting procedure, is required to be larger than $C \times$~the upper limit of the rigidity bin containing the rigidity of the event, where $C$ represents a coefficient. Since $\text{MDR}=1 / \Delta{ \eta} $, this cut allows to reject events with large associated deflection errors and significantly eliminates the spillover protons, as shown in figure  \ref{fig:mdrdef}. A coefficient of 10 is enough to remove all the spillover protons {\cite{Adriani2009}}. However, since the estimated MDR is about 1 TeV, $C=10$ restricts reconstructed energies to less than 100~GeV. Therefore, a coefficient of 6 is chosen to compromise the rejection power of spillover protons and the upper limit of detectable energy. Moreover, ``sub-bins'' are introduced in highest bins to increase the statistics for high energy antiprotons , i.e. each of the last three rigidity bins is divided into 20 sub-bins. A cut, $\text{MDR} > 6 \times $ the upper limit of the sub-bin containing the rigidity of the event, is used instead of original one in the last three bins. In the final calculation, the spillover contamination has to be precisely determined. 

\subsection{ToF system criteria}

The ToF system measures the time-of-flight of the incident particles thereby providing a velocity determination. This information can be used to reject albedo particles which will cause an upward going proton to be misidentified as a downward going antiproton. It also helps to identify low energy (anti)protons. The $dE/dx$ measurements can exclude heavier particles, low energy electrons and pions. Combined with the hit information in the two top ToF scintillators, multiparticle events or interactions above the tracker can be rejected. The detailed selection cuts are as follows:

\begin{itemize}

\item Events satisfying following requirements are selected to remove multiparticle events where particles traverse different paddles on the same scintillator:
\begin{itemize}
\item[(a)] no more than 1 hit paddle on S11, S12, S22, S21; 
\item[(b)] at least 1 hit paddle on S1 and  S2; 
\item[(c)] no more than 2 hit PMTs outside the reconstructed track on S11, S12; 
\item[(d)] if there is a hit paddle on S11, S12, S21 or S22, its PMTs must be associated to the track extrapolated from tracker or there must be TDC signals belonging to that hit paddle. 
\end{itemize}
No cut is applied on S3 as the particles interacting below the tracker whose rigidities have already been proper measured should be selected in the sample.

\item The y measurement given by the TDC signals on S11 or S22 must be within a 6 cm tolerance margin around the y coordinate extrapolated to that plane from the reconstructed track. This cut checks the consistency between the track measured by the TOF and the one measured by the tracker.

\item Events with $\beta$ consistent with the expectation within $5 \sigma$ for (anti)protons are selected to reject other kind particles up to a few GeV, as shown in figure~\ref{fig:betareciprig}.

\begin{figure}[!htbp]
\begin{center}
\includegraphics[width=0.85\textwidth]{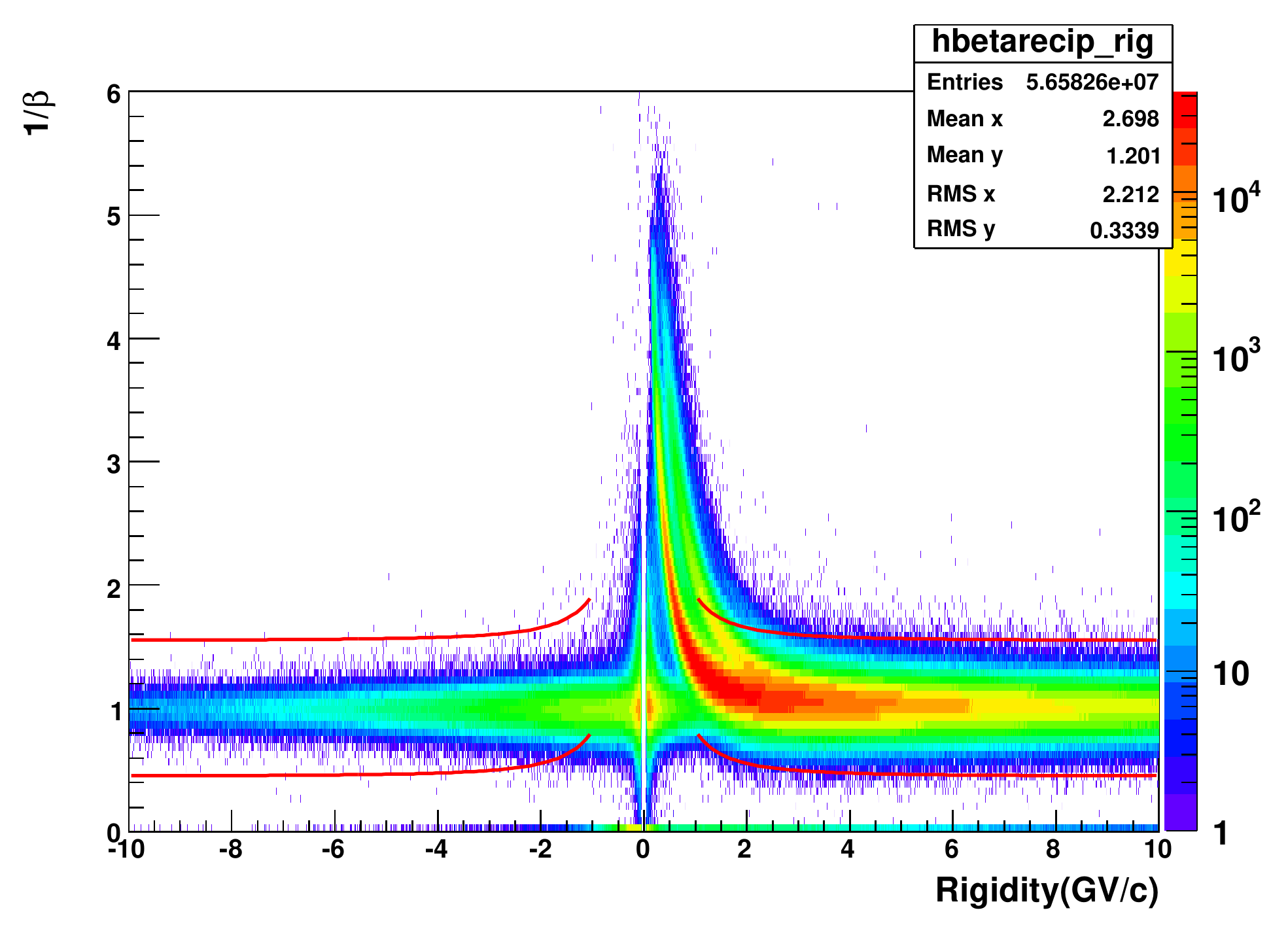}
\end{center}
\caption[$1/\beta$ distribution as a function of rigidity]{{\footnotesize $1/\beta$ distribution as a function of rigidity. Particles within the red lines are selected as (anti)protons candidates.}}
\label{fig:betareciprig}
\end{figure}

\item $dE/dx$ cuts in S1 and S2, similar to those used in the tracker, are applied to select singly-charged particles. 

\item Events with no more than 1 hit PMT outside the track on S11 and S12 and no large energy release in S11, S12, S21 and S22 are selected at low rigidities (below 14.6~GV) to reject particles interacting before the tracker, which are mainly pion background.
\end{itemize}

\subsection{Anticoincidence criteria}\label{sec:acsel}

The anticoincidence system (AC) is very useful to reduce events interacting inside the apparatus and producing secondaries. As mentioned in section \ref{sec:acinstrument}, these kind of events are always accompanied by multiple particles and can yield ``false'' triggers. In order to remove them, requirements are placed on AC, as shown below:
\begin{itemize}
\item Events with no signal in CARD scintillators are selected;
\item Events with no signal in CAT scintillator are selected.
\end{itemize}
No cut is put on the CAS scintillators since particles backscattered from the calorimeter can potentially be registered by the CAS scintillators but should not be rejected \cite{Orsi2006}.

\subsection{Calorimeter criteria} \label{sec:calocuts}
The main task of the calorimeter is to identify antiprotons from an electron background which is significantly more abundant. As discussed in section \ref{sec:calo}, the longitudinal and transverse segmentation of the calorimeter, combined with a $dE/dx$ measurement in each silicon strip, allows the rejection of electromagnetic showers. 

For an electron traversing matter, the radiation loss (so-called \textit{Bremsstrahlung}) exceeds the collision loss above a few tens of MeV and dominates the energy loss as the energy increases. A simplified description of an electromagnetic shower assumes an incident electron of energy $E_{0}$ will lose half its energy to a bremsstrahlung photon after one \textit{radiation length}, $X_{0}$, which corresponds to about 2 planes of the calorimeter. Each bremsstrahlung photon will after one radiation length produce an electron-positron pair, which in turn radiates another photon. This multiplication proceeds to a maximum depth until the energy of the produced secondaries reaches the critical energy $E_{c}$ below which ionization losses start to dominate. The maximum depth is given by $t_{max}=\ln{\left( E_{0}/E_{c} \right)}/ \ln{2}$~radiation lengths. Thus in the cascade the number of particles rises exponentially to a broad maximum and after that the shower decays slowly. The lateral spread of the shower depends mainly on the longitudinal depth and does not significantly depend on the energy of the primary electron. Multiple scatterings in the absorber have an important effect on the lateral spread, yielding two components in the shower: a narrow, strongly collimated central part due to the high-energy particles depositing most of the incident energy and a peripheral component spreading out as the shower penetrates deeper and low energy particles are created. The transverse spread is measured in a unit called the \textit{Moli\`{e}re radius} ($R_{M}$), defined as the average lateral spread of an electromagnetic shower initiated by an electron of energy $E_{c}$ when the electron traverses one $X_{0}$ of material. The electromagnetic shower is about 95\% laterally contained in 2$R_{M}$, which is about 1.8 cm (7.5 strips) for tungsten.

Unlike electromagnetic showers, a hadronic shower results from different inelastic hadronic interactions and consists of a wide variety of particles such as pions, protons and neutrons, with large fluctuations in multiplicity and energy loss between individual showers. On average, about half of the incident energy is acquired by the particles produced in inelastic interactions. The resulting secondaries thus have large transverse momentum and the hadronic shower tends to be more spread out laterally than an electromagnetic one. The longitudinal development of a hadronic cascade is described in units of \textit{(nuclear) interaction length}, $\lambda _{A}$, which has a value of 9.6 cm for tungsten at high energies, indicating that most incident hadrons will interact deeper in the calorimeter or even traverse the calorimeter without interacting.

\begin{figure}[!htbp]
\begin{center}
\includegraphics[width=10cm]{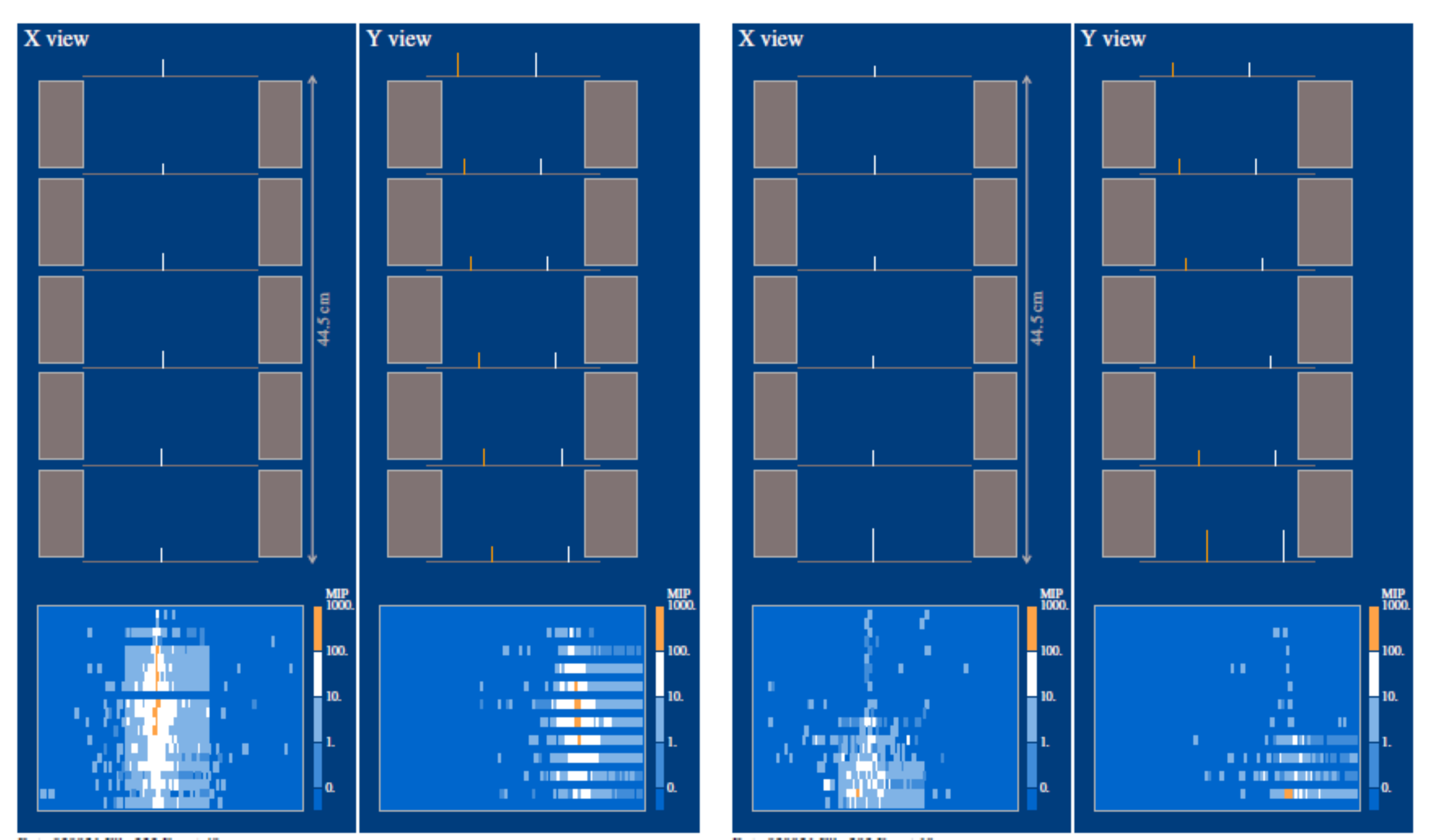}
\end{center}
\caption[An event display of an electron and proton recorded at CERN SpS facility]{{\footnotesize An event display of a 50 GeV electron (left) and proton (right) recorded at CERN SpS facility (taken from {\cite{Boezio2006}}). In the top part of the figure, the two views (X and Y) of the six silicon planes are shown inside the magnetic cavity. In the bottom part of the figure, the two views (X and Y) of the calorimeter are shown. The color scale shows the detected energy in each strip. Orange (blue) area corresponds to a high (low) energy deposit. Vacancies in some layers are due to that some strips were not used in the test. Evident topological and energetic differences between electromagnetic and hadronic showers can be seen.}}
\label{fig:shower}
\end{figure}

Examples of an electromagnetic shower and a hadronic shower are illustrated in figure \ref{fig:shower}. In order to separate hadrons and leptons, an energy dependent criteria has been developed based on both simulations and tests with particle beams {\cite{Boezio2006}}. Several calorimeter variables are used for the lepton/hadron separation. The total energy deposit in the calorimeter, referred to as $q_{tot}$, allows a powerful separation between leptons and hadrons. For electrons, since most showers are contained in the calorimeter, $q_{tot}$ is usually normally distributed for a given incident energy. For hadrons, the distribution of $q_{tot}$ is flat with a sharp peak at low energies for non-interacting particles. Therefore an energy-momentum match exists for electrons, i.e. the $q_{tot}$/rigidity satisfies a quasilinear relation, while for hadrons $q_{tot}$/rigidity assumes a lower value. Other calorimeter variables used in the analysis are described below.

\subsubsection{The starting point of the shower}
\begin{figure}[!htbp]
\begin{center}
\includegraphics[width=0.85\textwidth]{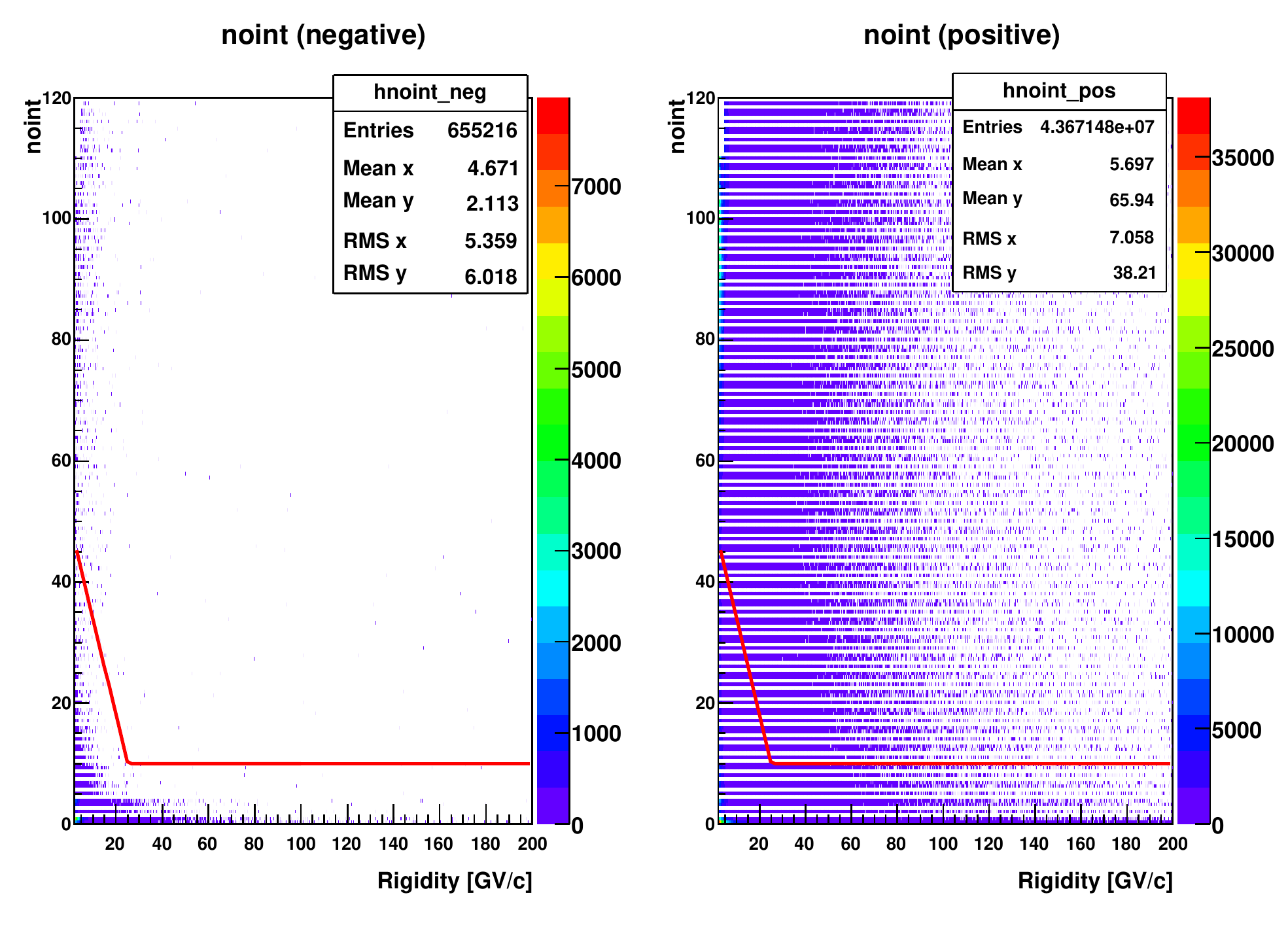}
\includegraphics[width=0.85\textwidth]{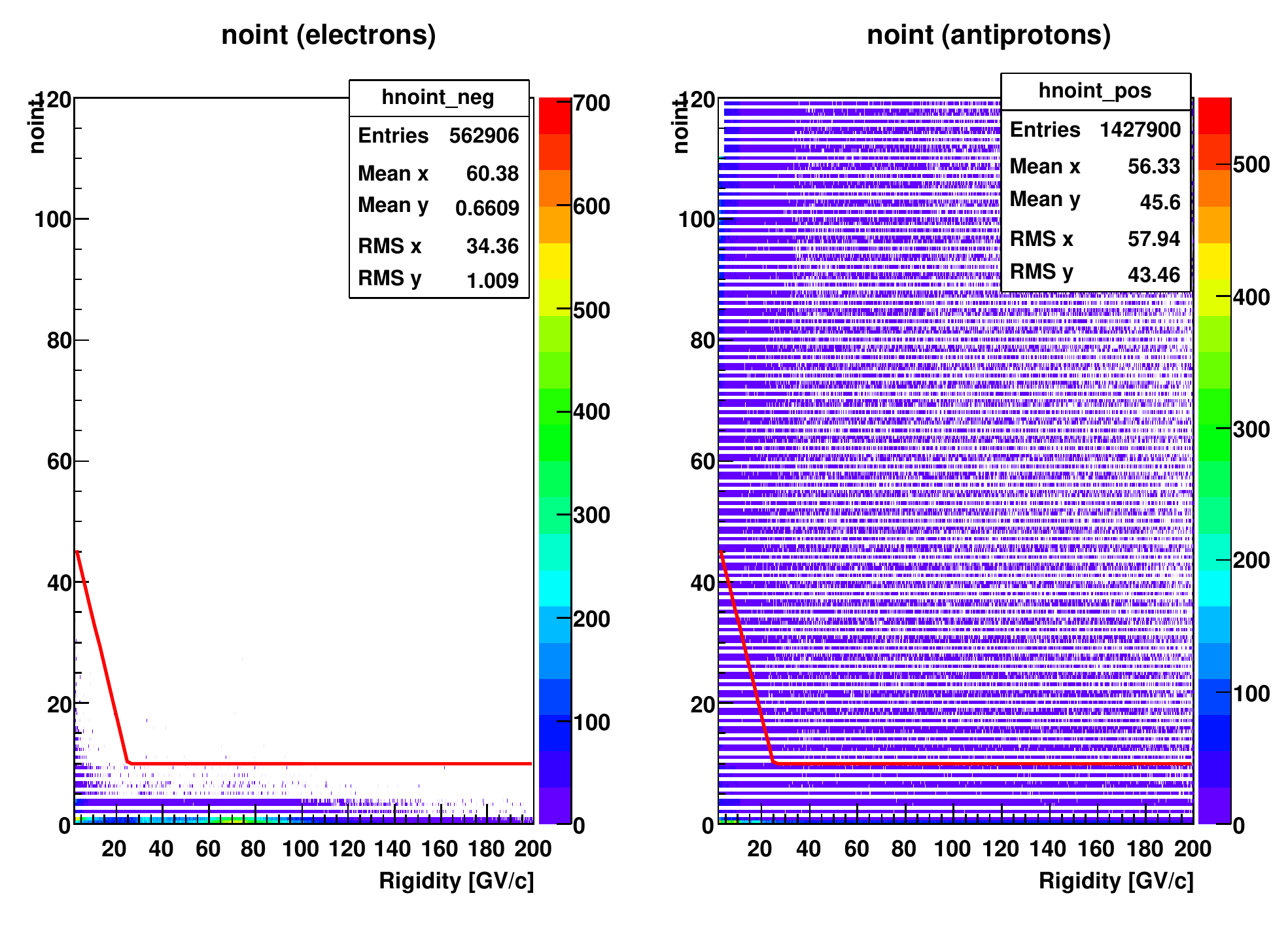}
\end{center}
\caption[Distribution of the variable \textit{noint}]{{\footnotesize The distribution of the variable \textit{noint}. \textit{Top}: \textit{noint} derived from flight data, both for negatively charged particles (mainly electrons) and positively charged particles (mainly protons). \textit{Bottom}: \textit{noint} derived from simulated electrons and antiprotons. The events above the red line are selected as (anti)protons.}}
\label{fig:noint}
\end{figure}

While a hadronic shower has a roughly uniform probability to start in any plane of the calorimeter, an electromagnetic shower is more likely to start in the first three planes. A variable used to characterize this difference is referred as \textit{noint}, given by
\begin{equation}
noint = \sum_{ j=1}^{ 2}  \sum_{i=1 }^{ 22} { \theta}_{ij } \cdot i,  
\end{equation}
where $\theta_{ij}=1$ if the $i$th plane of the $j$th view has strips registering energies compatible with a minimum ionizing particle within 4 mm from the reconstructed shower axis, otherwise $\theta_{ij}=0$. The variable \textit{noint} will increase as the interaction starts in deeper planes. Therefore it assumes low values for electromagnetic showers, and takes higher values for a non- or partially-interacting hadron. The distribution of \textit{noint} from flight data and simulation are shown in figure \ref{fig:noint}.

\begin{figure}[!htb]
\begin{center}
\includegraphics[width=0.85\textwidth]{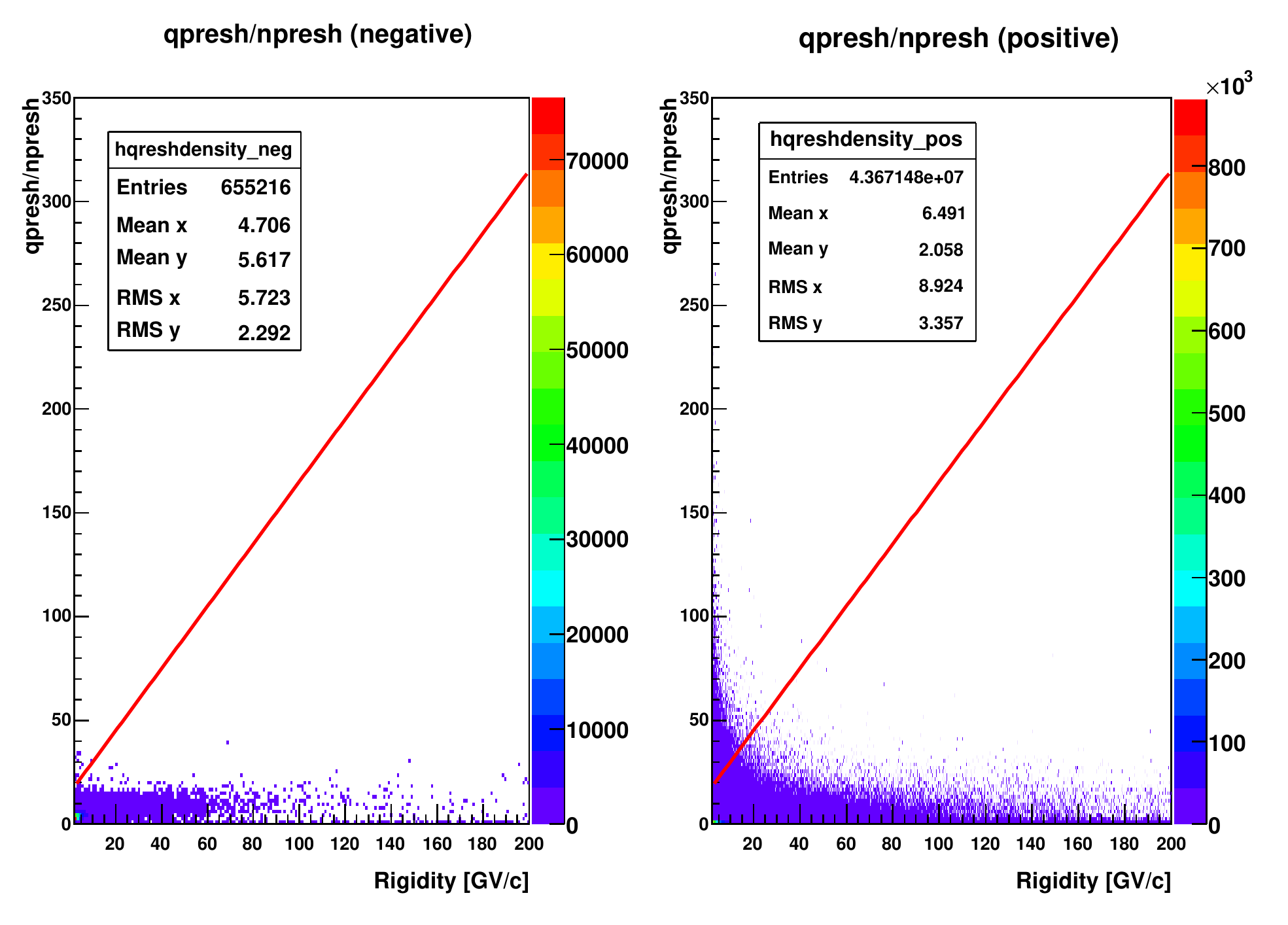}
\end{center}
\caption[The distribution of the variable $qpresh/npresh$]{{\footnotesize The distribution of the variable $qpresh/npresh$ derived from flight data. The negatively charged particles are mainly electrons, and the positively charged particles are mainly protons. The events above the red line are selected as (anti)protons.}}
\label{fig:qpreshdensity}
\end{figure}

Another variable which is sensitive to the starting point of the shower is the ratio of energy deposited in a cylinder of diameter 2 strips ($q_{presh}$) and the number of strips hit in the same cylinder ($n_{presh}$), i.e. $q_{presh}/n_{presh}$, shown in figure \ref{fig:qpreshdensity}. As an electron interacts immediately in the first planes, the average energy deposit in each strip is expected to be lower compared to a hadron which starts to interact deeper in the calorimeter or does not interact.

\subsubsection{The longitudinal and topological profile}
While the energy deposit of an electromagnetic shower decreases after the shower maximum and spreads out laterally, the hadronic showers deposit their energy approximately uniformly and any maximum lies deeper in the calorimeter. A quantity related to the longitudinal profile is defined as 
\begin{equation}
q_{core}=\sum_{ j=1}^{ 2}  \sum_{i=1 }^{ pl_{max}} {Qhit}_{ij } \cdot i, 
\end{equation}
where ${Qhit}_{ij }$ is the energy released in the \textit{j}th view of \textit{i}th plane within a cylinder of radius 2$R_{M}$ centered on the shower axis, and $pl_{max}$ is the calculated electromagnetic shower maximum for a given incident energy provided by the tracking system.  

Furthermore, for an electromagnetic shower, before achieving the shower maximum, the shower multiplication is expected to increase with shower depth and the shower particles should be collimated along the shower axis. A variable called $n_{core}$ is used to reveal this behavior, given by
\begin{equation}
n_{core}=\sum_{ j=1}^{ 2}  \sum_{i=1 }^{ {pl_{max}}} {Nhit}_{ij } \cdot i, 
\end{equation}
where ${Nhit}_{ij }$ is the number of hit strips in the \textit{j}th view of \textit{i}th plane within a cylinder of radius 2$R_{M}$ centered on the shower axis, and the plane number $pl_{max}$ is closest to the calculated electromagnetic shower maximum of the \textit{j}th view.

The energy density in the shower core weighted by the depth in the calorimeter, $q_{core}/n_{core}$, which is more sensitive to the shower difference between hadrons and electrons, is finally used to separate hadrons and leptons as shown in figure \ref{fig:qdensity_ncyl}.


\subsubsection{The lateral profile}
\begin{figure}[!htbp]
\begin{center}
\includegraphics[width=0.85\textwidth]{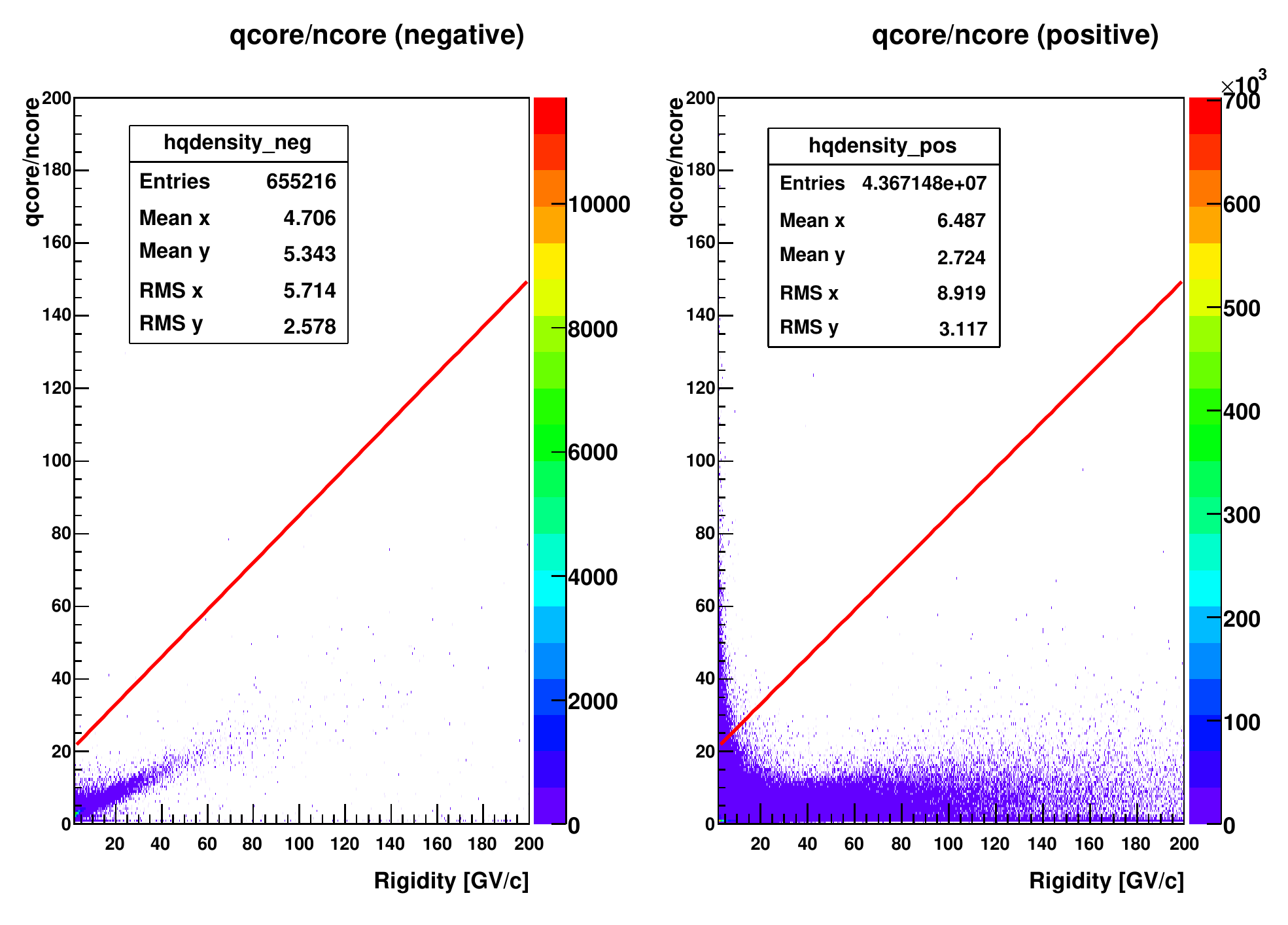}
\includegraphics[width=0.85\textwidth]{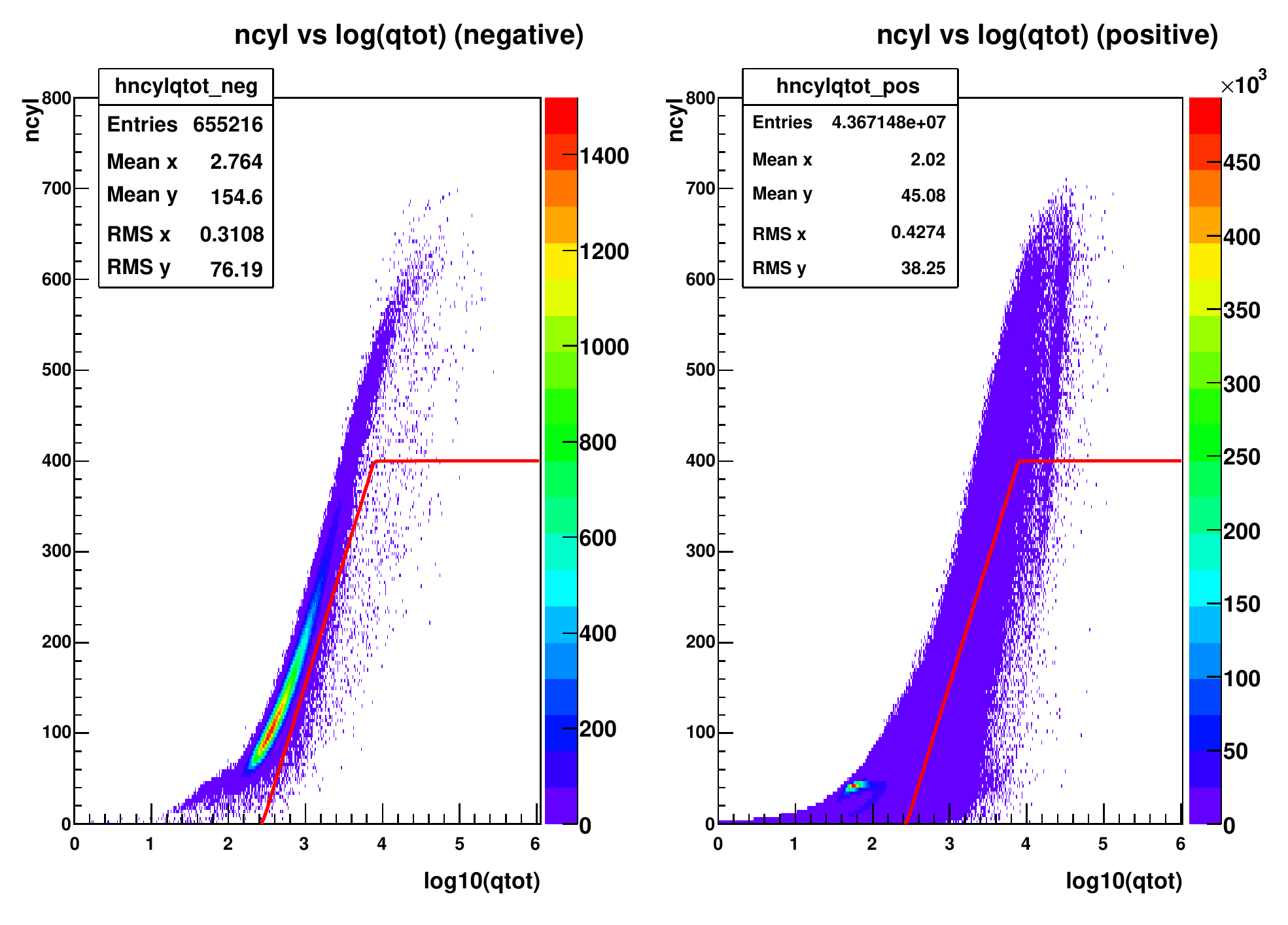}
\end{center}
\caption[The distribution of the variable $n_{cyl}$]{{\footnotesize The distribution of the variable $q_{core}/n_{core}$ (\textit{top}) and the variable $n_{cyl}$ (\textit{below}). The negatively charged particles are mainly electrons, and the positively charged particles are mainly protons. The events below the red line are selected as (anti)protons.}}
\label{fig:qdensity_ncyl}
\end{figure}

While the lateral development of a hadronic shower depends on the traverse momentum of produced secondaries, which usually carry about half of the incident energy and thus cause a wide lateral spread, an electromagnetic shower has a lateral spread due to the multiple scattering of low energy particles which is usually less broad. 
A variable called $n_{cyl}$, which is the number of strips hit in the cylinder of radius 2$R_{M}$ around the shower axis, is sensitive to the lateral profile. Due to the difference between the behaviour of leptons and hadrons in the calorimeter, $n_{cyl}$ assumes a higher value for electrons than for hadrons (shown in figure \ref{fig:qdensity_ncyl} as well).

\subsection{Selecting Galactic particles}\label{sec:selgalactic}

As discussed in section \ref{sec:CRintroduction}, the Earth's geomagnetic field prevents low energy particles from reaching the atomosphere. In order to select Galactic cosmic rays, only events with rigidities larger than the minimum value needed for a cosmic ray to penetrate the geomagnetic field and reach PAMELA are selected:
\begin{equation}
{rigbin}_{lowerlimit} > \text{cutoff}_{PAMELA} =1.3 \times SVC ,
\end{equation}
where ${rigbin}_{lowerlimit}$ means the lower limit of the rigidity bin. The Stoermer vertical cutoff (SVC), as defined in section \ref{sec:CRintroduction}, is estimated using the satellite position. A coefficient of 1.3 is used here to ensure a robust selection of Galactic particles. 

\section{Selection efficiencies} \label{sec:efficiencies}

In order to determine the antiproton flux, the number of antiproton candidates surviving all the selection criteria is compensated for selection efficiencies. The efficiency of a set of selection cuts can be calculated as the surviving fraction of events when applying those cuts to a selected sample of antiprotons. Efficiency samples can be obtained in different ways: using simulations, test beams or flight data. The purity of the sample can be controlled well by using the first two methods. However, conditions can change during the flight which make the test beam data less useful. For example, during PAMELA flight, the performance of the Viking VA1 chips (see figure \ref{fig:ladder}), responsible for the readout of tracker, has degraded with time. In some periods, a S11 PMT was not operational and the PMT high voltage levels of the ToF varied. These conditions are impossible to reproduce in a test with particle beams. They also complicate the simulation model. Therefore, the flight data itself is used which intrinsically include the detector performance over time. Since independent detectors have to be used to select the sample of particles and determine their rigidities, biased samples might be introduced if the response of different detectors are correlated. To better understand and estimate the efficiencies, simulations are therefore used to cross-check the results.

As it is impossible to select an unbiased, statistically significant antiproton sample from flight data, the efficiencies are derived from a proton sample with the assumption that protons and antiprotons behave identically in the detectors except for inelastic interactions. Therefore the only exception is the calorimeter, where the antiproton efficiency is corrected from simulations, taking into account that antiprotons and protons have different cross section for inelastic interactions. The efficiencies of the selection criteria discussed in section \ref{sec:selection} are presented below, in the order they were applied.

\subsection{Basic tracker cuts efficiency}
The basic tracker cuts are used to reconstruct a good quality track and give a reliable rigidity. Thus the efficiency sample of basic tracker cuts should be obtained without using the tracker system. In the low energy region, the rigidity can be obtained with the ToF system as $\beta m / \sqrt{1- {  \beta }^{ 2}}$. Due to the limited time resolution of the ToF system, this method is only applied below 1.7~GV/c \cite{HofverbergPhd}. For higher energy particles, the ToF rigidity resolution worsens since a small difference in the time of flight measurements produces a large relative difference in the reconstructed $\beta$. Therefore, the tracker system is the only detector which can determine the rigidity of particles for the rigidity range of interest in this work. Fortunately, the efficiency, depending on the energy released in silicon planes, the curvature of the track and the multiple scattering, is expected to be constant for relativistic particles. For particles with a certain charge, the energy deposited in silicon planes is proportional to ${\beta}^{-2}$. A particle deposits energy in the silicon planes and consequently creates a number of \textit{clusters}, where a cluster is defined as one or more strips in the sensitive silicon plane with a signal 7 standard deviations from the intrinsic noise of the channel. The number of created clusters naturally depends on the amount of energy released in the plane. For low velocity particles,  as the multiplicity of clusters increases while the velocity decreases, the probability that the tracking algorithm will find a unique track decreases. However, for relativistic particles where $\beta$ approaches unity, the probability is expected to be constant and thus the tracker efficiency also. The efficiency effected by the curvature of the track is also constant for relativistic particles since they produce nearly straight tracks. The last effect, characterized as the scattering angle ($\theta  \sim z/ \left ( p \beta   \right ) $) where $z$ represents the particle charge and $p$ the momentum, is negligible for singly-charged and $\beta  \simeq  1$ particles. Hence, for the basic tracker efficiency calculation, a sample is selected without rigidity determination. Once the basic tracker efficiency is obtained, the basic tracker cuts can be used to select efficiency samples for other selection criteria, allowing rigidity dependent efficiencies to be estimated. 

\subsubsection{Selecting an experimental proton sample} 
To select a clean proton sample, a set of cuts are applied to raw flight data as follows. In particular, in order to ensure the tracks of incident particles are inside the fiducial acceptance without using the spectrometer, the calorimeter is used to identify particle trajectories.

\begin{description}
\item[Single particle selection] The anticoincidence criteria described in section \ref{sec:acsel} is applied. The number of hit paddles on S11, S12, S21, S22 is required to be not larger than one. The number of hit paddles on both S1 and S2 should be at least one. 

\item[Charge one particle selection] 
The energy released on S1 should be less than 1.8 MIP.

\item[Downgoing and high energy particle selection]  \hfill \\
\begin{itemize}
\item $0.92 < \beta <1.10$. This cut selects downgoing particles whose velocity is positive and relativistic particles with $\beta$ close to 1, considering the resolution of $\beta$ is about 0.08.
\item Particles must cross both the last x-view plane and the last y-view plane of the calorimeter.
\end{itemize}

\item[Geometry constraint] 
Particle tracks are reconstructed inside the calorimeter iteratively. At each step, the hits in the calorimeter are fitted to define a single track, and the most distant points are rejected. The same fitting procedure is repeated until no hit can be found departing from the track further than a certain distance. The extrapolated tracks with good ${\chi}^{2}$ are required to be inside the acceptance, defined by the geometry of the 8 tracker planes and the 6 ToF planes with a 0.7 cm tolerance applied to each side of every plane. This is the minimum tolerance which guarantees no underestimation of the efficiency due to uncertainties in  of the calorimeter track fit and mechanical tolerances {\cite{SergioPrivate}}. This cut also reject pions which are low rigidity particles with highly curved trajectories and then enter the calorimeter with an inclined track. Their tracks back-propagated from the calorimeter therefore are not expected to be inside the acceptance.

\item[Electron rejection]

The energy deposited in the calorimeter strip closest to the track divided by the total energy deposit in calorimeter should be larger than 0.8 (this cut is referred to as CaloNotIntCut). This cut discards all the particles interacting in the calorimeter which naturally includes electrons since they interact immediately when they traverse the calorimeter.
\end{description}

\subsubsection{The efficiency of basic tracker selection}\label{sec:trkbas_eff}
\begin{figure}[!htbp]
\begin{center}
\includegraphics[width=0.85\textwidth]{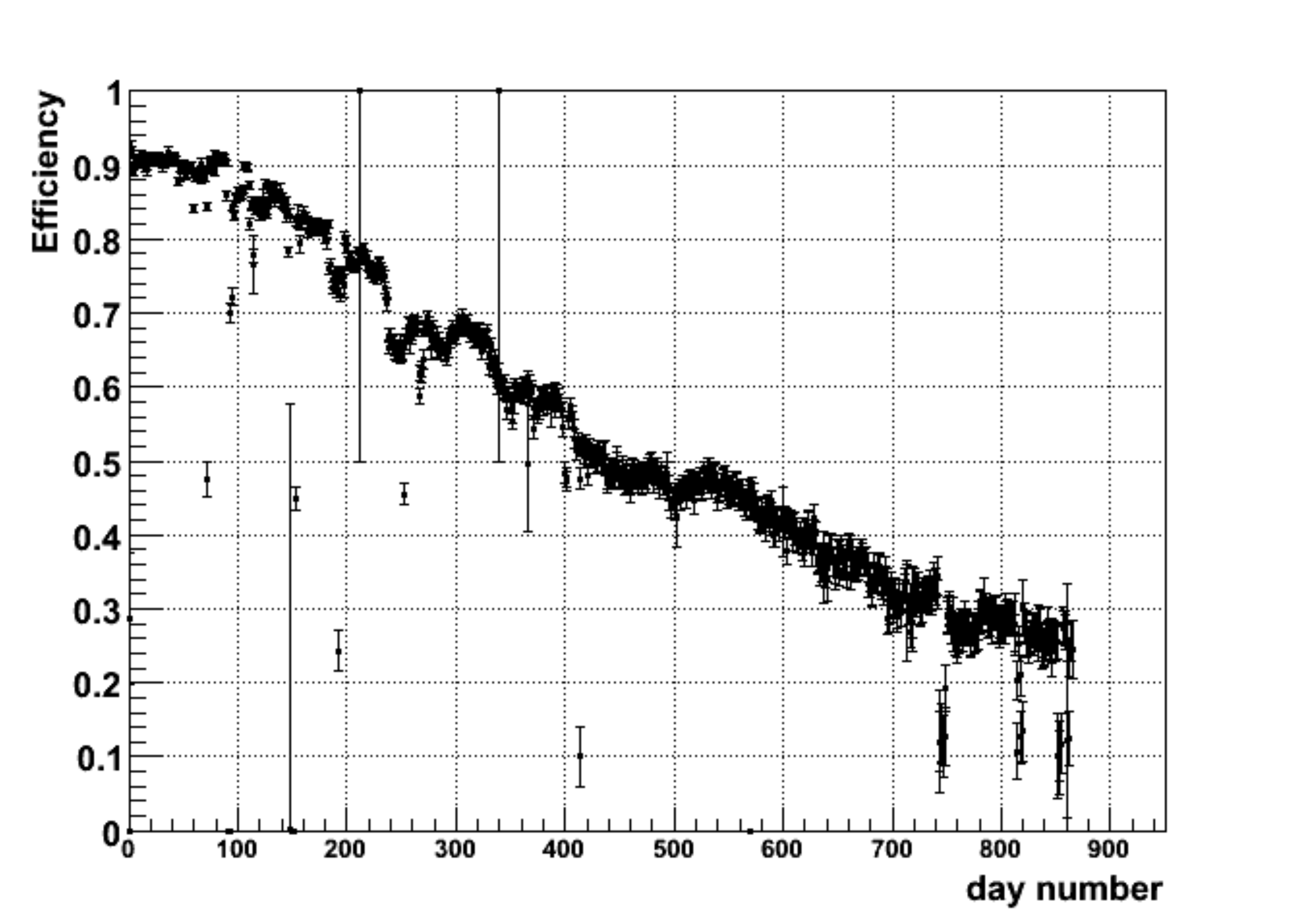}
\end{center}
\caption[Time evolution of the basic tracker cuts efficiency]{{\footnotesize Time evolution of the basic tracker cuts efficiency.}}
\label{fig:trkbasicdailyeff}
\end{figure}

After applying the cuts described above to flight data, a proton sample is selected. The efficiency is defined as the fraction of events in the sample passing the basic tracker cuts detailed in section \ref{sec:tracker_sel}. The efficiency calculated for each day on orbit is plotted in figure \ref{fig:trkbasicdailyeff}. The observed decrease is due to the tracker degradation, i.e. the efficiency falls gradually as the number of malfunctioning VA1 chips increases. Some other tracker performance issues can cause a short-term change of efficiency. For example, in 2006, from October 7th to October 11th (corresponding day numbers are 89 to 93 in figure \ref{fig:trkbasicdailyeff}), the power supply for the tracking system had a problem hence the tracking system was working only for a very short time. As a result, efficiencies for those days are zero or close to zero. In order to solve this problem, the power supply was changed to a redundant backup and DSP\footnote[1]{Digital Signal Processor - a part of the tracker electronics.} 7 was switched off until October 23rd, which means one layer was not working on the y view with a resulting drop in efficiency. The days where the tracker was switched off will be excluded from the flux reconstruction to avoid an overestimation of the live time and an underestimation of efficiency. 

\begin{figure}[!htbp]
\begin{center}
\includegraphics[width=0.85\textwidth]{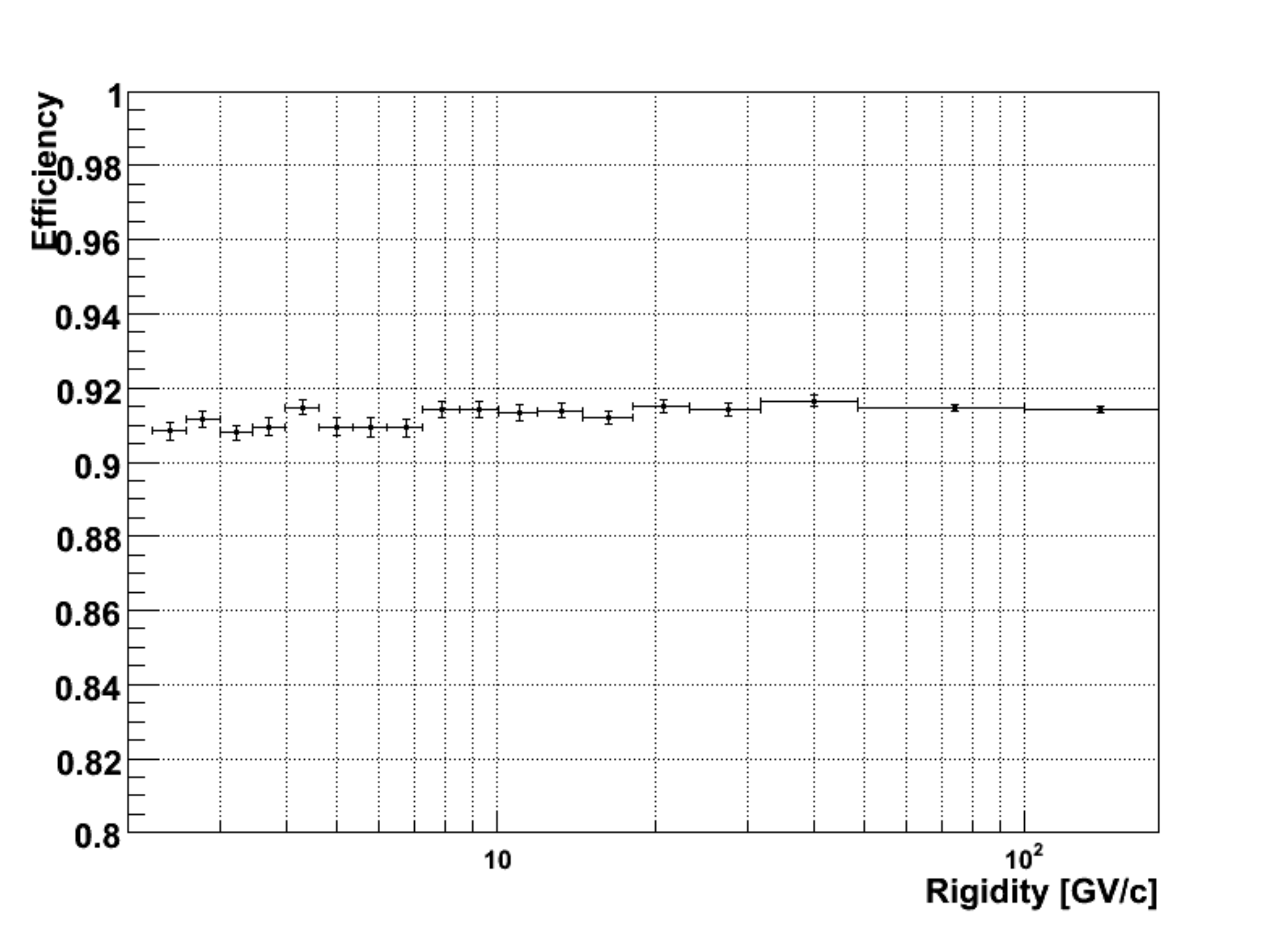}
\end{center}
\caption[The simulated basic tracker cuts efficiency for July 2006]{{\footnotesize The simulated basic tracker cuts efficiency for July 2006.}}
\label{fig:trkbasicsimeff}
\end{figure}

The simulated basic tracker efficiency for the tracker configuration with malfunctioning VA1 chips as in flight during July 2006 is 91.3\% with a variation less than 0.7\%, as presented in figure \ref{fig:trkbasicsimeff}. The simulated result is slightly higher than the value obtain from flight data for that month, which is $\left( 90.6 \pm 0.1\right)\%$. The discrepancy between simulation and flight data is expected because the simulation can not give complete information regarding the $\chi^{2}$. However, the result shows a rather constant efficiency (variation less than 0.7\%), which is consistent with the expectation that the efficiency is independent of rigidity.

The efficiency derived by selecting protons not interacting in the calorimeter might be biased since the interacting protons may cause particles to be back-scattered from the calorimeter and reduce the probability for the tracker algorithm to find the correct single track. However, this underestimation is estimated to be negligible \cite{Wu_LicThesis}.

\subsection{Additional tracker cuts efficiency}
Since the  efficiency of the basic tracker cuts has been estimated, these cuts can be used to select efficiency samples for the other selection cuts, the efficiencies of which can be referred to as relative efficiencies. For example, if the selection criteria applied to select candidates of a certain species are grouped as A, B, C, etc, the sample for C's efficiency should be obtained by applying A plus B and other necessary cuts to reject extra background. Here C's efficiency is a relative efficiency, while A's efficiency is called an absolute efficiency. 

Instead of checking that tracks extrapolated from the calorimeter fall within the acceptance, the efficiency sample for the additional tracker cuts is derived by considering tracks reconstructed by the tracker system, i.e. applying the basic tracker cuts and the geometrical selection described in section~\ref{sec:tracker_sel}.

\begin{figure}[!htbp]
\begin{center}
\includegraphics[width=0.85\textwidth]{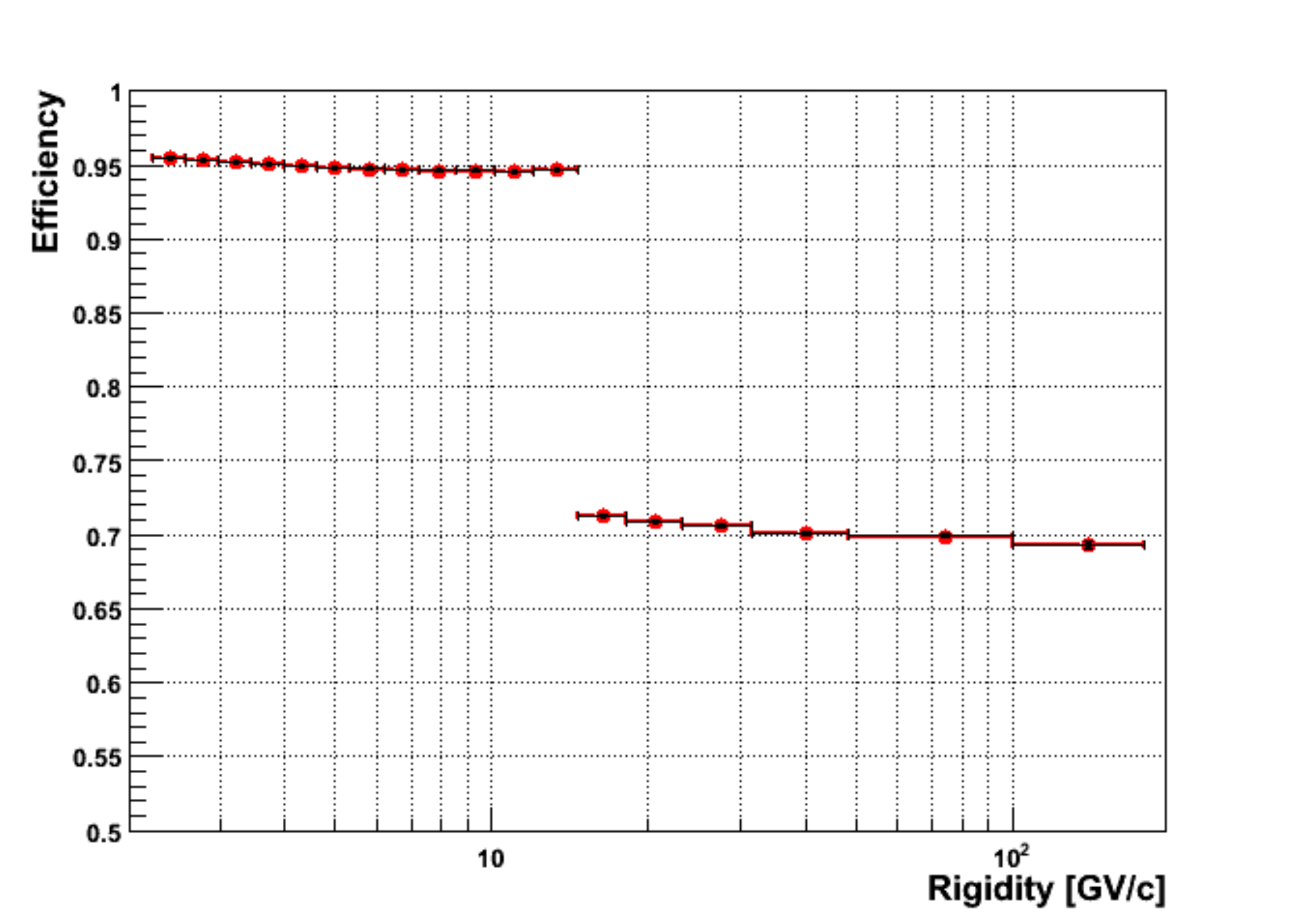}
\end{center}
\caption[Additional tracker efficiency derived by using and without using cuts on S2]{{\footnotesize Additional tracker efficiency integrated over the whole live time using flight data. The black (red) points are derived by using (without using) cuts on S2.}}
\label{fig:trkaddeff}
\end{figure}

\begin{figure}[!htbp]
\begin{center}
\includegraphics[width=0.85\textwidth]{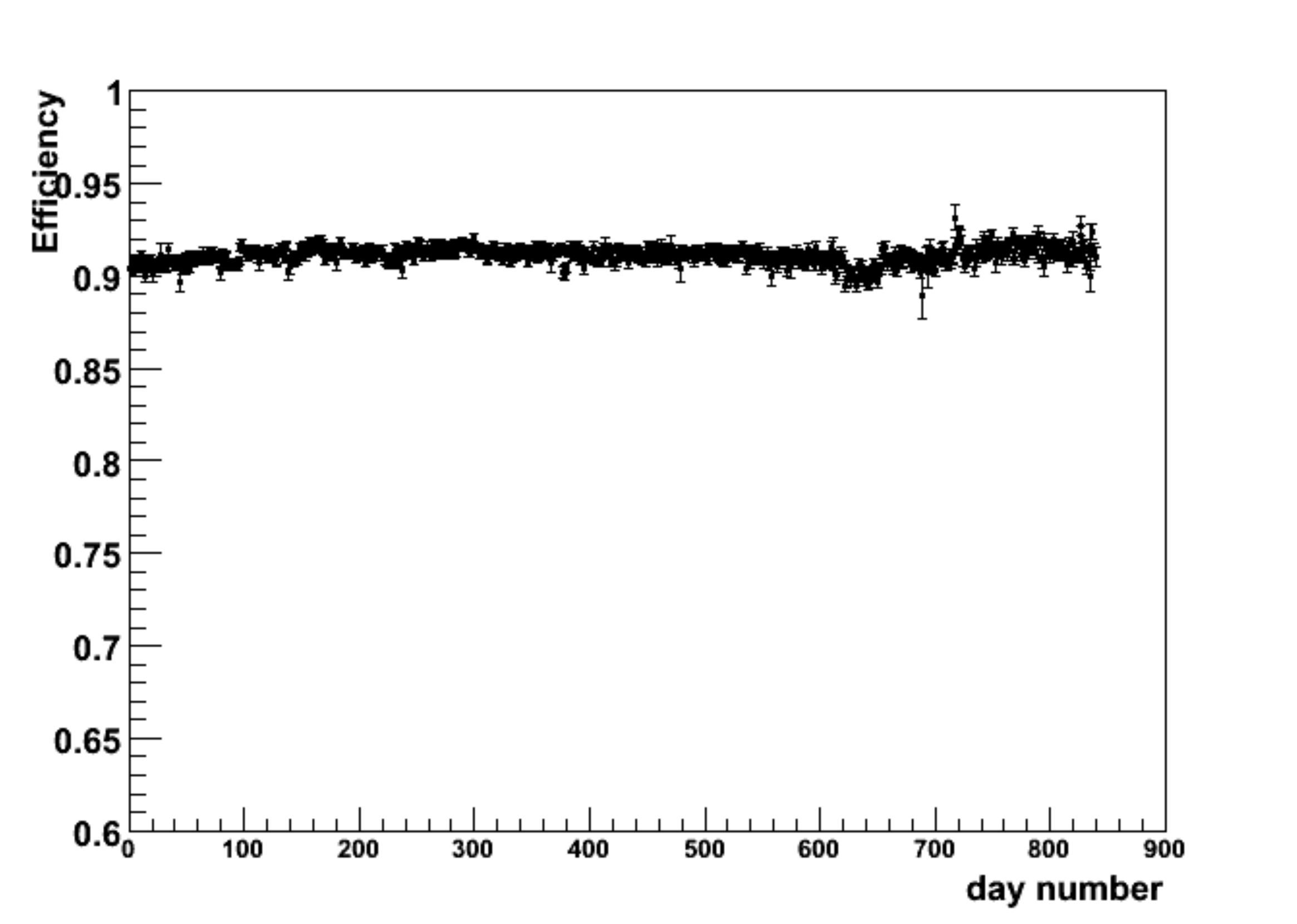}
\end{center}
\caption[Time evolution of the additional tracker cuts efficiency]{{\footnotesize Time evolution of the additional tracker cuts efficiency.}}
\label{fig:trkaddefftime}
\end{figure}

The rigidity dependent efficiency of the additional tracker cuts, integrated over the whole live time, is shown in figure \ref{fig:trkaddeff} (black points). A discontinuity occurs at 14.6 GV since different cuts are applied below and above this rigidity. However, for the rigidity range where the same cuts are employed, the dependence on rigidity is fairly constant. The slight decrease at high rigidities may be caused by delta rays. As the energy increases, more delta rays are generated thereby releasing more energy in the silicon planes and producing a higher hit multiplicity. These events are subsequently rejected by the delta ray cut (explained in \ref{sec:addtrkcuts}). The time variation of the overall efficiency is shown in figure \ref{fig:trkaddefftime}. The changes of the tracker performance are visible in the efficiency. 

A correlation between the ToF and the tracker exists, as the cut used to remove multiple tracks in the tracker, named TrkMultipleTracksCut, is based on the PMT information from S1 and S2. Therefore a cut using information from S2, which might be sensitive to TrkMultipleTracksCut, is removed from the sample selection criteria to derive the efficiency. The resulting efficiency is compared with the one produced using the S2 cut in figure \ref{fig:trkaddeff}. Evidently, the difference between the two cases is negligible and will be omitted in the calculation of the efficiency.

\subsection{ToF efficiency} \label{sec:tofeff}
A sample of protons has been derived without using the ToF system. Single down-going charge one protons are selected with good quality tracks inside the fiducial acceptance. The basic tracker cuts and the additional tracker cuts are both used to choose a sample of singly-charged, single particle events. The tracks of these particles must be inside the fiducial acceptance. The AC cuts are used to further reduce secondary events. Since the deflection of particles is required to be positive, the remaining background are electrons, positrons  and albedo singly-charged particles. Therefore the cut CaloNotIntCut is applied to reject the electrons and positrons. Events with positive deflections are required which means the remaining contamination is upward-going antiprotons and therefore negligible ($\sim 10^{-4}$ of protons). 

\begin{figure}[!htbp]
\begin{center}
\includegraphics[width=0.85\textwidth]{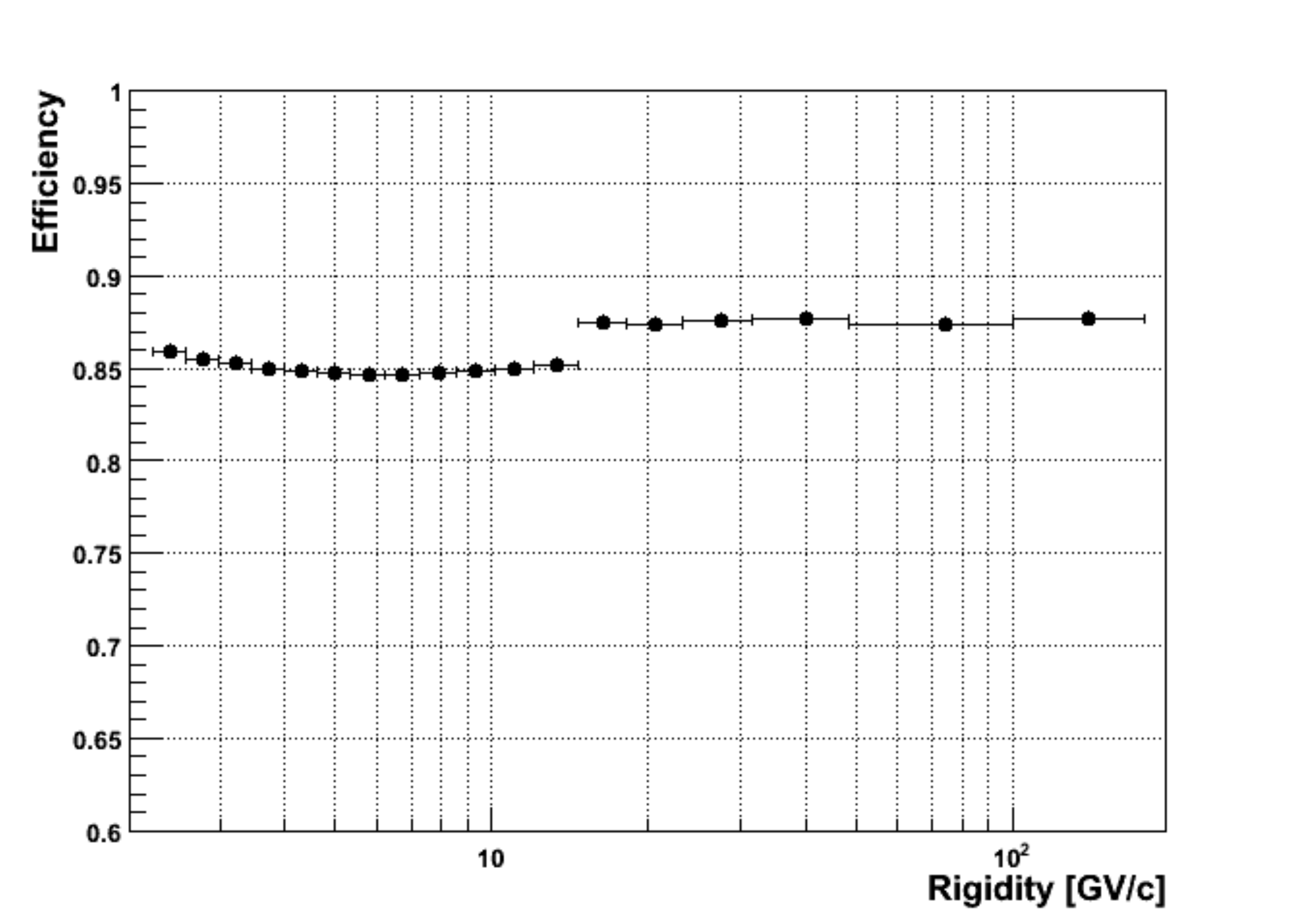}
\end{center}
\caption[ToF efficiency based on flight data and integrated over the whole live time]{{\footnotesize ToF efficiency based on flight data and integrated over the whole live time.}}
\label{fig:tofeff}
\end{figure}

\begin{figure}[!htbp]
\begin{center}
\includegraphics[width=0.85\textwidth]{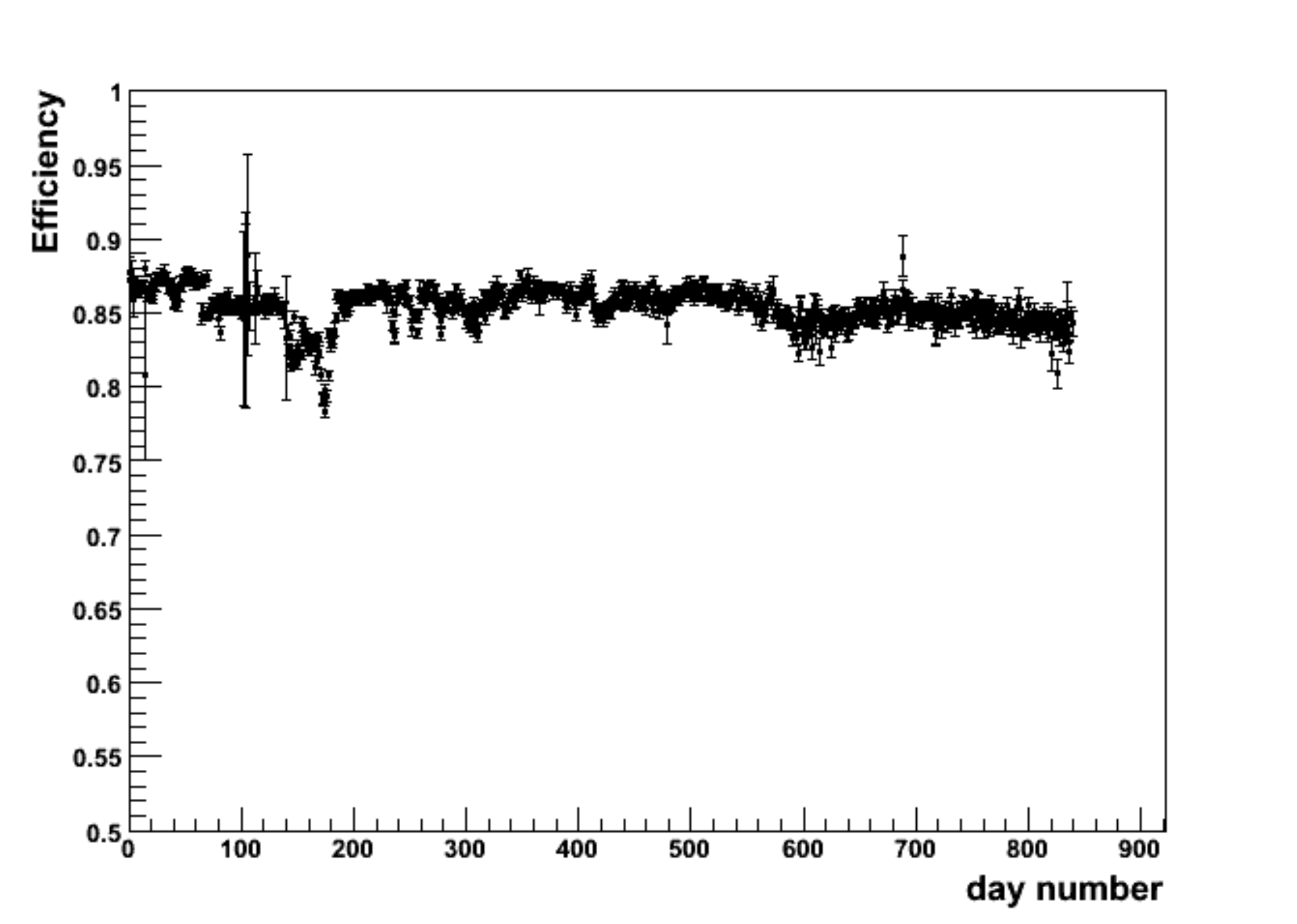}
\end{center}
\caption[Time evolution of the ToF efficiency]{{\footnotesize Time evolution of the ToF efficiency.}}
\label{fig:tofefftime}
\end{figure}

The rigidity dependent efficiency of the ToF cuts, integrated over the whole live time, is shown in figure \ref{fig:tofeff}.  A discontinuity occurs at 14.6 GV since different cuts are applied below and above this rigidity. However, for the rigidity range where the same cuts are employed, the efficiency does not vary more than 1\%. The time variation of overall efficiency is shown in figure \ref{fig:tofefftime}. A fluctuation can be observed which is caused by a known variation in the ToF performance. For example, around the day number of 20, 40, 180, drops of efficiency correspond to changes in the PMT high voltage settings. Around the day 70, a failure of one S11 PMT introduces a decrease of efficiency. A variation around the 200th day is due to a change in TDC threshold.

\begin{figure}[!htbp]
\begin{center}
\includegraphics[width=0.85\textwidth]{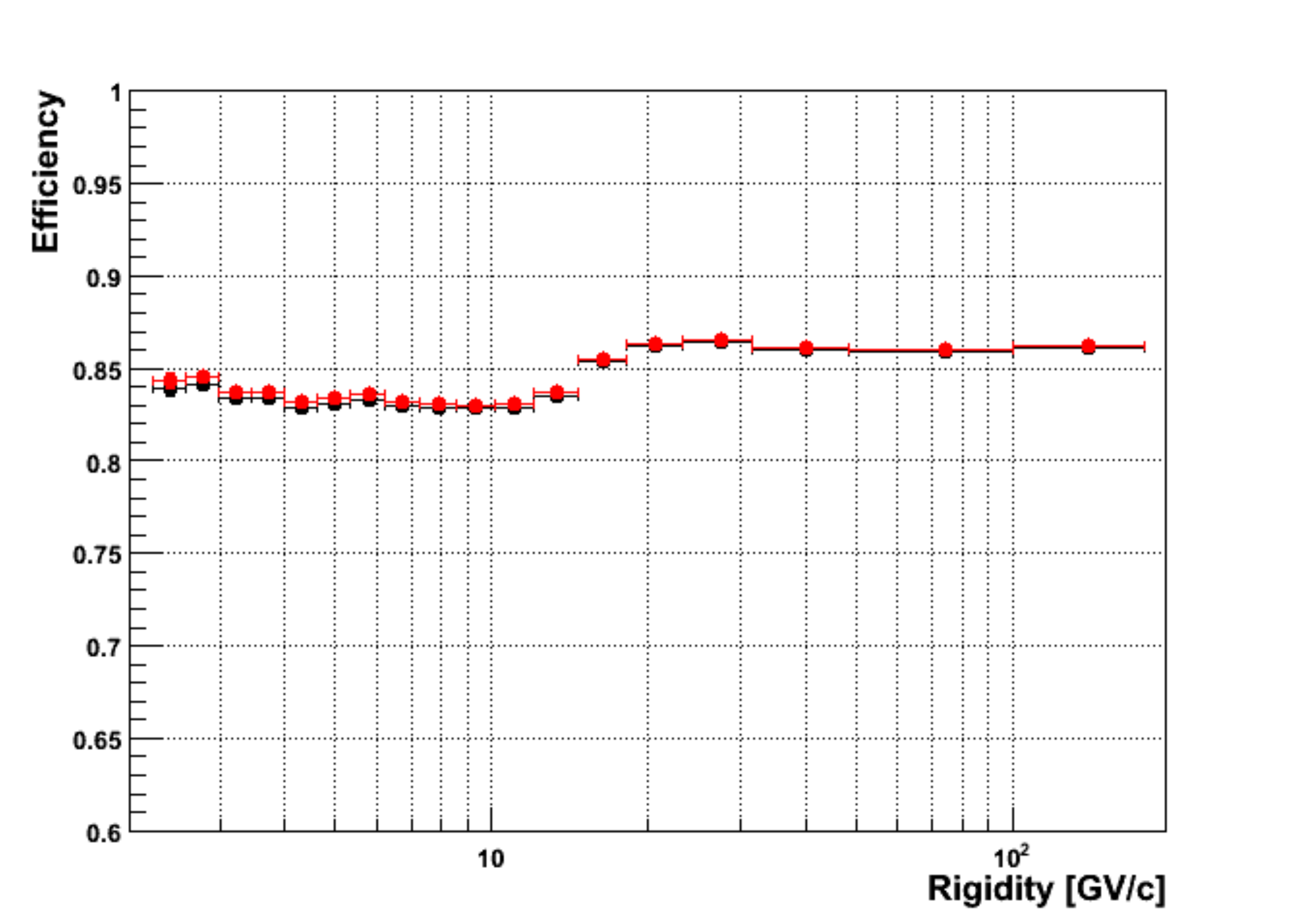}
\end{center}
\caption[Simulated ToF efficiency derived by using and without using AC cuts]{{\footnotesize Simulated ToF efficiency. The red (black) one is the efficiency derived by using (without using) AC cuts.}}
\label{fig:tofanticompare}
\end{figure}

The events which produce delta rays above the spectrometer, or are backscattered from the calorimeter while not interacting before S3, should be part of the proton sample. However, a ``no hit'' requirement on CAT and CARD removes a fraction of this kind of events, resulting in an overestimation of the ToF efficiency. In order to estimate the correlation between ToF and AC, the simulated efficiencies are derived by using the AC cuts and without using the AC cuts, called $\varepsilon_{tof\_ac}$ and  $\varepsilon_{tof\_noac}$ respectively. The results are shown in figure \ref{fig:tofanticompare}. As expected,  $\varepsilon_{tof\_ac}$ is higher than  $\varepsilon_{tof\_noac}$, since AC cuts remove a fraction of particles which should be included in the sample and will be rejected by the ToF cuts. The difference between the two efficiencies is less than 0.5\%. This is added to the ToF efficiency as a systematic error.

\subsection{Anticoincidence efficiency}
\begin{figure}[!htbp]
\begin{center}
\includegraphics[width=0.85\textwidth]{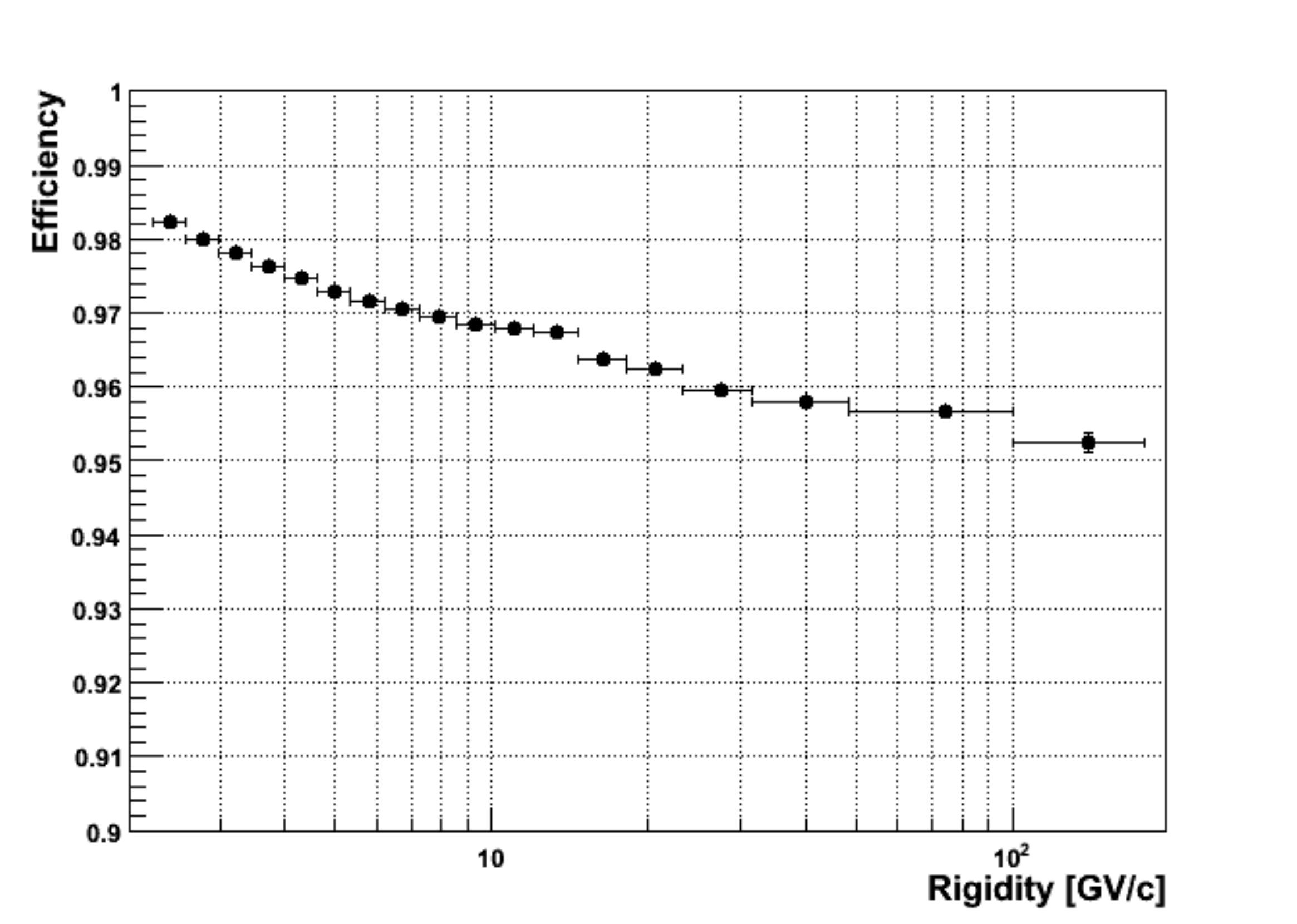}
\end{center}
\caption[AC efficiency based on flight data and integrated over the whole live time]{{\footnotesize AC efficiency based on flight data and integrated over the whole live time.}}
\label{fig:antieff}
\end{figure}

\begin{figure}[!htbp]
\begin{center}
\includegraphics[width=0.85\textwidth]{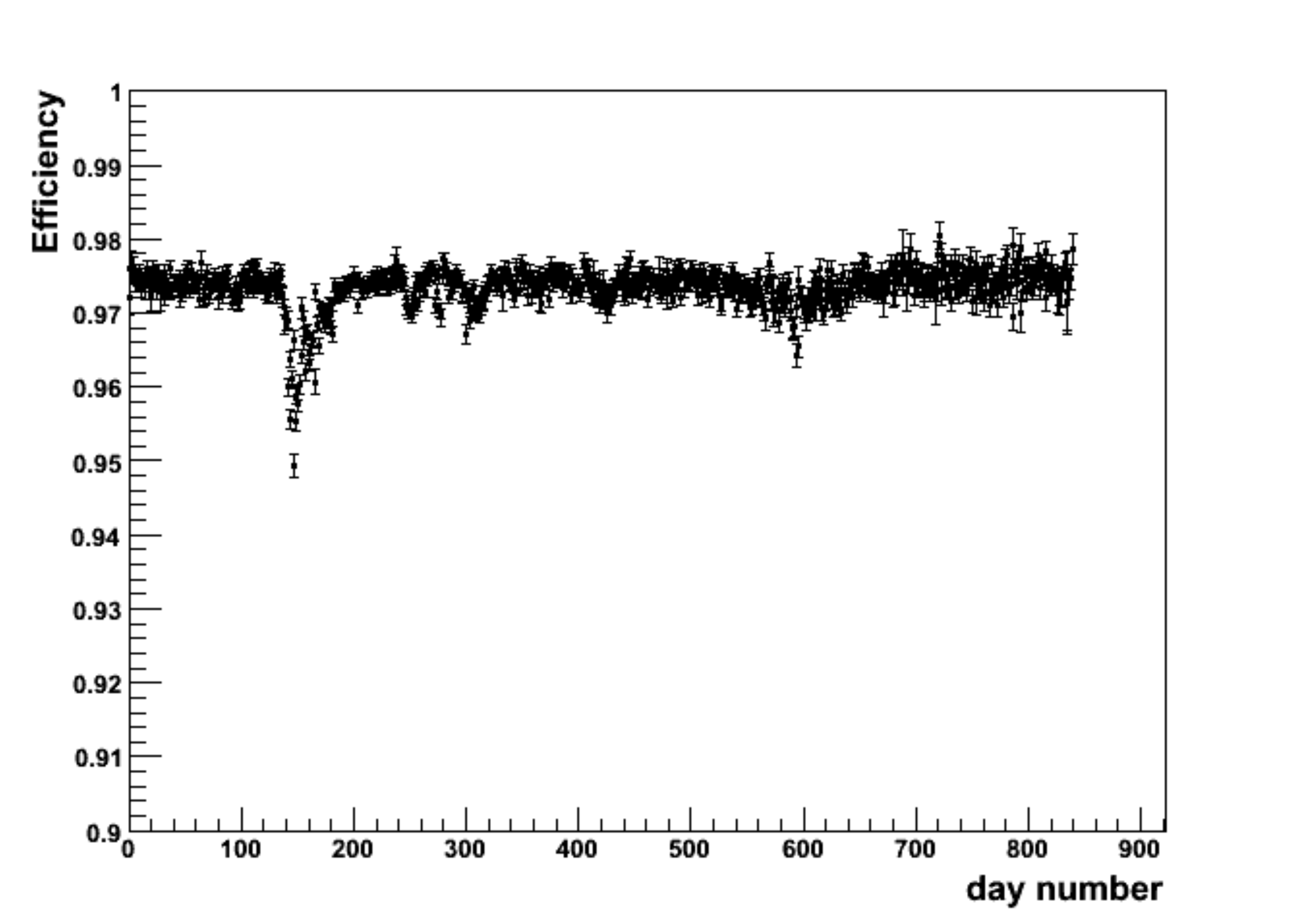}
\end{center}
\caption[Time evolution of the AC efficiency]{{\footnotesize Time evolution of the AC efficiency.}}
\label{fig:antiefftime}
\end{figure}

After applying the basic tracker cuts, the additional tracker cuts and the ToF cuts to flight data, the proton sample for estimation of the AC cuts efficiency is further cleaned by using the calorimeter cuts described in section \ref{sec:calocuts}. The AC efficiency is the fraction of events yielding a signal on CARD or CAT in the proton sample. Since all the particles interacting hadronically before S3 are removed in the sample, the particles hitting CARD or CAT are either delta rays produced above the tracker or particles back-scattered from the calorimeter. These kinds of particles are good candidates but are rejected by the anticoincidence system, therefore the selected candidates need to be corrected for this effect. The derived efficiency is about 96\% over the entire rigidity range, decreasing as the rigidity increases, as shown in figure \ref{fig:antieff}. There is no significant time dependence since the AC performance is stable (see figure \ref{fig:antiefftime}). However, a small drop can be observed correlated to the ToF performance. This drop is caused by the ToF system which has a lower rejection efficiency for delta rays or backscattered particles when selecting a proton sample. Thus more events in the sample will give a signal in the anticoincidence system, yielding a lower AC efficiency during those days. The correlation between ToF and AC has already been discussed in section \ref{sec:tofeff}.


\subsection{Calorimeter efficiency}
The calorimeter efficiency is derived for flight protons by calculating the fraction of events passing the calorimeter criteria over the sample obtained by applying the basic tracker cuts, the additional tracker cuts, the ToF cuts and the AC cuts. Since only positive deflection events are selected, the sample should only consist of positively charged particles, i.e. protons and a negligible number of positrons ($\sim 10^{-3}$ of protons).

\begin{figure}[!htb]
\begin{center}
\includegraphics[width=0.85\textwidth]{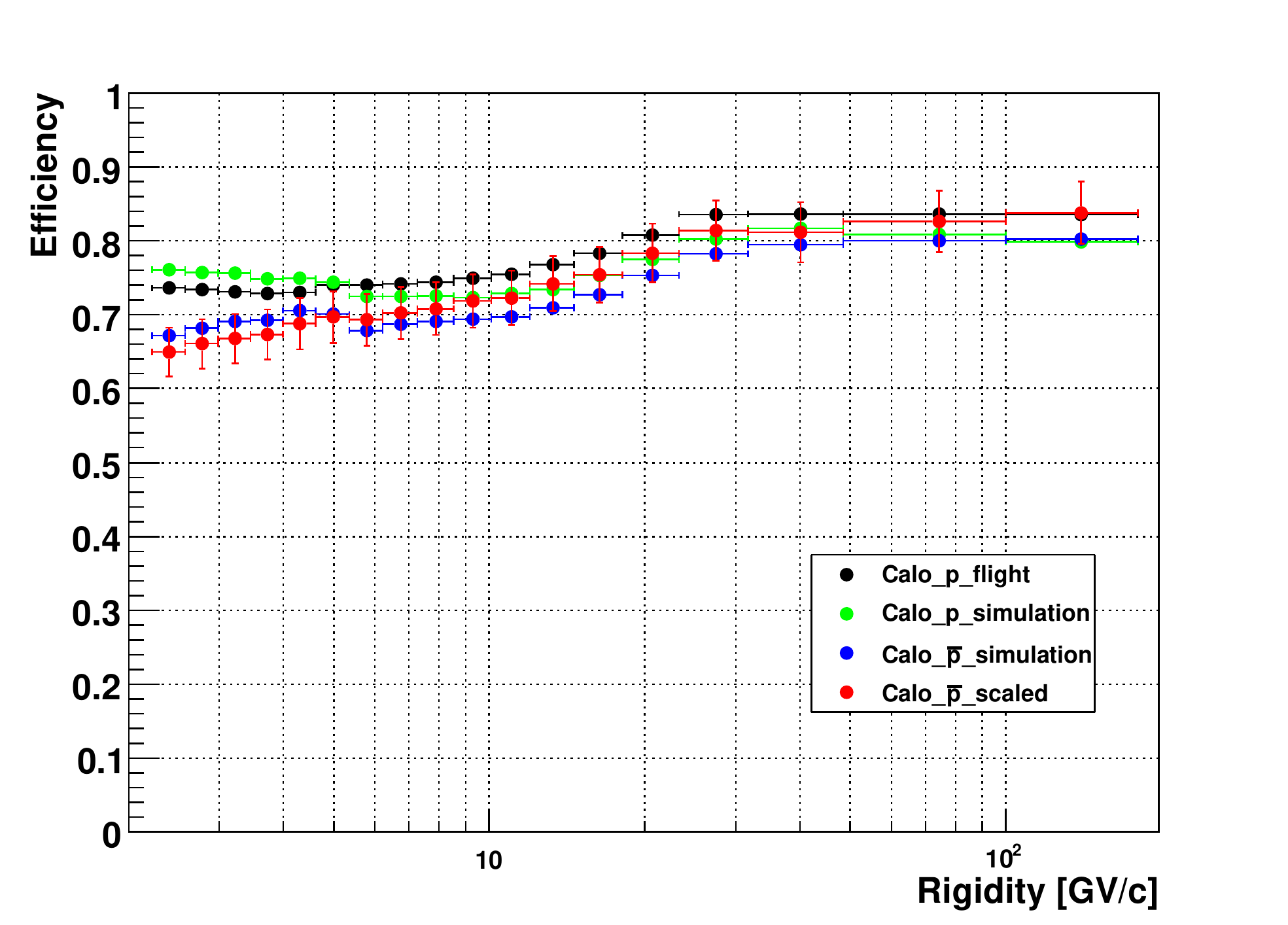}
\end{center}
\caption[The calorimeter efficiency evaluated from flight data and simulation]{\footnotesize The calorimeter efficiency for protons evaluated from flight data (black) and simulation (green). The blue points show the calorimeter efficiency for antiprotons from simulation. The red points are the calorimeter antiproton efficiency scaling the flight proton efficiency by the ratio of simulated antiproton efficiency and simulated proton efficiency.}
\label{fig:caloeff}
\end{figure}

Antiprotons and protons behave differently in the calorimeter mainly due to their interaction cross sections. At 2 GeV, the cross section for a $\bar{p}p$ interaction is about 40 mb larger than a $pp$ interaction \cite{PDGbookletWeb}. The difference decreases at high energy but can not be assumed approximately equal until approximately 100 GeV. Therefore a systematically lower calorimeter efficiency for antiprotons than for protons is expected until $\sim 100$ GeV. This is consistent with the simulated antiproton and proton efficiencies as presented in figure \ref{fig:caloeff}. Hence, the antiproton efficiency is estimated by scaling the flight proton efficiency by the ratio of simulated antiproton efficiency and proton efficiency. The scaling factor is derived from simulation, however, this factor may not be exactly equal to the discrepancy appearing during flight. To account for the discrepancy between flight efficiency and simulated antiproton efficiency, a conservative systematic uncertainty of 5\% is assigned to the antiproton calorimeter efficiency in the final calculation.

\subsection{MDR efficiency}
\begin{figure}[!htb]
\begin{center}
\includegraphics[width=0.85\textwidth]{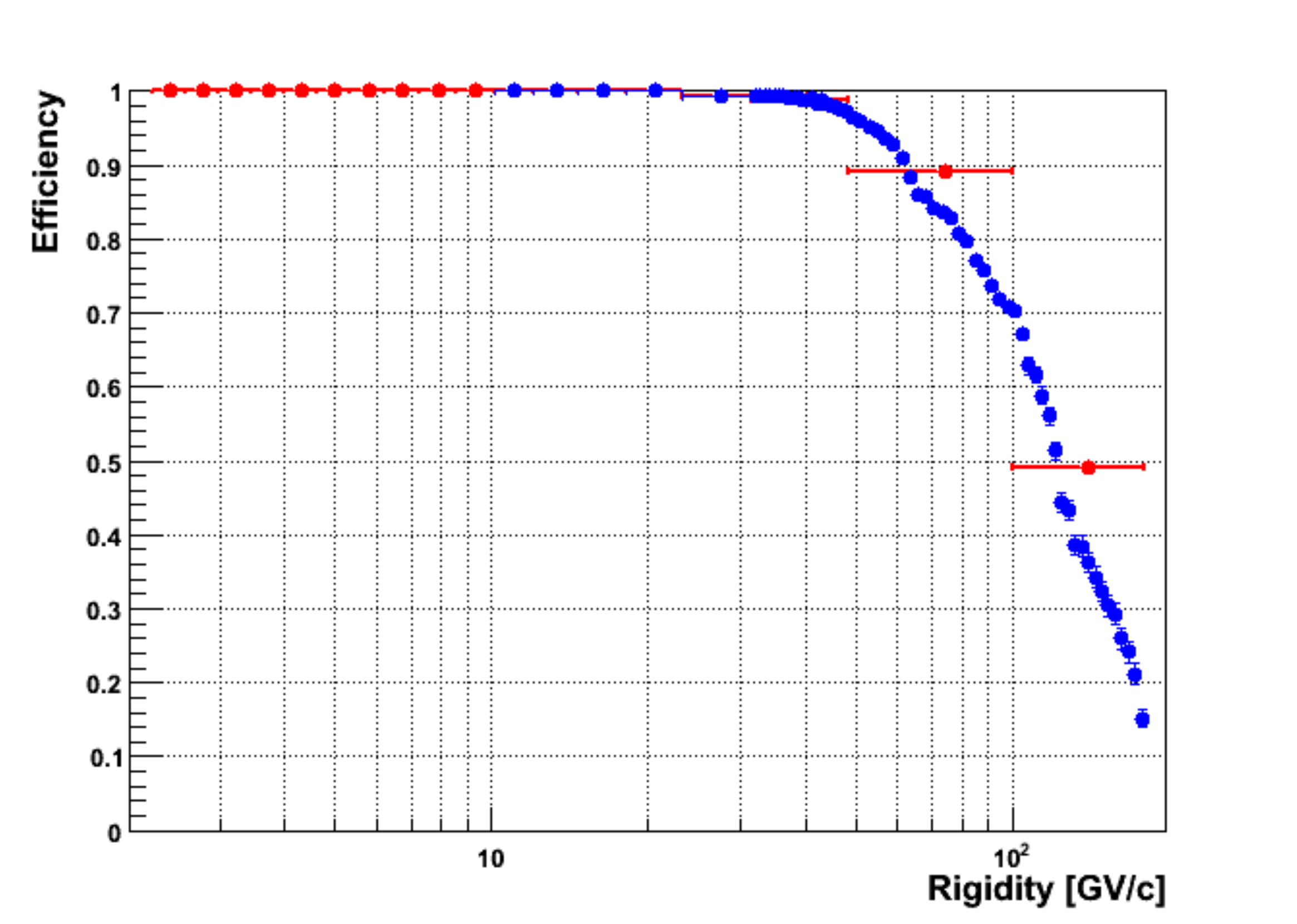}
\end{center}
\caption[The MDR efficiency for protons evaluated from flight data]{{\footnotesize The MDR efficiency for protons evaluated from flight data. The blue points show the efficiencies in sub bins. The red points show the mean efficiency in wider bins.}}
\label{fig:mdreff}
\end{figure}

The MDR efficiency is derived from the sample obtained by applying all the other cuts implemented in section \ref{sec:selection} apart from the MDR cut. As mentioned in section \ref{sec:tracker_sel}, in the last three bins the MDR cut requires a MDR larger than 6 times the upper limit of the sub-bin containing the rigidity of the event. An MDR efficiency is thus calculated for each sub-bin with a center rigidity $R_{j}$ and then a mean efficiency in bin $i$ (${\varepsilon}_{i}$) is determined by weighting the efficiency in the sub-bin $j$ (${\varepsilon}_{j}$) by a theoretical flux, which is $\sim R^{-2.7}$, in the following way:
\begin{equation}
\bar{{\varepsilon}_{i}} =  \frac{ \sum_{j=1 }^{20} R_{j}^{-2.7} \times {\varepsilon}_{j} \times \Delta R_{j}}{ \sum_{j=1 }^{20} R_{j}^{-2.7} \times \Delta R_{j}}.
\end{equation}
As shown in figure \ref{fig:mdreff}, the MDR efficiency decreases dramatically above $\sim$20~GV. 

\subsection{Trigger efficiency}
The PAMELA trigger efficiency is a product of the trigger efficiency for each ToF layer required in a particular trigger configuration. The trigger efficiency is calculated to exceed 0.997 with an error of the order $0.5 \times 10^{-4}$ \cite{SergioPrivate}. Compared to other factors discussed in this section, the trigger inefficiency is completely negligible and therefore will be omitted when calculating the total selection efficiency.

\subsection{Total selection efficiency}
\begin{figure}[!htbp]
\begin{center}
\includegraphics[width=0.85\textwidth]{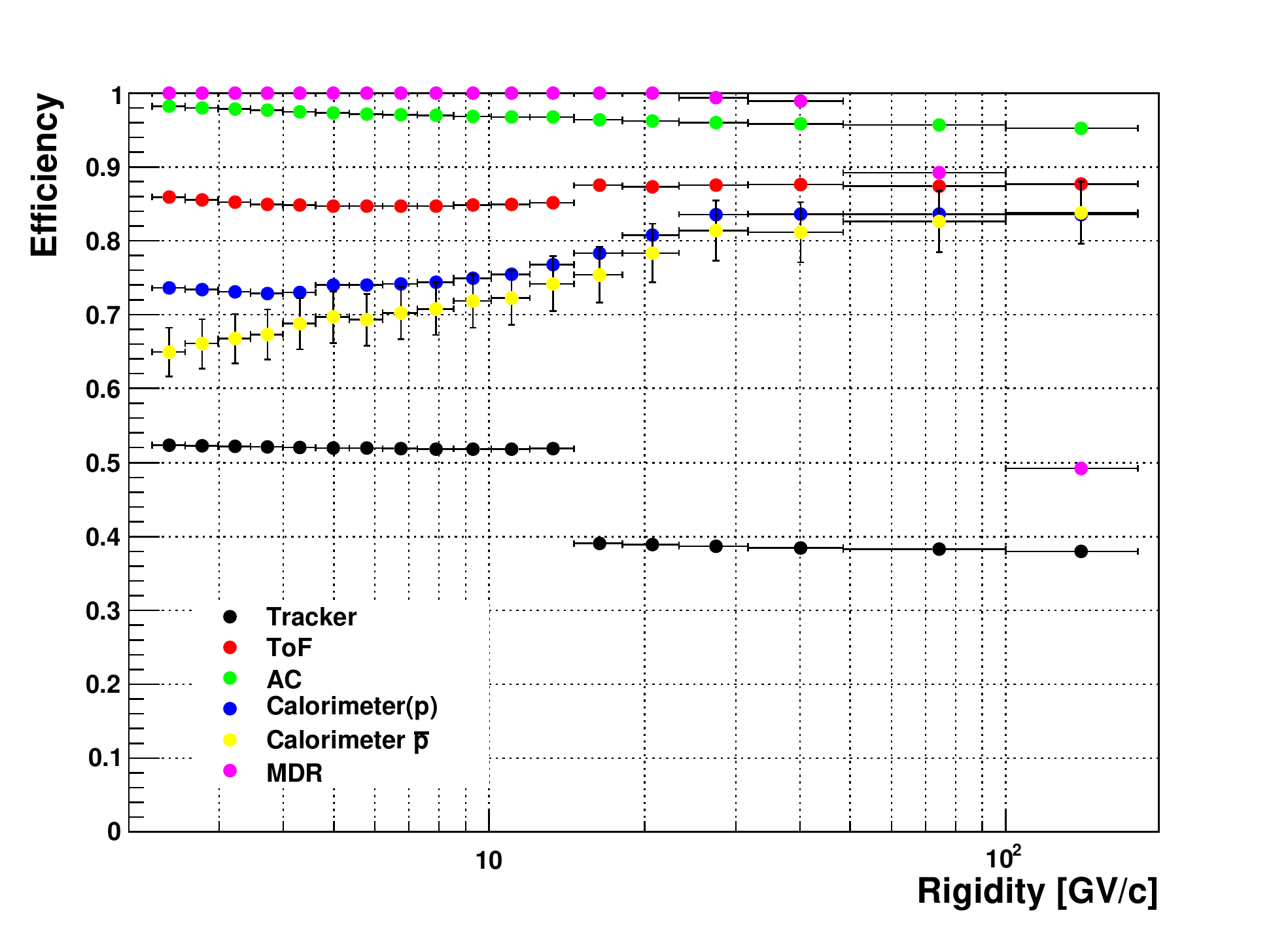}
\includegraphics[width=0.85\textwidth]{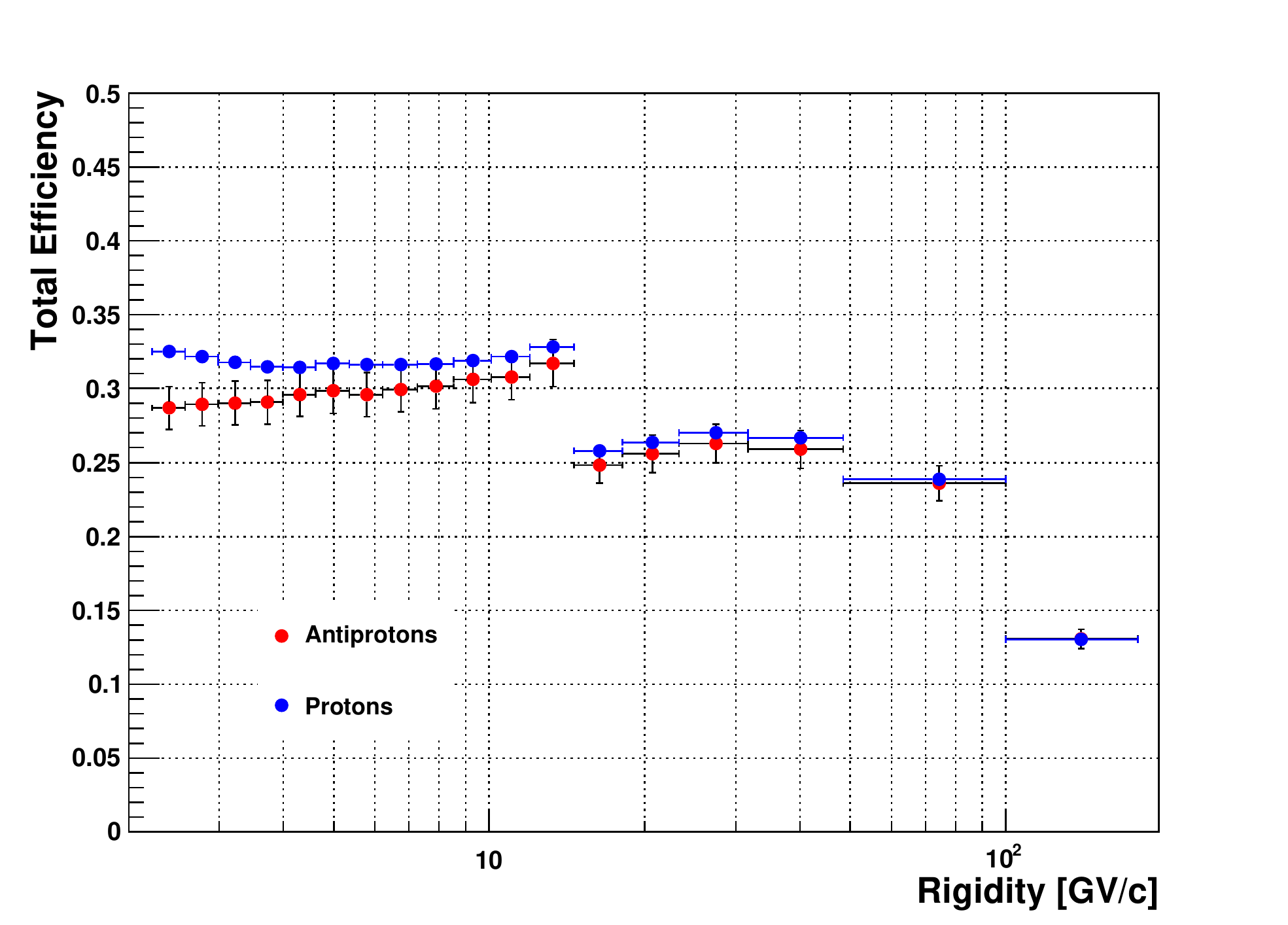}
\end{center}
\caption[All grouped efficiencies and the total efficiency integrated over the whole live time]{{\footnotesize Top: All grouped efficiencies integrated over the whole live time. Bottom: Total efficiency integrated over the whole live time for protons and antiprotons.}}
\label{fig:toteff}
\end{figure}

All grouped efficiencies discussed above are compared in figure \ref{fig:toteff}. The tracker efficiency here is defined as the total efficiency of the basic and additional tracker selection cuts, calculated by multiplying the basic and the additional tracker efficiencies. The selection efficiency is dominated by the tracker selection. Above 100 GV, the MDR selection also play an crucial role. After all individual efficiencies are derived and possible correlations are understood, the total antiproton and proton selection efficiencies can thus be calculated by multiplying all terms and associated errors, as shown in figure \ref{fig:toteff}.

\section{Correction factors} \label{sec:corrections}
Apart from selection efficiencies, several other correction factors, i.e. the geometrical factor and hadronic interaction losses in the instrument, the live time of measurements and transmission through the geomagnetic field, should be applied to the selected candidates to reconstruct the antiproton flux. Moreover, to derive the antiproton-to-proton flux ratio, differences between the corrections between antiprotons and protons are also addressed.

\subsection{ Geometrical factor and hadronic interaction losses}
\subsubsection{Geometrical constraint}
The flux intensity is generally a product of a count rate and a proportionality factor called the \textit{gathering power} of the detector. Under the hypothesis of an isotropic flux, the gathering power is expressed as a geometrical factor , defined by
\begin{equation}
G \left (  R   \right ) =  \int_{  \Omega }d \Omega  \int_{ S } dS \left | cos \theta   \right | f \left ( x, y,  \theta ,  \phi ,  R   \right ) , 
\end{equation}
where $R$ is the rigidity, $\Omega$ is the total solid angle, $S$ is a reference plane orthogonal to the z axis defined in figure \ref{fig:detectors}, and $f$ is a weighting function that is either 1 or 0 depending whether the trajectory of incident particle satisfies the acceptance requirements of the apparatus. The acceptance requirements are:
\begin{itemize}
\item the trajectory must cross at least one of the two layers in each plane of the ToF system: (S11 OR S12) AND (S21 OR S22) AND (S31 OR S32).
\item the particle must cross all the 6 planes of the tracking system.
\item the trajectory must be fully contained in the magnetic cavity without touching the walls of the cavity.
\end{itemize}

The geometrical factor is dependent on the rigidity of the incident particle. As lower rigidity particles are more deflected by the magnetic field towards the walls of the magnetic cavity, where they are absorbed before reaching the lower face of magnetic cavity, the geometrical factor is expected to decrease at low rigidities. At high rigidities where particle trajectories are approximately straight, the geometrical factor is expected to be constant. The geometrical factor presents the geometrical constraints of a particle telescope and does not depend on the particle species.

In order to calculate the geometrical factor, an approach based on the work by Sullivan \cite{Sullivan1971} has been performed with simulations. A set of particles are generated on a generation surface just above the scintillator plane S1, each with random parameters $ \left ( x, y,  \theta ,  \phi ,  R   \right ) $, in which $ \left (x, y   \right ) $ is the coordinate, the $ \left ( \theta ,  \phi \right ) $ is the incident direction and $R$ is the initial rigidity.

The measured PAMELA magnetic field in the spectrometer has been used in the simulation. For a given rigidity, the geometrical factor is given by:
\begin{equation}
G_{F}=\frac{n_{sel}}{n_{tot}}\cdot G_{gen}, 
\end{equation}
where $n_{sel}$ is the number of selected particles satisfying all the acceptance requirements, $n_{tot}$ is the total number of generated particles, and $G_{gen}$ is the gathering power of the generation surface with area $S_{gen}$. $G_{gen}$ can be expressed as:
\begin{equation}
G_{gen} =  \int_{  \Omega_{gen} }d \Omega  \int_{ S_{gen} } d S \left | cos \theta   \right | = S_{gen} \pi  \left ( 1-  { cos}^{2 }   { \theta }_{max }   \right ) , 
\end{equation}
where ${ \theta }_{max } $ is the maximum generation zenith angle. As only downward-going particles are of interest, the angular domain is limited to the downward hemisphere, characterized by $\pi /2 < \theta < \pi$.

\subsubsection{Interaction losses}
The geometrical factor discussed above depends only on the geometrical constraints of the instrument. However, due to the presence of material a particle traversing the apparatus may interact inside the acceptance. Since an incident particle entering the acceptance of PAMELA should always be accounted for in the calculation of the particle flux, inelastically interacting events, which are rejected by the selection criteria applied on flight data should be evaluated. In detail, two effects should be considered:
\begin{itemize}
\item the loss of particles which would traverse the acceptance cleanly if not interacting inside the acceptance.
\item the gain of particles which would not traverse the acceptance if not scattering inside the acceptance.
\end{itemize}
This correction is not accounted for in the selection efficiency calculation since particles which interact inside the instrument and produce secondaries are rejected. Therefore, the geometry of the apparatus and all physical processes are implemented in the simulation to estimate the effective acceptance which includes the correction due to inelastic interaction effect. An independent work has been performed by Bruno \cite{BrunoPrivate}. A sample of downgoing particles have been isotropically generated from a surface placed above the ``dome'' (the top container of the instrument). The starting point, direction and the area of generation surface were chosen to ensure no bias in the calculation of the fraction of in-acceptance events and include those events in the sample whose tracks were initially not contained in the acceptance but were deflected into the acceptance by scattering (mainly in the dome). The effective acceptance is calculated by:
\begin{equation}
A_{F}=\frac{n_{sel}}{n_{tot}}\cdot G_{gen}.
\end{equation}
Since the interaction processes are included in the simulation, the particle loss and gain due to interactions are naturally included in the effective acceptance.  

\begin{figure}[!htb]
\begin{center}
\includegraphics[width=0.85\textwidth]{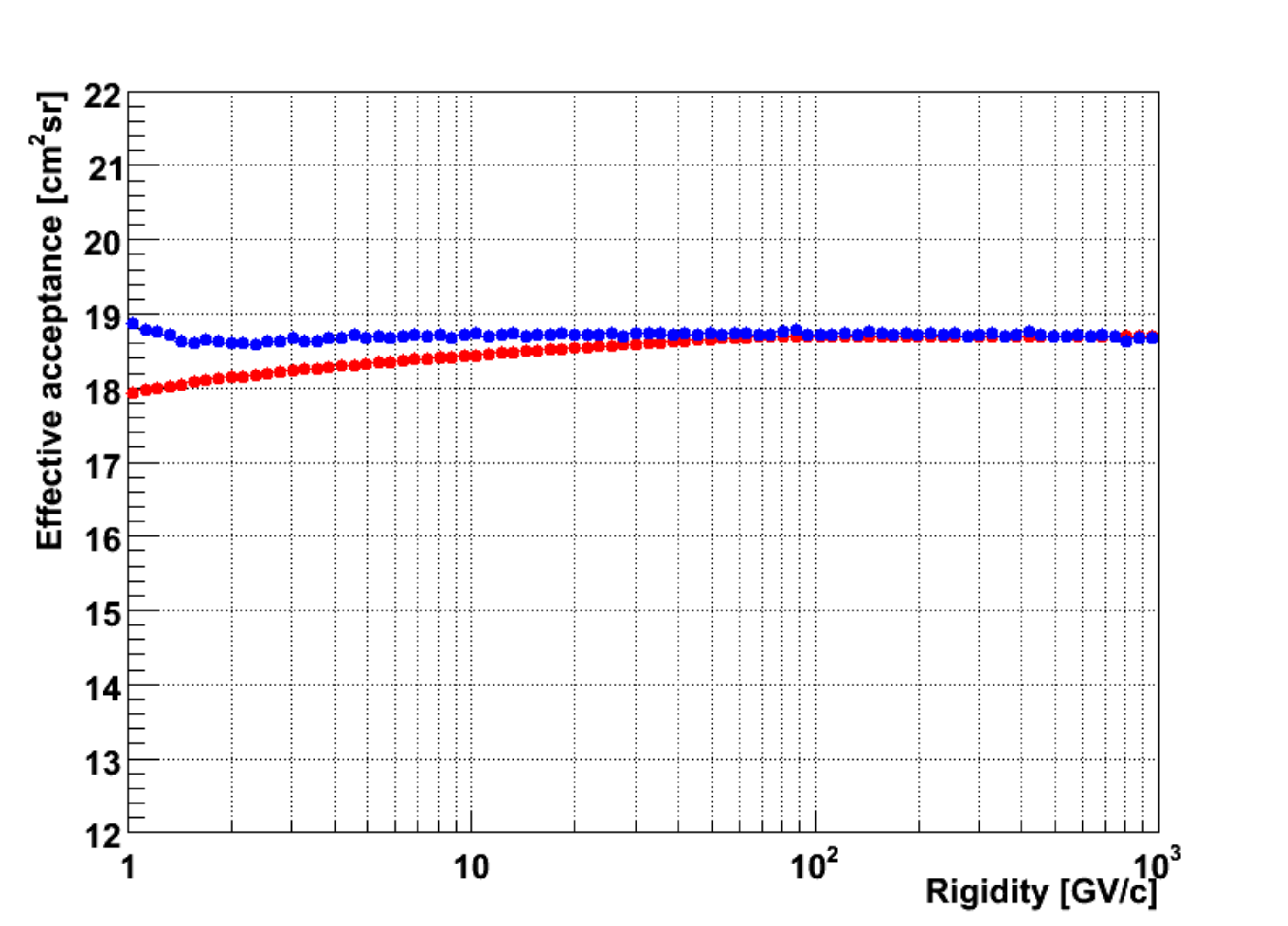}
\end{center}
\caption[The PAMELA effective acceptance for protons and antiprotons]{{\footnotesize The PAMELA effective acceptance for protons (blue) and antiprotons (red). From \cite{BrunoPrivate}.}}
\label{fig:acceptance}
\end{figure}

The results of the calculation are shown in figure \ref{fig:acceptance}, both for positively-charged particles and negatively-charged particles {\cite{BrunoPrivate}. Below 1 GeV, the geometry significantly constrains the acceptance. However, above that energy, the shape of the effective acceptance mainly reflects the cross sections of inelastic interactions for protons and antiprotons. The inelastic cross section for protons increases with energy from 1 GeV to 2 GeV and remains constant between 2 GeV and 1000 GeV, resulting a decreasing effective acceptance and an almost unchanged acceptance below and above 2 GeV. For antiprotons, the inelastic cross sections increases dramatically until 100 GeV, resulting in an increasing effective acceptance below that energy. The difference between protons and antiprotons is due to the difference in inelastic cross section for these two species. The hadronic generator FLUKA is employed in GPAMELA simulation code \cite{BrunoPhd}, which was developed by the PAMELA collaboration based on the GEANT package \cite{Brun1994} version 3.21. In order to simulate the hadronic physical processes and to generate secondary cascades from hadron-nucleus interactions, different models are used in FLUKA generator depending on the energy of  the cosmic ray hadrons. An error of 5\% is added to the effective acceptance considering systematic uncertainties related to the hadronic generator. 

\subsection{Transmission through the geomagnetic field and live time}

As discussed in section \ref{sec:selgalactic}, only particles satisfying the following requirement are accepted:
\begin{equation*}
{rigbin}_{lowerlimit} > \text{cutoff}_{\text{PAMELA}} =1.3 \times SVC.
\end{equation*}
Galactic particles not satisfying the rigidity requirement are rejected and should be accounted for when deriving the flux. 

This correction was calculated together with the live time, which is defined as the time when the experiment is operational and ready for a new trigger. Contrary to the live time, the time when the instrument is switched off or is reading out and processing data is called the dead time. The live time corrected by transmission through the geomagnetic field is calculated by:
\begin{equation}
T_{live} \left ( bin  \right ) = \int_{0 }^{{bin}_{lowerlimit}/1.3 } f \left( SVC \right) d R ,
\end{equation}
where $T_{live} \left( bin \right)$ represents the live time spent for $1.3 \cdot SVC$ lower than ${bin}_{lowerlimit}$ (the low edge of the bin), and $f \left( SVC \right)$ is the cutoff distribution weighted by relative live time. Assuming that the particle flux is isotropic, the loss of particles can be compensated for by multiplying the measured particle intensity with the inverse of the corrected live time. The result is presented in figure \ref{fig:livetime}. Below about 20 GV a continuous increase of live time with rigidity can be seen. 

\begin{figure}[!htb]
\begin{center}
\includegraphics[width=0.85\textwidth]{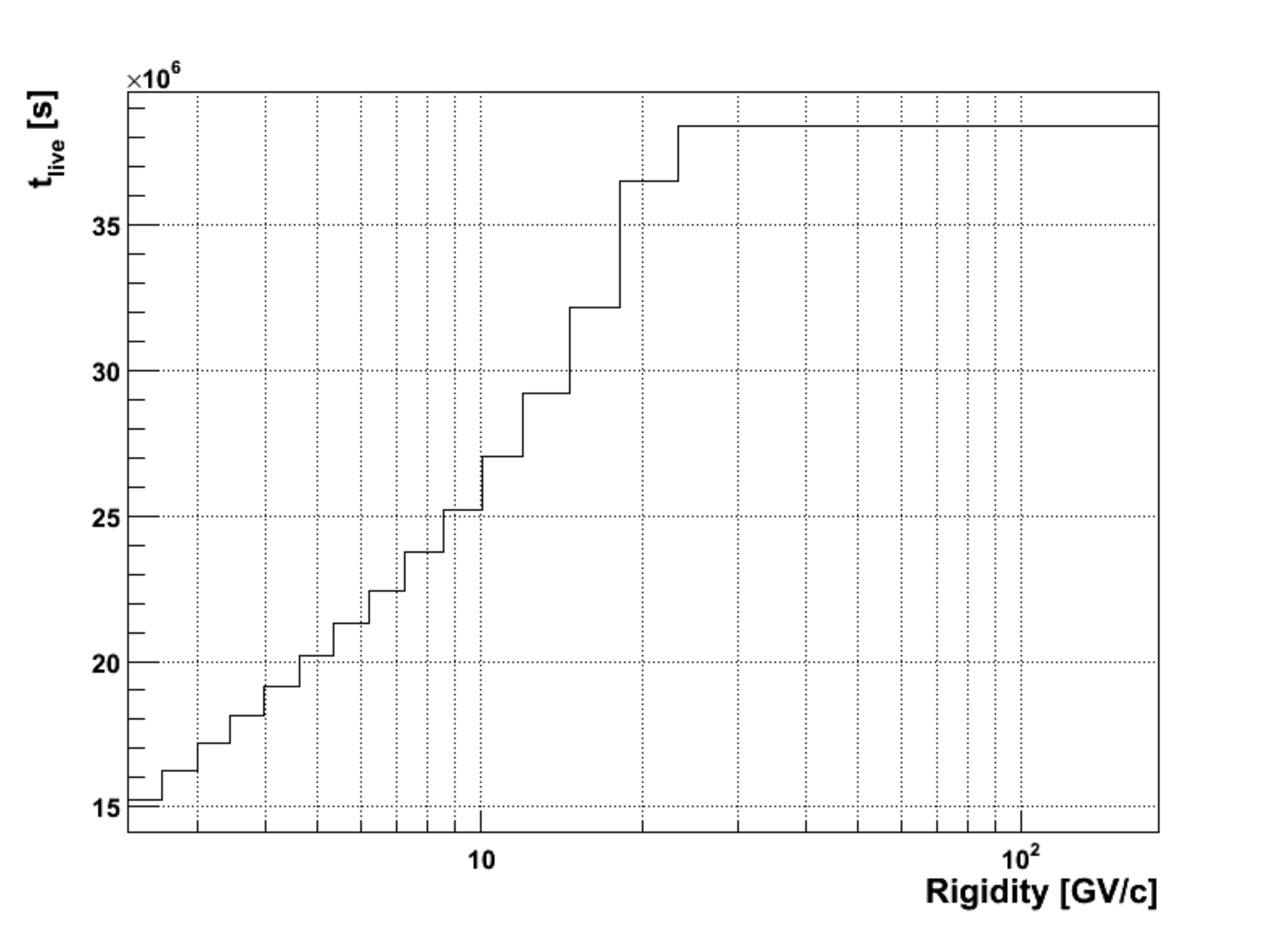}
\end{center}
\caption[The live time]{{\footnotesize The live time spent for PAMELA cutoff $ \leq {bin}_{lowerlimit}$.}}
\label{fig:livetime}
\end{figure}

\section{Antiproton flux and antiproton-to-proton flux ratio} \label{sec:antiproton_results}
The raw number of antiprotons and proton candidates surviving all the selection criteria are presented in in table \ref{tab:candidates}. These candidates are selected by accounting for the time when the tracker was not operating satisfactorily or the statistics is too low, as discussed in section \ref{sec:trkbas_eff}. The days of detector failure are labelled as ``non-operational'' days and are excluded in the data analysis to ensure stable efficiencies. For the rigidity range considered in this thesis, i.e. between 2.23~GV/c and 180~GV/c, the pion contamination was estimated to be negligible {\cite{BrunoPrivate}}, as well as the electron contamination {\cite{Adriani2009}}. Possible proton contamination due to the spillover effect was studied using both simulation and flight data. The antiproton selection criteria except the MDR cut is applied to simulated protons to reproduce the spillover observed in real flight data, and the proton contamination after applying the MDR cut is then determined. The systematic uncertainty due to the proton contamination only exists in the highest energy bins and was estimated to be $\sim$20\% for the rigidity bin 48.5-100~GV/c and $\sim$30\% for the bin 100-180~GV/c.

\begin{table}[!htbp]
\centering
\begin{tabular}{c | c | c}
\hline
{ Rigidity GV/c} & { Antiprotons} & { protons} \\
\hline
{ 2.23 - 2.58  } & { 49 } & { 1701576} \\
{ 2.58 - 2.99  } & { 77 } & { 1644062 } \\
{ 2.99 - 3.45  } & { 78 } & { 1485846 } \\
{ 3.45 - 3.99  } & { 96} & { 1376729} \\
{ 3.99 - 4.62  } & { 102} & { 1246995 } \\
{ 4.62 - 5.36  } & { 107} & { 1130713 } \\
{ 5.36 - 6.23  } & { 108} & { 993868 } \\
{ 6.23 - 7.27  } & { 98} & { 873283  } \\
{ 7.27 - 8.53  } & { 85 } & { 762115  } \\
{ 8.53 - 10.1  } & { 94} & { 671588  } \\
{ 10.1 - 12.0  } & { 105} & { 567089  } \\
{ 12.0 - 14.6  } & { 78 } & { 527378 } \\
{ 14.6 - 18.1  } & { 64 } & { 359288  } \\
{ 18.1 - 23.3  } & { 55 } & { 337924  } \\
{ 23.3 - 31.7  } & { 41 } & { 272516  } \\
{ 31.7 - 48.5  } & { 36   } & { 194277  } \\
{ 48.5 - 100.0 } & { 22  } & { 106888   } \\
{ 100.0 - 180.0} & {  3  } & { 13596    } \\
\end{tabular}
\caption[Selected candidates after excluding ``non-operational'' days]{The antiproton and proton candidates after excluding ``non-operational'' days.}
\label{tab:candidates}
\end{table} 

In order to construct the antiproton flux at the top of the PAMELA payload, the raw number should be corrected for selection efficiencies, hadronic interaction losses and the geometrical factor, transmission through the geomagnetic field and measurement live time. The selection efficiencies discussed in section \ref{sec:efficiencies} show a time dependence, especially the tracker efficiency. For a bin $i$, if the raw number in each day $N_{j}$ is corrected with a daily efficiency, ${\varepsilon}_{ j}$, the systematic effects due to the time variation of detector performance can be minimized. However, as there are a small number of antiproton candidates per day, if an event is selected by chance in a day when the efficiency is low and thus no candidate is expected to be selected, the corrected number will be overestimated. Therefore, instead of using a daily efficiency, the correction is done on the basis of a `bunch' of several days to eliminate this overestimation by increasing the statistics, as well as to reduce the effect of time dependent efficiencies. A value of 60 days is chosen to be appropriate for one bunch in final calculation. 

The resulting flux is presented in table \ref{tab:pbarflux}. The range of an energy bin is converted from the range of rigidity bin in table \ref{tab:candidates}, assuming a singly-charged particle and by using $E_{k}=\sqrt{p^{2}c^{2}+m^{2}c^{4}}- mc^{2}$, where $E_{k}$ is the kinetic energy, $m$ the mass of the particle, $p$ the momentum and $c$ the velocity of light. Data points are centered in each bin according to a technique developed by Lafferty and Wyatt \cite{Lafferty1995}. For a bin with lower limit $E_{l}$ and upper limit $E_{u}$, the weighted center $E_{c}$ is determined as the abscissa value at which the measured spectrum is equal to the expectation average value of the ``true'' spectrum, which can be expressed as
\begin{equation}
f \left ( E_{c}  \right ) =  \frac{ 1}{ E_{u}-E_{l}} \int_{ E_{l}}^{E_{u} }f \left ( E  \right ) dE,
\end{equation}
where $f \left ( E \right)$ is a theoretical model of antiproton spectrum \cite{Donato2001, Ptuskin2006}.

Since for antiprotons and protons the discrepancy between correction factors only exists for the effective acceptance and the calorimeter efficiency due to the difference of hadronic cross sections, the $\bar{\text{p}}$/p ratio can be calculated as
 \begin{equation}
R\left ( bin  \right )=\frac{ N_{\bar{p}} \left ( bin  \right )/ \left( {\varepsilon  }_{\bar{p}calo}\left ( bin  \right ) \times G_{\bar{p}}  \right) }{N_{p} \left ( bin  \right ) / \left( {\varepsilon  }_{pcalo}\left ( bin  \right ) \times G_{p}  \right) },
\end{equation}
where $N_{\bar{p}}$ is the total number of antiproton candidates, $N_{p}$ the total number of proton candidates, ${\varepsilon  }_{\bar{p}calo}$ the calorimeter efficiency for antiprotons, ${\varepsilon  }_{pcalo}$ the calorimeter efficiency for protons, $G_{\bar{p}}$ the acceptance for antiprotons and $G_{p}$ the acceptance for protons. The resulting ratio, which increases with energy up to $\sim 10$ GeV and then flattens, is presented in table \ref{tab:pbarflux}.

\begin{table}[!htb]
\centering
\begin{tabular}{c | c | c|c}
\hline
Energy  & Energy mean & $\bar{\text{p}}$ flux& $\bar{\text{p}}$/p flux ratio\\
(GeV)& (GeV)&$m^{-2}sr^{-1}s^{-1}{GeV}^{-1} $& \\
\hline
{ 1.48 - 1.81  } & { 1.61 } & {$ 1.84_{ - 0.29 -0.04}^{ +0.29 +0.04} \times 10^{-2}$} & {$ 3.31_{ - 0.48 -0.29}^{ +0.48 +0.29} \times 10^{-5}$} \\
{ 1.81 - 2.20  } & { 2.03 } & {$ 2.50_{ - 0.31 -0.06}^{ +0.31 +0.06} \times 10^{-2}$} &{$ 5.36_{ - 0.61 -0.47}^{ +0.61 +0.47} \times 10^{-5}$} \\
{ 2.20 - 2.64  } & { 2.42 } & {$ 2.03_{ - 0.25 -0.05}^{ +0.25 +0.05} \times 10^{-2}$} & {$ 5.90_{ - 0.67 -0.52}^{ +0.67 +0.52} \times 10^{-5}$} \\
{ 2.64 - 3.16  } & { 2.90 } & {$ 2.15_{ - 0.24 -0.05}^{ +0.24 +0.05} \times 10^{-2}$} &{$ 7.64_{ - 0.78 -0.67}^{ +0.78 +0.67} \times 10^{-5}$} \\
{ 3.16 - 3.78  } & { 3.47 } & {$ 1.68_{ - 0.19 -0.04}^{ +0.19 +0.04} \times 10^{-2}$} & {$ 8.90_{ - 0.88 -0.78}^{ +0.88 +0.78} \times 10^{-5}$} \\
{ 3.78 - 4.50  } & { 4.14 } & {$ 1.49_{ - 0.16 -0.03}^{ +0.16 +0.03} \times 10^{-2}$} &{$ 1.03_{ - 0.10 -0.09}^{ +0.10 +0.09} \times 10^{-4}$} \\
{ 4.50 - 5.36  } & { 4.93 } & {$ 1.06_{ - 0.11 -0.03}^{ +0.11 +0.03} \times 10^{-2}$} &{$ 1.19_{ - 0.12 -0.11}^{ +0.12 +0.11} \times 10^{-4}$} \\
{ 5.36 - 6.39  } & { 5.87 } & {$ 7.72_{ - 0.84 -0.16}^{ +0.84 +0.16} \times 10^{-3}$} & {$ 1.22_{ - 0.13 -0.11}^{ +0.13 +0.11} \times 10^{-4}$} \\
{ 6.39 - 7.64  } & { 7.00 } & {$ 5.30_{ - 0.62 -0.12}^{ +0.62 +0.12} \times 10^{-3}$} & {$ 1.18_{ - 0.13 -0.11}^{ +0.13 +0.11} \times 10^{-4}$} \\
{ 7.64 - 9.21  } & { 8.40 } & {$ 5.00_{ - 0.57 -0.11}^{ +0.57 +0.11} \times 10^{-3}$} & {$ 1.47_{ - 0.16 -0.13}^{ +0.16 +0.13} \times 10^{-4}$} \\
{ 9.21 - 11.1  } & { 10.1 } & {$ 3.60_{ - 0.38 -0.08}^{ +0.38 +0.08} \times 10^{-3}$} &{$ 1.94_{ - 0.19 -0.17}^{ +0.19 +0.17} \times 10^{-4}$} \\
{ 11.1 - 13.7  } & { 12.3 } & {$ 2.03_{ - 0.26 -0.05}^{ +0.26 +0.05} \times 10^{-3}$} &{$ 1.54_{ - 0.18 -0.14}^{ +0.18 +0.14} \times 10^{-4}$} \\
{ 13.7 - 17.2  } & { 15.3 } & {$ 1.42_{ - 0.20 -0.03}^{ +0.20 +0.03} \times 10^{-3}$} & {$ 1.96_{ - 0.25 -0.17}^{ +0.25 +0.17} \times 10^{-4}$} \\
{ 17.2 - 22.4  } & { 19.6 } & {$ 6.74_{ - 1.00 -0.12}^{ +1.00 +0.12} \times 10^{-4}$} &{$ 1.79_{ - 0.24 -0.16}^{ +0.24 +0.16} \times 10^{-4}$} \\
{ 22.4 - 30.1  } & { 26.2 } & {$ 2.53_{ - 0.42 -0.05}^{ +0.42 +0.05} \times 10^{-4}$} &{$ 1.64_{ - 0.26 -0.15}^{ +0.26 +0.15} \times 10^{-4}$} \\
{ 30.1 - 47.6  } & { 38.0 } & {$ 1.32_{ - 0.25 -0.03}^{ +0.25 +0.03} \times 10^{-4}$} &{$ 1.96_{ - 0.33 -0.17}^{ +0.33 +0.17} \times 10^{-4}$} \\
{ 47.6 - 99.2  } & { 67.4 } & {$ 2.36_{ - 0.51 -0.06}^{ +1.16 +0.06} \times 10^{-5}$}&{$ 1.71_{ - 0.47 -0.18}^{ +0.47 +0.18} \times 10^{-4}$} \\
{ 99.2 - 179.1 } & { 128.9} & {$ 3.9_{ - 2.2 -0.3}^{ +6.1 +0.3} \times 10^{-6}$} & {$ 1.6_{ - 1.2 -0.2}^{ +1.8 +0.2} \times 10^{-4}$} \\
\end{tabular}
\caption[The antiproton flux and $\bar{\text{p}}$/p flux ratio measured by PAMELA]{The antiproton flux and $\bar{\text{p}}$/p flux ratio measured by PAMELA. The first set of errors refer to the $1\sigma$ statistical errors and the second set are systematic errors. }
\label{tab:pbarflux}
\end{table}

The statistical and systematic errors are shown separately in table \ref{tab:pbarflux}. For both the antiproton flux and the $\bar{\text{p}}$/p ratio, statistical errors dominate over the entire energy range. Figure \ref{fig:fluxcompare} and figure \ref{fig:ratiocompare} compare the antiproton flux and the $\bar{\text{p}}$/p ratio derived in this work respectively with other experimental data. PAMELA measurements are consistent with other measurements but with significantly better statistics. The results derived in this work are also compared with those officially published by the PAMELA Collaboration \cite{Adriani2010_pbar} extending to lower energies. The agreement is excellent\footnote[2]{The published results used a different method to estimate the total efficiency. Loose selections were also used in the rigidity range 6.23-14.6~GV once it became clear that pion contamination was minimal.}.

The majority of cosmic ray antiprotons are generally believed to be produced secondarily by interactions of cosmic ray protons with the ISM. Due to the kinematic constraints on the antiproton production, a peak around 2~GeV is expected to appear in the antiproton flux and is confirmed by results shown here. The  antiproton spectrum is a useful tool to constrain propagation models. Different models may give different predictions on antiproton fluxes. For example, the diffusion reacceleration models usually produce too few antiprotons \cite{Moskalenko2002, Trotta2011}. Some models expect slightly different antiproton spectra which could not be discriminated by antiproton data published before PAMELA (e.g. as in \cite{Ptuskin2006}). Thanks to  the more accurate antiproton data provided by PAMELA, stronger constraints may be able to be placed on cosmic ray propagation models. In the next chapter, the antiproton flux and the $\bar{\text{p}}$/p ratio measured by PAMELA will be used to further study cosmic ray propagation models.
 
\begin{figure}[!htb]
\begin{center}
\includegraphics[angle=90, width=350pt]{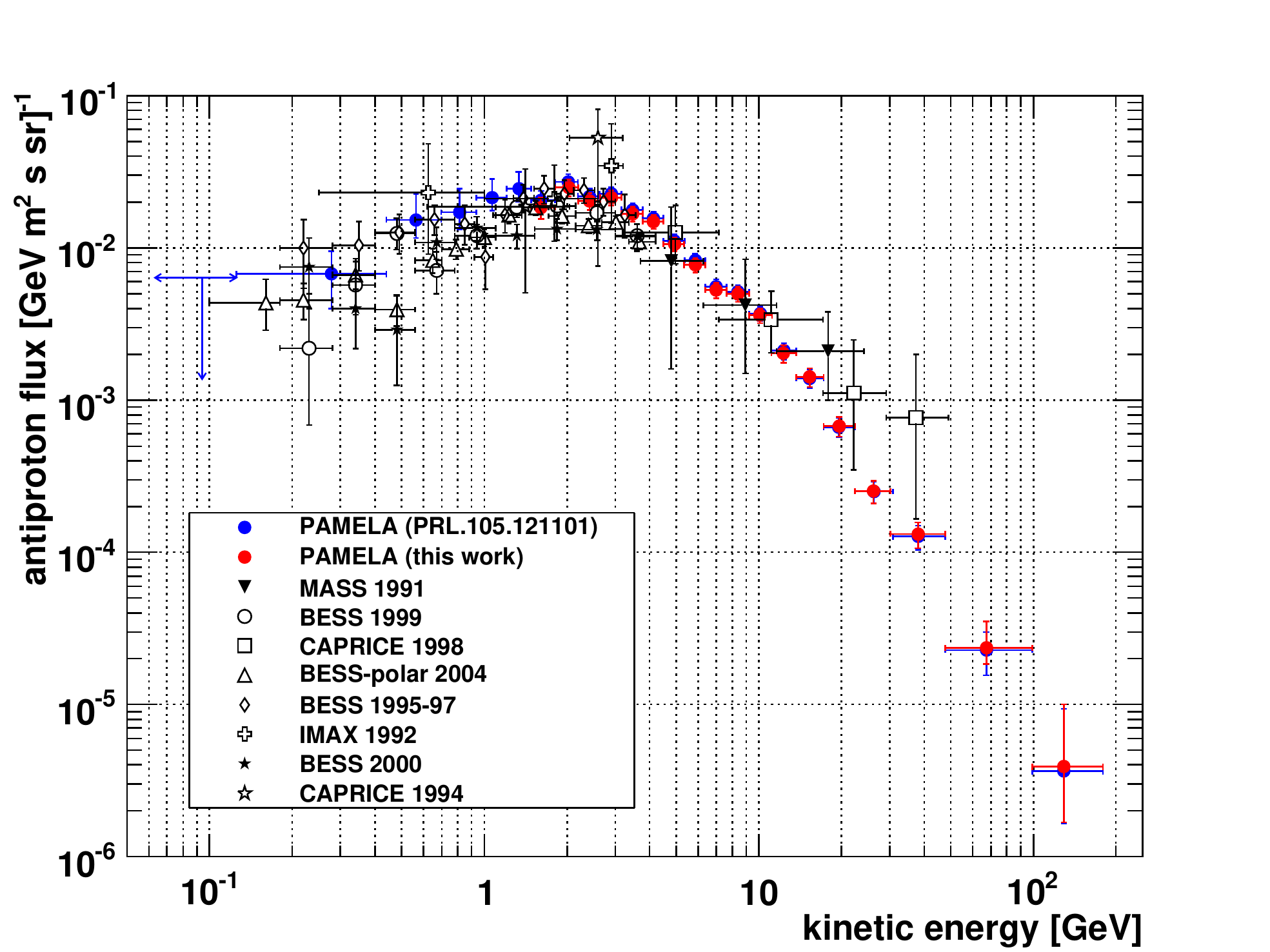}
\end{center}
\caption[The antiproton flux measured by PAMELA compared with the results from other experiments]{\footnotesize The $\bar{p}$ flux measured by PAMELA compared with the results from other experiments (see {\cite{Boezio2001}} (and references therein) \cite{Maeno2001} and \cite{Abe2008}). PAMELA measurements related to this work are shown in red symbols. PAMELA measurements published in \cite{Adriani2010_pbar} are shown in blue symbols. }
\label{fig:fluxcompare}
\end{figure}

\begin{figure}[!htb]
\begin{center}
\includegraphics[angle=90, width=350pt]{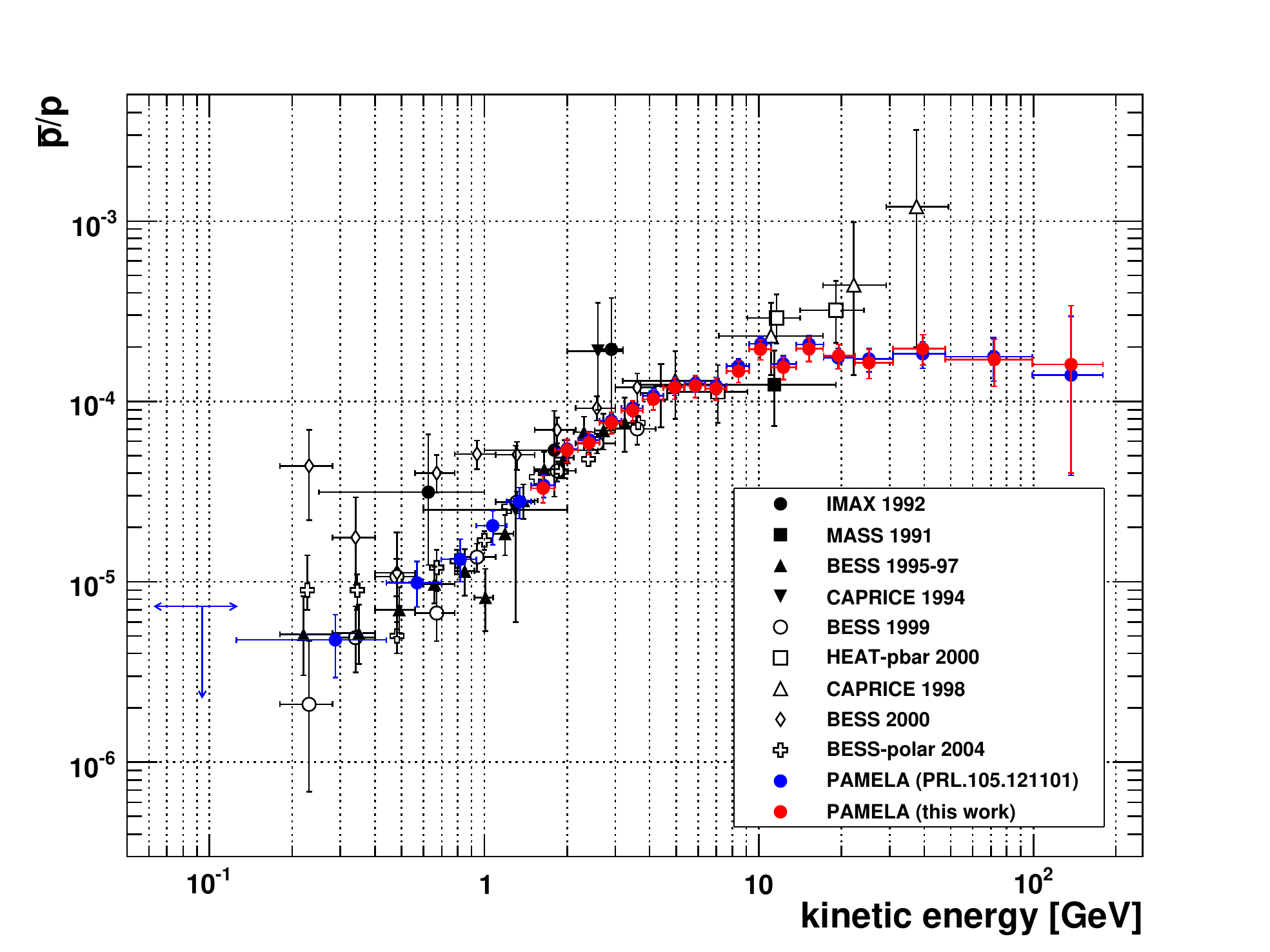}
\end{center}
\caption[The $\bar{\text{p}}$/p ratio measured by PAMELA compared with the results from other experiments]{\footnotesize The $\bar{\text{p}}$/p ratio measured by PAMELA compared with the results from other experiments (see references in \cite{Adriani2009}). PAMELA measurements related to this work are shown in red symbols. PAMELA measurements published in \cite{Adriani2010_pbar} are shown in blue symbols. }
\label{fig:ratiocompare}
\end{figure}

\section{Discussion}
Before the PAMELA experiment was carried out, the CAPRICE1998 measurement indicated an increasing trend in the antiproton flux above 10~GeV and therefore a possible presence of a primary antiproton contribution. The much more precise PAMELA results disfavor this trend and generally agree with the pure secondary production calculations. However, in contrast to antiproton flux, PAMELA observed a clear rise above 10~GeV in the positron fraction disagreeing with the prediction of secondary positron production. A considerable number of interpretations, including astrophysical or exotic primary sources, have been proposed to explain both the antiproton data and the unexpected positron fraction. Nearby Pulsars have been proposed as a good candidate for the positron excess since electron-positrons pairs are expected to be produced in pulsars but no antiprotons (e.g. \cite{Hooper2009_pulsar, Malyshev2009, Profumo2012}). Another explanation suggested for the steep increase in the positron fraction is provided by positrons and electrons as secondary products of hadronic interactions inside aged SNRs \cite{Blasi2009_positron}. This model also predicts an antiproton flux compatible with the PAMELA data but predicts a harder antiproton component beyond the 100~GeV energy region which strongly differs from the conventional antiproton production created only from spallation in the ISM \cite{Blasi2009_pbar}. The dark matter scenario, as discussed in section \ref{sec:darkmatter}, may also result in anomalous features in spectrum of cosmic ray antiparticles. Both antiprotons and positrons are generally believed to be the final states of dark matter annihilation, however, no excess over the standard secondary background is seen in the antiproton spectrum and this has therefore placed strong constraints on dark matter properties. The dominant final states of dark matter annihilation or decay are leptons instead of quarks \cite{Yin2009, Cirelli2009}. Candidates like the Kaluza-Klein particles which annihilate mostly to leptons, provide the possibility to reproduce the observed electron and positron spectrum \cite{Hooper2009b, Okada2009}. Dark matter annihilation channels of charged gauge bosons (W, Z) can only be accommodated in PAMELA data with $>10$~TeV dark matter mass \cite{Cirelli2009}, unless electrons and antiprotons have significant different boost factors (for example in \cite{Donato2009}). Nevertheless, a robust estimation of the cosmic ray propagation is a foundation of the investigation on possible primary sources and will be studied in next chapter.

%% file: Phd-Ch-Model_Constraints.tex
\chapter{Constraints on transport and acceleration models}
\label{chapt:model_constraints}


As discussed in the last chapter, the cosmic ray antiprotons measured by PAMELA can be considered as an important means to test propagation models and dark matter properties. In previous works published in the literature (see references in section \ref{sec:CRstatus}), cosmic ray propagation is usually studied using data from different experiments. However, inconsistencies might exist between data sets and thus introduce systematic errors in the final results. Since PAMELA provides measurements of a variety of cosmic ray species, this systematic effect can be potentially reduced. The work presented in this chapter examines whether  PAMELA antiproton and proton data can provide strong and reliable constraints on the source and propagation parameters. In addition, the value of the upcoming PAMELA B/C ratio is demonstrated by employing the B/C ratio from other experiments covering a comparable energy range to PAMELA. The unprecedented accuracy of the PAMELA B/C ratio is expected to give even better constraints. 

The propagation and source parameters under study are summarized in section \ref{sec:paras_sum}. The data used in this work are shown in section \ref{sec:data}. Two statistical approaches, i.e. the $\chi^{2}$ minimization method (see section \ref{sec:chisquare_analysis}) and the Bayesian method (see section \ref{sec:bayes_analysis}), are used to constrain the propagation and acceleration models. Moreover, the Bayesian analysis allows us to test hypotheses, for which the electron flux and the positron fraction are calculated as a consistency check and as an input for future dark matter searches.


\section{Summary of studied parameters} \label{sec:paras_sum}

Different propagation models are studied in this work (as also defined in chapter \ref{chapt:propagation}):
\begin{itemize}
\item the plain diffusion models (PD);
\item the diffusion reacceleration models (DR);
\item the diffusion convection models (DC);
\item the diffusion reacceleration convection models (DRC).
\end{itemize}

The public numerical package GALPROP (version 54.0.572) \cite{GALPROPWebsite} is used. In GALPROP,  the source abundance of protons is normalized based on the propagated proton spectrum at solar position for an energy of 100~GeV, referred to as $N_{p}$. The normalizations of other nuclei are then scaled by their source abundances relative to that for protons. Therefore, instead of the absolute normalization abundances of the injection spectra for different cosmic ray species, $N_{p}$ is used as a free parameter to characterize the source term. Other free parameters related to the source term and the propagation processes were described in section \ref{sec:paradescription}. A summary of these model parameters follows here: 

\begin{itemize}
\item $D_{0}$: a free normalization of the diffusion coefficient at a reference rigidity 4~GV;
\item $\delta$: the spectral index of the diffusion coefficient;
\item $v_{A}$: the Alfv\'en speed characterizing the reacceleration effect;
\item $dV/dz$: the derivative of convection velocity;
\item $\nu$: the injection spectrum index;
\item $N_{p}$: the normalization of the propagated proton spectrum at a reference kinetic energy of 100~GeV. 
\end{itemize}

For stable cosmic ray species, a degeneracy exists between $D_{0}$ and the halo size of the Galaxy, $z_{h}$. Therefore, these two parameters cannot be constrained simultaneously by using stable secondary-to-primary ratios. In this work, $z_{h} = 4$~kpc is assumed in agreement with earlier GALPROP studies of $\rm^{10}Be$, $\rm^{26}Al$, $\rm^{36}Cl$, $\rm^{54}Mn$, and the B/C ratio~\cite{Strong2001,Moskalenko2001a} and to ease the comparison of results. Other GALPROP parameters are held at the conventional reacceleration configuration~\cite{Strong2001, Ptuskin2006}, tuned to reproduce the ACE isotopic abundances of~\cite{Weidenbeck2001}. For studies of the B/C ratio, the nuclear chain starts from $^{28}$Si since all primary elements from Si down to C have an important effect on the B/C ratio. For studies only including proton and antiproton data, the nuclear chain starts from $^{4}$He since primary helium nuclei and proton interactions with the ISM are dominant in the secondary production of protons and antiprotons. 

To reproduce solar modulation, the force-field approximation described in section~\ref{sec:CRintroduction} is used in this analysis. The modulation potential $\Phi$ is chosen to follow that reported by each experiment, i.e. 500~MV for PAMELA \cite{Adriani2011} and HEAO3 \cite{Engelmann1990}, 325~MV for ACE-CRIS \cite{George2009}, 850~MV for CREAM-1 \cite{Ahn2008}, 400~MV for Spacelab-2 \cite{Swordy1990} and 450~MV for AMS01\cite{Aguilar2010}. It is worth to note that the derived values of $\Phi$ suffer from some uncertainties since their determination is phenomenological and depends on the choices of interstellar spectra (see equation \ref{eq:forcefield}). To take into account these uncertainties, the solar parameters are free in the Bayesian analysis. The prior probability distributions (simply called priors), stating our initial knowledge on the parameters before seeing the data, are specified as described in section \ref{sec:priors}. Unlike the model parameters,  the solar parameters can be considered as nuisance parameters which are include in the parameter scan but are not of primary interest.

\section{Data} \label{sec:data}

\begin{figure}[!htb]
\begin{center}
\includegraphics[width=0.85\textwidth]{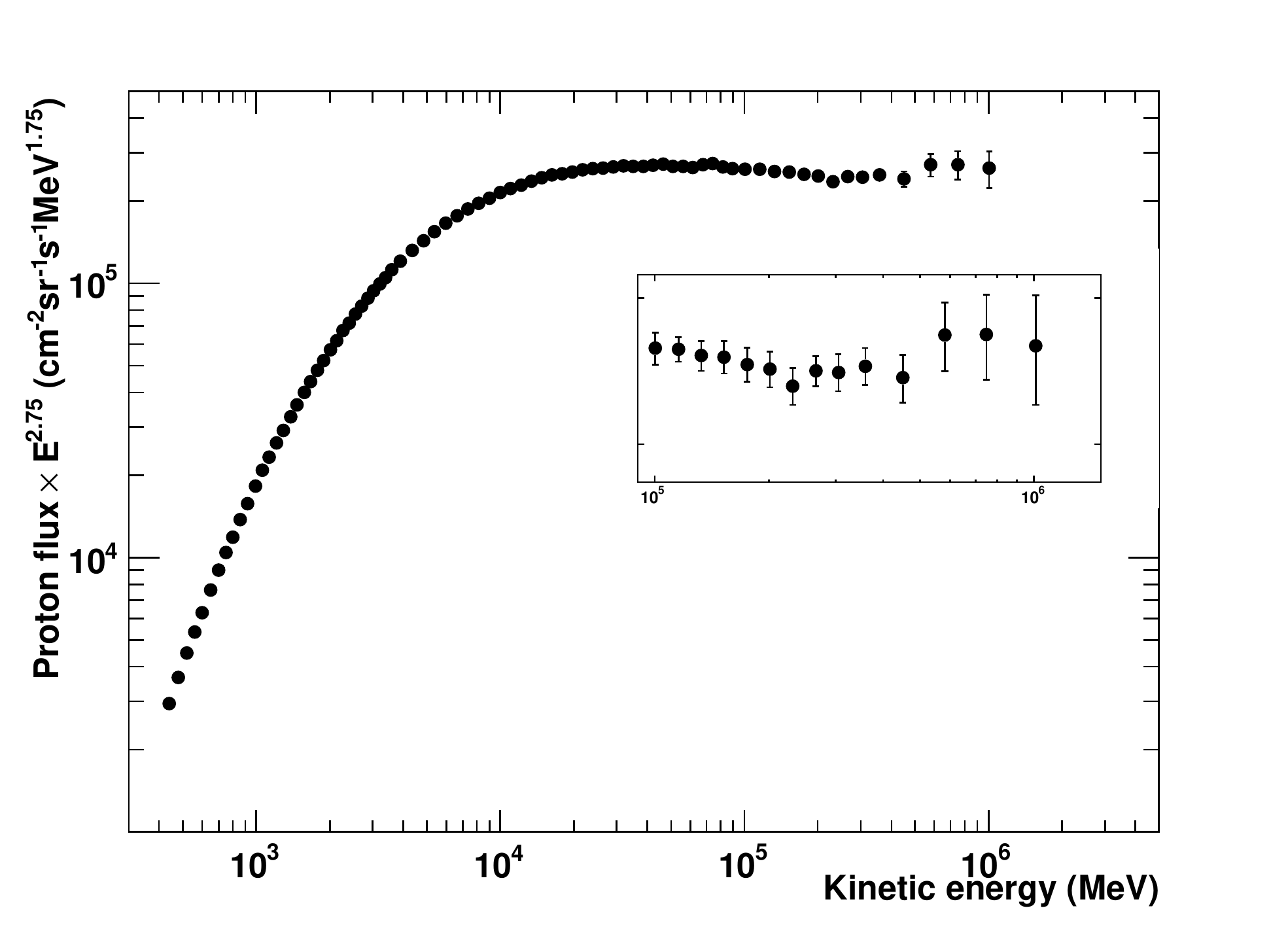}
\end{center}
\caption[Proton flux measured by PAMELA]{\footnotesize Proton flux measured by PAMELA above 1~GeV/n \cite{Adriani2011}.}
\label{fig:PAMELA_proton}
\end{figure}

As mentioned, PAMELA achieves significantly better statistics and extends the energy range compared to previous experiments, especially on antiparticles, e.g., the antiproton flux and the $\bar{\text{p}}$/p ratio, as shown in figure \ref{fig:fluxcompare} and \ref{fig:ratiocompare}. These measurements cover the energy range from 60~MeV to 180~GeV. Such precise measurements enable us to put better constraints on cosmic ray transport parameters. In the analyses presented in this chapter, the published PAMELA data on the $\bar{\text{p}}$/p ratio \cite{Adriani2010_pbar} are used to be consistent with other studies in the literature. Together with the antiproton data, we also use the proton flux measured by PAMELA with great accuracy from 400~MeV to 1.2~TeV~\cite{Adriani2011}, as shown in figure \ref{fig:PAMELA_proton}, to improve constraints on the primary injection spectrum. A hardening in the spectra around 200~GeV can be seen. Some interpretations were proposed to explain this hardening, for example dispersion in the source injection spectra \cite{Yuan2011} or the neutral atoms presented during the acceleration process in a shock \cite{Blasi2012}. This spectrum hardening, however, is not modeled in this work in order to focus on the propagation processes described in chapter \ref{chapt:propagation}.

We propose to exclusively use PAMELA data to study cosmic ray propagation for several reasons. Usually in order to constrain propagation parameters, it is necessary to combine data sets from a variety of experiments to cover a wide enough energy range. As pointed out in \cite{Trotta2011}, errors might be underestimated for an experiment. In order to compensate the systematic discrepancies in the reported uncertainties from data sets, one can therefore introduce a set of nuisance parameters to rescale the reported errors. These rescaling factors increase the computational time for parameter space scans and hinder model selection. These difficulties can be avoided by using only PAMELA data, which additionally allows the use of a smaller parameter set. Another unavoidable problem caused by incorporating data sets from various experiments is that the modulation potential $\Phi$ based on the assumed interstellar spectrum of cosmic ray species may differ between experiments. Relatively poorly understood solar physics makes studies of cosmic ray transport in the Galaxy more difficult. Using only PAMELA data decreases uncertainties on derived propagation parameters by including $\Phi_{\text{PAMELA}}$ as a nuisance parameter, though a potential bias might still arise from the simplified approximation of solar modulation used. 

\begin{figure}[!htb]
\begin{center}
\includegraphics[width=0.85\textwidth]{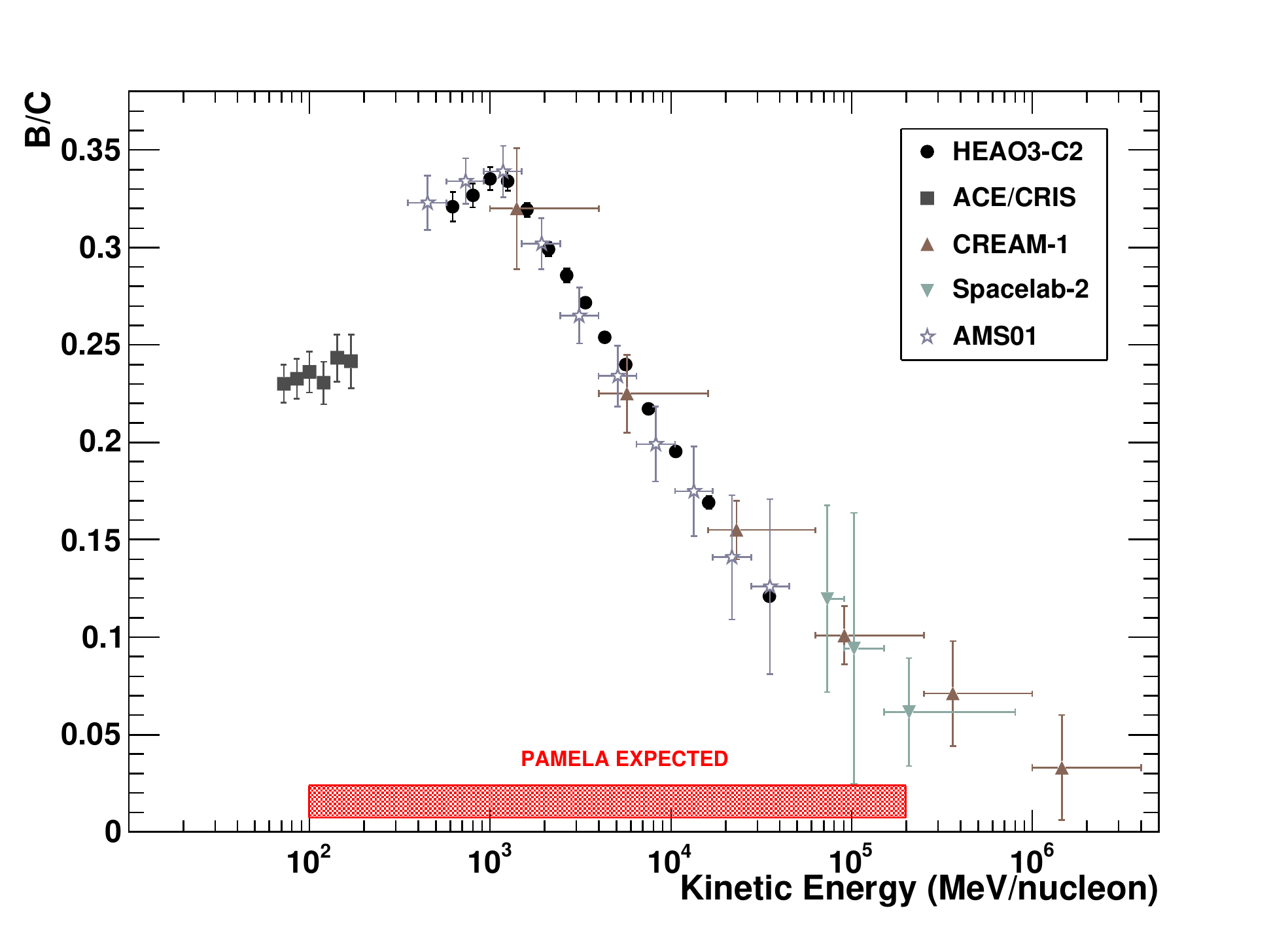}
\end{center}
\caption[B/C ratio measured by previous experiments]{\footnotesize B/C ratios measured by HEAO3 \cite{Engelmann1990}, ACE-CRIS \cite{George2009}, CREAM-1 \cite{Ahn2008}, Spacelab-2 \cite{Swordy1990} and AMS01 \cite{Aguilar2010}. PAMELA is expected to measure the B/C ratio from 100~MeV/n to 200~GeV/n (the red band).}
\label{fig:BCRatio_data}
\end{figure}

The B/C ratio, one of the quantities most sensitive to the propagation parameters, is expected to be measured from 100~MeV/n to 200~GeV/n by the PAMELA experiment in near future. PAMELA is also able to measure hydrogen and helium isotopes, i.e. $^{2}$H and $^{3}$He, over an energy range from 100~MeV/n to 700~MeV/n and 900~MeV/n respectively \cite{Formato2011_ICRC, BoezioPrivate}. These isotopes are believed to be produced during interactions of $^{4}$He. The ratios $^{2}$H/$^{4}$He and $^{3}$He/$^{4}$He have been shown to be as constraining as the B/C ratio \cite{Coste2011}. However, all these ratios from PAMELA are not available yet, therefore the B/C ratio measured by previous experiments~\cite{Swordy1990, Engelmann1990, George2009, Ahn2008, Aguilar2010} (see figure \ref{fig:BCRatio_data}) are used in this work to constrain the transport parameters. The energy range covered by the B/C data sets chosen in the analysis is comparable to the energy range that PAMELA will provide. The ratios of $^{2}$H/$^{4}$He and $^{3}$He/$^{4}$He are not included in the analyses for three reasons. Firstly, they are less accurately measured by individual experiment than the B/C ratio because of the detetors' isotopic separation ability. Secondly, including more data from various experiments will increase the uncertainties due to discrepancies between data sets. Thirdly, a larger number of modulation parameters need to be dealt with. 

Different combinations of data sets are used in the analysis presented in this chapter and are labeled as follows:
\begin{itemize}
\item[``-o'':] only the antiproton flux;
\item[``-a'':] only the B/C ratio;
\item[``-b'':] a combination of the $\bar{\text{p}}$/p ratio and the proton spectrum;
\item[``-c'':] a combination of the B/C ratio, the $\bar{\text{p}}$/p ratio and the proton spectrum.
\end{itemize}

\section{$\chi^{2}$ minimization approach} \label{sec:chisquare_analysis}

To understand how well a model reflects the observed data, a commonly performed statistical test is the $\chi^{2}$ `goodness of fit' test. Assuming we have a number of $N$ data points ($X_{i}$, $Y_{i}$) with Gaussian distributed errors, the $\chi^{2}$ test statistic is defined as:
\begin{equation} \label{eq:chisquare}
\chi^{2}\left( \mathbf{\Theta}\right) = \sum_{i=1}^{N}\frac{\left( f\left( X_{i}, \mathbf{\Theta}\right) - Y_{i} \right)^{2}}{\sigma_{i}^{2}},
\end{equation}
where $f\left( X_{i}, \mathbf{\Theta} \right)$ is the theoretical value at abscissa $X_{i}$, and $\sigma_{i}$ are the uncertainties on the measurements $Y_{i}$. Since the theoretical expectation depends on the vector of physical parameters $\mathbf{\Theta}$, the best-fit parameters can be extracted by minimizing the $\chi^{2}$. Usually the reduced $\chi^{2}$ is used, i.e. $\chi^{2}$ divided by the number of degrees of freedom (d.o.f), where the d.o.f is equal to the number of data points minus the number of free parameters. 

GALPROP is interfaced with the minimization library MINUIT~\cite{James1975} for this analysis. The MINUIT package provides several algorithms to perform a minimization of a multi-parameter function, such as the the Nelder-Mead SIMPLEX algorithm \cite{Nelder1965} and the MIGRAD algorithm based on the variable metric method \cite{Fletcher1970}. An error matrix is calculated as a by-product of MIGRAD and the parabolic errors are derived. More general asymmetric errors can be further produced by the MINOS method. In this work, $f\left( X_{i}, \mathbf{\Theta}\right)$ corresponds to the cosmic ray fluxes and/or flux ratios given by GALPROP at kinetic energy $X_{i}$. The $\chi^{2}$ function is derived from the GALPROP prediction of $f\left( X_{i}, \mathbf{\Theta}\right)$ and the experimental data described in section \ref{sec:data}. The efficient MIGRAD method is used to evaluate the best-fit parameter values and MINOS is used to derive the uncertainties of the parameters. If MIGRAD fails to converge then the minimizer is switched to the slower SIMPLEX method.

\subsection{Analysis and results for an unmodified diffusion coefficient} \label{sec:chi2_unmodified_diff}

The antiproton data measured by PAMELA and the B/C ratio reported by other experiments are used separately to check their constraining power for propagation parameters. Instead of the $\bar{\text{p}}$/p ratio, the antiproton flux is used since once the proton flux is fixed, the antiproton flux is more sensitive than the $\bar{\text{p}}$/p ratio to the propagation parameters. The source injection index is fixed at $\nu=2.3$ and the propagated proton spectrum is normalized as $N_{p}=4.69\times 10^{-9}$\,cm$^{-2}$\,sr$^{-1}$\,s$^{-1}$\,MeV$^{-1}$ at 100~GeV to fit the PAMELA proton data. The simplest model, i.e. PD, is studied. Results obtained from the antiproton flux (PD-o model) are $D_{0}= 3.88\pm0.14 \times 10^{28}$~cm$^{2}$/s and $\delta=0.479\pm0.024$, which are close but less constraining than those obtained from the B/C ratio (PD-a model), $D_{0}= 5.57\pm0.04 \times 10^{28}$~cm$^{2}$/s and $\delta=0.490\pm0.008$. By fitting the antiproton flux, the best-fit PD-o model gives a reduced $\chi^{2}$ of 0.70, which means that the data are overfitted by the PD-o model. The DR-o model, with $v_{A}=14\pm^{10}_{14}$~km\,s$^{-1}$ and diffusion coefficients consistent with the PD-o model, does not change the value of $\chi^{2}$. The large error on $v_{A}$ indicates that reacceleration is not constrained by only using the measured antiproton spectrum. The convection velocity $dV/dz$ is converged at zero, since the model with convection (DC-o) always increases the $\chi^{2}$ value compared to the PD-o model and is disfavored. The PD-o model is sufficient here to reproduce the antiproton data due to two possible reasons:  (1) the low energy antiprotons are primarily produced through ``tertiary'' processes coming from inelastic scattering of high energy antiprotons, therefore the antiproton flux may not be very sensitive to the other low energy processes, i.e. reacceleration and convection; (2) the antiproton flux is dependent on the injection spectrum of primaries since antiproton production is significantly influenced by the kinetic energy of their progenitors, i.e. protons and helium nuclei. 

\begin{figure}[!htb]
\begin{center}
\includegraphics[width=0.85\textwidth]{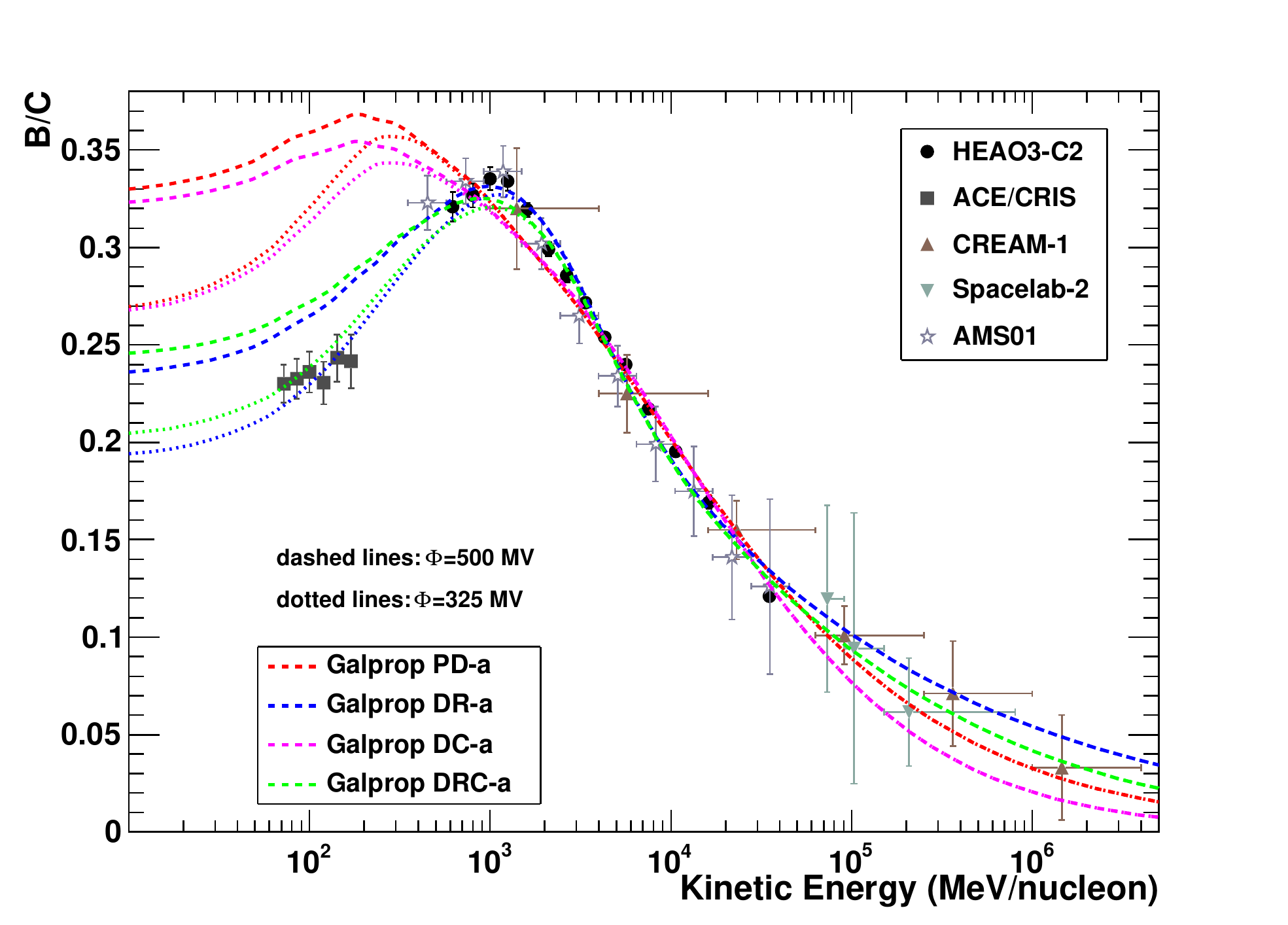}
\end{center}
\caption[The B/C ratio compared with various models in the $\chi^{2}$ study]{\footnotesize The B/C ratio for the best-fit parameters of PD-a, DR-a, DC-a and DRC-a models as listed in table \ref{tab:chi2analysis}.}
\label{fig:BC_fits_a}
\end{figure}

Tighter constraints are placed on transport parameters by using the B/C ratio, which are summarized in table \ref{tab:chi2analysis} as well as the corresponding $\chi^{2}$/d.o.f. From the reduced $\chi^{2}$, PD-a and DC-a models can not reproduce the B/C ratio. Large deviations at low energy can be seen in figure \ref{fig:BC_fits_a} from the comparison of the data with the best-fit PD-a and DC-a models. The reacceleration process (DR and DRC models) can explain the B/C ratio adequately. The DRC-a model is found to best fit the B/C data. Compatible results are obtained by fixing the injection spectrum index $\nu=2.5$, which indicates that the sensitivity of the B/C ratio to the injection spectrum is weak. This is expected since the boron nuclei have almost the same energy/nucleon as their primaries. 

\begin{figure}[tbp]
\begin{center}
\includegraphics[width=0.85\textwidth]{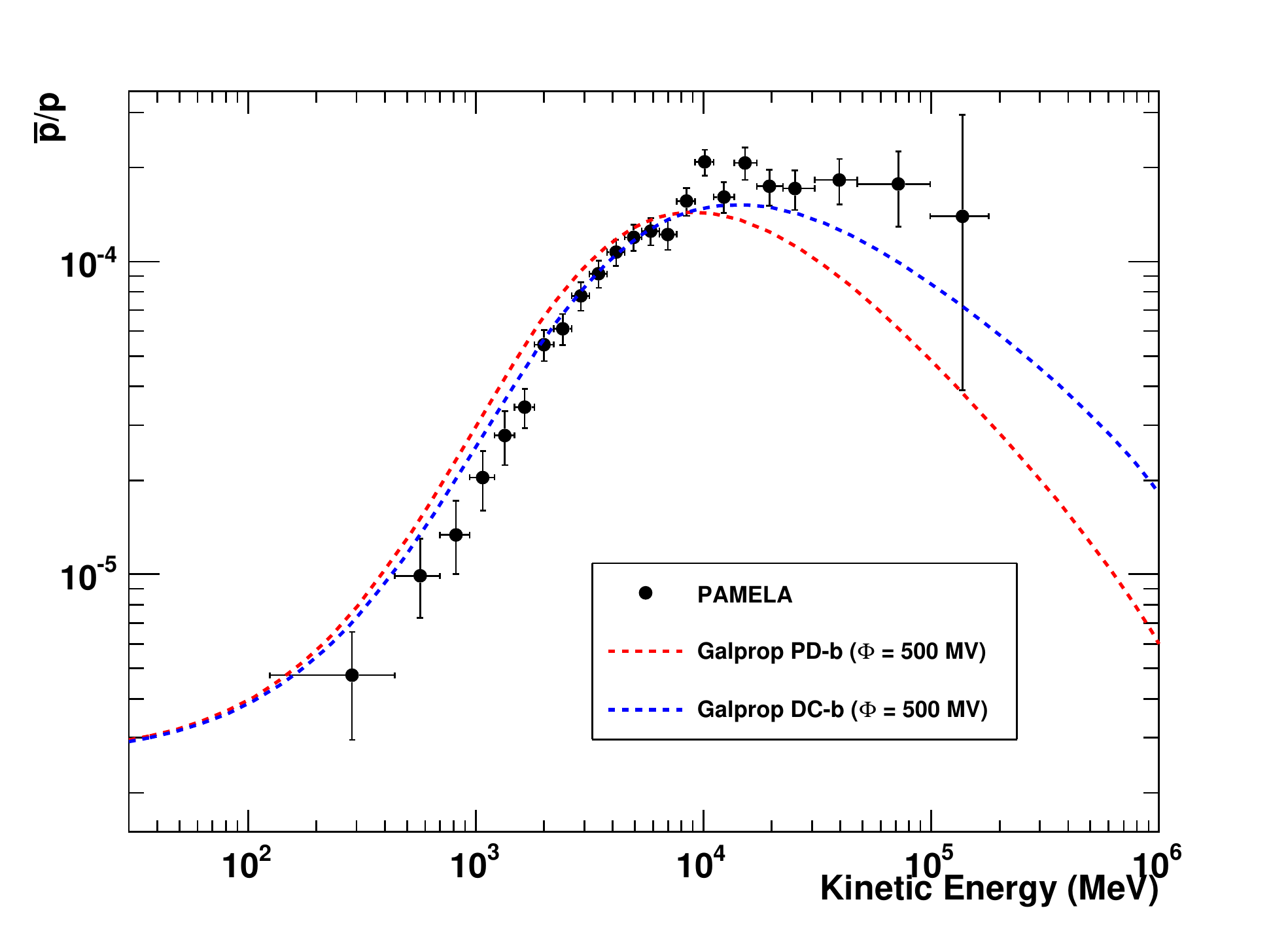}
\includegraphics[width=0.85\textwidth]{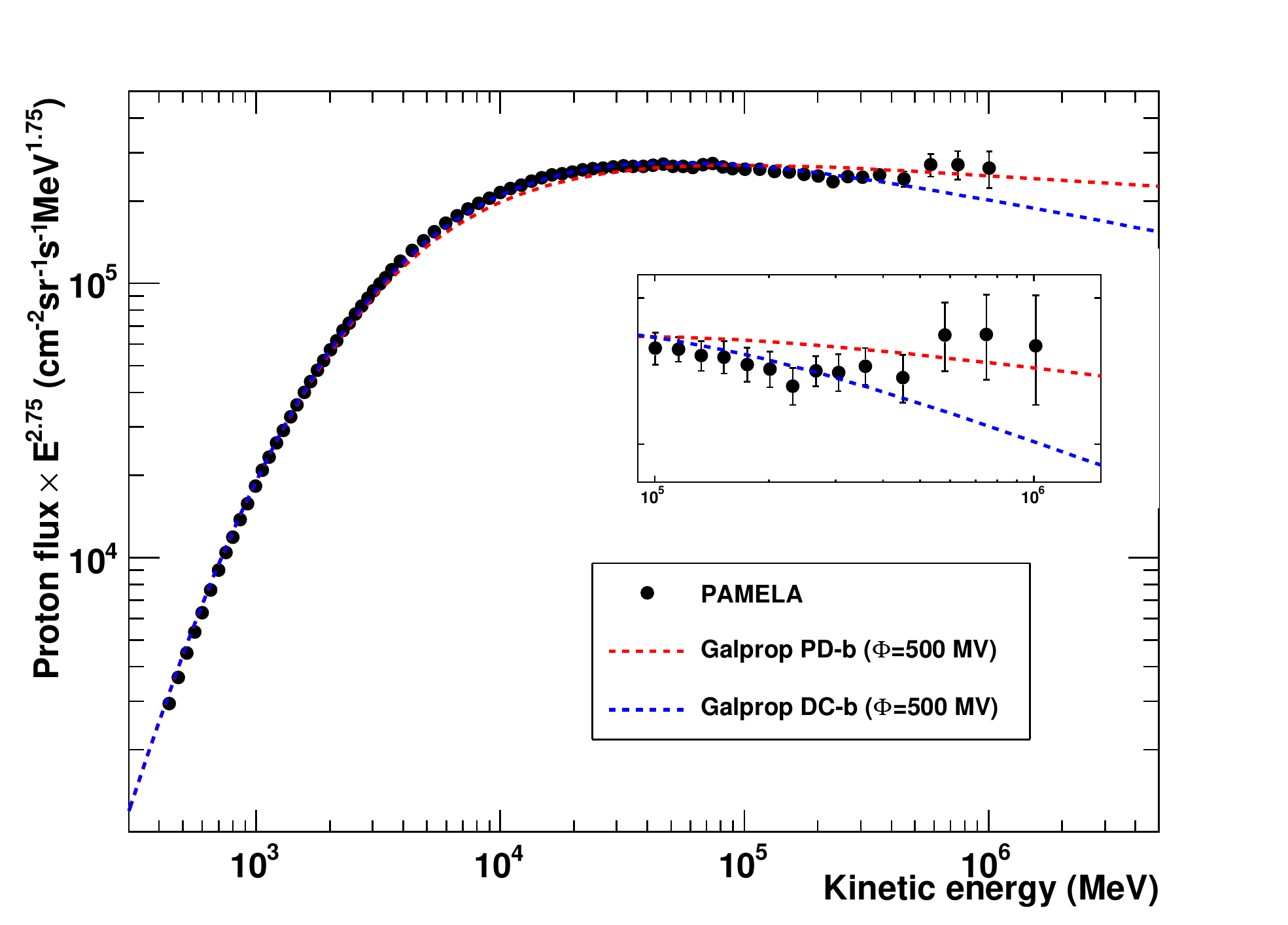}
\end{center}
\caption[The $\bar{\text{p}}$/p ratio and the proton flux compared with various models in the $\chi^{2}$ study]{\footnotesize The $\bar{\text{p}}$/p ratio (top) and the proton spectrum (bottom) for the best-fit parameters of PD-b and DC-b models as listed in table \ref{tab:chi2analysis}.}
\label{fig:ppbar_fits_b}
\end{figure}

In order to obtain complementary information on transport and source parameters, a simultaneous fit to both the secondary-to-primary ratios and the primary fluxes is necessary. Using only PAMELA data, the proton flux is combined with the $\bar{\text{p}}$/p ratio to estimate the parameters. When only the data of antiproton spectrum were fitted, the source parameters were fixed and the simplest model, PD-o, seems sufficient to describe the antiproton data. However, if the source parameters are varied, the derived values of transport parameters will be considerably changed. Therefore a simultaneous fit on the $\bar{\text{p}}$/p ratio and the proton flux may allow us to constrain all the processes. The results listed in table \ref{tab:chi2analysis} show that parameters for both PD-b and DC-b models can be constrained. However, the reliability of these best-fit parameters needs to be questioned. A rather high spectral index of the diffusion coefficient $\delta=0.84\pm0.04$ appears in the PD-b model, which can reproduce the proton spectrum but not the $\bar{\text{p}}$/p ratio, as shown in figure \ref{fig:ppbar_fits_b}. This bias on the estimated parameters is due to the dominant weight of the proton flux since it is more precisely measured than the $\bar{\text{p}}$/p ratio. The DC-b model, with the reduced $\chi^{2}$ close to 1, gives comparable best-fit parameters to that for the DC-a model and can generally fit both the proton flux and the $\bar{\text{p}}$/p ratio. Contrary to the indication from the B/C ratio, reacceleration is disfavored by fitting simultaneously the $\bar{\text{p}}$/p ratio and the proton flux, i.e. $v_A \to 0$ for DR-b and DRC-b models. Generally, the derived best-fit propagation parameters have errors at least twice larger than the ones obtained from the B/C ratio, showing that the combination of $\bar{\text{p}}$/p ratio and the proton flux is not as constraining as the B/C ratio.

\begin{figure}[!htb]
\begin{center}
\includegraphics[width=0.65\textwidth]{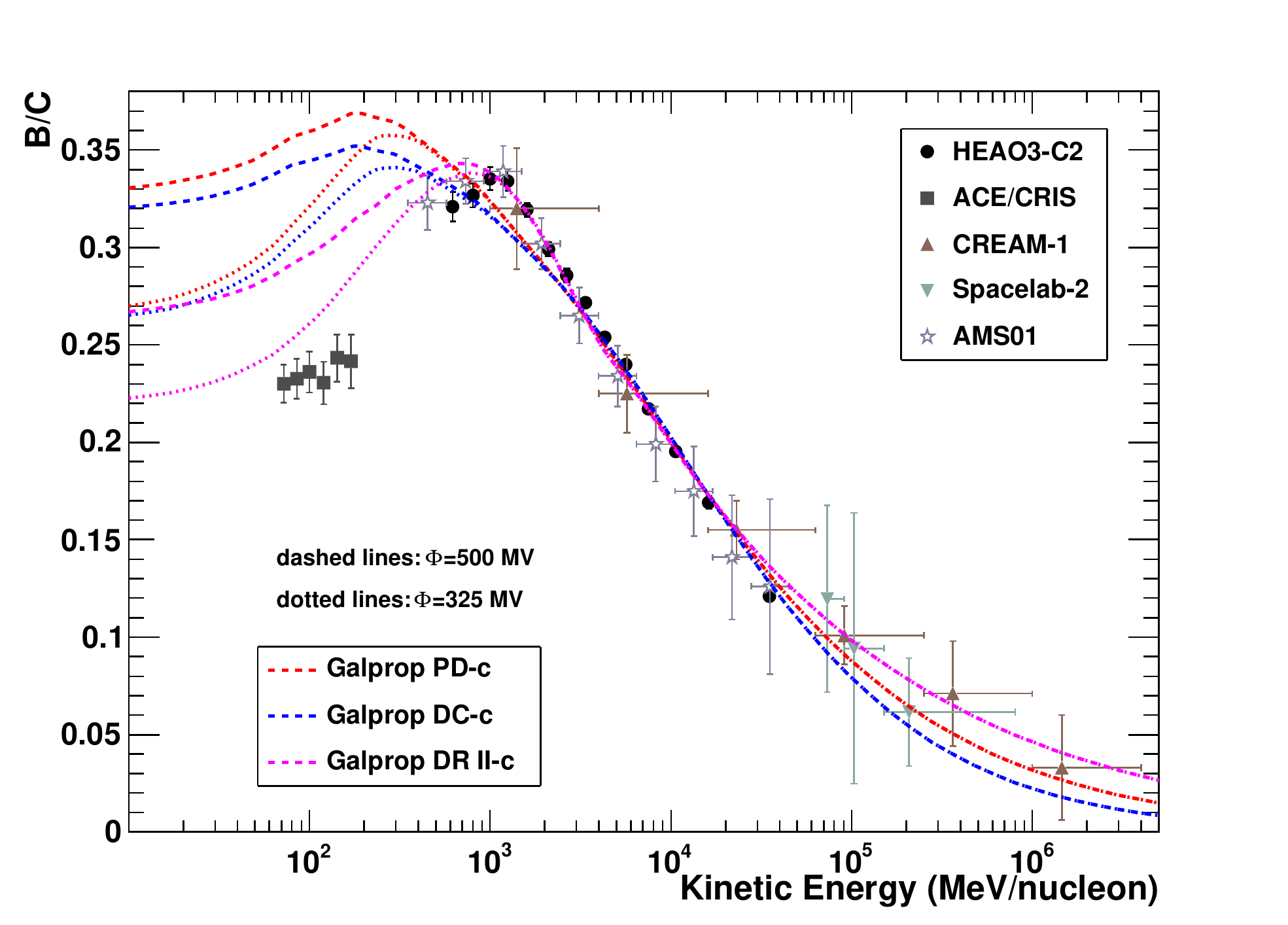}
\includegraphics[width=0.65\textwidth]{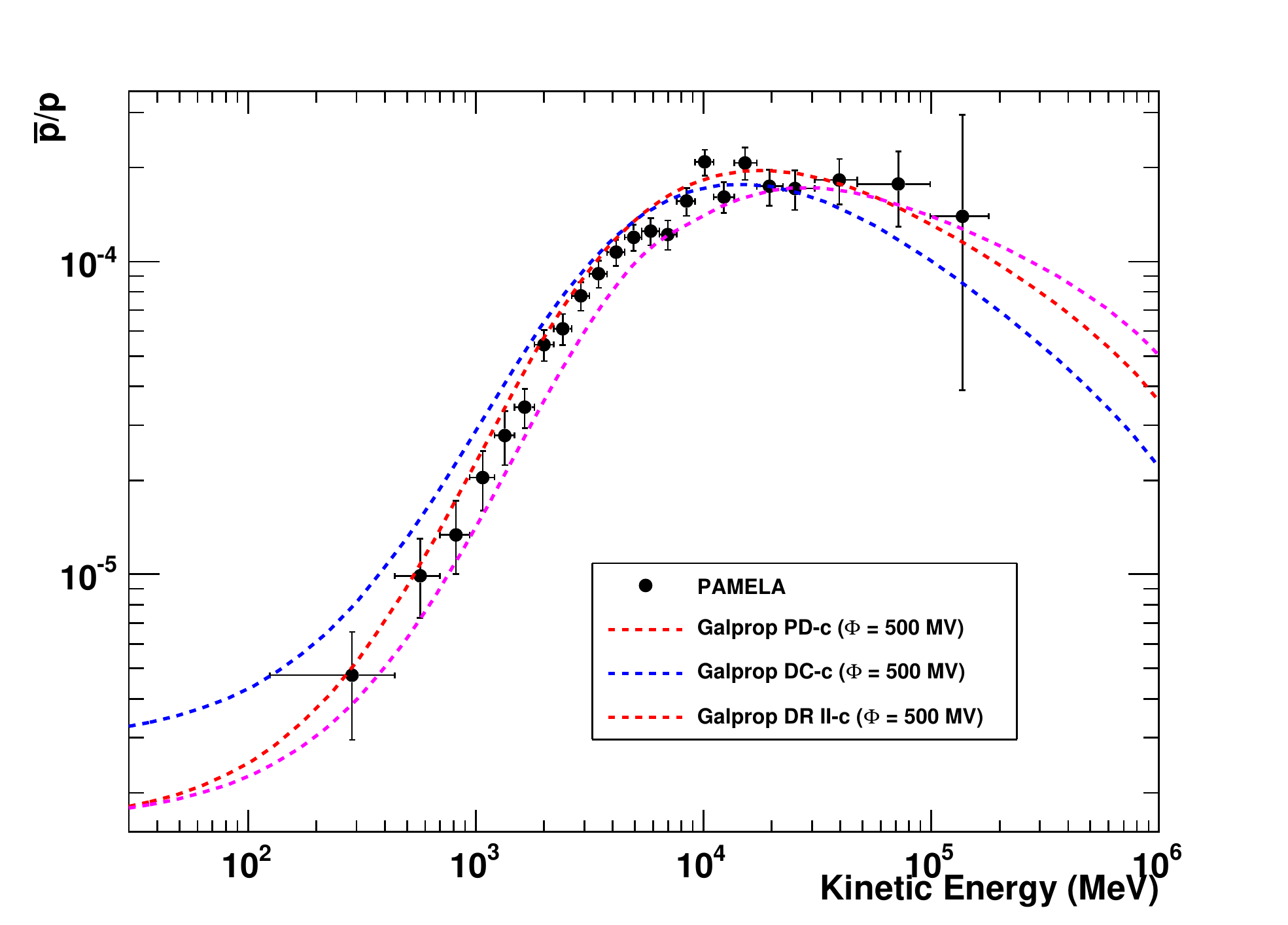}
\includegraphics[width=0.65\textwidth]{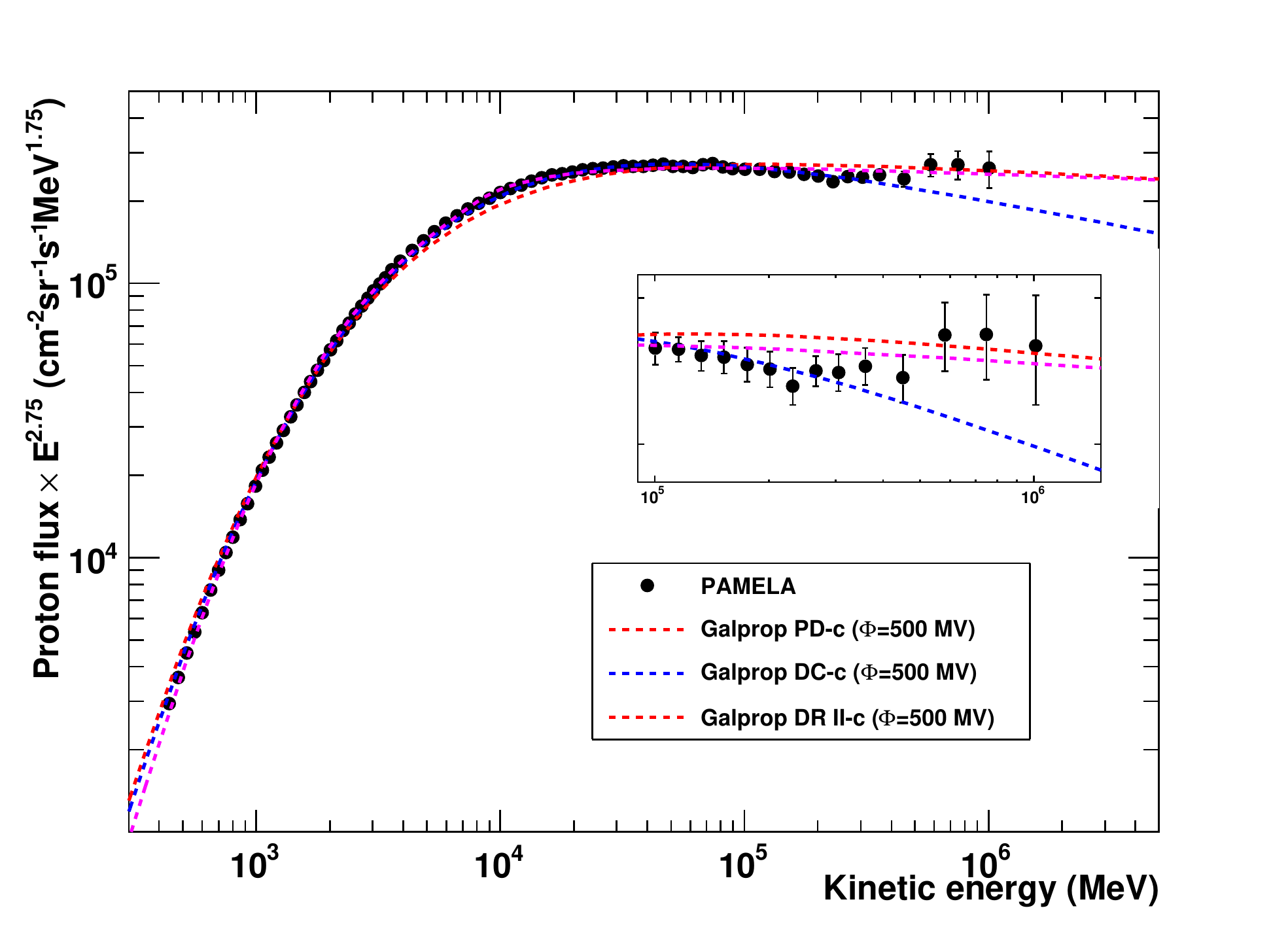}
\end{center}
\caption[The B/C ratio, the $\bar{\text{p}}$/p ratio and the proton flux compared with various models in the $\chi^{2}$ study]{\footnotesize The B/C ratio (top), the $\bar{\text{p}}$/p ratio (middle) and the proton spectrum (bottom) for the best-fit parameters of PD-c, DC-c and DR II-c models as listed in table \ref{tab:chi2analysis}.}
\label{fig:bcpbarp_fits_c}
\end{figure}

Since the estimated parameters might not be reliable by fitting only the $\bar{\text{p}}$/p ratio and the proton flux, the B/C ratio is also included in the simultaneous fit. The B/C ratio is more constraining than the $\bar{\text{p}}$/p ratio and therefore will decrease the bias on the transport parameters. The best-fit PD-c and DC-c models can reproduce all the data except for the low energy B/C ratio, as shown in figure \ref{fig:bcpbarp_fits_c}. Unlike the PD-b model, the PD-c model gives a lower spectral index of the diffusion coefficient $\delta=0.495\pm0.007$ which provides a satisfactory fit on the $\bar{\text{p}}$/p ratio. Reacceleration, which is expected to explain the low energy B/C ratio, is still disfavored since it conflicts with the proton flux. One way to solve this disagreement is to introduce an unphysical adhoc break in the injection spectrum, as did in \cite{Moskalenko2002, Trotta2011}. The same configuration, referred to as ``DR II'', is also tested here to fit the B/C ratio, the $\bar{\text{p}}$/p ratio and the proton spectrum. The best-fit parameters of this DR II-c model are also given in table \ref{tab:chi2analysis}. However, the predicted B/C ratio of this model is still higher than the data below 1~GeV (see figure \ref{fig:bcpbarp_fits_c}). This is because that a higher $v_{A}$ (as the best-fit value of the DR-a model) than the one obtained for the DR II-c model, is needed to explain the B/C data. In order to account for the low energy B/C ratio, a nonlinear diffusion coefficient will be further studied in section \ref{sec:modified_diff}.

\begin{sidewaystable}
\begin{center}
\begin{tabular}{c | c | c | c | c | c | c | c}
\hline
\hline
Model &  $D_{0} \cdot 10^{-28}$ & $\delta$  & $v_{A} $  & $\text{d}V/\text{d}z$ & $\nu$ &$N_{p}\cdot 10^{9}$&$\chi^{2}/$d.o.f \\
 &\!(cm$^{2}$/s)\!&  & \!(km/s)\! & \!\!(km/s/kpc)\!\! &&\!\!(cm$^{2}$/sr/s/MeV)\!\!& \\
\hline
\hline
&&&&&&&\\
PD-a&\!$3.57\pm0.05$\!&\!$0.490\pm0.009$\!&---&---&[2.3]&[4.69]&13.6\\[2mm]
PD-b&\!$3.13\pm0.11$\!&\!$0.84\pm0.05$\!&---&---&\!$1.97\pm0.04$\!&\!$4.76\pm0.04$\!&2.47\\[2mm]
PD-c&\!$3.56\pm0.04$\!&\!$0.495\pm0.008$\!&---&---&\!$2.303\pm0.008$\!&\!$4.74\pm0.04$\!&5.94\\[2mm]
&&&&&&&\\
DR-a&\!$6.68\pm0.13$\!&\!$0.304\pm0.006$\!&\!$34.7\pm1.0$\!&---&  [2.3]& [4.69]&2.13 \\[2mm]
DR-b&\!$3.13\pm0.11$\!&\!$0.84\pm0.05$\!&\!$0\pm^{0.5}_{0}$\!&---&\!$1.97\pm0.04$\!&\!$4.76\pm0.04$\!&2.50\\[2mm]
DR-c&\!$3.56\pm0.04$\!&\!$0.495\pm0.008$\!&\!$0\pm^{0.6}_{0}$\!&---&\!$2.303\pm0.008$\!&\!$4.74\pm0.04$\!&5.98\\[2mm]
DR II-c&\!$5.29\pm0.09$\!&\!$0.368\pm0.007$\!&\!$28.7\pm0.7$\!&---&\!$1.869\pm0.10$/$2.413\pm0.009$\!&\!$4.59\pm0.04$\!&1.68\\[2mm]
&&&&&&&\\
DC-a&\!$2.13\pm0.13$\!&\!$0.648\pm0.020$\!&---&\!$10.6\pm^{1.3}_{1.2}$\!&[2.3]&[4.69]&11.5\\[2mm]
DC-b&\!$2.65\pm0.16$\!&\!$0.630\pm^{0.029}_{0.028}$\!&---&\!$11.9\pm^{2.2}_{1.9}$\!&\!$2.30\pm0.04$\!&\!$4.72\pm0.04$\!&1.02\\[2mm]
DC-c&\!$2.29\pm0.08$\!&\!$0.622\pm0.012$\!&---&\!$9.7\pm0.8$\!&\!$2.309\pm0.008$\!&\!$4.66\pm0.04$\!&3.98\\[2mm]
&&&&&&&\\
DRC-a&\!$3.5\pm^{0.6}_{0.5}$\!&\!$0.438\pm^{0.023}_{0.028}$\!&\!$40.0\pm^{2.3}_{1.9}$\!&\!$31\pm^{10}_{8}$\!&  [2.3]& [4.69]&1.58 \\[2mm]
DRC-b&\!$2.65\pm0.16$\!&\!$0.63\pm0.03$\!&\!$0\pm^{1.7}_{0}$\!&\!$11.9\pm^{2.2}_{1.9}$\!&\!$2.30\pm0.04$\!&\!$4.72\pm0.04$\!&1.01\\[2mm]
DRC-c&\!$2.29\pm0.08$\!&\!$0.622\pm0.012$\!&\!$0\pm^{1.5}_{0}$\!&\!$9.7\pm0.8$\!&\!$2.309\pm0.008$\!&\!$4.66\pm0.04$\!&4.01\\[2mm]
&&&&&&&\\
\hline
\end{tabular}
\caption[The best-fit parameters for PD, DR, DC and DRC models in the $\chi^{2}$ study]{\footnotesize The best-fit parameters for PD, DR, DC and DRC models by fitting only B/C ratio (labelled as a), by fitting the $\bar{\text{p}}$/p ratio plus the proton spectrum (labelled as b), and by fitting simultaneously the B/C ratio,  the $\bar{\text{p}}$/p ratio and the proton spectrum (labelled as c). The value in the square brackets correspond to the fixed value of the parameter. The break on the injection index used in the DR II-c model is placed at 100~GeV.}
\label{tab:chi2analysis}
\end{center}
\end{sidewaystable}

\subsubsection{Comparison with previous studies}

Except for the biased values obtained for PD-b and DR-b, the spectral index $\delta$ of the diffusion coefficient is well constrained between 0.3 and 0.65 for all models considered. Whereas the PD models favour a Kraichnan turbulence spectrum of $\delta = 0.5$, the DC models favour a slightly higher value of $\delta$ between 0.62 and 0.65. The Kolmogorov spectrum of turbulence, $\delta=1/3$, is only recovered for the DR-a model and DR II-c, in agreement with earlier studies (e.g., \cite {Moskalenko2002, Trotta2011}). Including observations of primary nuclei, i.e. the proton flux, tends to disfavour reacceleration unless a break is introduced in the injection spectrum as done in the DR II-c model. The Galactic wind velocities obtained for the DC models are around 10~km\,s$^{-1}$\,kpc$^{-1}$ and in good agreement with other studies, e.g., \cite{Moskalenko2002}. The same is valid for the Alfv\'en velocities of about 35-40~km\,s$^{-1}$ for the DR-a and DRC-a models. 

\subsection{Analysis and results for a modified diffusion coefficient} \label{sec:modified_diff}

The discrepancy between the low energy B/C observations and the models which are compatible with other data, could be explained by nonlinear MHD wave effects, i.e. the spectral wave density decreases from small to large wave numbers \cite{Ptuskin2006}. This turbulence dissipation effect can result in a low energy dependence of the cosmic ray diffusion coefficient, as mentioned in section \ref{sec:CRstatus}. A spatial diffusion coefficient $D_{xx}=D_{0}\beta^{\eta} \left( \rho / \rho_{0} \right)^{\delta}$ is adopted to account for this effect, where $\eta$ is added as a free parameter in the fit. Unlike introducing an artifical break on the diffusion coefficient at a rigidity 3~GV or 4~GV as used in \cite{Strong1998, Moskalenko2002}, or on the injection spectrum at 10~GV used in \cite{Moskalenko2002, Trotta2011}, this approach is physically motivated.

\begin{figure}[!h]
\begin{center}
\includegraphics[width=0.65\textwidth]{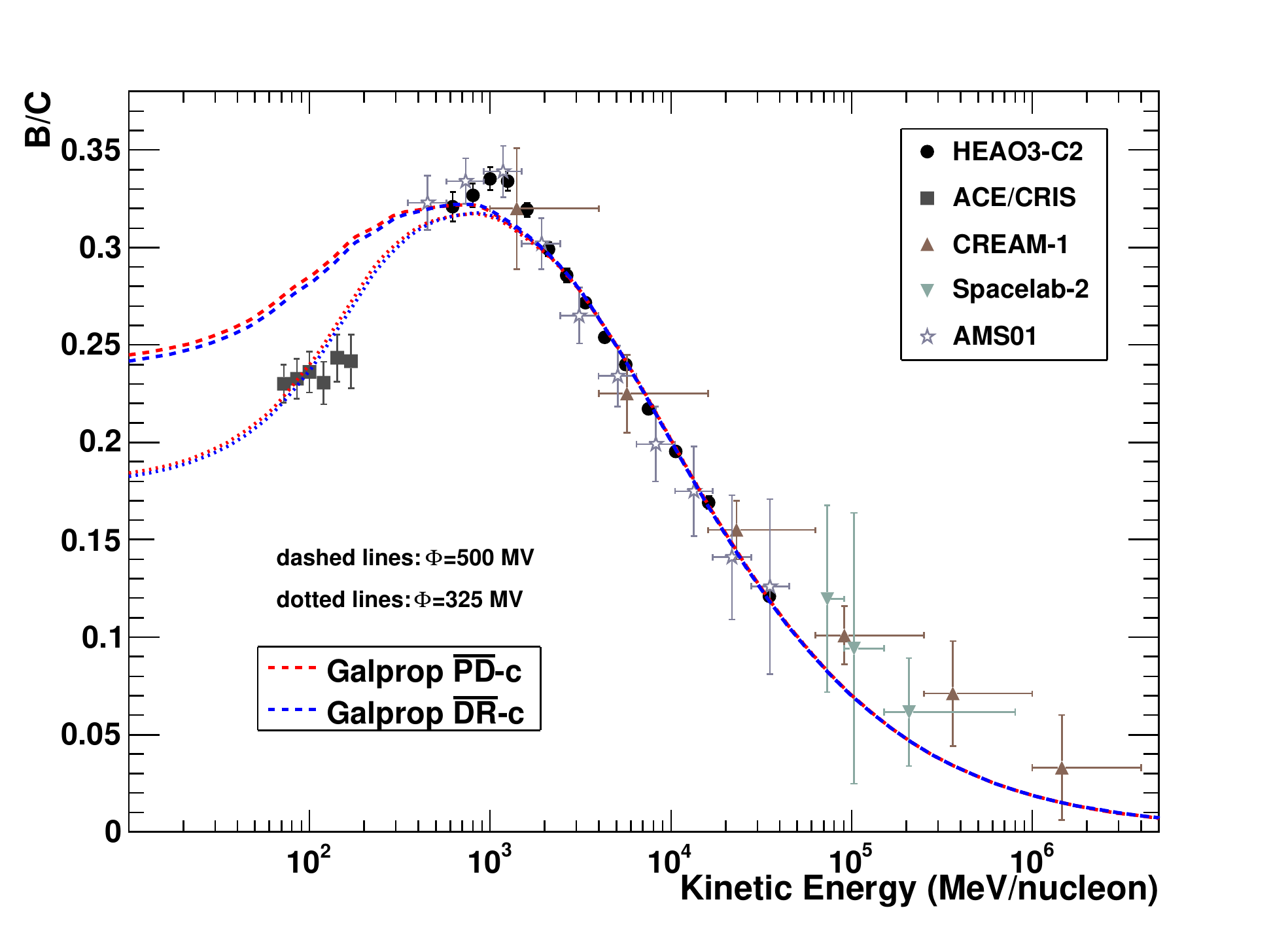}
\includegraphics[width=0.65\textwidth]{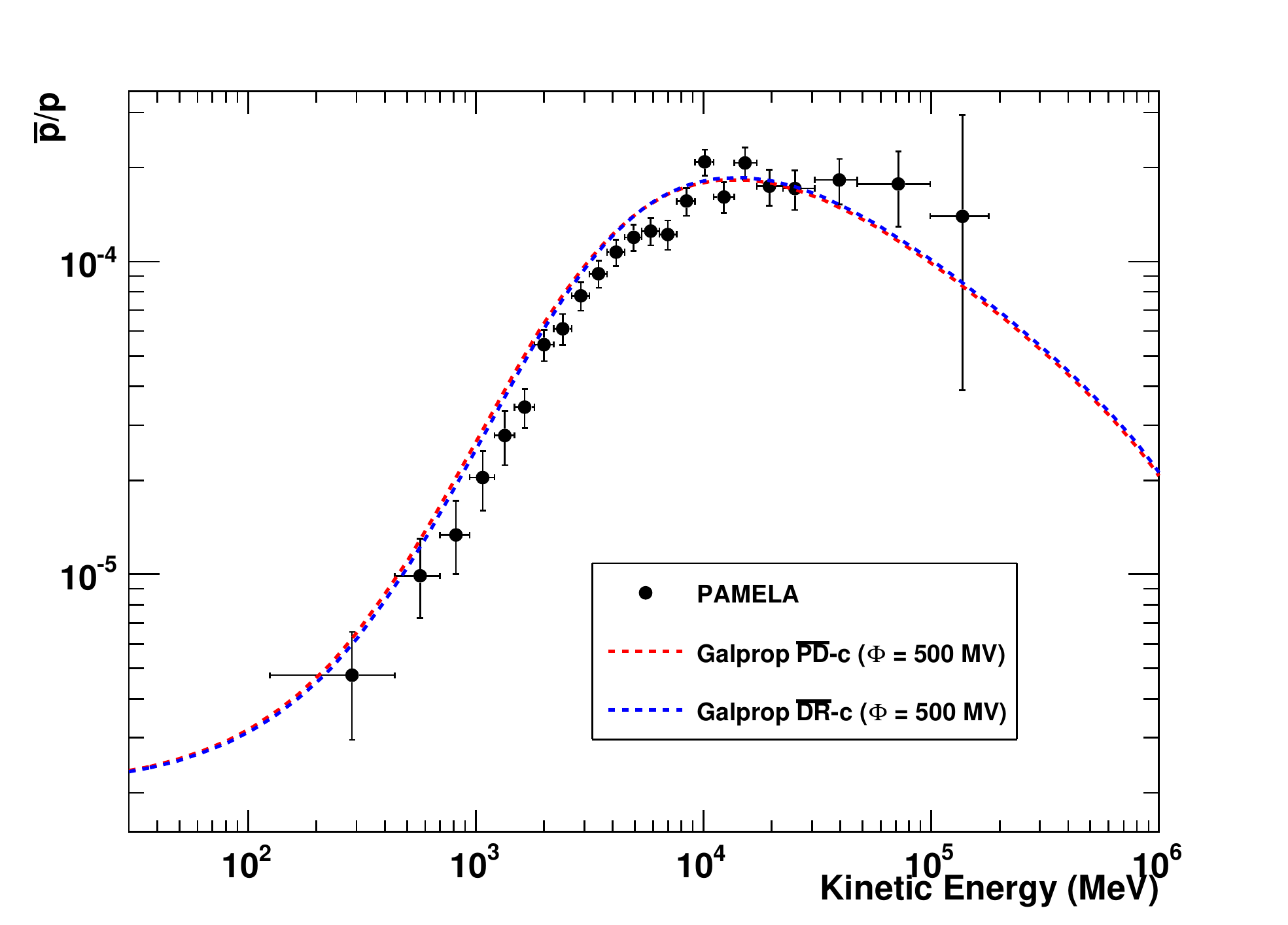}
\includegraphics[width=0.65\textwidth]{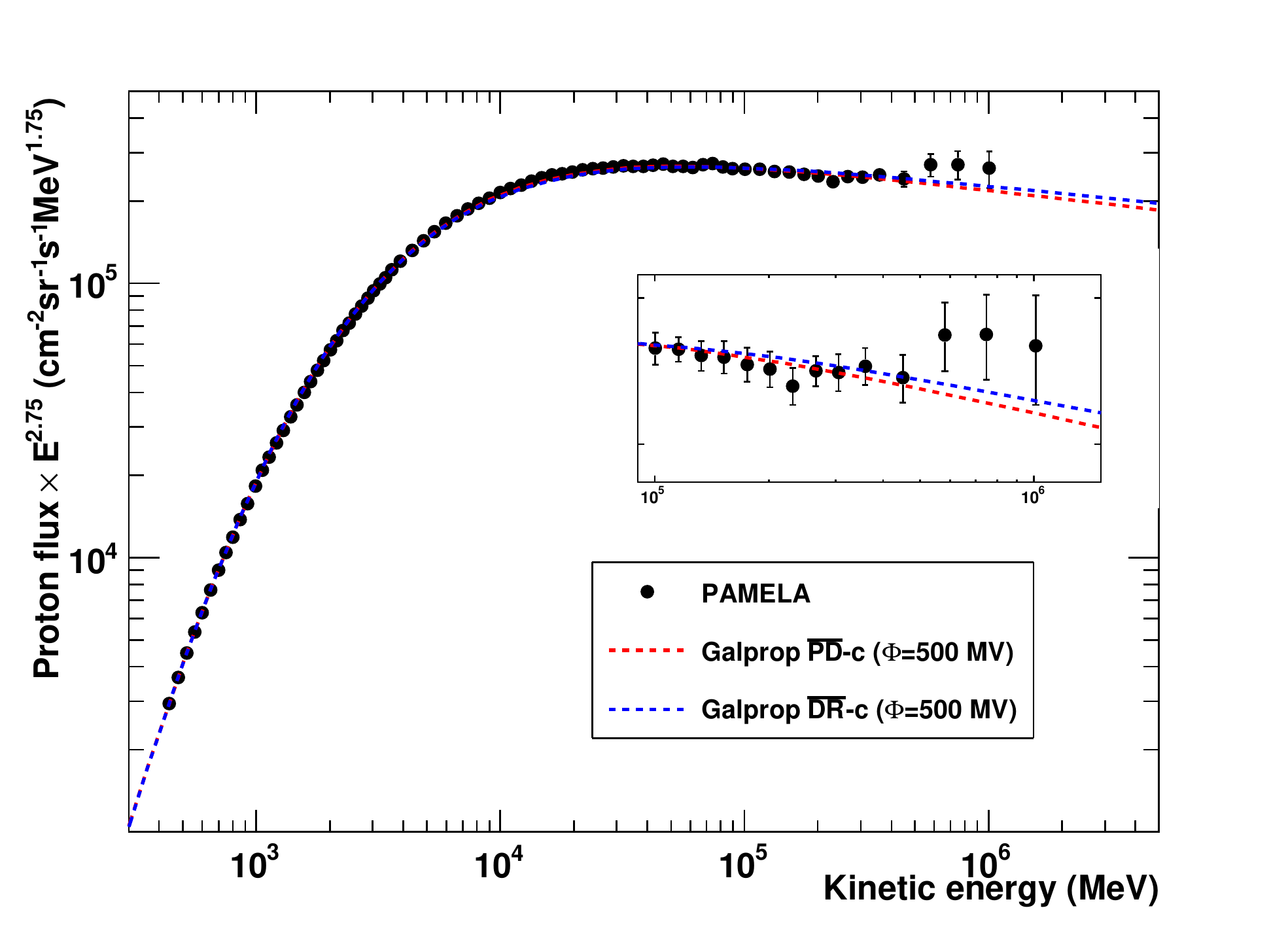}
\end{center}
\caption[The B/C ratio, the $\bar{\text{p}}$/p ratio and the proton flux compared with various models in the $\chi^{2}$ study]{\footnotesize The B/C ratio (top), the $\bar{\text{p}}$/p ratio (middle) and the proton spectrum (bottom) for the best-fit parameters of $\rm \overline{PD}$-c and $\rm \overline{DR}$-c models as listed in table \ref{tab:chi2analysis_eta}.}
\label{fig:bcpbarp_fits_eta}
\end{figure}

Models including $\eta$ to parameterize low energy MHD physics are referred as $\rm \overline{PD}$, $\rm \overline{DR}$, $\rm \overline{DC}$ and $\rm \overline{DRC}$. The same combinations of data sets as used in the previous section, are employed to study these models. All the results are summarized in table \ref{tab:chi2analysis_eta}. The effect of $\eta$ competes with effects of convection and reacceleration. If $\eta$ is included in the fit of the B/C ratio, no convection is favored. To fit both the $\bar{\text{p}}$/p ratio and the proton flux, the values of the reduced $\chi^{2}$ are much smaller than 1.0. This unexpectedly good fit may be caused by the possible correlated and/or overestimated errors, or the errors are not Gaussian distributed, which then would result in a significant deviation from the $\chi^{2}$ distribution and make the $\chi^{2}$ minimization approach inappropriate in this specific case. A more precise study is beyond the scope of this thesis, but will be addressed in a future work. The smaller values of the reduced $\chi^{2}$ indicates that models including a low energy dependence in the diffusion coefficient are preferred to explain the $\bar{\text{p}}$/p ratio and the proton flux. But using only the $\bar{\text{p}}$/p ratio and the proton flux is not enough to constrain models including $\eta$. In the simultaneous fit of the B/C ratio, the $\bar{\text{p}}$/p ratio and the proton flux, the convection velocity is converged to zero due to degeneracy between $\eta$ and $dV/dz$ and the dominant effect of $\eta$ at low energy.

The $\rm \overline{PD}$-c model is generally consistent with all the data except for a slight overprediction of the $\bar{\text{p}}$/p ratio below 10~GeV. The best-fit values of $\delta=0.621\pm0.009$ and $\eta=-1.75\pm0.10$ are close to the values given in \cite{Ptuskin2006}, i.e. $\delta=0.60$ and $\eta=-2$. Reacceleration can be invoked to fit the data. However, since the required $v_{A}$ is very weak and the reduced $\chi^{2}$ for model $\rm \overline{DR}$-c is compatible with the one for model $\rm \overline{PD}$-c, reacceleration seems not to be necessary to explain the data. This can also be illustrated in figure \ref{fig:bcpbarp_fits_eta}. Nevertheless, $\eta$ is found to dominate over other competing processes and may result in too many degenerated parameters being constrained at low energy.


\begin{sidewaystable}
\begin{center}
\begin{tabular}{c | c | c | c | c | c | c | c|c}
\hline
\hline
Model &  $D_{0} \cdot 10^{-28}$ & $\delta$  & $v_{A} $  & $\text{d}V/\text{d}z$ & $\nu$ &$N_{p}\cdot 10^{9}$& $\eta$&$\chi^{2}/$d.o.f \\
 &\!(cm$^{2}$/s)\!&  & \!(km/s)\! & \!\!(km/s/kpc)\!\! &&\!\!(cm$^{2}$/sr/s/MeV)\!\!& &\\
\hline
\hline
&&&&&&&&\\
$\rm \overline{PD}$-a&\!$2.70\pm0.05$\!&\!$0.642\pm0.012$\!&---&---&[2.3]&[4.69]&$-1.86\pm0.13$&2.00\\[2mm]
$\rm \overline{PD}$-b&\!$3.64\pm^{0.16}_{0.15}$\!&\!$0.46\pm0.04$\!&---&---&\!$2.39\pm0.04$\!&\!$4.64\pm0.04$\!&$-2.30\pm0.21$&0.37\\[2mm]
$\rm \overline{PD}$-c&\!$2.80\pm^{0.05}_{0.04}$\!&\!$0.621\pm0.010$\!&---&---&\!$2.236\pm0.009$\!&\!$4.61\pm0.04$\!&$-1.75\pm0.11$&1.23\\[2mm]
&&&&&&&&\\
$\rm \overline{DR}$-a&\!$4.3\pm0.3$\!&\!$0.470\pm0.027$\!&\!$25.3\pm^{2.1}_{2.2}$\!&---&  [2.3]& [4.69]&$-0.26\pm^{0.20}_{0.22}$&0.81 \\[2mm]
$\rm \overline{DR}$-b&\!$3.64\pm^{0.16}_{0.15}$\!&\!$0.46\pm0.04$\!&\!$0\pm^{3}_{0}$\!&---&\!$2.39\pm0.04$\!&\!$4.64\pm0.04$\!&$-2.30\pm0.20$&0.36\\[2mm]
$\rm \overline{DR}$-c&\!$2.86\pm0.05$\!&\!$0.617\pm0.010$\!&\!$6.9\pm^{1.4}_{1.7}$\!&---&\!$2.225\pm0.010$\!&\!$4.62\pm0.04$\!&$-1..72\pm0.10$&1.19\\[2mm]
&&&&&&&&\\
$\rm \overline{DC}$-a&\!$2.70\pm0.05$\!&\!$0.642\pm0.012$\!&---&\!$0\pm^{0.22}_{0}$\!&[2.3]&[4.69]&$-1.86\pm0.13$&2.06\\[2mm]
$\rm \overline{DC}$-b&\!$3.20\pm^{0.26}_{0.33}$\!&\!$0.44\pm0.04$\!&---&\!$6\pm^{5}_{3}$\!&\!$2.46\pm^{0.06}_{0.05}$\!&\!$4.63\pm0.04$\!&$-2.4\pm0.4$&0.31\\[2mm]
$\rm \overline{DC}$-c&\!$2.80\pm^{0.05}_{0.04}$\!&\!$0.621\pm0.010$\!&---&\!$0\pm^{0.17}_{0}$\!&\!$2.236\pm0.009$\!&\!$4.61\pm0.04$\!&$-1.75\pm0.11$&1.23\\[2mm]
&&&&&&&&\\
$\rm \overline{DRC}$-a&\!$4.3\pm0.3$\!&\!$0.470\pm0.027$\!&\!$25.3\pm2.2$\!&\!$0\pm^{2.9}_{0}$\!&  [2.3]& [4.69]& $-0.26\pm^{0.20}_{0.22}$&0.84 \\[2mm]
$\rm \overline{DRC}$-b&\!$3.20\pm0.28$\!&\!$0.44\pm0.04$\!&\!$0$\!&\!$6\pm^{5}_{3}$\!&\!$2.46\pm^{0.06}_{0.05}$\!&\!$4.63\pm0.04$\!&$-2.4\pm0.4$&0.30\\[2mm]
$\rm \overline{DRC}$-c&\!$2.69\pm^{0.17}_{0.20}$\!&\!$0.630\pm^{0.018}_{0.016}$\!&\!$11\pm^{3}_{4}$\!&\!$2.5\pm^{3.1}_{2.5}$\!&\!$2.224\pm0.010$\!&\!$4.63\pm0.04$\!&$-1.63\pm^{0.16}_{0.14}$&1.19\\[2mm]
&&&&&&&&\\
\hline
\end{tabular}
\caption[The best-fit parameters for $\rm \overline{PD}$, $\rm \overline{DR}$, $\rm \overline{DC}$ and $\rm \overline{DRC}$ models in the $\chi^{2}$ study] {\footnotesize The best-fit parameters for $\rm \overline{PD}$, $\rm \overline{DR}$, $\rm \overline{DC}$ and $\rm \overline{DRC}$ models by fitting only B/C ratio (labelled as a), by fitting the $\bar{\text{p}}$/p ratio plus the proton spectrum (labelled as b), and by fitting simultaneously the B/C ratio,  the $\bar{\text{p}}$/p ratio and the proton spectrum (labelled as c). The value in the square brackets correspond to the fixed value of the parameter.}
\label{tab:chi2analysis_eta}
\end{center}
\end{sidewaystable}

\subsection{What did we learn from the $\chi^{2}$ study?}

From the $\chi^{2}$ study, the following conclusions can be drawn:
\begin{itemize}
\item The antiproton data alone is not enough to constrain different propagation processes.
\item Reacceleration can explain the B/C data but produces too many protons at energies of a few GeV. A break in the injection spectrum is included in some studies to fit the data of the proton spectrum. If this break is not introduced, however, in order to fit all the data, including the B/C ratio, the $\bar{\text{p}}$/p ratio and the proton spectrum, reacceleration is always disfavored, i.e. $v_{A} \to 0$.
\item Except for the low energy B/C ratio ($<1$~GeV/n) reported by ACE-CRIS, the PD and DC models can generally explain the high energy B/C ratio, the $\bar{\text{p}}$/p ratio and the proton spectrum.
\item A low energy dependence applied to the diffusion coefficient can fit all the data. But other important processes at low energy, such as convection and reacceleration, are not allowed to be studied since the low energy dependence of the diffusion coefficient dominates over these effects at low energy. 
\item The estimated parameters might be biased in the simultaneous fit to both the secondary-to-primary ratios and the primary fluxes due to the very precise proton data. The parameters characterizing the low energy precesses could also be biased due to the solar modulation which is simply modeled by an effective parameter, used in the force-field approximation.
\item The statistical uncertainties on parameters depend on the parameters under study and the data used. The errors on $N_{p}$ are constant to be about 1\% since this parameter is normalized to the measured proton flux. By fitting all the data simultaneously, parameters other than $N_{p}$ have errors less than 10\%, which are at least twice precise than those estimated by fitting PAMELA proton and $\bar{\text{p}}$/p data. However, if the parameter $\eta$ is included in the fit, the values of $v_{A}$ and $dV/dz$ are not constrained very well, i.e. uncertainties are generally larger than 25\%.

\end{itemize}

\section{Bayesian approach} \label{sec:bayes_analysis}

As discussed in the last section,  a simultaneous fit to both primary and secondary cosmic ray data might bias the parameters. This was also argued in \cite{Coste2011}. Indeed, primary fluxes are more prone to systematics and are more sensitive to solar modulation than secondary-to-primary ratios. This bias can be reduced by specifying priors on the source parameters and taking into account the uncertainties on the solar modulation potentials. For these reasons a Bayesian method is used, allowing parameters to be estimated based on prior knowledge and information contained in the likelihood, i.e. the probability to observe the data measured for a particular model assumption. 

Given the observed data set $\mathbf{D}$ and the parameters $\mathbf{\Theta}$ under study in a hypothesis (model) $H$, Bayes' theorem states that: 
\begin{equation}
P \left( \mathbf{\Theta}|\mathbf{D}, H \right) = \frac{P\left(\mathbf{D}|\mathbf{\Theta}, H\right) P \left(\mathbf{\Theta} | H\right)}{P\left(\mathbf{D} | H\right)},
\end{equation}
where $P \left( \mathbf{\Theta}|\mathbf{D}, H \right)$ is the posterior probability density function (p.d.f.) of the parameters, $P\left(\mathbf{D}|\mathbf{\Theta}, H\right) \equiv L\left(  \mathbf{\Theta} \right)$ is the likelihood, $P \left(\mathbf{\Theta} | H \right)$ is the prior, and $P\left(\mathbf{D} |H\right)$ is the Bayesian evidence. 

The posterior probability distributions of the transport and source parameters are derived by Bayesian inference which can naturally produce credible regions in the parameter space. This will help us to understand the uncertainties and correlations between the parameters. Furthermore, the Bayesian evidence offers a useful tool to select models \cite{Berkhof2003, Trotta2008}. The evidence is a normalization constant and is defined as: 
\begin{equation} \label{eq:evidence}
Z = \int  P \left( \mathbf{D}|\mathbf{\Theta}, H \right) P \left(\mathbf{\Theta}|H\right)\mathrm{d} \mathbf{\Theta}.
\end{equation}
The evidence is independent of the parameters and therefore it is usually neglected in parameter estimation. However, when comparing alternative models, the evidence is the key ingredient to choose which one is better. A model which depends on fewer free parameters and fits better the data will have a larger evidence. A comparison between two competing models $H_{0}$ and $H_{1}$ can be performed by comparing their respective posterior probabilities as follows:
\begin{equation}
\frac{P \left(H_{1}|\mathbf{D} \right)}{P \left(H_{0}|\mathbf{D} \right)}=\frac{P \left(\mathbf{D}|H_{1} \right) P \left(H_{1} \right)}{P \left(\mathbf{D}|H_{0} \right) P \left(H_{0} \right)} = B_{10} \frac{P \left(H_{1} \right)}{P \left(H_{0} \right)},
\end{equation}
where $P \left(H_{1} \right) / P \left(H_{0} \right)$ is a \textit{priori} probability ratio for models $H_{0}$ and $H_{1}$ and usually can be set to unity, $B_{10}$ is called the Bayes factor and is defined as the ratio of two models evidences. Given the observations $\mathbf{D}$, if $B_{10}>1$, model $H_{1}$ is favored versus model $H_{0}$, and vice versa. While the $\chi^{2}$ method addresses the goodness of fit, the Bayesian approach provides a model selection criterion.

The main difficulty of the Bayesian approach is its very expensive computation cost on the calculation of the posterior distribution, and especially the Bayesian evidence. To perform sufficiently fast Bayesian analysis, a publicly available package, MultiNest \cite{Feroz2008, Feroz2009}, which implements a nested sampling algorithm was integrated with GALPROP to study cosmic ray propagation models. Compared to traditional Markov Chain Monte Carlo (MCMC) techniques (see e.g. \cite{Mackay2003}), MultiNest is highly efficient which reduces the computation time by a factor of $\sim100$. Using a log-likelihood function, $\mathrm{ln} L \left( \mathbf {\Theta} \right)= -1/2 \chi^{2}$, MultNest directly produces the evidence and the posterior distribution. Once the samples of posterior distribution $f(\mathbf{\Theta})$ in n-dimension parameter space are generated, it is able to estimate the one-dimensional (1D) marginal probability $P \left( \Theta_{i} |\mathbf{D}, H \right)$ for the parameters of interest $\Theta_{i}$ by integrating $f(\mathbf{\Theta})$ over all other parameters, as:
\begin{equation}
P \left( \Theta_{i} |\mathbf{D}, H \right) = \int P \left( \mathbf{\Theta}|\mathbf{D}, H \right) \mathrm{d} \Theta_{1} ... \mathrm{d} \Theta_{i-1} \mathrm{d} \Theta_{i+1}...\mathrm{d} \Theta_{n}.
\end{equation}

Distinguishing from the frequentist confidence interval which indicates how frequently the observed interval contains the parameters, the Bayesian credible interval states the degree of belief that the parameters lie inside the interval. The two-tail symmetric $\alpha$\% credible interval [ $\Theta_{i}^{-}$, $\Theta_{i}^{+}$ ] can be obtained by:
\begin{equation}
\int_{-\infty}^{\Theta_{i}^{-}} P \left( \Theta_{i} |\mathbf{D}, H \right) \mathrm{d} \Theta_{i} = \frac{1-\alpha \%} {2} = \int_{\Theta_{i}^{+}}^{\infty} P \left( \Theta_{i} |\mathbf{D}, H \right) \mathrm{d} \Theta_{i} .
\end{equation}
The integration can be calculated by counting a fraction of $(1-\alpha \%)/2$ of the number of samples falling outside each side of the interval. Two-dimensional (2D) marginal posterior p.d.f.s are defined in a similar way. The $\alpha$\% credible regions are produced by finding out the contours in which the integration of the 2D marginal posterior density equals to $\alpha$\%. The best-fit parameters which maximize the likelihood function is also given by MultiNest as a by-product.

\subsection{Models and priors} \label{sec:priors}

In this section, only the DR and DRC models were studied in the framework of Bayesian inference. This choice was made for several reasons. Firstly, without the need of an arbitrary break on the diffusion coefficient, the reacceleration process which is expected when relativistic particles scatter on magnetic turbulence, well describes the secondary-to-primary ratios. This can be seen from the $\chi^{2}$ study where the DR and DRC models give much smaller $\chi^{2}$ compared with the rather high values for the PD and DC models, as shown in table \ref{tab:chi2analysis}. Secondly, since the DR model has been studied widely in the literature, it is natural to choose it as a reference case. Thirdly, in order to understand if convection can better explain the data and to study the correlation between each process, the DRC model is also studied. 

The solar modulation potentials are included as nuisance parameters in the Bayesian analysis to diminish systematic effects due to uncertainties on the modulation potentials. Since solar modulation mainly affects cosmic ray nuclei with energies below a few GeV, the low energy dependence of diffusion coefficient which was proved to be dominant over other processes may not allow any useful information to be extracted. Therefore, only standard models with $\eta$ equal to unity are studied here. 

Based on our current knowledge, priors are specified for the free parameters listed in section \ref{sec:paras_sum} to restrict parameters in physically reasonable regions. The propagation and source parameters characterizing a model are of interest. As shown in table \ref{tab:priors}, the prior on each transport parameter is uniform to assign equal probabilities on all the possible values within the prior range. The source parameters are assumed to follow a Gaussian distribution with an expected mean. The solar modulation parameters also adopt Gaussian priors, for which the mean values are chosen to be the estimated value given by each experiment. As shown in equation \ref{eq:evidence}, the evidence of a model depends on the priors for the parameters. If the likelihoods get higher values at lower prior probability regions, the evidence will be suppressed. This will increase our confidence in model rejection. 

\begin{table*}[!t]
\begin{center}
\scalebox{0.9}{
\begin{tabular}{l | l | c | c }
\hline
\hline
 & Parameters & Prior range & Prior type \\
\hline
\multirow{4}{*}{Propagation parameters} & $D_{0} $ ($10^{28}$~cm$^{2}$/s) & [0.5, 15] & Uniform\\
&$\delta$& [0.1, 1.0] & Uniform\\
&$v_{A} $ (km/s) & [0, 100] & Uniform\\
& $\text{d}V/\text{d}z$ (km/s/kpc) & [0, 50] & Uniform\\
&&&\\
\multirow{2}{*}{Source parameters} & $\nu$ & [1.7, 2.9] & $N$(2.3, 0.2)\\
& $N_{p}$ ($10^{-9}$~cm$^{2}$/sr/s/MeV)& [4.54, 4.84] &$N$(4.69, 0.05)\\
&&&\\
\multirow{6}{*}{Modulation parameters} & $\Phi_{\textrm{HEAO3}}$~(MV) & [350, 650] & $\mathcal{N}(500,\,50)$\\
& $\Phi_{\textrm{ACE/CRIS}}$~(MV) &[226, 424] & $\mathcal{N}(325,\,33)$\\
& $\Phi_{\textrm{CREAM-1}}$~(MV)  &[595, 1105] & $\mathcal{N}(850,\,85)$\\
& $\Phi_{\textrm{Spacelab-2}}$~(MV) & [280, 520] & $\mathcal{N}(400,\,40)$\\
& $\Phi_{\textrm{AMS01}}$~(MV)  &[315, 585] & $\mathcal{N}(450,\,45)$\\
& $\Phi_{\textrm{PAMELA}}$~(MV)  &[350, 650] & $\mathcal{N}(500,\,50)$\\
&&&\\
\hline
\end{tabular}
}
\caption[Priors for propagation model parameters and solar modulation parameters in the Bayesian study] {\footnotesize Priors for propagation model parameters and solar modulation parameters. The notation $\mathcal{N}(\mu,\,\sigma)$ is used to represent a Gaussian distribution with mean $\mu$ and strandard deviation $\sigma$.}
\label{tab:priors}
\end{center}
\end{table*} 

\subsection{Results}

\begin{figure}[!htb]\label{fig:DR_BC_pdf}
\begin{center}
\includegraphics[width=0.9\textwidth]{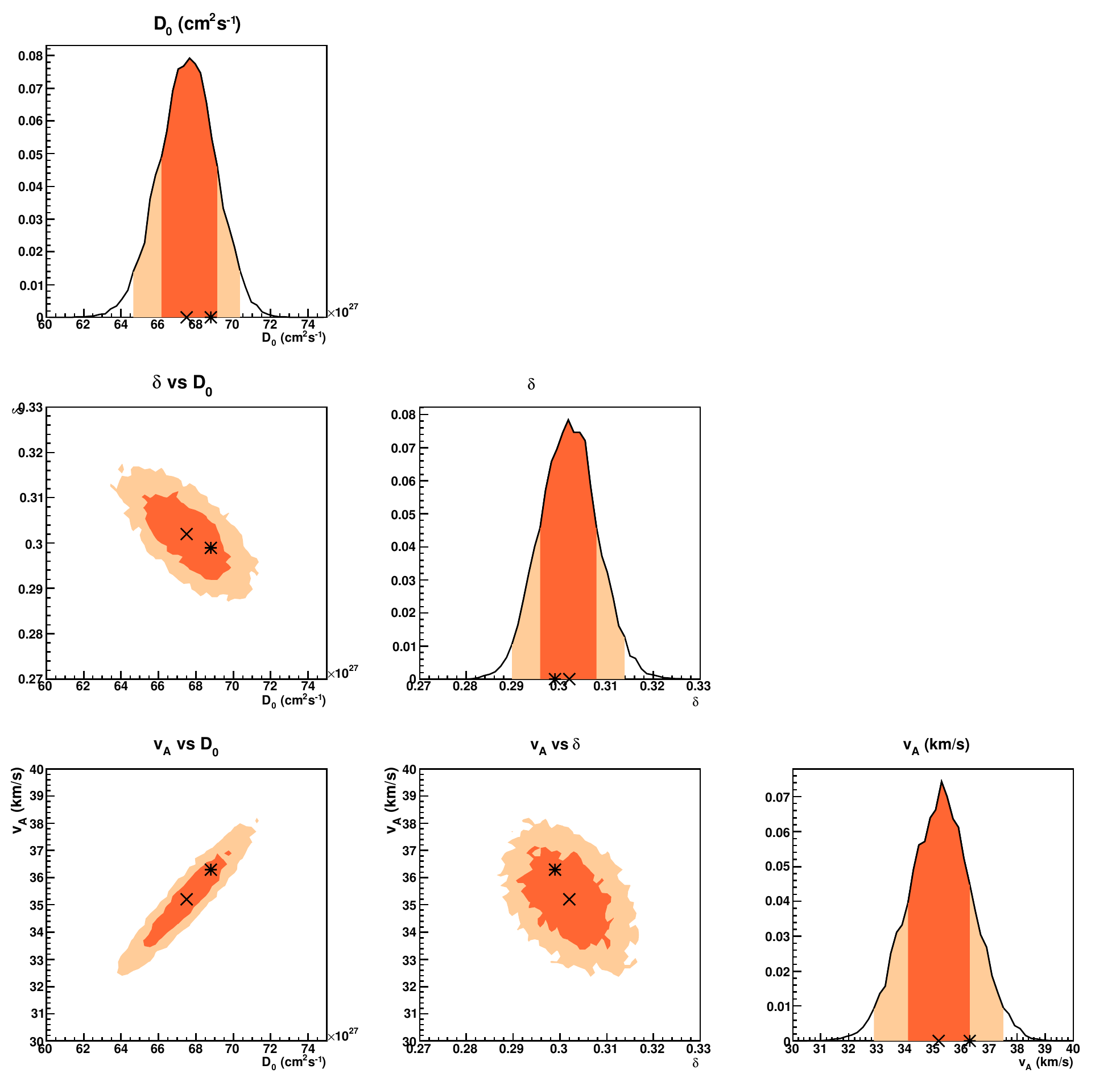}
\end{center}
\caption[The posteriors of the propagation parameters for the DR-a model in the Bayesian study]{\footnotesize The 1D (diagonal) and 2D (off-diagonal) marginalized posterior p.d.f.s of the propagation parameters for the DR-a model (as shown in table \ref{tab:bayesian_results}). The dark/light orange color represents the 68\%/95\% credible interval. The cross is the posterior mean, the star the best fit.} 
\label{fig:DR_a_pdf}
\end{figure}

\begin{figure}[!htbp]
\begin{center}
\includegraphics[angle=90,width=0.9\textwidth]{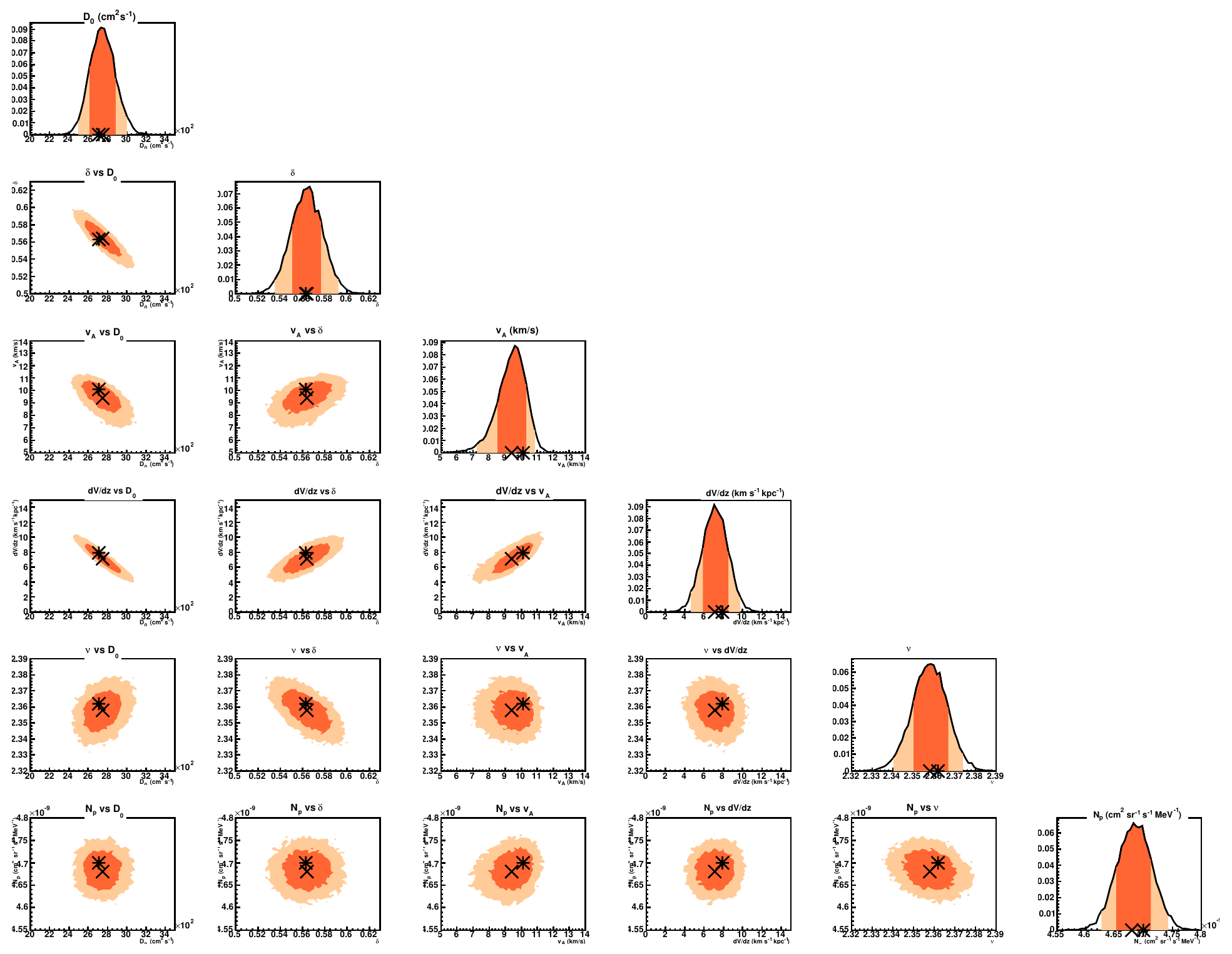}
\end{center}
\caption[The posteriors of the source and propagation parameters for the DRC-c model in the Bayesian study]{\footnotesize The 1D (diagonal) and 2D (off-diagonal) marginalized posterior p.d.f.s of the propagation parameters for the DRC-c model (as shown in table \ref{tab:bayesian_results}). The dark/light orange color represents the 68\%/95\% credible interval. The cross is the posterior mean, the star the best fit.} 
\label{fig:DRC_c_pdf}
\end{figure}

Identical data sets as employed in the $\chi^{2}$ study are used. For all the studied models, the constraints on parameters and the best-fit parameters maximizing the likelihood are summarized in table \ref{tab:bayesian_results} and the marginal posterior p.d.f.s for the model parameters are produced. Examples of the posterior p.d.f.s and the 68\% and 95\% credible intervals (dark and light orange, respectively) for the DR-a and DRC-c models are given in figures \ref{fig:DR_a_pdf} and \ref{fig:DRC_c_pdf}. The posterior p.d.f.s and the 68\% and 95\% contours are also computed for other models but are not shown here.

\begin{sidewaystable}
\begin{center}
\begin{tabular}{c | c | c | c | c | c | c | c}
\hline
\hline
Model &  $D_{0} \cdot 10^{-28}$ & $\delta$  & $v_{A} $  & $\text{d}V/\text{d}z$ & $\nu$ &$N_{p}\cdot 10^{9}$& log-evidence/$\chi^{2}/$d.o.f \\
 &\!(cm$^{2}$/s)\!&  & \!(km/s)\! & \!\!(km/s/kpc)\!\! &&\!\!(cm$^{2}$/sr/s/MeV)\!\! &\\
\hline
\hline
&&&&&&&\\
\multirow{2}{*}{DR-a}&\!$6.75\pm0.15$ \!&\!$0.302\pm0.007$\!&\!$35.2\pm1.2$\!&---&  [2.3] & [4.69]&$-50.48\pm0.08$ \\[2mm]
&6.88 & 0.299 & 36.3 &---&[2.3] & [4.69]&2.36\\
\hline
\multirow{2}{*}{DR-b}&\!$3.72\pm0.16$\!&\!$0.42\pm0.04$\!&\!$1.7\pm1.2$\!&---&\!$2.45\pm0.04$\!&\!$4.67\pm0.03$\!&$-34.60\pm0.10$\\[2mm]
&3.76&0.39&0.26&---&2.48&4.67&0.29\\
\hline
\multirow{2}{*}{DR-c}&\!$3.51\pm0.05$\!&\!$0.501\pm0.008$\!&\!$2.7\pm1.4$\!&---&\!$2.368\pm0.009$\!&\!$4.68\pm0.03$\!&$-184.21\pm0.12$\\[2mm]
&3.52&0.502&4.1&---&2.367&4.69&2.34\\
&&&&&&&\\
\hline
\hline
\multirow{2}{*}{DRC-a}&\!$3.3\pm0.5$\!&\!$0.446\pm0.021$\!&\!$39.7\pm2.0$\!&\!$33\pm8$\!&  [2.3]& [4.69]&$-39.76\pm0.08$ \\[2mm]
&3.21 & 0.45 & 39.5 &32.8& [2.3]& [4.69]&1.70\\
\hline
\multirow{2}{*}{DRC-b}&\!$3.3\pm0.4$\!&\!$0.40\pm0.04$\!&\!$3.3\pm2.2$\!&\!$6\pm5$\!&\!$2.50\pm0.05$\!&\!$4.67\pm0.03$\!&$-34.94\pm0.10$\\[2mm]
&3.6&0.40&2.9&1&2.48&4.65&0.30\\
\hline
\multirow{2}{*}{DRC-c}&\!$2.75\pm0.13$\!&\!$0.564\pm0.015$\!&\!$9.4\pm1.0$\!&\!$7.1\pm1.4$\!&\!$2.358\pm0.009$\!&\!$4.68\pm0.03$\!&$-171.78\pm0.12$\\[2mm]
&2.71&0.563&10.1&7.9&2.362&4.70&2.13\\
&&&&&&&\\
\hline
\end{tabular}
\caption[The posterior mean with standard deviation and the best-fit parameteres maximizing the likelihood for all the propagation and source parameters for DR and DRC models in the Bayesian study]{\footnotesize The posterior mean with standard deviation (\textit{the first row}) and the best-fit parameteres maximizing the likelihood (\textit{the second row}) for all the propagation and source parameters for DR and DRC models by using only B/C ratio (labelled as a), by using the $\bar{\text{p}}$/p ratio plus the proton spectrum (labelled as b), and by using the B/C ratio,  the $\bar{\text{p}}$/p ratio and the proton spectrum (labelled as c). The derived evidences are also given. The value in the square brackets correspond to the fixed value of the parameter.}
\label{tab:bayesian_results}
\end{center}
\end{sidewaystable}

\begin{table*}[!t]
\begin{center}
\scalebox{0.9}{
\begin{tabular}{c | c | c | c | c | c | c}
\hline
\hline
Model & \!$\Phi_{HEAO3}$\! & \!$\Phi_{ACE-CRIS}$\! & \!$\Phi_{CREAM-1}$\!  & \!$\Phi_{Spacelab-2}$\! & \!$\Phi_{AMS01}$\! & \!$\Phi_{PAMELA}$\! \\
 &\! (MV) \!& \!(MV)\! & \!(MV)\! & \!(MV)\! &\!(MV)\!&\!(MV)\! \\
\hline
\hline
&&&&&&\\
\multirow{2}{*}{DR-a}&\!$510\pm50$\!&\!$342\pm21$\!&\!$850\pm90$\!&\!$400\pm40$\!& \!$470\pm50$\!&--- \\[2mm]
&591 & 367 & 945 & 438 & 568&---\\
\hline
\multirow{2}{*}{DR-b}&---&---&---&---&---&\!$641\pm7$\!\\[2mm]
&--- & --- & --- & ---&---&\!649\!\\
\hline
\multirow{2}{*}{DR-c}&\!$490\pm50$\!&\!$227.2\pm1.2$\!&\!$850\pm90$\!&\!$400\pm40$\! & \!$450\pm50$\!&\!$640\pm7$\! \\[2mm]
&415 &226 & 851 & 373 & 447&650\\
&&&&&&\\
\hline
\hline
\multirow{2}{*}{DRC-a}&\!$510\pm50$\!&\!$296\pm21$\!&\!$850\pm90$\!&\!$400\pm40$\!& \!$460\pm50$\!&--- \\[2mm]
&503 & 272 & 884 & 377& 569&---\\
\hline
\multirow{2}{*}{DRC-b}&---&---&---& --- &---&\!$622\pm16$\!\\[2mm]
&--- & --- & --- & ---& ---&644\\
\hline
\multirow{2}{*}{DRC-c}&\!$500\pm50$\!&\!$227.4\pm1.4$\!&\!$850\pm90$\!&\!$400\pm40$\!& \!$450\pm50$\!&\!$640\pm9$\! \\[2mm]
&453 & 226 & 862 & 464 & 407&649\\
&&&&&&\\
\hline
\end{tabular}}
\caption[The posterior mean with standard deviation and the best-fit parameteres maximizing the likelihood for the solar modulation parameters for DR and DRC models from the Bayesian study]{\footnotesize The posterior mean with standard deviation (\textit{the first row}) and the best-fit parameteres maximizing the likelihood (\textit{the second row}) for the solar modulation parameters for DR and DRC models by using only B/C ratio (labelled as a), by using the $\bar{\text{p}}$/p ratio plus the proton spectrum (labelled as b), and by using the B/C ratio,  the $\bar{\text{p}}$/p ratio and the proton spectrum (labelled as c).}
\label{tab:bayesian_results_phi}
\end{center}
\end{table*} 

\begin{figure}[!htbp]
\begin{center}
\includegraphics[angle=90,width=0.95\textwidth]{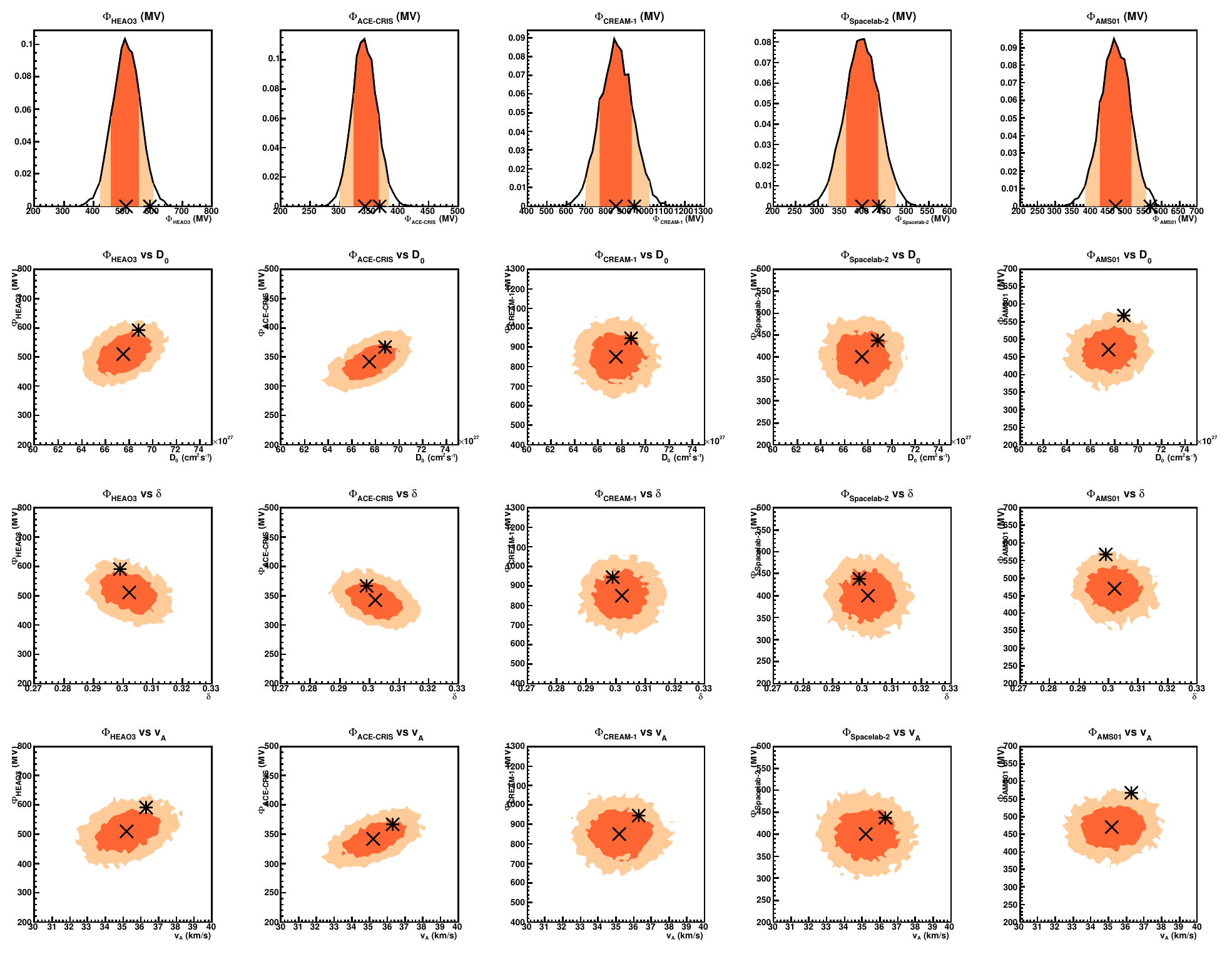}
\end{center}
\caption[The posteriors of solar modulation parameters for the DR-a model in the Bayesian study]{\footnotesize The 1D marginalized posterior p.d.f.s of the solar modulation parameters for the DR-a model (as shown in table \ref{tab:bayesian_results}) and the relation between the solar modulation parameters and the model parameters. The abscissae of the 2D contours are propagation parameters while the ordinates are the modulation parameters. The units of the propagation parameters follow the same in figure \ref{fig:DR_BC_pdf}. The dark/light orange color represents the 68\%/95\% credible interval. The cross is the posterior mean, the star the best fit.}
\label{fig:BC_DR_phi}
\end{figure}

A negative correlation between the normalization $D_{0}$ and spectral index $\delta$ of the diffusion coefficient is seen for all the models. This can be inferred from equation \ref{eq:diff_coeffieient}. In order to keep a roughly constant diffusion coefficient to reproduce the secondary-to-primary ratios, a larger $D_{0}$ leads to a smaller $\delta$. The relationship between the diffusion parameters ($D_{0}$ or $\delta$) and other propagation parameters are model dependent. For instance, the correlation between $\delta$ and the Alfv\'en velocity $v_{A}$ is negative in the DR-a model but positive in the DRC-c model. If the source parameters are also under study, the correlation between the injection index $\nu$ and spectral index of the diffusion coefficient $\delta$ can be found to be negative, as well as the correlation between $\nu$ and the normalization of the propagated proton spectrum $N_{p}$. The former one is because a flatter injection spectrum needs to be diffuse more to hold the propagated cosmic ray slope that is observed. The latter one is naturally obtained to fit the data, i.e. a flatter injection spectrum (lower value of $\nu$) will deviate more from the data if it has to fit a lower value of normalization for the propagated proton spectrum ($N_{p}$).

An example of the 1D marginalized posterior p.d.f.s of solar modulation and the correlations between solar modulation parameters and the model parameters is shown for the DR-a model in figure \ref{fig:BC_DR_phi}. In this case, mainly $\Phi_{HEAO3}$ and $\Phi_{ACE-CRIS}$ are correlated with the model parameters. The reason is that below  300~MeV/n the data are only from ACE-CRIS and above this energy the most accurate data sensitive to solar modulation are from HEAO3. The values of $\Phi_{HEAO3}$ or $\Phi_{ACE-CRIS}$ are positively correlated with $D_{0}$ and $v_{A}$, and negatively correlated with $\delta$. This indicates that for a DR model, less modulation needs a smaller $\delta$ and more reacceleration.

Comparing the results in table \ref{tab:bayesian_results} and table \ref{tab:chi2analysis}, the constraints from the Bayesian analysis are consistent with the ones from the $\chi^{2}$ method for the DR-a model and the DRC-a model. Only the DR-a model prefers the Kolmogorov spectrum of turbulence of $\delta=1/3$. For the DR-b model and DRC-b model, the bias on the diffusion parameters due to the dominant weight of the primary proton flux is diminished mainly after freeing solar modulation parameters and by applying Gaussian priors on the source and modulation parameters. For these two models reacceleration is still not favored since $v_{A}$ is estimated to be nearly zero. This is also the case for the DR-c model. The Bayesian results for the DR-c model do not deviate significantly from the ones derived from the $\chi^{2}$ analysis. For the DRC-c model, while the $\chi^{2}$ study converges at $v_{A} \to 0$, weak reacceleration with $v_{A} = 9.4\pm1.0$~km\,s$^{-1}$ is allowed with Bayesian method. This is possibly because the best-fit $\Phi$ for PAMELA is estimated as 649~MV in the Bayesian analysis, i.e. higher than the one used in the $\chi^{2}$ study (500~MV). Increased solar activity will suppress the overproduction of  the proton flux  around a few GeV caused by reacceleration.

\subsection{Model selection}
When selecting between two competing models $H_{0}$ and $H_{1}$, the evaluated Bayes factors indicate the strength of evidence. Empirical thresholds on the logarithm of the Bayes factors are $\mathrm{ln} B_{10}=\Delta \mathrm{ln} Z = 1.0,\, 2.5,\,5.0$, representing weak, moderate and strong evidence (see e.g. \cite{Trotta2008}). 

\begin{table*}[!t]
\begin{center}
\begin{tabular}{  c | c }
\hline
\hline
 Data &$\Delta \mathrm{ln} Z$ \\
\hline
B/C & $10.72\pm0.11$\\
 $\bar{\text{p}}$/p + p&$-0.35\pm0.14$ \\
B/C+ $\bar{\text{p}}$/p + p& $12.43\pm0.17$\\
\hline
\end{tabular}
\caption{The difference in log-evidence between the DRC model and the DR model in the Bayesian study.}
\label{tab:evidence_diff}
\end{center}
\end{table*} 

Assuming $H_{0}$ is the DR model and $H_{1}$ is the DRC model, for each combination of data sets the differences in log-evidence between the two models are shown in table \ref{tab:evidence_diff}. By incorporating the proton flux with the $\bar{\text{p}}$/p ratio, $\Delta \mathrm{ln} Z \to 0$ (i.e. $\mathrm{ln} B_{01} \to 0$) thus no model is favored. As indicated from the $\chi^{2}$ study, the antiproton data can be sufficiently described by the PD model. The addition of reacceleration and convection processes is not expected to improve the description of the data. By using only B/C data or combining all the data together, $\Delta \mathrm{ln} Z >5.0$. These large values of $\Delta \mathrm{ln} Z$ strongly support the DRC model over the DR model. To explain all the data, the DRC model is therefore selected as the ``best'' one. 

\begin{figure}[!htb]
\begin{center}
\includegraphics[width=0.95\textwidth]{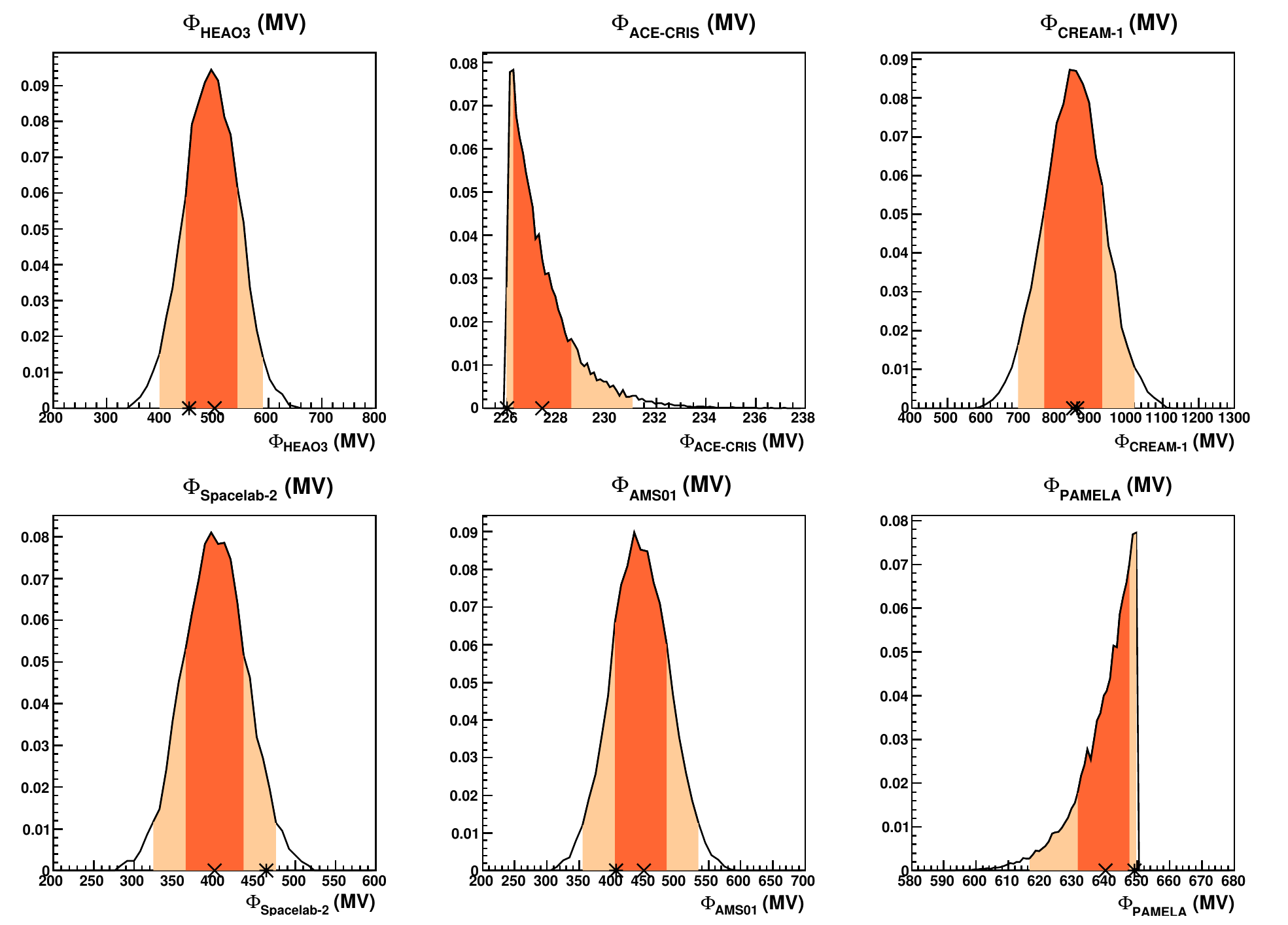}
\end{center}
\caption[The posteriors of solar modulation parameters for the DRC-c model in the Bayesian study]{\footnotesize The 1D marginalized posterior p.d.f.s of the solar modulation parameters for the DRC-c model (as shown in table \ref{tab:bayesian_results}). The cross is the posterior mean, the star the best fit.}
\label{fig:DRC_c_phi}
\end{figure}

\begin{figure}[!htbp]
\begin{center}
\includegraphics[width=0.65\textwidth]{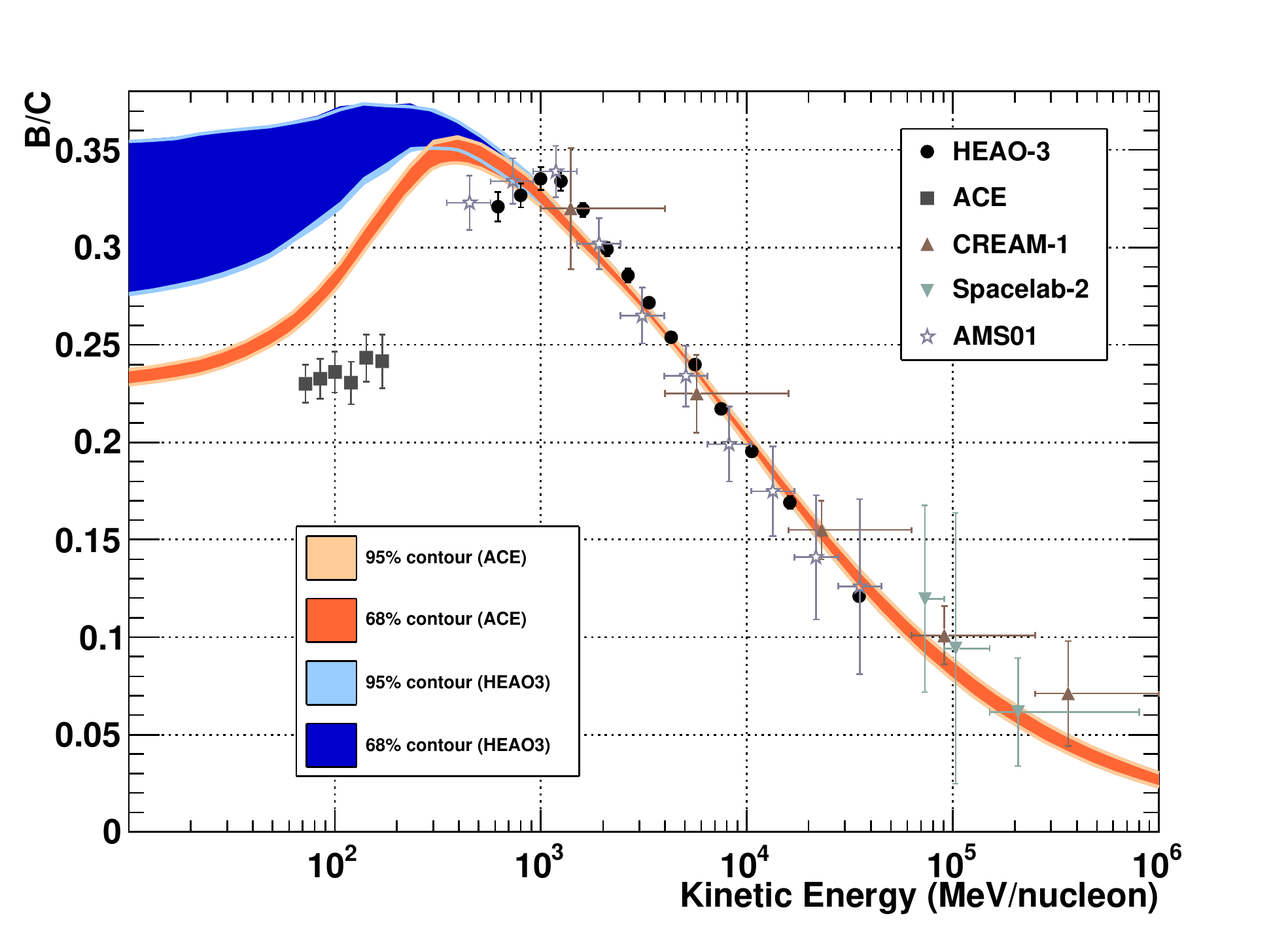}
\includegraphics[width=0.65\textwidth]{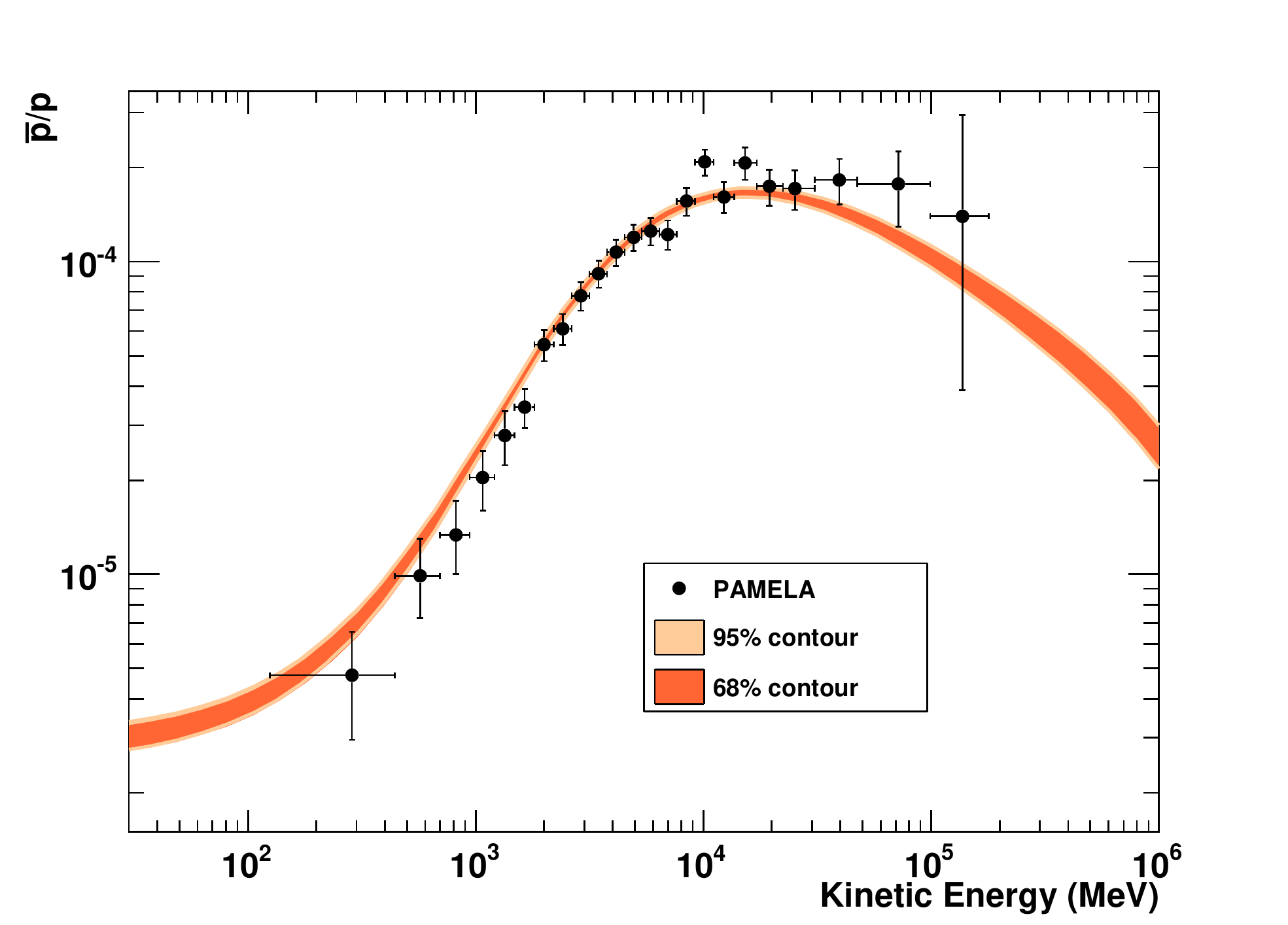}
\includegraphics[width=0.65\textwidth]{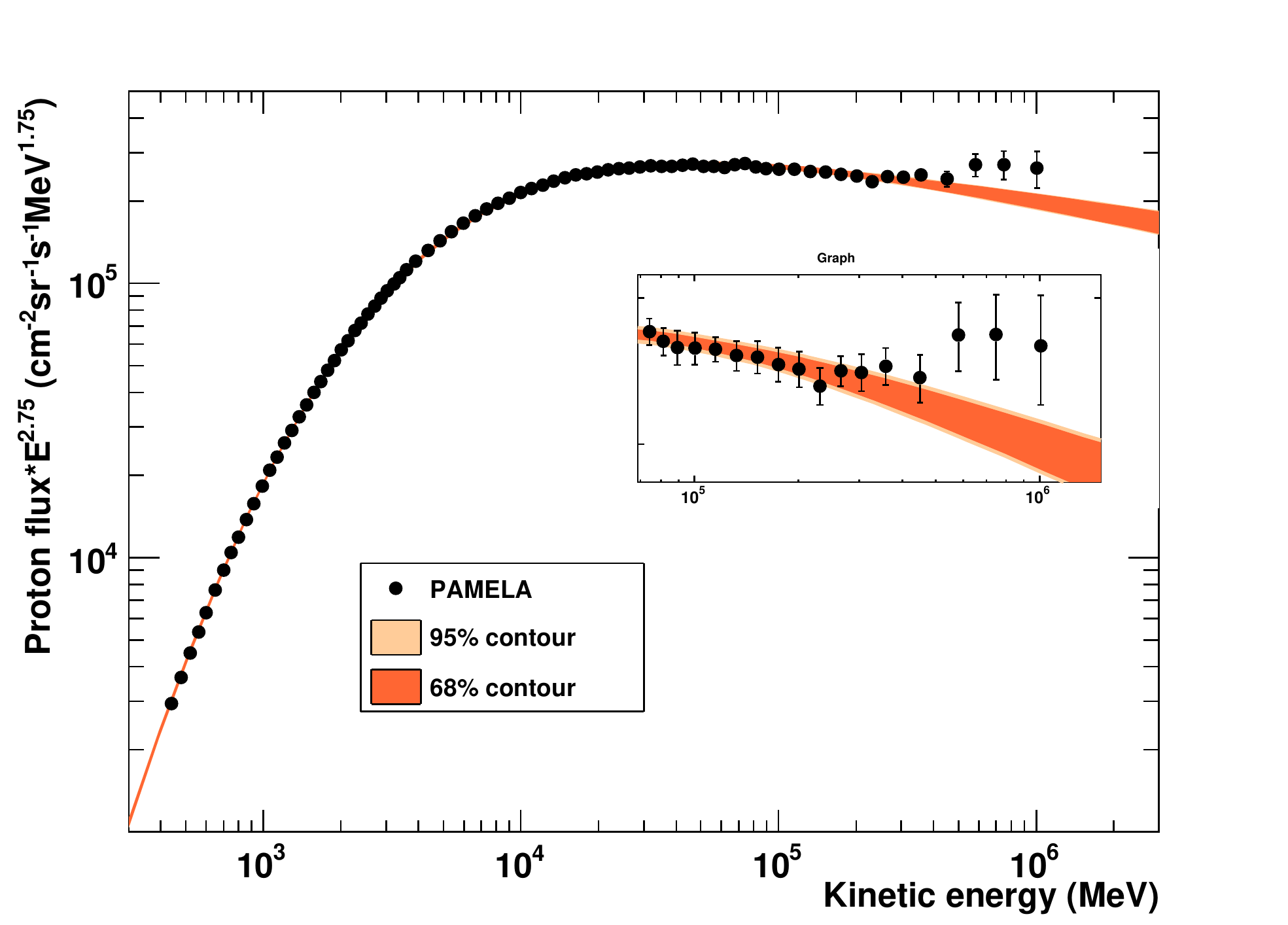}
\end{center}
\caption[The B/C ratio, the $\bar{\text{p}}$/p ratio and the proton spectrum compared with the DRC-c model in the Bayesian study]{\footnotesize The B/C ratio (top), the $\bar{\text{p}}$/p ratio (middle) and the proton spectrum (bottom) for the best model (DRC-c) as listed in table \ref{tab:bayesian_results}. The dark/light orange (or blue) color represents the 68\%/95\% credible interval. The blue and orange colors in the top figure are plotted with the posterior values of $\Phi_{ACE-CRIS}$ and $\Phi_{HEAO3}$, respectively. The middle and bottom figures are plotted with the posterior values of $\Phi_{PAMELA}$.}
\label{fig:bestmodel_bcpbarp}
\end{figure}

The predictions of this DRC-c model for the fitted cosmic ray spectra and ratios are shown in figure \ref{fig:bestmodel_bcpbarp}. This model can not reproduce the B/C ratio below $\sim 1$~GeV. As can been seen in table \ref{tab:bayesian_results_phi} and figure \ref{fig:DRC_c_phi}, the best fit value of $\Phi_{ACE-CRIS}=226$~MV in this model, as well as the posterior mean, are close to the lower limit value of $\Phi_{ACE-CRIS}$ specified in its prior. This indicates that the value of $\Phi_{ACE-CRIS}$ lower than 226~MV is favored and would allow to better fit the ACE-CRIS data. The discrepancy has already been seen in the $\chi^{2}$ study, which can only be recovered by either adding a break in the injection spectrum or considering a low energy dependence in the diffusion coefficient. However, a break in the injection spectrum is difficult to explain physically. If this break is introduced, it either underestimates the antiproton flux (see \cite{Moskalenko2002, Trotta2011}), or overpredicts the B/C ratio at low energy (with lower value of  $v_{A}$ than that in  \cite{Moskalenko2002, Trotta2011}) as mentioned in section \ref{sec:chi2_unmodified_diff}. Taking into account a nonlinear diffusion coefficient can satisfactorily explain all the data, however, it has strong correlation with reacceleration and convection, as indicated in the $\chi^{2}$ study. Including modulation potentials as nuisance parameters in the Bayesian analysis will increase the correlations at low energy. In order to concentrate the effort on understanding the processes of reacceleration and convection, the nonlinear diffusion coefficient is therefore not studied with the Bayesian method. Besides these two explanations, several effects may also be responsible for the discrepancy. The dominant weight of the proton spectrum on the fitting could be the most important reason. Even reacceleration with $v_{A}=9.4\pm1.0$~km\,s$^{-1}$ is allowed to fit the proton spectrum as a compensation for higher solar modulation potential ($\Phi_{\text{PAMELA}}=640\pm9$~MV) than the value fixed in the $\chi^{2}$ study (500~MV), it is still too weak to account for the rapid decreasing of boron flux below around 1~GeV. Using the force-field approximation to model the solar modulation may be another problem since it is too simplified and adopting a different $\Phi$ may significantly change the estimated values of other parameters. The systematic inconsistencies between data sets could also cause a bias on the constraints since data set from different experiments were employed in this work. 

\subsection{The electron spectrum and the positron fraction}

In addition to the cosmic ray nuclei, PAMELA also measures precisely the electron and positron components in the cosmic radiation. PAMELA reported the electron absolute spectrum between 1~GeV and 625~GeV \cite{Adriani2011_electron} as well as the positron fraction between 1.5~GeV and 100~GeV \cite{Adriani2009_positron, Adriani2010_positron}. These data were not used in the fitting procedure described in previous sections, and therefore allow a cross check with the prediction of electrons and positrons from the best model. In a traditional scenario, cosmic ray electrons originate from SNRs and positrons are mainly secondary production created by cosmic ray protons interacting with the ISM. The positron fraction measured by PAMELA increases with energy above 10~GeV, conflicting with the trend predicted by secondary production. Primary sources such as pulsars and dark matter may give an extra contribution. Therefore, a precise and reliable determination of the electron flux and positron flux (or positron fraction) play important roles in studying primary sources.   

\begin{figure}[!htbp]
\begin{center}
\includegraphics[width=0.85\textwidth]{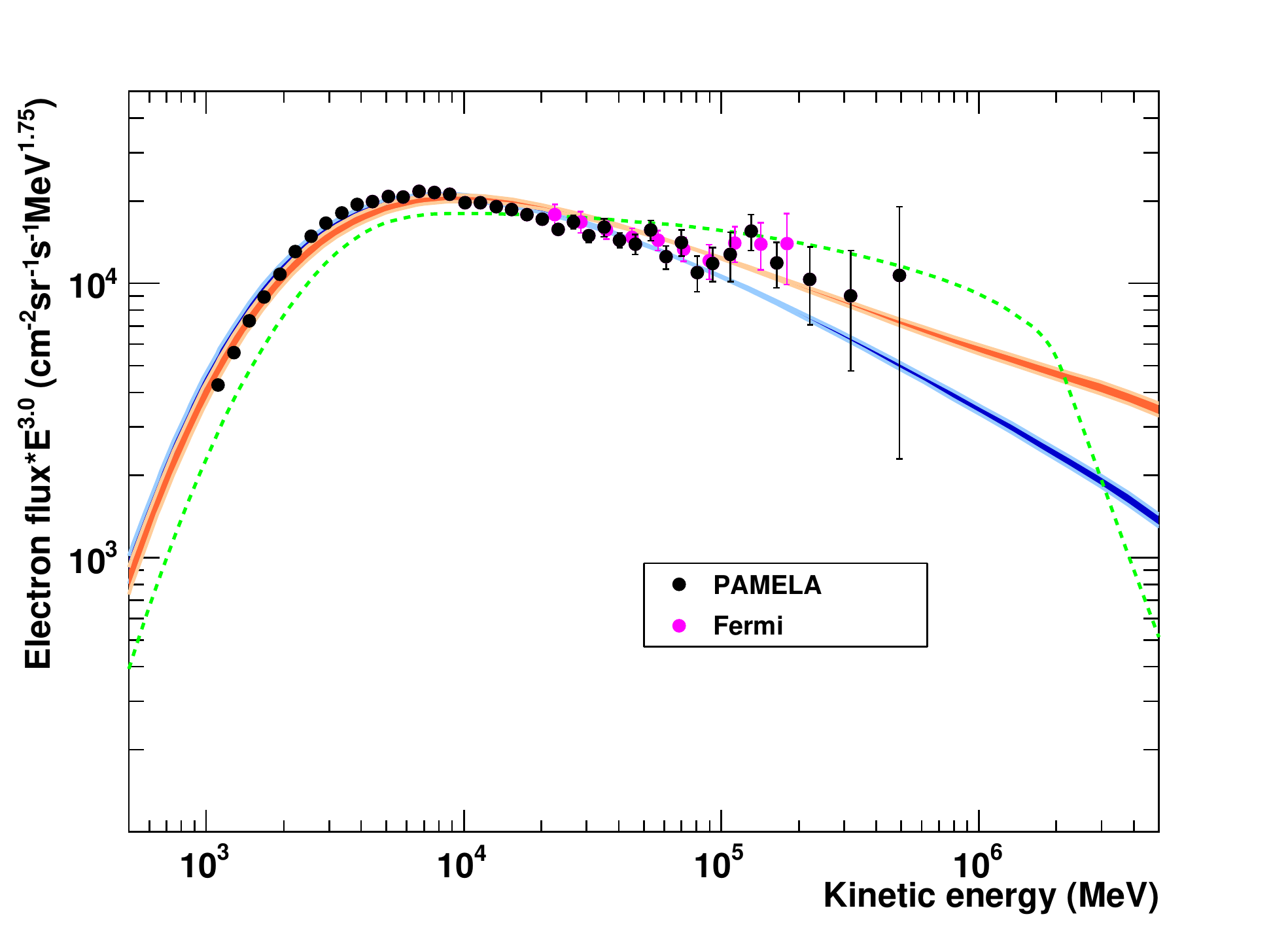}
\includegraphics[width=0.85\textwidth]{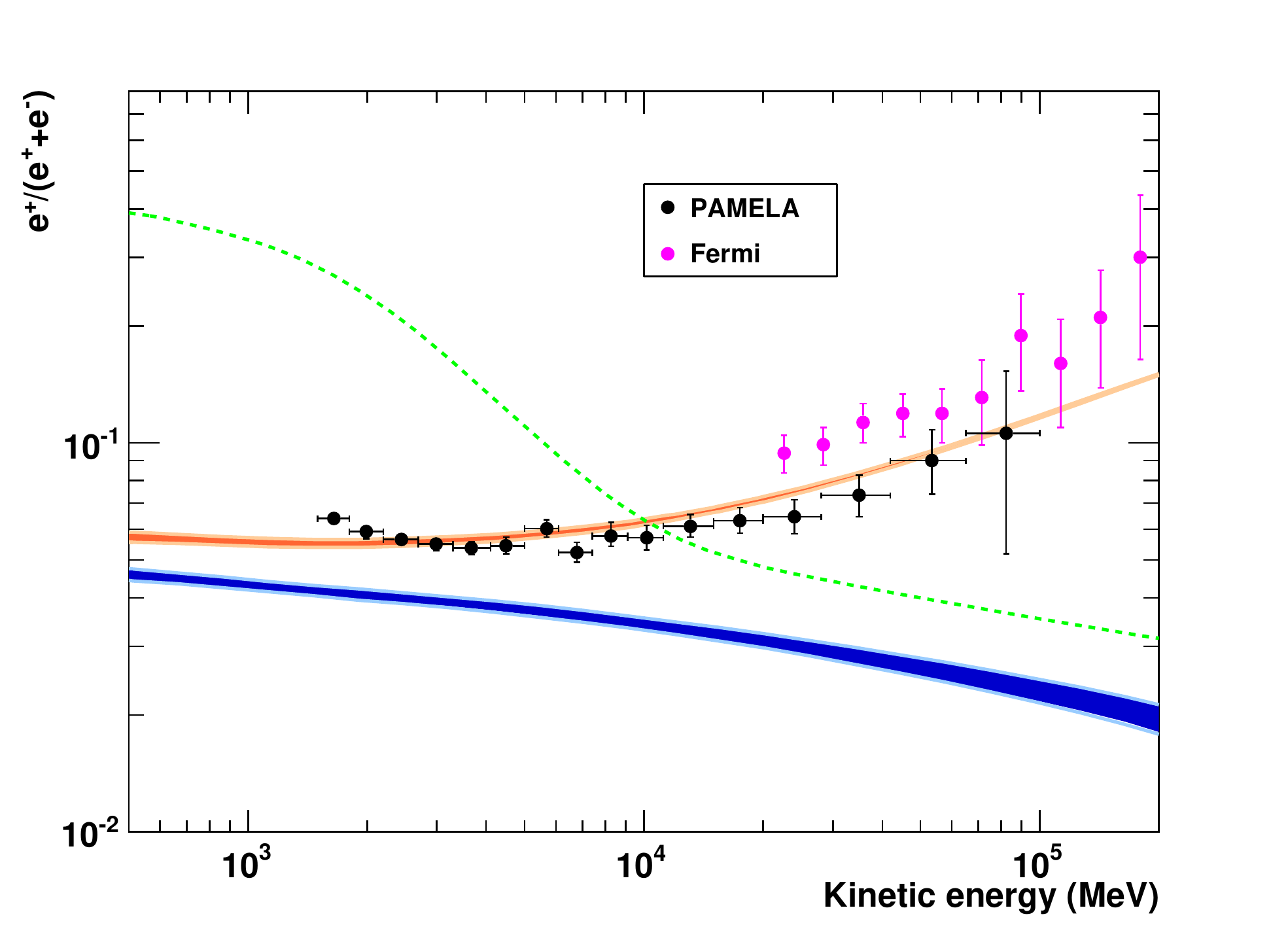}
\end{center}
\caption[The electron spectrum and the positron fraction predicted from the DRC-c model in the Bayesian study]{\footnotesize The electron spectrum (top) and the positron fraction (bottom) for the best model (DRC-c) as listed in table \ref{tab:bayesian_results}. The dark/light color represents the 68\%/95\% credible interval. The blue bands are calculated for the single primary component model with $e^{-}$ injection slope=2.72. The orange bands are calculated for the two primary components model for which one component has an $e^{-}$ injection slope=2.68 and another one has an $e^{\pm}$ injection slope=2.1. The green dash line is plotted using the best-fit parameters derived in \cite{Trotta2011}. Data points are the measurements from PAMELA \cite{Adriani2011_electron, Adriani2010_positron} and Fermi \cite{Ackermann2012}.}
\label{fig:electron_positron}
\end{figure}

\begin{figure}[tb]
\begin{center}
\includegraphics[width=0.85\textwidth]{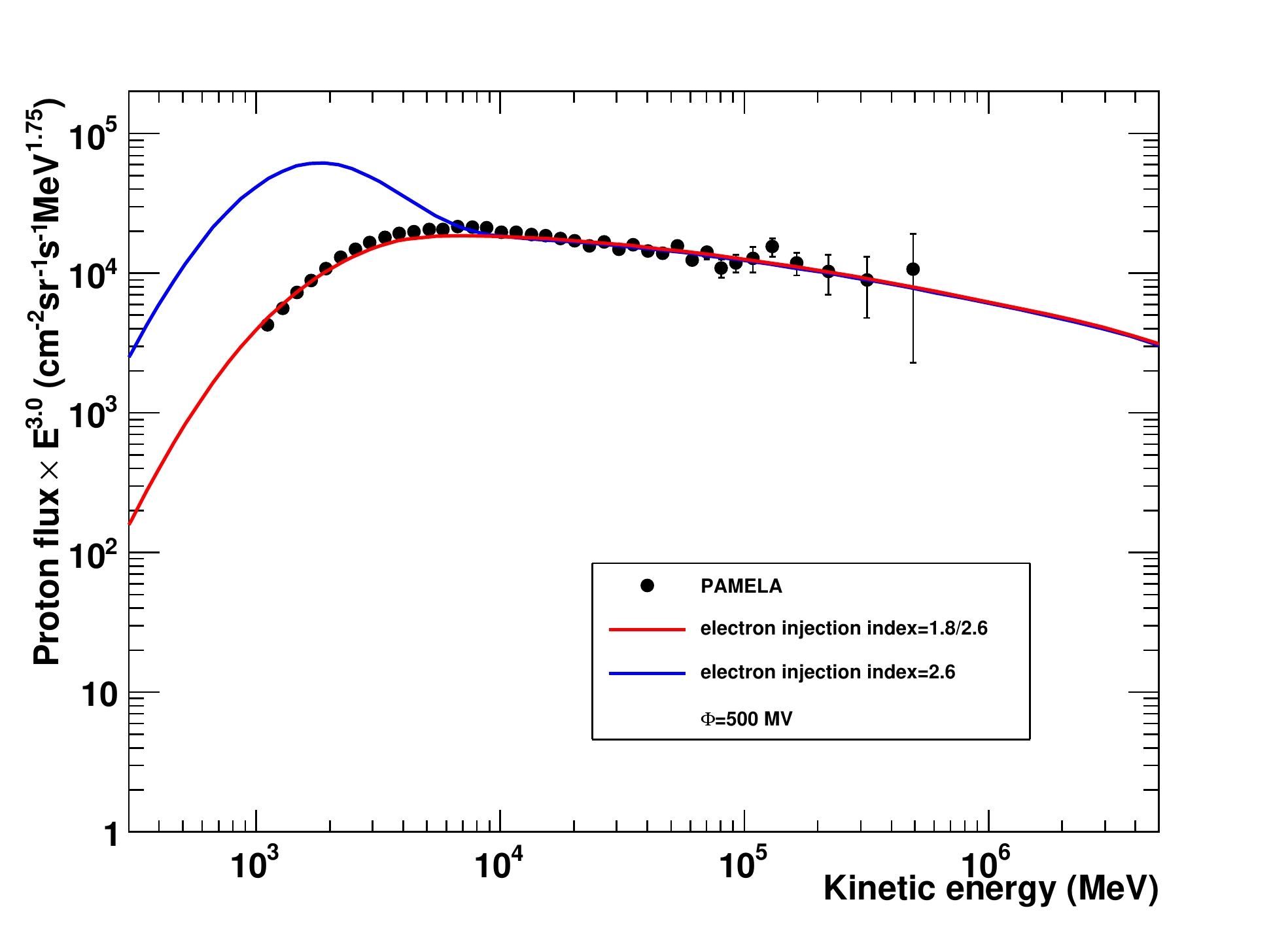}
\end{center}
\caption[The electron spectrum and the positron fraction predicted from the DR II-c model in the $\chi^{2}$ study]{\footnotesize The electron spectrum for the DR II-c model as listed in table \ref{tab:chi2analysis}. The electron injection spectrum is modeled with a broken power law with index 1.8/2.6 below/above 4~GV (red) and a single power law with index 2.6 (blue). Data points are the measurements of the electron flux from PAMELA \cite{Adriani2011_electron}.}
\label{fig:electron_DRII_c}
\end{figure}

The primary electron injection spectrum is normalized to the PAMELA electron data at 70~GeV and is tuned with a power law injection index 2.72 to fit the electron spectrum measured by PAMELA. The electron spectrum and the positron fraction are calculated based on the best model and compared with the PAMELA data, as shown with blue color in figure \ref{fig:electron_positron}. The theoretical calculation of the electron spectrum agrees well with the PAMELA data. It is noted that no significant break in the electron injection spectrum at 4~GV, as adopted in \cite{Trotta2011} which requires a strong reacceleration and introduces a break on the primary proton injection index at 100~GeV, is necessary here to describe the data. The same configuration was tested using the $\chi^{2}$ minimization method to fit the B/C ratio, the $\bar{\text{p}}$/p ratio and the proton flux. As shown in figure \ref{fig:electron_DRII_c}, using the best-fit parameters obtained for this model (see DR II-c in table \ref{tab:chi2analysis}), a broken power law with index 1.8/2.6 below/above 4~GV is needed in the electron injection spectrum to fit the electron data. Otherwise an anomalous bump arises around 1~GeV, which is caused by a combination effect of reacceleration and energy loss. In the best model (DRC-c model as listed in table \ref{tab:bayesian_results}) under study here, the reacceleration is weak and a break is not required in the electron injection spectrum. 

However, the DRC-c model (with an electron injection index 2.72) predicts an evident lower positron fraction than the PAMELA data. The discrepancy below 10~GeV can be due to, for example, a charge-sign dependent solar modulation. But above 10~GeV, the prediction which considers only secondary positrons produced in cosmic ray spallation, shows an opposite trend of the positron fraction with PAMELA data. Clearly additional components are required to interpret the raising observed by PAMELA. Since no obvious deviation from the modeled flux is observed in the electron data, positron production from the nearby astrophysical sources (e.g. pulsars) and/or from the exotic sources (e.g. dark matter) may contribute to the extra component in positron spectrum. Both the positrons and the electrons are expected to be produced in equal amounts from these extra primary sources. To reproduce both the electron spectrum and the positron fraction measured by PAMELA, two primary components are considered here, employing the same injection indices as used in \cite{Adriani2011_electron}, i.e. a standard primary component with injection index 2.69 only contributing for electrons and an extra component with injection index 2.1 producing equal amount of electrons and positrons. The agreement between this model (plotted in orange color) and the data can be seen in figure \ref{fig:electron_positron}. Invoking pulsars or dark matter as the extra component need more realistic treatment concerning the nature of the source, for example the source distribution, the mass of the dark matter particle, etc. 

\subsection{What did we learn from the Bayesian study?}
From the Bayesian study, the following conclusions can be drawn:
\begin{itemize}
\item The bias caused by the dominance of very precise proton data is reduced in the Bayesian analysis by specifying priors on source parameters and solar modulation parameters, but still could possibly exist. 
\item Strong reacceleration is required to explain the B/C data but still yields to too many protons below a few~GeV, as concluded in the $\chi^{2}$ study. When including the proton spectrum to the fitting procedure, unlike $v_{A} \to 0$ in the $\chi^{2}$ study, weak reacceleration is allowed in the Bayesian analysis since uncertainties on modulation parameters are taken into account.
\item Negative correlations are found between parameters $D_{0}$ and $\delta$, $\nu$ and $\delta$, as well as $\nu$ and $N_{p}$, as expected. Correlations between other parameters are model dependent.
\item To fit only the current PAMELA data of the $\bar{\text{p}}$/p ratio and the proton spectrum, reacceleration and convection can not be constrained.
\item When fitting only the B/C data or all the data, the DRC model is found to perform better than the DR model based on the Bayesian evidences. However, they still cannot reproduce the low energy B/C data reported by ACE-CRIS due to a prior limited on $\Phi_{ACE-CRIS}$.
\item The source and propagation parameters are not able to be well constrained by fitting PAMELA proton and $\bar{\text{p}}$/p data alone, i.e. errors on $v_{A}$ and $dV/dz$ are comparable to their posterior means. To simultaneously fit all the data, the errors on parameters range from 2\% on $\delta$ to 20\% on $dV/dz$. The uncertainties on $\delta$ are small since $\delta$ is well constrained by the high energy data. At low energies, multiple physical processes shape the cosmic ray spectra and therefore the estimation of the parameters characterizing low energy processes is hindered by their degeneracy. 
\end{itemize}

\section{Conclusion}

Cosmic ray propagation models have been studied using both the $\chi^{2}$ minimization method and the Bayesian method. Different combinations of data sets have been used to test their constraining ability. Using only the antiproton and proton data from PAMELA does not allow to constrain propagation models and therefore B/C data provided by other experiments were also included in the analysis. The B/C data is sensitive to propagation parameters but insensitive to source parameters. Models with strong reacceleration or with a nonlinear diffusion coefficient can well reproduce the B/C data. Other models fail to reproduce the B/C ratio in the low energy range. In order to constrain both the propagation and the source parameters, a simultaneous fit on the B/C ratio, the $\bar{\text{p}}$/p ratio and the proton spectrum have been performed. However, in the $\chi^{2}$ study, reacceleration models are disfavored in the simultaneous fit since reacceleration produce too many protons at low energy. 

The uncertainties due to solar modulation and the bias caused by the very precise proton data in the fitting procedure are taken into account in the Bayesian analysis by specify priors on the modulation and the source parameters. The models accounting for the low energy dependence of the diffusion coefficient were not studied using the Bayesian method since this effect is dominant over other low energy processes, i.e. reacceleration and convection. Including solar modulation parameters as nuisance parameters in the Bayesian analysis increases the number of correlated parameters and no strong constraints are expected to be obtained. 

From the estimated evidences in the Bayesian study, the diffusion model with weak reacceleration and convection is selected as the best one to explain all the data. However, a deviation between the low energy part of B/C data and the theoretical calculations is seen. More reacceleration or an extremely low modulation potential (about 100~MV) for ACE-CRIS could solve this problem. But since protons are much more precisely measured than other species, a small deviation caused by an increase of $v_{A}$ can significantly increase the $\chi^{2}$ value (or decrease the likelihood) and therefore will be excluded in the fitting procedure. Regarding using $\Phi$ $\sim 100$~MV for the ACE-CRIS experiment, this modulation level is too low to characterize the solar activity although the B/C data observed by ACE-CRIS using in this work is taken during 1997-1998 solar minimum period. A low energy dependence in the diffusion coefficient may also account for the low energy B/C data, as studied with $\chi^{2}$ method. Moreover, the inconsistencies between data sets from different experiments could result in the disagreement between the data and the model prediction.It is also worth to stress that no model studied in this work can account for the feature of the PAMELA proton spectrum at $\sim 200$\,GeV, since a source dispersion as suggested in \cite{Yuan2011} or an acceleration model as proposed in \cite{Blasi2012} is not taken into account. The electron flux and the positron fraction were also produced by this model and compared with the PAMELA data. The model with a standard primary component of electrons predicts a lower positron fraction compared to the PAMELA data. Especially above 10~GeV the positron fraction measured by PAMELA increases with energy but the model decreases with energy. But by adding an extra primary component, both the electron and positron fraction can be reproduced. Nevertheless, this model can reproduce most of the observations except for the B/C ratio below 1~GeV. More robust constraints will be studied in the future by using upcoming PAMELA nuclei data. 

\vspace{10 mm}
This study not only improves our understanding on the cosmic ray acceleration and propagation mechanisms, but will also provide a useful tool to study the astrophysical sources (e.g. pulsars) or search for exotic contributions (e.g. dark matter). Cosmic ray antimatter (e.g. $\bar{\text{p}}$  and e$^{+}$) are produced as secondary products in the ISM. They maybe also produced as the final state of dark matter annihilation and then propagated in the Galaxy before reaching Earth. An accurate and reliable estimation of the propagation will help us to discriminate whether the final state includes quarks. It also allows us to constrain the mass and the cross sections of dark matter, since different properties on dark matter are expected to give different amount of contribution in the antimatter spectrum. Similarly, the diffuse gamma ray emissions can also be studied based on cosmic ray propagation since they are also expected to be produce by cosmic ray proton interactions in ISM, and will provide another channel of dark matter search. The cosmic ray $e^{\pm}$ pairs can also be generated from pulsars. The fast energy loss of electrons indicates that only $e^{\pm}$ produced locally can reach Earth. Based on the determination of cosmic ray propagation, properties of nearby pulsars such as the injection spectrum and the energy conversion factor can therefore also be constrained.


%% file: Phd-Ch-Outlook.tex
\chapter{Discussion and outlook} 
\label{chapt:outlook}


Studying the propagation of Galactic cosmic rays can help us understand cosmic ray astrophysical sources, the properties of the Galaxy including the Galactic magnetic field, the interstellar medium and the nuclear interactions happening therein. However, since the discovery of cosmic rays one century ago, questions regarding the details of their acceleration and propagation mechanisms are still under debate. Different acceleration models predict different values of the injection spectral index. Whether and how other possible processes such as reacceleration and convection play a role in cosmic ray propagation is still not certain. The diffusion index $\delta$ related to the spectrum of magnetic turbulence differs from 0.2 to 0.9 in the literature. 

Studies on cosmic ray acceleration and propagation, rely on accurate measurements on cosmic ray nuclei over a broad energy range. The determination of source and propagation parameters, which plays an important role for us to understand relevant mechanisms, are based on fitting the data of secondary-to-primary ratios and primary fluxes.  Therefore the precision and reliability of the parameters are limited by the uncertainties and energy ranges of cosmic ray species measured and possible systemic effects existing in different experiment. To precisely measure light nuclei cosmic ray fluxes (e.g., protons, helium nuclei, antiprotons) and secondary-to-primary ratios (e.g., B/C, $\bar{\text{p}}$/p, $^{2}$H/$^{4}$He and $^{3}$He/$^{4}$He) over a wide energy range, the satellite-borne experiment PAMELA was launched in 2006 and has been studying a variety of cosmic ray species for almost six years. The absence of atmospheric overburden and the long live time makes the PAMELA measurements more accurate than those from balloon-borne experiments. These observations provide a wealth of opportunities to further study the acceleration and propagation mechanisms of Galactic cosmic rays. 

Among all the scientific objectives PAMELA are designed for, the measurement of cosmic ray antiprotons is one of the primary task through which not only the propagation mechanism but also exotic sources can be investigated. In the first part of this work, the cosmic ray antiprotons were identified with the PAMELA instrument over backgrounds presented by  other cosmic ray components and particles produced by cosmic rays interacting with the experiment materials. The selection efficiencies of each individual detector and other correction factors concerning the geometrical factor, hadronic interaction losses, the live time of measurements and transmission through the geomagnetic field were estimated. Finally the antiproton energy spectrum and the antiproton-to-proton flux ratio were reconstructed between 1.5~GeV and 180~GeV. 

In the second part of this thesis, different propagation models were studied in this work, by using the PAMELA antiproton data including a low energy part down to $\sim$100~MeV (not described in this thesis), the PAMELA proton data from 400~MeV to 1.2~TeV, and the B/C ratio reported by previous experiments but with comparable energy range expected by PAMELA. The GALPROP code which solves the cosmic ray transport equation numerically, was employed to simulate cosmic ray propagation. Analyses have been performed relying on statistical methods, i.e. the $\chi^{2}$ minimization method and Bayesian inference. In previous studies, statistical analyses were mainly carried out by using semi-analytical models (e.g. most recently \cite{Putze2010, Maurin2010, Putze2011}) or were only performed for the numerical diffusion reacceleration model with a break in the injection spectrum index \cite{Trotta2011}.  In this thesis different GALPROP models such as the plain diffusion (PD) model, the diffusion reacceleration (DR) model, the diffusion convection (DC) model and the diffusion reacceleration convection (DRC) model are studied for the first time based on statistical analyses. Models without an artificial break on the injection spectrum and on the diffusion coefficient are the main focus of the analyses. 

Different combinations of data sets are used to constrain models in this work. Using only PAMELA data is expected to minimize uncertainties due to inconsistencies between data sets, however, current PAMELA data (the $\bar{\text{p}}$/p ratio and the proton flux) has been proved to be not enough to constrain propagation parameters. Stronger and more reliable constraints are allowed by a simultaneous fit including the PAMELA data as well as the B/C ratio from other experiments.

The goodness of fit of each model was studied in the $\chi^{2}$ study. Only models considering a low energy dependence in diffusion coefficient due to nonlinear MHD waves can describe simultaneously the B/C data as well as the $\bar{\text{p}}$/p ratio and the proton spectrum. However, since the effect of nonlinear diffusion coefficient dominates at low energy, other processes (i.e. reacceleration and convection) are not possible to be studied. Models with a linear diffusion coefficient either cannot fit the B/C ratio below 1~GeV (PD and DC models) or generate too many protons at a few GeV (DR and DRC models). To reduce the uncertainties on the results due to solar modulation and the possible bias due to the dominance of the PAMELA proton spectrum in the fit the Bayesian analysis which specifies priors on the source parameters and the solar modulation parameters is used. The p.d.f.s of different parameters and the correlations between them are also able to be studied. From the $\chi^{2}$ study, the DR and DRC models can explain the B/C ratio well but cannot fit the data of the proton flux which is more prone to solar modulation and systematic effects. These two models are studied in the Bayesian analysis by considering priors are specified on the solar modulation and source parameters which can reduce the uncertainties due to solar modulation and the possible bias caused by the dominant proton data in the fit. Based on the Bayesian evidences, the DRC model favors a Karaichnan turbulence and has been proved to be better than the DR model when describing all the data. The credible interval of the parameters and the fluxes have been shown based on the posterior samples produced in the Bayesian method. Only weak reacceleration and convection are allowed in the DRC model. The B/C ratio above 1~GeV, the proton and antiproton data can all be reproduced by the DRC model. The electron flux and positron fraction can also be accommodated in this model if an additional primary component of electrons and positrons is taken into account. However, the predicted B/C ratio is still higher than the data below 1~GeV.


Several effects, including inconsistencies between data sets, solar modulation and the dominance of the proton data in the fitting procedure, might influence the reliability of the results and result in misinterpretation:
\begin{itemize}

\item The B/C ratio data used in this work are from experiments other than PAMELA. Systematic discrepancies may exist between data sets and result in a bias in the model constraints.

\item Solar modulation is a main factor affecting the spectra of cosmic ray nuclei and electrons below 10~GeV and tens of GeV, respectively. The simplified force-field approximation depending on a single parameter, the modulation potential $\Phi$, is used in this work to model the solar modulation and may bias the results. A more realistic solar modulation should include a charge-sign dependency.

\item Including the proton spectrum in the fit provides constraints on the source parameters. However, since the proton spectrum is measured more accurate than data of other species, it has a dominant weight in the fit. Any systematic bias in the proton data may therefore significantly bias the results. 

\end{itemize}

The forthcoming secondary-to-primary ratios measured by PAMELA, including the B/C ratio between 100~MeV and 200~GeV, the $^{2}$H/$^{4}$He ratio between 100~MeV/n and 700~MeV/n, and the $^{3}$He/$^{4}$He ratio between 100~MeV/n and 900~MeV/n, are expected to allow better and more robust constraints on transport parameters. The B/C ratio provided by PAMELA is expected to be more precise than previous published data and can hopefully help clarify the longstanding issue concerning the value of the spectral index of the diffusion coefficient. Degeneracy between diffusion and other low energy processes, i.e. reacceleration and convection, is therefore expected to be broken. Incorporating secondary-to-primary ratios and primary fluxes exclusively from PAMELA, the bias due to data set inconsistencies and solar modulation uncertainties can be reduced. Moreover, when using data from a single experiment, more realistic modulation models, which require a number of parameters to describe the modulation effects  \cite{Potgieter2008, Gast2009, Bobik2012}, will be easier to treat and further decrease the uncertainties due to solar modulation. The bias caused by the very precise proton data is difficult to eliminate. One simple way is increasing the errors on the proton data but some features in the proton spectrum may disappear and the corresponding source information may be lost. Additionally the model test ability of the data will be lost. The reliability of the results can also be checked by seeing whether the results are consistent by fitting only the secondary-to-primary ratios and by fitting the secondary-to-primary ratios plus the primary fluxes.



Observations in other channels, for example electrons, positrons and gamma rays, will give a consistency check on the acceleration and propagation models. PAMELA has measured the electron flux up to 625~GeV and positrons up to 300~GeV \cite{Rossetto2012Phd}. The $\gamma$-ray diffuse emission has been recently measured by the Fermi-Large Area Telescope between 100~MeV and 10~GeV \cite{Abdo2009}. The statistical techniques applied here can also be adopted to other channels. Multi-messenger observations and statistical analyses on the measurements allow us to obtain complementary information on cosmic ray acceleration and propagation models.

The AMS02 experiment, successfully installed on the International Space Station (ISS) in May last year, will also measure the energy spectra for a wide range of cosmic ray species in near furture. It has an acceptance of 0.5~m$^{2}$\,sr \cite{Schael2011_TEVPA} which is two orders of magnitude larger than that of PAMELA (21.5~cm$^{2}$\,sr), and is expected to operate on ISS for more than 10 years. The large acceptance and the long lifetime of AMS02 will drastically increase the statistics of the measurements than that have been achieved by all the previous experiments. The data with unprecedented accuracy may dramatically improve the model constraints and may potentially allow us to understand the acceleration and propagation mechanisms.  

Furthermore, based on accurate and reliable constraints on the cosmic ray propagation models, primary contributions causing from nearby pulsars or dark matter can be probed indirectly through anomalous antimatter components in cosmic rays (mainly antiprotons and positrons) and gamma rays, created during dark matter particle annihilation. The dark matter contribution can be extracted with respect to the astrophysical background of cosmic ray antimatter and gamma ray diffuse emission. The dark matter interpretation has been proposed to explain the positron excess above 10~GeV first observed by PAMELA and then confirmed by Fermi. Additional information on whether the dark matter annihilations result in hadronic final states will be given after the antiproton data at high energy available from AMS02. Information on the dark matter cross section and mass can also be constrained by the spectra of cosmic ray antimatter and gamma rays. 


Using upcoming data to improve the constraints on acceleration and propagation models and to investigate properties of primary sources such as nearby pulsars or dark matter will be an important future task and a development of this thesis.


%% file: Phd-Acknowledgements.tex
\chapter*{Acknowledgements}
\label{chapt:acknowledgements}

First, I would like to express my deepest gratitude to my supervisor, Professor Mark Pearce for introducing me to the research area of cosmic rays, giving me the opportunity to work with PAMELA and guiding me throughout my PhD study. Without your encouragement and help, this work would not have been performed. I would also like to thank my co-supervisor Professor Bengt Lund-Jensen for your kind help especially during the difficult time when I started my PhD.

A special and deep thanks goes to Antje Putze for all the useful discussions and valuable advice on my studies concerning cosmic ray propagation. This thesis would not have been finished without your contribution and help. My sincere gratitude also gives to William Gillard. Without your valuable suggestions and skilled help on Galprop, the work would have been much harder. The benefits I got during the time spent in Grenoble were remarkable.

My grateful thanks also goes to Mirko Boezio for your enlightening suggestions on my work of antiproton measurements with PAMELA, to Per Carlson for giving rewarding ideas, and to Petter Hofverberg, Alessandro Bruno, Emiliano Mocchiutti, Wolfgang Menn and Nico De Simone for answering all my questions about data analysis and detector performance with lots of patience. Thanks to Elena Vannuccini, Sergio Ricciarini, Massimo Bongi, Nico Mori and Paolo Papini. I really learnt a lot and enjoyed the time when I was in Florence. Thanks to all my colleagues in the PAMELA Collaboration for all the interesting discussions.

Thanks to my office roommates Christoffer, Merlin and Shabnnam who make our office a pleasant academic atmosphere with interesting discussions. I would like to thank Cecilia (who left our group two years ago) and Jenela for your care when I was sick and in hospital. Thanks for Stefen for lending me the book about how to breath correctly. Thanks for Laura for your laughter and being here during all my PhD period. Thanks for Elena for helping me to back up my thesis when my laptop was broken. Thanks for Mozsi for all the spontaneous sweets and strange food which we all enjoyed. Thanks for Oscar for showing me the very beautiful photos taking by him. Thanks to everyone in my group at KTH for bringing much pleasure during the work. 

Funding from the Swedish National Space Board is gratefully acknowledged. Support from China Scholarship Council is also acknowledged, which partially funded this work.

I am also grateful to my colleagues in the Physics department of China university of Geoscience for all your support. Thanks to my friends Xiaogai,  Weidong, Jiajia, Lingquan, Jing, Ameng, Changrong, Hui and Wei. Life in Stockholm is much more enjoyable with your company and support. 

My last appreciation goes to my family. Thanks for your encouragement, your sharing, your blessing. You always give me courage to move on. Thanks to my mother for your endless love. I will always miss you! Thanks my husband for all your understanding and love, for helping me go through my hardest time. Everything that you have done is greatly appreciated!

%% file: Phd-References.tex
\label{chapt:biblo}